\documentclass[12pt]{article}
\pdfoutput=1

\usepackage{jheppub}
\makeatletter
\def\@fpheader{\relax}
\makeatother

\usepackage{amsthm,amsmath,amssymb,mathtools}
\usepackage{mathrsfs}
\usepackage{bm}
\usepackage{bbm}
\usepackage{float}
\usepackage{caption}
\usepackage{subcaption}
\usepackage[most]{tcolorbox}
\usepackage[mathscr]{euscript}
\usepackage{dsfont}
\usepackage[mathscr]{euscript}
\usepackage{mathrsfs}
\usepackage{hyperref}
\usepackage{tikz}
\usetikzlibrary{shapes.geometric}
\usetikzlibrary{arrows.meta}
\usetikzlibrary{decorations.markings}

\renewcommand\mod{\, \text{mod}\ }

\newcommand\norm[1]{\left\lVert#1\right\rVert}

\newenvironment{nohyphens}{%
  \hyphenpenalty=10000
  \exhyphenpenalty=10000
  \sloppy %
}{\par}

\title{Modular factorization of superconformal indices}

\author[a]{Vishnu Jejjala}
\author[b]{\!\!, Yang Lei}
\author[a]{\!\!, Sam van Leuven}
\author[c]{\!\!, Wei Li}

\affiliation[\,a]{Mandelstam Institute for Theoretical Physics, School of Physics, NITheCS, and CoE-MaSS, \\ University of the Witwatersrand, Johannesburg 2050, South Africa}
\affiliation[\,b]{Institute for Advanced Study \& School of Physical Science and Technology,  \\ Soochow University, Suzhou 215006, P.R.~China}
\affiliation[\,c]{Institute of Theoretical Physics,  Chinese Academy of Sciences, \\100190 Beijing, P.R.~China}

\emailAdd{v.jejjala@wits.ac.za}
\emailAdd{leiyang@suda.edu.cn}
\emailAdd{sam.vanleuven@wits.ac.za}
\emailAdd{weili@mail.itp.ac.cn}

\abstract{\begin{nohyphens}
Superconformal indices of four-dimensional $\mathcal{N}=1$ gauge theories factorize into holomorphic blocks.
We interpret this as a modular property resulting from the combined action of an $SL(3,\mathbb{Z})$ and $SL(2,\mathbb{Z})\ltimes \mathbb{Z}^2$ transformation.
The former corresponds to a gluing transformation and the latter to an overall large diffeomorphism, both associated with a Heegaard splitting of the underlying geometry.
The extension to more general transformations leads us to argue that a given index can be factorized in terms of a family of holomorphic blocks parametrized by modular (congruence sub)groups.
We find precise agreement between this proposal and new modular properties of the elliptic $\Gamma$ function.
This leads to our conjecture for the ``modular factorization'' of superconformal lens indices of general $\mathcal{N}=1$ gauge theories.
We provide evidence for the conjecture in the context of the free chiral multiplet and SQED and sketch the extension of our arguments to more general gauge theories.
Assuming the validity of the conjecture, we systematically prove that a normalized part of superconformal lens indices defines a non-trivial first cohomology class associated with $SL(3,\mathbb{Z})$.
Finally, we use this framework to propose a formula for the general lens space index.
\end{nohyphens}
}

\date{}

\begin{document} 

\maketitle

\section{Introduction}\label{sec:intro}

Indices and partition functions of two-dimensional (supersymmetric) CFTs are well-known to possess modular properties. 
Within the context of string theory, such properties have played a central role in the study of black holes. 
For example, modularity leads to an asymptotic formula for the entropy of CFT states, the Cardy formula, which matches exactly with the Bekenstein--Hawking entropy of an associated supersymmetric black hole \cite{Strominger:1996sh,Dijkgraaf:1996it}. 
Various generalizations, including near-BPS and near-extremal black holes, typically still involve this key ingredient.
A closely related application is quantum gravitational physics in AdS$_3$. 
Here, modularity implies an expression for the elliptic genus as an average over the modular group, which can be beautifully interpreted in terms of the gravitational path integral \cite{Dijkgraaf:2000fq,Manschot:2007ha}. 
The latter is notoriously difficult to compute from first principles, which reflects the power of modularity.

Recent work has revisited the study of supersymmetric black holes in AdS$_{d>3}$ spaces from the perspective of the dual CFT \cite{Benini:2015eyy,Hosseini:2017mds,Cabo-Bizet:2018ehj,Choi:2018hmj,Benini:2018ywd}.\footnote{See \cite{Zaffaroni:2019dhb} for a review and an extensive collection of early references.}
In particular, various asymptotic limits of the superconformal index have been shown to reproduce the Bekenstein--Hawking entropy exactly, improving the earlier efforts of \cite{Kinney:2005ej,Chang:2013fba}.\footnote{Further progress aimed at understanding the associated microstates can be found in \cite{Murthy:2020rbd,Agarwal:2020zwm,Imamura:2021ytr,Gaiotto:2021xce,Murthy:2022ien,Lee:2022vig,Imamura:2022aua,Choi:2022ovw,Chang:2022mjp,Choi:2022caq}.}
Because $d$ is larger than three, one does not expect to have modularity as an available tool.
However, in the context of AdS$_{5}$ black holes, surprisingly, an $SL(3,\mathbb{Z})$ modular-like property turns out to either feature explicitly in or underlie the original works \cite{Cabo-Bizet:2018ehj,Choi:2018hmj,Benini:2018ywd} and various follow-ups \cite{Honda:2019cio,ArabiArdehali:2019tdm,Kim:2019yrz,Cabo-Bizet:2019osg,Lezcano:2019pae,Lanir:2019abx,Cabo-Bizet:2019eaf,ArabiArdehali:2019orz,Cabo-Bizet:2020nkr,Benini:2020gjh,GonzalezLezcano:2020yeb,Cabo-Bizet:2020ewf,Amariti:2021ubd,Cassani:2021fyv,ArabiArdehali:2021nsx,Aharony:2021zkr,Ardehali:2021irq,Colombo:2021kbb,Choi:2021rxi,Cabo-Bizet:2021plf}.
This property is associated to the elliptic $\Gamma$ function and was first proposed in a purely mathematical context \cite{Felder_2000}.
For concreteness, we state it here:
\begin{equation}\label{eq:Y3-prop-intro}
    \Gamma(z;\tau,\sigma)=e^{-i\pi Q(z;\tau,\sigma)}\Gamma\left(\tfrac{z}{\sigma};\tfrac{\tau}{\sigma},-\tfrac{1}{\sigma}\right)\Gamma\left(\tfrac{z}{\tau};\tfrac{\sigma}{\tau},-\tfrac{1}{\tau}\right)\,,
\end{equation}
while leaving a detailed discussion to the main text.
In the physical context, the elliptic $\Gamma$ function arises as the main building block of four-dimensional $\mathcal{N}=1$ gauge theory indices \cite{Dolan:2008qi}.
The relevance of the modular property to the asymptotics of the superconformal index was emphasized in \cite{Benini:2018ywd,ArabiArdehali:2019tdm,Gadde:2020bov,Goldstein:2020yvj,Jejjala:2021hlt,Choi:2021rxi}.

The $SL(3,\mathbb{Z})$ modular property of the elliptic $\Gamma$ function has been previously invoked in the physics literature.
It made an early appearance in \cite{Spiridonov:2012ww} in the context of anomaly matching conditions for Seiberg dual theories.
More relevant to the present work is the physical interpretation of Nieri and Pasquetti \cite{Nieri:2015yia}.
Based on a factorization property of the superconformal index of large class of $\mathcal{N}=1$ gauge theories \cite{Yoshida:2014qwa,Peelaers:2014ima}, the authors showed that the superconformal lens indices $\mathcal{I}_{L(p,1)}$ of those theories can be factorized into so-called holomorphic blocks $\mathcal{B}_S$.
Schematically:
\begin{equation}\label{eq:nieri-pasq-result}
    \mathcal{I}_{L(p,1)}\cong \sum \norm{\mathcal{B}_S}^2_{f_{p}}\,.
\end{equation}
They further proposed that the factorization reflects a Heegaard-like splitting of the underlying geometry:
\begin{equation}\label{eq:l(p,1)-split-intro}
    L(p,1)\times S^1 \cong \left(D_2\times T^2\right)_S\overset{f_{p}}{\sqcup}\left(D_2\times T^2\right)_S\,.
\end{equation}
The holomorphic blocks are interpreted as partition functions on the $\left(D_2\times T^2\right)_S$ geometries.
These geometries are glued with an appropriate $SL(3,\mathbb{Z})$ element combined with orientation reversal, denoted by $f_{p}$, which acts on $T^3=\partial D_2\times T^2$.
The subscript $S$ indicates the action of an element inside $SL(2,\mathbb{Z})\ltimes \mathbb{Z}^2$, the group of large diffeomorphsims of $D_2\times T^2$.\footnote{This observation is left unmentioned in previous works, but forms the basis of the present work.}

A main ingredient in the proposal consists of a set of modular properties of the elliptic $\Gamma$ function.
This includes \eqref{eq:Y3-prop-intro}, which features as a special case of \eqref{eq:nieri-pasq-result}: it reflects the factorization of the $S^3\times S^1$ index of a free chiral multiplet.
This provides a remarkable physical interpretation of the modular property of the elliptic $\Gamma$ function.
In general, we note that the factorization property \eqref{eq:nieri-pasq-result} is rather distinct from the properties of ordinary modular forms.
Indeed, the property involves three functions that in general do not stand on an identical footing, although for the chiral multiplet they do.
Furthermore, there is a combined action of $SL(3,\mathbb{Z})$ and $SL(2,\mathbb{Z})\ltimes \mathbb{Z}^2$, where the former relates the variables of the holomorphic blocks while the latter is an overall transformation between the left and right hand sides.
For the original modular property in~\eqref{eq:Y3-prop-intro}, the relevant $SL(3,\mathbb{Z})$ element exchanges $\tau$ and $\sigma$, while the $SL(2,\mathbb{Z})\ltimes \mathbb{Z}^2$ transformation acts like an $S$-transformation on the third argument of the $\Gamma$ functions on the right hand side.

Recently, a proposal was made for the modular interpretation of the factorization property in the inspiring work of Gadde \cite{Gadde:2020bov}.
Key to this insight is again the foundational mathematical work \cite{Felder_2000}.
There, it was already observed that the elliptic $\Gamma$ function fits into a $1$-cocycle $X_g$ for $g\in \mathcal{G}$ with $\mathcal{G}=SL(3,\mathbb{Z})\times \mathbb{Z}^{3}$, where the $\mathbb{Z}^{3}$ factor contains large gauge transformations associated to a line bundle over $T^3$.
Technically, $X_g$ is an element of $H^1(\mathcal{G},N/M)$, the first group cohomology of $\mathcal{G}$ valued in the space $N/M$ of meromorphic functions modulo phases, which satisfies a defining $1$-cocycle condition:
\begin{equation}\label{eq:1-cocycle-felder-intro}
    X_{g_1g_2}(\boldsymbol{\rho})\cong X_{g_1}(\boldsymbol{\rho})X_{g_2}(g_{1}^{-1}\boldsymbol{\rho})\,,
\end{equation}
where $\boldsymbol{\rho}$ is shorthand notation for the moduli associated to the $D_2\times T^2$ geometries, on which $\mathcal{G}$ acts.
The equality in \eqref{eq:1-cocycle-felder-intro} holds up to multiplication by functions in $M$.
This generalizes the notion of a (weak) Jacobi form, such as the elliptic genus, which can be thought of as an element of $H^0(\mathcal{J},N/M)$ for $\mathcal{J}=SL(2,\mathbb{Z})\times \mathbb{Z}^{2}$.\footnote{We review this mathematical framework in Section \ref{ssec:mod-group-cohomology}.}
In the physical context, \eqref{eq:1-cocycle-felder-intro} corresponds to a property of the collection of (normalized) lens indices of a free chiral multiplet \cite{Gadde:2020bov}.
Gadde proposes an extension to general $\mathcal{N}=1$ gauge theories by constructing a ``normalized part of the lens index'' $\hat{\mathcal{Z}}^{\alpha}_{g}(\boldsymbol{\rho})$ with $g\in \mathcal{G}$.
Based on holomorphic block factorization of the physical lens index, \cite{Gadde:2020bov} argues that $\hat{\mathcal{Z}}^{\alpha}_{g}(\boldsymbol{\rho})$ similarly satisfies a $1$-cocycle condition.
An important part of the conjecture is that $\hat{\mathcal{Z}}^{\alpha}_{g}(\boldsymbol{\rho})$ furnishes a non-trivial cohomology class and that local trivializations, or ``locally exact'' expressions, are related to the holomorphic block factorization.

The combined $SL(3,\mathbb{Z})$ and $SL(2,\mathbb{Z})\ltimes \mathbb{Z}^2$ action in the factorization property \eqref{eq:nieri-pasq-result} can be understood from the $1$-cocycle condition.
In particular, for the case of the $S^3\times S^1$ index, one focuses on an order three element $Y^3=1$ in $SL(3,\mathbb{Z})$, which, using the $1$-cocycle condition, implies the following equation for the normalized part of the index $\hat{\mathcal{Z}}^{\alpha}_{Y}(\boldsymbol{\rho})$ \cite{Gadde:2020bov}:
\begin{equation}\label{eq:Y3-prop-gen-intro}
    \hat{\mathcal{Z}}^{\alpha}_{Y}(\boldsymbol{\rho})\hat{\mathcal{Z}}^{\alpha}_{Y}(Y^{-1}\boldsymbol{\rho})\hat{\mathcal{Z}}^{\alpha}_{Y}(Y^{-2}\boldsymbol{\rho})=e^{i\pi \mathcal{P}(\boldsymbol{\rho})}\,,
\end{equation}
where $\mathcal{P}(\boldsymbol{\rho})$ turns out to capture the 't Hooft anomalies of the theory.
With some work, this property can be translated into the factorization of the physical index.\footnote{This connection is not obvious. Indeed, the three functions appearing in \eqref{eq:Y3-prop-gen-intro} appear on an equal footing, whereas in the factorization property \eqref{eq:nieri-pasq-result} this is not the case. We return to the translation between the two in Section \ref{ssec:coh-perspective}.}

In a previous paper \cite{Jejjala:2021hlt}, we proposed a generalization of \eqref{eq:Y3-prop-gen-intro} to more general order three elements and suggested an interpretation in terms of new factorization properties of the index.
In this work, we turn the logic around: we first present physical arguments for a (modular) family of factorization properties of a given superconformal lens index.
We prove our proposal for the free chiral multiplet and SQED, and sketch how these arguments can be extended to more general $\mathcal{N}=1$ gauge theories.
The primary tool we use in the proof consists of new modular properties of the elliptic $\Gamma$ function that generalize \eqref{eq:Y3-prop-intro} in multiple directions.
These properties are derived in Appendix \ref{app:mod-props-Gamma} without making use of the fact that the elliptic $\Gamma$ function is part of a $1$-cocycle, as opposed to our previous work \cite{Jejjala:2021hlt}.
Assuming the validity of our proposal, we are able to provide a systematic and rigorous proof of the $1$-cocycle condition for $\hat{\mathcal{Z}}^{\alpha}_{g}(\boldsymbol{\rho})$.
In particular, our approach supplies a physical interpretation of the fact that $\hat{\mathcal{Z}}^{\alpha}_{g}(\boldsymbol{\rho})$ defines a non-trivial cohomology class.
We now give a more detailed summary of the remainder of this paper.

\paragraph{Summary:}

In Section \ref{sec:heegaard-splitting} we review the Heegaard-like splitting of (secondary) Hopf surfaces with topology $L(p,q)\times S^1$, as represented in \eqref{eq:l(p,1)-split-intro}, including a mapping between the complex structure moduli of the Hopf surface and the $D_2\times T^2$ geometries.
We emphasize certain ambiguities in the Heegaard splitting of a Hopf surface.
If a given Hopf surface admits a Heegaard splitting in terms of some gluing transformation $f_{(p,q)}$
\begin{equation}\label{eq:l(p,1)-split-2-intro}
    L(p,q)\times S^1 \cong D_2\times T^2\overset{f_{(p,q)}}{\sqcup}D_2\times T^2\,,
\end{equation}
it also admits the Heegaard splittings:
\begin{equation}\label{eq:l(p,1)-split-3-intro}
    L(p,q)\times S^1 \cong \left(D_2\times T^2\right)_h\overset{f_{(p,q)}'}{\sqcup}\left(D_2\times T^2\right)_{\tilde{h}}\,,
\end{equation}
with $f_{(p,q)}'=hf_{(p,q)}\tilde{h}^{-1}$ and $h,\tilde{h}\in H\equiv SL(2,\mathbb{Z})\ltimes \mathbb{Z}^2$, the large diffeomorphism group of $D_2\times T^2$.
The subscripts indicate the action of these elements on the respective moduli of the $D_2\times T^2$ geometries, generalizing \eqref{eq:l(p,1)-split-intro} to arbitrary large diffeomorphisms in $H$.

In Section \ref{ssec:towards-conjecture}, we tentatively propose a generalization of the factorization property \eqref{eq:nieri-pasq-result} to reflect the ambiguities in the Heegaard splitting.
Schematically:
\begin{equation}\label{eq:tent-prop}
    \mathcal{I}_{L(p,q)}\cong \sum \norm{\mathcal{B}_h}^2_{f_{(p,q)}'}\,.
\end{equation}
We stress that the left hand side is independent of $(h,\tilde{h})$: any two Heegaard splittings with gluing transformations $f$ and $f'=hf\tilde{h}^{-1}$ lead to the same Hopf surface, and therefore each factorization to the same compact space index.
We continue in Section \ref{ssec:hol-blocks} with a comprehensive review of the original holomorphic block factorization of lens indices, as written in \eqref{eq:nieri-pasq-result}.
In the process, we promote an observation of \cite{Nieri:2015yia} to a consistency condition: the lens index should not depend on the boundary conditions imposed to compute an individual holomorphic block.
This condition constrains the proposal \eqref{eq:tent-prop}, as we will show in Section \ref{ssec:consistency-cond}.
In particular, we find that only certain pairs of large diffeomorphisms $(h,\tilde{h})$ are compatible with the condition, which depend on $f$.
This subset of large diffeomorphisms, which we denote by $S_f\subset H\times H$, can be parametrized in terms of modular (congruence sub)groups.
This motivates our conjecture for the \emph{modular factorization of lens indices} in Section \ref{ssec:mod-fact-conjecture}.
For example, we find that the $L(p,\pm 1)\times S^1$ indices for any $p\geq 0$ can be factorized respectively in terms of two $SL(2,\mathbb{Z})$ families of holomorphic blocks.\footnote{As is common, one defines $L(0,\pm 1)\cong S^2\times S^1$.}

In Section \ref{ssec:geom-int-univ-blocks}, we discuss a geometric interpretation of the compatible diffeomorphisms $(h,\tilde{h})\in S_f$.
We find that they parametrize all the ways in which the ``time circle'' inside a given Hopf surface can be embedded into the $D_2\times T^2$ geometries of the Heegaard splitting, such that the associated gluing transformation fixes this cycle.
Such an embedding is indeed labeled by two large diffeomorphisms $(h,\tilde{h})\in H\times H$ and the associated gluing transformation is given by $f'=hf\tilde{h}^{-1}$, as in \eqref{eq:l(p,1)-split-3-intro}.
The condition effectively solves $\tilde{h}$ in terms of $h$ (or vice versa) and leads precisely to the modular subset $S_f\subset H\times H$.\footnote{In the most general case, $h$ itself cannot be entirely arbitrary in $H$, but sits in a congruence subgroup. We will describe this in detail in Section \ref{ssec:consistency-cond}.}
At the level of a lens index, we thus conclude that its factorization is only compatible with those pairs of holomorphic blocks which are defined with respect to a common time circle.
All in all, this provides the geometric rationale for the modular factorization of four-dimensional indices.
Note that the origin of the modular structure is rather distinct from the $SL(2,\mathbb{Z})$ modularity of 2d torus partition functions, and in particular relies on holomorphic block factorization.

For a given lens index, we have so far described the set of compatible holomorphic blocks.
It will also be of interest to instead fix a pair of holomorphic blocks and consider the set of indices which can be factorized in terms of them.  
As a consequence of the above, the relevant indices are labeled by a subset of gluing transformations, namely those which fix the time circle.
This subset of $SL(3,\mathbb{Z})$ is large enough to glue (secondary) Hopf surfaces of arbitrary topology.
The chosen holomorphic blocks provide the unique pair in terms of which all lens indices associated to the subset can be factorized.
To factorize indices associated to other subsets of $SL(3,\mathbb{Z})$, one requires an alternative pair of holomorphic blocks.
Hence, we obtain a patchwise picture for the holomorphic block factorization of lens indices: the lens index associated to any element in $SL(3,\mathbb{Z})$ can be factorized in terms of some pair of holomorphic blocks, but not all elements can be factorized in terms of the same blocks.
The maximal patch inside $SL(3,\mathbb{Z})$ that can be factorized in terms of a common pair of holomorphic blocks is in bijection with $SL(2,\mathbb{Z})\ltimes \mathbb{Z}^2$.

Finally, in Section \ref{ssec:evidence} we show remarkable agreement between our physical arguments and a family of modular properties obeyed by the elliptic $\Gamma$ function, which vastly generalize \eqref{eq:Y3-prop-intro}.
For example, the generalization relevant for the superconformal index of a free chiral reads:
\begin{align}\label{eq:ZS23-hol-blocks-intro}
    \begin{split}
        \Gamma(z;\tau,\sigma)&=e^{-i\pi Q_{\mathbf{m}}(z;\tau,\sigma)}\Gamma\left(\tfrac{z}{m\sigma+n};\tfrac{\tau-\tilde{n}(k\sigma+l)}{m\sigma+n},\tfrac{k\sigma+l}{m\sigma+n}\right)\Gamma\left(\tfrac{z}{m\tau+\tilde{n}};\tfrac{\sigma-n(\tilde{k}\tau+\tilde{l})}{m\tau+\tilde{n}},\tfrac{\tilde{k}\tau+\tilde{l}}{m\tau+\tilde{n}}\right)\,,
    \end{split}
\end{align}
and may be compared to \eqref{eq:Y3-prop-intro}.
A detailed derivation of the modular properties, including \eqref{eq:ZS23-hol-blocks-intro}, is contained in Appendix \ref{app:mod-props-Gamma}.\footnote{A subset of these results appears implicitly in \cite[Theorem 3.8]{Felder_2008}.}
These properties form the basis for a proof of modular factorization in the context of the free chiral multiplet and SQED.
We also point out how these proofs can be extended to more general $\mathcal{N}=1$ gauge theories.

In Section \ref{sec:gen-modularity}, we review relevant aspects of group cohomology.
We then show how the modular factorization of lens indices can be used to systematically prove that $\hat{\mathcal{Z}}^{\alpha}_g(\boldsymbol{\rho})$ satisfies a $1$-cocycle condition for $\mathcal{G}$ such as in \eqref{eq:1-cocycle-felder-intro}.
The strategy of the proof follows the original mathematical work \cite{Felder_2000}, which can be viewed as a proof for the example of the free chiral multiplet.
We also show that $\hat{\mathcal{Z}}^{\alpha}_{g}(\boldsymbol{\rho})$ defines a non-trivial cohomology class.
This will follow from the connection between holomorphic block factorization and locally exact expressions for $\hat{\mathcal{Z}}^{\alpha}_{g}(\boldsymbol{\rho})$, mentioned before, and the fact that a given set of holomorphic blocks can only be used to factorize a strict subset of $SL(3,\mathbb{Z})$ of the form $SL(2,\mathbb{Z})\ltimes \mathbb{Z}^2$.
Finally, we return to the perspective of our previous work \cite{Jejjala:2021hlt}, showing how the modular factorization can also be obtained from relations in $SL(3,\mathbb{Z})$ that generalize $Y^3=1$ referred to above.

In Section \ref{sec:gen-lens-index}, we turn to an application of the $1$-cocycle condition: a concrete formula for the $L(p,q)\times S^1$ index for $q>1$.
Up until now, such a formula has been absent in the literature.
We test our formula in context of the free chiral multiplet by subjecting it to various consistency checks.
Finally, we end the paper in Section \ref{sec:sum-future} with a discussion of future directions, including implications for supersymmetric AdS$_5$ black holes and modular properties of indices.

In Appendix \ref{app:defs}, we collect the definitions and properties of special functions appearing in the main text. 
In Appendix \ref{app:hopf-surfaces}, we review of secondary Hopf surfaces with topology $L(p,q)\times S^1$ and their Heegaard splitting.
In Appendix \ref{app:lens-indices}, we collect the contour integral expressions of lens indices for general gauge theories.
In Appendix \ref{app:mod-props-Gamma}, we derive various modular properties of the elliptic $\Gamma$ function.
For the reader's convenience, Table \ref{tab:notation} supplies a glossary of notation.

\newpage

\vspace*{\fill}

{\small
\begin{table}[H]
\centering
\begin{tabular}{c|c|c}
$\mathcal{G}$     & gluing group: $SL(3,\mathbb{Z})\ltimes \mathbb{Z}^{3r}$ & \eqref{eq:def-mathcalG} \\
$H$     &   large diffeomorphisms of $D_2\times T^2$: $SL(2,\mathbb{Z})_{13}\ltimes \mathbb{Z}^2 $ & \eqref{eq:def-H}\\
$\mathcal{H}$     &  $H$ with large gauge symmetries: $ SL(2,\mathbb{Z})_{13}\ltimes \mathbb{Z}^{2(r+1)}$ & \eqref{eq:defn-calH}\\
$f$  & gluing element $f=g\,\mathcal{O}$ with $g\in\mathcal{G}$ and $\mathcal{O}$ orientation reversal & \eqref{eq:Mg-defn-lambda-mu}\\
$S_f$     &  set of compatible $(h,\tilde{h})$ for a given $f$ & \eqref{eq:h-ht-Lens-gen}\\
$F$ &subgroup of $\mathcal{G}$ that fixes time circle: $SL(2,\mathbb{Z})_{12}\ltimes \mathbb{Z}^{2}$ & \eqref{eq:def-F1}\\
$F_h$ &
$h^{-1}F\mathcal{O}h\mathcal{O}$ with $h \in H$
& \eqref{eq:Zf-Bh}\\
$\mathcal{M}_{(p,q)}(\hat{\boldsymbol{\rho}})$ & Hopf surface of topology $L(p,q)\times S^1$& \eqref{eq:notation-hopf-surface} \\
$\hat{\boldsymbol{\rho}}$ & moduli of the Hopf surface & \eqref{eq:hat-rho}\\
$M_f(\boldsymbol{\rho}, \tilde{\boldsymbol{\rho} })$ &
Heegaard splitting of Hopf surface
& \eqref{eq:notation-Mg-split} \\
$\boldsymbol{\rho}$ & $(\vec{z}; \tau, \sigma)
$, moduli of left $D_2\times T^2$ geometry
& \eqref{eq:rho-homog}\\
$\tilde{\boldsymbol{\rho}}$ & $(\vec{\tilde{z}}; \tilde{\tau}, \tilde{\sigma})$, moduli of right $D_2\times T^2$ geometry & \eqref{eq:gluing-condition} \\
$\mathcal{I}_{(p,q)}(\hat{\boldsymbol{\rho}})$    &  supersymmetric partition function on $\mathcal{M}_{(p,q)}(\hat{\boldsymbol{\rho}})$  & \eqref{eq:notation-lens-index} \\
$\mathcal{Z}_f(\boldsymbol{\rho})$    &  $\mathcal{I}_{(p,q)}(\hat{\boldsymbol{\rho}})$ with Heegaard splitting $M_f(\boldsymbol{\rho}, \tilde{\boldsymbol{\rho} })$ & \eqref{eq:defn-Zf-notation} \\
$\mathcal{Z}^{\alpha}_f(\boldsymbol{\rho})$    &  contribution to $\mathcal{Z}_f(\boldsymbol{\rho})$ at Higgs branch vacuum $\alpha$ &  \eqref{eq:lens-higgs-form} \\
$Z_f(\boldsymbol{\rho})$     &  free chiral multiplet partition function & \eqref{eq:Zchiralmultiplet} \\
$\hat{\mathcal{Z}}^{\alpha}_g(\boldsymbol{\rho})$    & normalized part index at vacuum $\alpha$ & \eqref{eq:defn-hatZ}\\
$\mathcal{B}^{\alpha}(\boldsymbol{\rho})$    & partition function on $D_2\times T^2(\boldsymbol{\rho})$ at vacuum $\alpha$%
& above \eqref{eq:equiv-hol-block-facts} \\
$\mathcal{B}^{\alpha}_{h}(\boldsymbol{\rho})$    & $\mathcal{B}^{\alpha}(h\boldsymbol{\rho})$ with $h\in \mathcal{H}$  & \eqref{eq:defn-Bh} \\
$\mathcal{C}^{\alpha}(\boldsymbol{\rho})$    & partition function on $D_2\times T^2(\boldsymbol{\rho})$ at vacuum $\alpha$%
&  %
\\
& %
\qquad \qquad \qquad \qquad \qquad with opposite b.c.\ from $\mathcal{B}^{\alpha}(\boldsymbol{\rho})$ &   above \eqref{eq:equiv-hol-block-facts} %
\end{tabular}
\caption{Summary of notation, with equations in which they are first defined.  }
\label{tab:notation}
\end{table}
}

\vspace*{\fill}

\newpage

\section{Heegaard splitting of \texorpdfstring{$\bm{L(p,q)\times S^1}$}{L(p,q)xS1}}\label{sec:heegaard-splitting}

In this section, we review the Heegaard splitting of lens spaces $L(p,q)$ and discuss the generalization to (secondary) Hopf surfaces of topology $L(p,q)\times S^1$.
We end with a discussion of ambiguities in the Heegaard splitting of a Hopf surface, which will play a central role in the remainder of this paper.

\subsection{Topological aspects}\label{ssec:top-aspects}

The Heegaard splitting of a general smooth three-manifold $M_3$ is the statement that $M_3$ is obtained from the gluing of two genus $g$ handlebodies $H_g$ and $H_g'$:\footnote{See Chapter 1 of \cite{saveliev1999} for a pedagogical review.}
\begin{equation}\label{eq:3d-heegaard-split}
    M_3\cong  H_g\overset{f}{\sqcup}H_g'\,,
\end{equation}
where the boundary $\Sigma_g=\partial H_g$ is identified with $\Sigma_g'=\partial H_g'$ up to the action of an orientation reversing diffeomorphism $f$.
We will be interested in the lens space $L(p,q)$, which can be defined as a quotient of the three-sphere $S^3$ viewed as a subset of $\mathbb{C}^2$:\footnote{The minus sign in the phase is conventional in the physics literature \cite{Lin:2005nh,Benini:2011nc,Razamat:2013opa,Closset:2013vra}. It reflects the standard choice of supercharge used to define the associated lens index (see Appendix \ref{app:lens-indices}).}
\begin{equation}\label{eq:lens-quotient}
    (z_1,z_2)\sim (e^{\frac{ 2\pi i q}{p}  }z_1,e^{-\frac{2\pi i}{p}  }z_2)\qquad \Longleftrightarrow \qquad (z_1,z_2)\sim (e^{\frac{ 2\pi i }{p}  }z_1,e^{-\frac{2\pi is}{p}  }z_2)\,,
\end{equation}
with $\gcd(p,q)=1$.
For later convenience, we have introduced an equivalent description in terms of $s=q^{-1}\mod p$.
Both $q$ and $s$ are defined $\mod p$.
We note that $L(1,0)= S^3$ and the fundamental group $\pi_1(L(p,q))=\mathbb{Z}_p$.

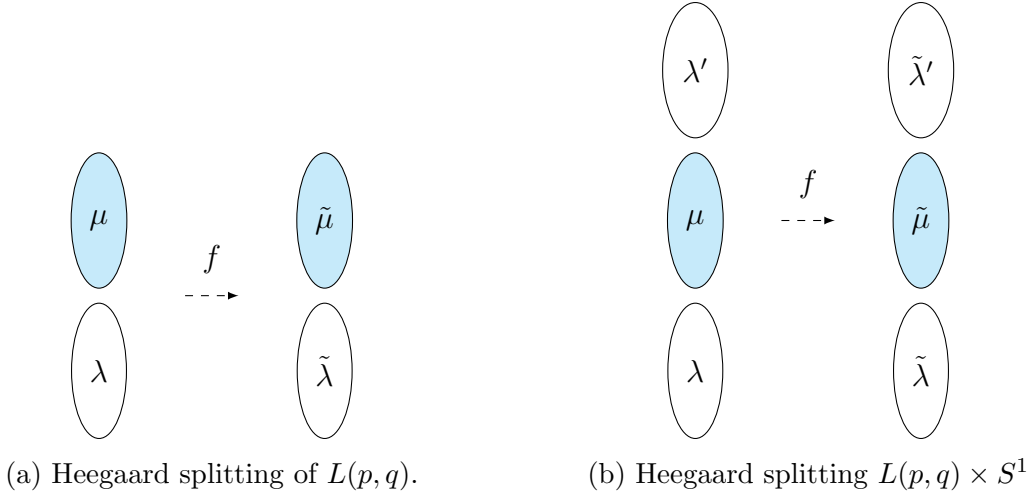
\begin{figure}[t]
     \centering
     \begin{subfigure}[t]{0.49\textwidth}
         \centering
         \begin{tikzpicture}
    \node (rho1) at (1,-1) {};
    \node (rho2) at (2,-1) {};
    \node[ellipse,
    draw = black,
    text = black,
    fill = cyan!20,
    minimum width = 0.2cm, 
    minimum height = 1.8cm] (e2) at (0,0) {$\mu$};
    \node[ellipse,
    draw = black,
    text = black,
    minimum width = 0.2cm, 
    minimum height = 1.8cm] (e3) at (0,-2) {$\lambda$};
    \node[ellipse,
    draw = black,
    text = black,
    fill = cyan!20,
    minimum width = 0.2cm, 
    minimum height = 1.8cm] (f2) at (3,0) {$\tilde{\mu}$};
    \node[ellipse,
    draw = black,
    text = black,
    minimum width = 0.2cm, 
    minimum height = 1.8cm] (f3) at (3,-2) {$\tilde{\lambda}$};
    \draw [-latex][dashed] (rho1) -- node[above=1.5mm] {$f$} (rho2) ;
    \end{tikzpicture}
        \caption{Heegaard splitting of $L(p,q)$.}
         \label{fig:heeg-Lpq}
     \end{subfigure}
     \hfill
     \begin{subfigure}[t]{0.49\textwidth}
         \centering
    \begin{tikzpicture}
    \node (rho1) at (1,-2) {};
    \node (rho2) at (2,-2) {};
    \node[ellipse,
    draw = black,
    text = black,
    minimum width = 0.2cm, 
    minimum height = 1.8cm] (e1) at (0,0) {$\lambda'$};
    \node[ellipse,
    draw = black,
    text = black,
    fill = cyan!20,
    minimum width = 0.2cm, 
    minimum height = 1.8cm] (e2) at (0,-2) {$\mu$};
    \node[ellipse,
    draw = black,
    text = black,
    minimum width = 0.2cm, 
    minimum height = 1.8cm] (e3) at (0,-4) {$\lambda$};
    \node[ellipse,
    draw = black,
    text = black,
    minimum width = 0.2cm, 
    minimum height = 1.8cm] (f1) at (3,0) {$\tilde{\lambda}'$};
    \node[ellipse,
    draw = black,
    text = black,
    fill = cyan!20,
    minimum width = 0.2cm, 
    minimum height = 1.8cm] (f2) at (3,-2) {$\tilde{\mu}$};
    \node[ellipse,
    draw = black,
    text = black,
    minimum width = 0.2cm, 
    minimum height = 1.8cm] (f3) at (3,-4) {$\tilde{\lambda}$};
    \draw [-latex][dashed] (rho1) -- node[above=1.5mm] {$f$} (rho2) ;
    \end{tikzpicture}
         \caption{Heegaard splitting $L(p,q)\times S^1$}
         \label{fig:heeg-LpqS1}
     \end{subfigure}
        \caption{On the left, we depict two solid tori $D_{2}\times S^1$ with their cycles identified by $f=g\,\mathcal{O}$ with $g\in SL(2,\mathbb{Z})$ as in \eqref{eq:3d-lens-space-sl2-defn}. Similarly, on the right we depict $D_{2}\times T^2$ geometries with their cycles identified by $f=g\,\mathcal{O}$ with $g\in SL(3,\mathbb{Z})$ as in \eqref{eq:Mg-defn-lambda-mu}.}
        \label{fig:solid-2tori-gluing}
\end{figure}

Every lens space admits a genus $1$ Heegaard splitting \cite{saveliev1999}.
The relevant (large) diffeomorphisms $f$ are classified by $SL(2,\mathbb{Z})$, the mapping class group of $T^2$, and in the following we consider $f$ to be an element of $SL(2,\mathbb{Z})$ combined with orientation reversal.
Let us denote the non-contractible and contractible cycles on either $H_1=D_2\times S^1$ by $(\lambda,\,\mu)$ and $(\tilde{\lambda},\,\tilde{\mu})$, respectively.
The gluing transformation $f$ identifies these cycles as:
\begin{equation}\label{eq:3d-lens-space-sl2-defn}
    \begin{pmatrix}\mu &\; \lambda\end{pmatrix}=\begin{pmatrix}\tilde{\mu}&\; \tilde{\lambda} \end{pmatrix} f^{-1}\,,
\end{equation}
where for $L(p,q)$ the transformation $f$ is given by:
\begin{equation}
    f=g\,\mathcal{O}\,,\qquad g=\begin{pmatrix} 
        -s & -r \\
        -p & -q
    \end{pmatrix}\,,\quad  \mathcal{O}= \begin{pmatrix} 
        -1 & 0 \\
        0 & 1
    \end{pmatrix}\,,\quad qs-pr=1 \,.
\end{equation}
This description realizes $L(p,q)$ as a torus fibration over an interval with a $(1,0)$ cycle shrinking on one endpoint and a $(q,p)$ cycle on the other (see Appendix \ref{sapp:LpqS1}).
The slightly awkward convention for the entries of $g$ will facilitate comparison with the literature on indices.
See Figure \ref{fig:heeg-Lpq} for an illustration.
When $g$ is the identity matrix, the manifold is $S^2\times S^1$, denoted as $L(0,-1)$.

Clearly, the description of the lens space in terms of $f$ is redundant when compared to the quotient definition \eqref{eq:lens-quotient}.
These redundancies can be fixed with symmetries of the Heegaard splitting.
For example, two lens spaces are diffeomorphic if their Heegaard splittings are related through:
\begin{equation}
    f'=f^{-1}\qquad \text{or}\qquad  f'=\pm \mathcal{O}f\mathcal{O}\,.
\end{equation}
The first transformation exchanges $q\leftrightarrow s$, while the second maps $q\to -q$ and $s\to -s$.
In addition, consider the group of large diffeomorphisms of a solid torus.
This is the subgroup of $SL(2,\mathbb{Z})$ that preserves the contractible cycle $\mu$.
It is usually denoted by $\Gamma_{\infty}$ and corresponds to the integer shifts $\lambda\to \lambda+k\mu$.
The action on either solid torus should not change the topology, so that the manifolds associated to $f$ and $f'$ are diffeomorphic when:
\begin{equation}\label{eq:large-diffeos-M3}
    f'= \gamma f\tilde{\gamma}^{-1}\,,\qquad \gamma,\tilde{\gamma}\in \Gamma_{\infty}\,.
\end{equation}
These are generated by $(q,r)\to (q+p,r+s)$ and $(s,r)\to (s+p,r+q)$.
Taken together, we see that $f$ modulo the ambiguities implies that $L(p_1,q_1)$ and $L(p_2,q_2)$ are diffeomorphic if:
\begin{equation}\label{eq:Homeomorphic-condition}
    p_1=p_2\,,\qquad q_1=\pm q_2^{\pm 1}\mod p_1\,,
\end{equation}
as consistent with the quotient definition.
It turns out that this is also a necessary condition \cite{reidemeister1935homotopieringe}.%

Let us now proceed with the four-manifolds $L(p,q)\times S^1$.
In this case, the Heegaard-like splitting glues two $D_2\times T^2$ geometries along their boundary $T^3$ (see Figure \ref{fig:heeg-LpqS1}).
In general, the gluing map takes its value in $SL(3,\mathbb{Z})$, and the Heegaard splitting is defined through the identification:
\begin{equation}\label{eq:Mg-defn-lambda-mu}
    \begin{pmatrix}\lambda'&\; \mu &\; \lambda\end{pmatrix}=\begin{pmatrix}\tilde{\lambda}'&\; \tilde{\mu} &\; \tilde{\lambda}\end{pmatrix}\,f^{-1}\,,\qquad f=g \, \mathcal{O}\,,\quad g\in SL(3,\mathbb{Z})\,,\quad  \mathcal{O}= \begin{pmatrix} 
        1 & 0 & 0\\
        0 & -1 & 0 \\
        0 & 0 & 1
    \end{pmatrix}\,,
\end{equation}
where $\mu$ and $\tilde{\mu}$ indicate the contractible cycles.
To understand how a general $SL(3,\mathbb{Z})$ transformation realizes $L(p,q)\times S^1$, let us first describe the group $SL(3,\mathbb{Z})$ in some detail.
This group is generated by the elementary matrices $\{T_{ij}\}$ with $1\leq i\neq j\leq 3$, which are defined as $3\times 3$ matrices that differ from the identity matrix by the entry $1$ at the position $ij$.
These obey the following relations:
\begin{equation}\label{eq:sl3-relns}
\begin{aligned}
    & T_{ij}T_{kl}=T_{kl}T_{ij} \quad (i\neq l , j\neq k)\,,\qquad T_{ij}T_{jk}=T_{ik}T_{jk}T_{ij} \,,\qquad (T_{ij}T^{-1}_{ji}T_{ij})^4=\mathbbm{1} \,.
\end{aligned}
\end{equation}
Note that there are three obvious $SL(2,\mathbb{Z})$ subgroups in $SL(3,\mathbb{Z})$:
\begin{equation}\label{eq:defn-SL2ij}
    SL(2,\mathbb{Z})_{ij}\equiv \langle T_{ij},S_{ij} \rangle \,,\quad S_{ij}\equiv T_{ij}T^{-1}_{ji}T_{ij}\,,
\end{equation}
where we take $j>i$ and $S_{ij}$ and $T_{ij}$ correspond to the usual $S$ and $T$ generators.
Similar to the three-dimensional case, the large diffeomorphisms of $D_2\times T^2$ consist of those $SL(3,\mathbb{Z})$ transformations that fix $\mu$ \cite{Gadde:2020bov}.
Explicitly, they are given by:
\begin{equation}\label{eq:def-H}
\begin{aligned}
H\equiv SL(2,\mathbb{Z})_{13}\ltimes \mathbb{Z}^2 \,, \quad \textrm{with} \quad \mathbb{Z}^2 &= \langle T_{21}\,,T_{23}\rangle\,.
\end{aligned}
\end{equation}
A general element inside $H$ has the property that its $12$ and $32$ entries vanish, and its $22$ entry is equal to 1:
\begin{equation}\label{eq:h-shape}
h=\begin{pmatrix}
		* & 0 & * \\
		* & 1 & * \\
		* & 0 & *
		\end{pmatrix} \qquad \textrm{for} \quad h\in H\subset SL(3,\mathbb{Z})\,.
\end{equation}
This subgroup will play an important role in this paper.

Similar to the three-dimensional case, the manifolds associated to $f$ and $f'$ are diffeomorphic if they are related by any of the following relations:
\begin{equation}\label{eq:heeg-syms-top}
    \begin{alignedat}{2}
        f'&=hf\tilde{h}^{-1}\,,\qquad  f'=f^{-1}=\mathcal{O}g^{-1}\,,\qquad  f'=\mathcal{O}f^{-1}\mathcal{O}=g^{-1}\,\mathcal{O}\,,
    \end{alignedat}
\end{equation}
where $h,\tilde{h}\in H$ and the last transformation combines inversion with conjugation by $\mathcal{O}$.
We can use the first relation to show that the gluing of two $D_2\times T^2$ geometries with $f'=g'\,\mathcal{O}$ for general $g'\in SL(3,\mathbb{Z})$ produces a manifold diffeomorphic to $L(p,q)\times S^1$ for some $(p,q)$.
In particular, one can always find $h,\tilde{h}\in H$ for some $(p,q)$ such that:
\begin{equation}\label{eq:gSL2-from-gSL3}
    f'=hf\tilde{h}^{-1}\,,\qquad f=g_{(p,q)}\,\mathcal{O}\,,\quad g_{(p,q)}\equiv \begin{pmatrix}
		1 & 0 & 0 \\
		0 & -s & -r \\
		0 & -p & -q
		\end{pmatrix}\in SL(2,\mathbb{Z})_{23}\,.
\end{equation}
In other words, this means that there always exists a basis on the $D_2\times T^2$ geometries such that the gluing leaves invariant a non-contractible cycle.
From the preceding discussion we know that such a gluing produces a geometry with topology $L(p,q)\times S^1$, where we note that $g_{(p,q)}$ by itself still redundantly encodes $L(p,q)$.

\subsection{Hopf surfaces}\label{ssec:hopf-surfaces}

In this section, we summarize how manifolds with topology $L(p,q)\times S^1$ and $D_2\times T^2$ can be endowed with complex structure moduli and subsequently extend the Heegaard splitting to include a mapping of the moduli.
We refer to Appendix \ref{app:hopf-surfaces} for a more detailed discussion.

A primary Hopf surface is a complex manifold with topology $S^3\times S^1$.
It is defined as a quotient of $\mathbb{C}^2\setminus \lbrace(0,0)\rbrace$ by the $\mathbb{Z}$-action:
\begin{equation}\label{eq:hopf-surface-ids}
    (z_1,z_2)\sim (\hat{p}z_1,\hat{q}z_2)\,,\qquad  0<|\hat{p}|\leq |\hat{q}|<1\,,
\end{equation}
with $\hat{p}=e^{2\pi i\hat{\sigma}}$ and $\hat{q}=e^{2\pi i\hat{\tau}}$ representing the complex structure moduli.\footnote{The complex parameters $\hat{p}$ and $\hat{q}$ are not to be confused with the integers $p$ and $q$ defining the lens space $L(p,q)$.}
A secondary Hopf surface is defined as the lens quotient \eqref{eq:lens-quotient} of a primary Hopf surface.
As such, it has topology $L(p,q)\times S^1$.
We will denote these complex manifolds uniformly by:
\begin{equation}\label{eq:notation-hopf-surface}
    \mathcal{M}_{(p,q)}(\hat{\boldsymbol{\rho}})\,,
\end{equation}
where the primary Hopf surface has $(p,q)=(1,0)$ and we denote the moduli by:
\begin{equation}\label{eq:hat-rho}
   \hat{\boldsymbol{\rho}}\equiv(\hat{z}_a; \hat{\tau},\hat{\sigma})\,.
\end{equation}
The (real) holonomies $\hat{z}_{1,\ldots,r}$ along $S^1$ parametrize a rank $r$ vector bundle over the Hopf surface associated to a rank $r$ global symmetry.
With respect to $\hat{\boldsymbol{\rho}}$, the following transformations yield an identical Hopf surface:
\begin{align}\label{eq:syms-lens-geom}
    \begin{split}
         \hat{z}_a&\to \hat{z}_a+1\,,\quad \hat{\tau}\to \hat{\tau}+1\,,\quad \hat{\sigma}\to \hat{\sigma}+1\,,\\
         \hat{\tau}&\to \hat{\tau}-\tfrac{1}{p}\quad \text{and} \quad \hat{\sigma}\to \hat{\sigma} +\tfrac{q}{p}\,,\\
        \hat{\tau}&\to \hat{\tau}-\tfrac{s}{p}\quad \text{and} \quad\hat{\sigma}\to \hat{\sigma} +\tfrac{1}{p}\,,
    \end{split}
\end{align}
where we recall that $s=q^{-1}\mod p$.
The Hopf surface is also invariant under:
\begin{equation}\label{eq:syms-lens-geom-3}
    q\to q+p\,,\quad s\to s+p\,,
\end{equation}
and finally:
\begin{equation}\label{eq:syms-lens-geom-2}
    \hat{\tau}\leftrightarrow \hat{\sigma} \quad \text{and} \quad q\leftrightarrow s\,,
\end{equation}
both of which follow from the lens quotient in \eqref{eq:lens-quotient}.
Similar to the previous section, we also define:
\begin{equation}
    \mathcal{M}_{(0,-1)}(\hat{\boldsymbol{\rho}})\cong S^2\times T^2\,.
\end{equation}
In this case, $\hat{\sigma}$ can be interpreted as the modular parameter of the $T^2$, while $\hat{\tau}$ captures twists of $S^2$ as one-cycles along either of the $T^2$ cycles (see Appendix \ref{sapp:S2T2}).
In addition, $\hat{z}_a$ can be viewed as a complex holonomy, capturing two real holonomies along both cycles of the $T^2$.
In this case, the group of large diffeomorphisms and gauge transformations is given by $\mathcal{H}=SL(2,\mathbb{Z})\ltimes \mathbb{Z}^{2(1+r)}$ \cite{Closset:2013sxa}.
We will describe its action on $\hat{\boldsymbol{\rho}}$ momentarily.

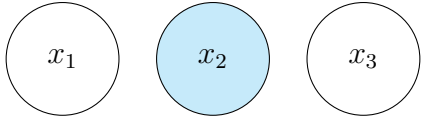
\begin{figure}[t]
    \centering
    \begin{tikzpicture}
    \node[ellipse,
    draw = black,
    text = black,
    minimum width = 1.5cm, 
    minimum height = 1.5cm] (e1) at (0,0) {$x_1$};
    \node[ellipse,
    draw = black,
    text = black,
    fill = cyan!20,
    minimum width = 1.5cm, 
    minimum height = 1.5cm] (e2) at (2,0) {$x_2$};
    \node[ellipse,
    draw = black,
    text = black,
    minimum width = 1.5cm, 
    minimum height = 1.5cm] (e3) at (4,0) {$x_3$};
    \end{tikzpicture}
    \caption{Schematic depiction of $D_2\times T^2$ with the contractible cycle shaded. Its complex structure moduli $(\tau,\sigma)$ are represented by the homogeneous moduli $x_i$.}
    \label{fig:solid-3torus}
\end{figure}

Finally, the $D_2\times T^2$ geometry can be endowed with complex structure moduli and holonomies in exactly the same way as $\mathcal{M}_{(0,-1)}(\hat{\boldsymbol{\rho}})$ \cite{Longhi:2019hdh}.
To distinguish from the closed manifolds, we denote these moduli by:
\begin{equation}\label{eq:rho-homog}
   \boldsymbol{\rho}\equiv(z_a;\tau,\sigma)=\left(\frac{Z_a}{x_1};\frac{x_2}{x_1},\frac{x_3}{x_1}\right) \,,
\end{equation}
where the second equality substitutes the projective moduli $(\vec{z}; \tau, \sigma)$ in terms of homogeneous moduli $(\vec{Z};x_1,x_2,x_3)$.
The $x_i$ can be thought of as complexifications of the cycle lengths of $T^3$; we associate $x_2$ to the contractible cycle \cite{Felder_2000,Gadde:2020bov}.
See Figure \ref{fig:solid-3torus}.

In order to write the Heegaard splitting of a Hopf surface, we must first define the action of the gluing transformation on the moduli $\boldsymbol{\rho}$.
The full gluing group consists of both the large diffeomorphisms and gauge transformations of (the rank $r$ vector bundle over) $T^3$ \cite{Gadde:2020bov}.
This group is given by:
\begin{equation}\label{eq:def-mathcalG}
    \mathcal{G}\equiv SL(3,\mathbb{Z})\ltimes \mathbb{Z}^{3r}\,.
\end{equation}
The subgroup $SL(3,\mathbb{Z})\subset \mathcal{G}$ acts on $\boldsymbol{\rho}$ by left matrix multiplication\footnote{This contrasts with the action of $SL(3,\mathbb{Z})$ on the cycles by right multiplication, cf.\ \eqref{eq:Mg-defn-lambda-mu}.} on the vector $\textbf{x}=(x_1,x_2,x_3)$, and the $t^{(a)}_{i}$ generators of each $\mathbb{Z}^3$ factor shift $Z_a$ by $x_i$ \cite{Felder_2000,Gadde:2020bov}: 
\begin{equation}\label{eq:calG-action}
\begin{aligned}
g \in SL(3,\mathbb{Z}): \qquad \textbf{x} &\mapsto g \cdot \textbf{x} \,, \\
t^{(a)}_i \in \mathbb{Z}^3_a: \qquad Z_a&\mapsto Z_a+x_i\,.
\end{aligned}
\end{equation}
For completeness, we collect the mixed relations satisfied by $T_{ij}$ and $t_i^{(a)}$ here, suppressing $a$, which together with \eqref{eq:sl3-relns} fully specifies the relations in the group $\mathcal{G}$:
\begin{equation}\label{eq:Z3-relns}
    \begin{aligned}
    T_{ij}t_k=t_kT_{ij}\,, \quad (i\neq k)\,,\qquad T_{ij}t_i=t_it_j^{-1}T_{ij} \,,\qquad t_it_j=t_jt_i\,.
    \end{aligned}
\end{equation}
The subgroup of large diffeomorphisms and gauge transformations of $D_2\times T^2$ is denoted by $\mathcal{H}$ and takes the form:
\begin{equation}\label{eq:defn-calH}
    \mathcal{H} =SL(2,\mathbb{Z})_{13}\ltimes \mathbb{Z}^{2(1+r)}\,,\qquad \mathbb{Z}^{2(1+r)}=\langle T_{21},T_{23},t_{1}^{(a)},t_{3}^{(a)}\rangle \,,
\end{equation}
which contains $H$, the group of large diffeomorphisms, defined in Section \ref{ssec:top-aspects}.
The action of $\mathcal{H}$ on $\boldsymbol{\rho}$ is obtained from \eqref{eq:calG-action} by viewing $\mathcal{H}\subset \mathcal{G}$.
This group also acts on $\mathcal{M}_{(0,-1)}(\hat{\boldsymbol{\rho}})\cong S^2\times T^2$, as mentioned above, and its action on the moduli $\hat{\boldsymbol{\rho}}$ is identical to the action on $\boldsymbol{\rho}$.

We can now state the Heegaard splitting of a general Hopf surface:
\begin{equation}\label{eq:notation-Mg-split}
    \mathcal{M}_{(p,q)}(\hat{\boldsymbol{\rho}})\cong M_{f}(\boldsymbol{\rho},\tilde{\boldsymbol{\rho}})\equiv   D_2 \times T^2 (\boldsymbol{\rho})\overset{f}{\sqcup}D_2 \times T^2 (\tilde{\boldsymbol{\rho}})\,,
\end{equation}
where $f=g_{(p,q)}\,\mathcal{O}$ with $g_{(p,q)}\in SL(2,\mathbb{Z})_{23}$ as in \eqref{eq:gSL2-from-gSL3}, as derived in Appendix \ref{app:hopf-surfaces}.\footnote{We turn to the Heegaard splitting of a Hopf surface for $g\in\mathcal{G}$ general in Section \ref{ssec:ambig-heegaard}.}
In addition, $\boldsymbol{\rho}$ and $\tilde{\boldsymbol{\rho}}$ capture the moduli of the two $D_2\times T^2$ geometries and are related via the gluing condition:
\begin{equation}\label{eq:gluing-condition}
    \tilde{\boldsymbol{\rho}}=f^{-1}\boldsymbol{\rho}\,.
\end{equation}
Finally, the moduli of the Hopf surface $\hat{\boldsymbol{\rho}}$ are related to $\boldsymbol{\rho}$ as follows:
\begin{equation}\label{eq:p-moduli}
   \boldsymbol{\rho}=(z_a;\tau,\sigma)=\begin{cases}
    (\hat{z}_a;\hat{\tau},\hat{\sigma})\,, \; &\text{for}\;p=r=0\,,\,q=s= -1\,,\\
    (\hat{z}_a;\hat{\tau}+s\hat{\sigma},p\hat{\sigma})\,, \; &\text{for}\;p\neq 0 \,.   \end{cases}
\end{equation}
Matching the holonomies requires us to set the imaginary part of $z_a$ to zero.
For later convenience, we note that the gluing condition implies:
\begin{equation}\label{eq:p-moduli-tilde}
   \tilde{\boldsymbol{\rho}}=(\tilde{z}_a;\tilde{\tau},\tilde{\sigma})=\begin{cases}
    (\hat{z}_a;-\hat{\tau},\hat{\sigma})\,, \; &\text{for}\;p=r= 0\,,\,q=s= -1\,,\\
    (\hat{z}_a;\hat{\sigma}+q\hat{\tau},p\hat{\tau})\,, \; &\text{for}\;p\neq 0 \,. 
    \end{cases}
\end{equation}

\subsection{Ambiguities in the Heegaard splitting}\label{ssec:ambig-heegaard}

\begin{figure}[t]
     \centering
     \begin{subfigure}[t]{0.49\textwidth}
         \centering
         \begin{tikzpicture}
    \node (rho1) at (0,1.5) {$\boldsymbol{\rho}$};
    \node (rho2) at (3,1.5) {$\tilde{\boldsymbol{\rho}}$};
    \draw [-latex] (rho2) -- node[above=1.5mm] {$S_{23}\mathcal{O}$} (rho1) ;
    \node[ellipse,
    draw = black,
    text = black,
    minimum width = 0.2cm, 
    minimum height = 1.8cm] (e1) at (0,0) {$x_1$};
    \node[ellipse,
    draw = black,
    text = black,
    fill = cyan!20,
    minimum width = 0.2cm, 
    minimum height = 1.8cm] (e2) at (0,-2) {$x_2$};
    \node[ellipse,
    draw = black,
    text = black,
    minimum width = 0.2cm, 
    minimum height = 1.8cm] (e3) at (0,-4) {$x_3$};
    \node[ellipse,
    draw = black,
    text = black,
    minimum width = 0.2cm, 
    minimum height = 1.8cm] (f1) at (3,0) {$x_1$};
    \node[ellipse,
    draw = black,
    text = black,
    fill = cyan!20,
    minimum width = 0.2cm, 
    minimum height = 1.8cm] (f2) at (3,-2) {$x_3$};
    \node[ellipse,
    draw = black,
    text = black,
    minimum width = 0.2cm, 
    minimum height = 1.8cm] (f3) at (3,-4) {$x_2$};
    \draw [-latex][dashed] (f1) -- (e1) ;
    \draw [-latex][dashed] (f2) -- (e3) ;
    \draw [-latex][dashed] (f3) -- (e2) ;
\end{tikzpicture}
        \caption{$M_{S_{23}\mathcal{O}}(\boldsymbol{\rho},\tilde{\boldsymbol{\rho}})$}
         \label{fig:MS23}
     \end{subfigure}
     \hfill
     \begin{subfigure}[t]{0.49\textwidth}
         \centering
         \begin{tikzpicture}
    \node (rho1) at (0,1.5) {$S_{13}\boldsymbol{\rho}$};
    \node (rho2) at (3,1.5) {$S_{13}\tilde{\boldsymbol{\rho}}$};
    \draw [-latex] (rho2) -- node[above=1.5mm] {$S_{12}^{-1}\mathcal{O}$} (rho1) ;
    \node[ellipse,
    draw = black,
    text = black,
    minimum width = 0.2cm, 
    minimum height = 1.8cm] (e1) at (0,0) {$x_3$};
    \node[ellipse,
    draw = black,
    text = black,
    fill = cyan!20,
    minimum width = 0.2cm, 
    minimum height = 1.8cm] (e2) at (0,-2) {$x_2$};
    \node[ellipse,
    draw = black,
    text = black,
    minimum width = 0.2cm, 
    minimum height = 1.8cm] (e3) at (0,-4) {$-x_1$};
    \node[ellipse,
    draw = black,
    minimum height = 1.8cm] (f1) at (3,0) {$x_2$};
    \node[ellipse,
    draw = black,
    text = black,
    fill = cyan!20,
    minimum width = 0.2cm, 
    minimum height = 1.8cm] (f2) at (3,-2) {$x_3$};
    \node[ellipse,
    draw = black,
    text = black,
    minimum width = 0.2cm, 
    minimum height = 1.8cm] (f3) at (3,-4) {$-x_1$};
    \draw [-latex][dashed] (f1) -- (e2) ;
    \draw [-latex][dashed] (f2) -- (e1) ;
    \draw [-latex][dashed] (f3) -- (e3) ;
\end{tikzpicture}
         \caption{$M_{S_{12}^{-1}\mathcal{O}}(S_{13}\boldsymbol{\rho},S_{13}\tilde{\boldsymbol{\rho}})$}
         \label{fig:MS12}
     \end{subfigure}
        \caption{Illustration of \eqref{eq:main-geom-equiv} for the example $\tilde{\boldsymbol{\rho}}=\mathcal{O}S_{23}^{-1}\boldsymbol{\rho}$ and $h=\tilde{h}=S_{13}$. We have written the homogeneous moduli inside the relevant cycles of the $D_2\times T^2$ geometries as in Figure \ref{fig:solid-3torus}. The arrows indicate the identification of cycles. Clearly, both Heegaard splittings represent the same Hopf surface $\mathcal{M}_{(1,0)}(\hat{\boldsymbol{\rho}})$ with $\boldsymbol{\rho}=\hat{\boldsymbol{\rho}}$.}
        \label{fig:geom-equiv}
\end{figure}
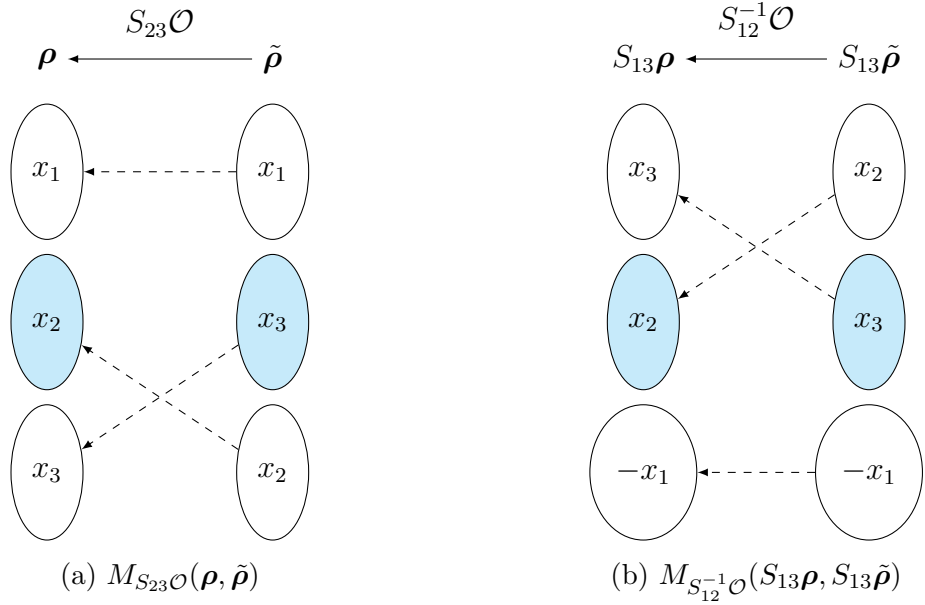

As we have seen in Section \ref{ssec:top-aspects}, the gluing transformation $f$ encodes the topology of $L(p,q)\times S^1$ redundantly. 
This leads to ambiguities in the Heegaard splitting.
For the Hopf surfaces, these ambiguities arise when combining the action $f\to f'$ with an action on the moduli $(\boldsymbol{\rho},\tilde{\boldsymbol{\rho}})$ such that the gluing condition is preserved: 
\begin{equation}\label{eq:combined-geom-action}
    \begin{alignedat}{2}
       f'&= hf\tilde{h}^{-1}\,,&\qquad (\boldsymbol{\rho}',\tilde{\boldsymbol{\rho}}')&=(h\boldsymbol{\rho},\tilde{h}\tilde{\boldsymbol{\rho}})\,,\qquad h,\tilde{h}\in\mathcal{H}\,,\\
\textrm{or}\quad       f'&= f^{-1}\,,&\qquad (\boldsymbol{\rho}',\tilde{\boldsymbol{\rho}}')&=(\tilde{\boldsymbol{\rho}},\boldsymbol{\rho})\,,\\
\textrm{or}\quad         f'&= \mathcal{O}f^{-1}\mathcal{O}\,,&\qquad (\boldsymbol{\rho}',\tilde{\boldsymbol{\rho}}')&=(\mathcal{O}\tilde{\boldsymbol{\rho}},\mathcal{O}\boldsymbol{\rho})\,,
    \end{alignedat}
\end{equation}
where the action of $h$ on $\boldsymbol{\rho}$ is defined by \eqref{eq:calG-action}.
It follows that a Hopf surface with Heegaard splitting $M_f(\boldsymbol{\rho},\tilde{\boldsymbol{\rho}})$, as defined in \eqref{eq:notation-Mg-split}, also admits a Heegaard splitting for any of these transformations:
\begin{align}\label{eq:main-geom-equiv}
    \begin{split}
         \mathcal{M}_{(p,q)}(\hat{\boldsymbol{\rho}})\cong  M_{f'}(\boldsymbol{\rho}',\tilde{\boldsymbol{\rho}}')\,.
    \end{split}
\end{align}
Here, $\boldsymbol{\rho}'$ is understood to be a function of $\boldsymbol{\rho}$, which in turn is related to $\hat{\boldsymbol{\rho}}$ as in \eqref{eq:p-moduli}.
This makes our claim \eqref{eq:l(p,1)-split-2-intro} in the introduction explicit.
For illustration, we present an example in Figure \ref{fig:geom-equiv} for $h=\tilde{h}=S_{13}$ and $f=S_{23}\,\mathcal{O}$.
As a consequence of \eqref{eq:gSL2-from-gSL3}, the expressions \eqref{eq:main-geom-equiv} relate the Hopf surfaces that were associated in Section \ref{ssec:hopf-surfaces} to $g_{(p,q)}\in SL(2,\mathbb{Z})_{23}\subset SL(3,\mathbb{Z})$ to any $g'\in SL(3,\mathbb{Z})$.

We stress that in general one should view the combined action \eqref{eq:combined-geom-action} as distinct from a large diffeomorphism, such as the $SL(2,\mathbb{Z})$ action on the complex structure of a two-torus.
Indeed, the large diffeomorphisms (and gauge transformations) of a general Hopf surface certainly do not include $\mathcal{H}\times \mathcal{H}$.
Instead, the action reflects ambiguities in the Heegaard splitting of the Hopf surface.

\bigskip
In establishing \eqref{eq:main-geom-equiv}, we have assumed that $\hat{\boldsymbol{\rho}}$ does not transform under the transformations \eqref{eq:combined-geom-action}. 
We now show that the symmetries of a Hopf surface $\mathcal{M}_{(p,q)}(\hat{\boldsymbol{\rho}})$, namely the transformation  \eqref{eq:syms-lens-geom} on $\hat{\boldsymbol{\rho}}$, the transformation \eqref{eq:syms-lens-geom-3} on $(p,q)$, and the transformation \eqref{eq:syms-lens-geom-2} on $\hat{\boldsymbol{\rho}}$ and $(p,q)$, can also be incorporated by a subset of transformations \eqref{eq:combined-geom-action}.
Namely, we want to show that: 
\begin{align}
    \begin{split}
         \mathcal{M}_{(p,q)}(\hat{\boldsymbol{\rho}})\cong \mathcal{M}_{(p,q')}(\hat{\boldsymbol{\rho}}') \cong  M_{f'}(\boldsymbol{\rho}',\tilde{\boldsymbol{\rho}}')\,,
    \end{split}
\end{align}
where $\mathcal{M}_{(p,q')}(\hat{\boldsymbol{\rho}}')$ is related to $ \mathcal{M}_{(p,q)}(\hat{\boldsymbol{\rho}})$ by any of the above symmetries, and $(\boldsymbol{\rho}',\tilde{\boldsymbol{\rho}}')$ are related to $\hat{\boldsymbol{\rho}}'$ in the same way as $(\boldsymbol{\rho},\tilde{\boldsymbol{\rho}})$ are related to $\hat{\boldsymbol{\rho}}$, as in \eqref{eq:p-moduli}.

For example, the first relation in \eqref{eq:combined-geom-action} with $h,\tilde{h}\in \langle T_{21},T_{31}\rangle$ can be used to derive the shift symmetries in \eqref{eq:syms-lens-geom}.
To see this, we first observe that:
\begin{align}\label{eq:largediffeo-matrixrelation}
    \begin{split}
        T_{21}\,f\,T_{21}^{-q}T_{31}^{-p}&=f\,,\qquad T_{21}^{-s}T_{31}^{-p}\,f\,T_{21}=f\,, \\
        T_{31}\,f\,T_{21}^{r}T_{31}^{s}&=f \,,\qquad T_{21}^{r}T_{31}^{q}\,f\,T_{31}=f\,,
    \end{split}
\end{align}
with $f=g_{(p,q)}\mathcal{O}$ as in \eqref{eq:gSL2-from-gSL3}.
Since $f$ is invariant, we need only consider the action on the moduli $(\boldsymbol{\rho},\tilde{\boldsymbol{\rho}})$.
Using the relation with $\hat{\boldsymbol{\rho}}$ one easily checks that the action on $(\boldsymbol{\rho},\tilde{\boldsymbol{\rho}})$ is equivalent to the shift symmetries, establishing the claim.
In addition, acting with $h=\tilde{h}=t_1^{(a)}$ also leaves $f$ invariant; this corresponds to the symmetry of the Hopf surface (plus vector bundle) under $\hat{z}_a\to\hat{z}_a+1$.

Similarly, the equivalence in \eqref{eq:syms-lens-geom-2} is the statement that:
\begin{equation}
    \mathcal{M}_{(p,q)}(\hat{z}_i;\hat{\tau},\hat{\sigma})\cong \mathcal{M}_{(p,s)}(\hat{z}_i;\hat{\sigma},\hat{\tau})\,.
\end{equation}
This is reproduced by the inversion on the second line of \eqref{eq:combined-geom-action}.
Finally, the fact that the Hopf surface only depends on $q,s\mod p$ follows from the first line of the combined action \eqref{eq:combined-geom-action}, now taking $h=T_{23}$ or $\tilde{h}=T_{23}$.
These actions change $f$ by the shifts $(s,r)\to (s+p,r+q)$ and $(q,r)\to (q+p,r+s)$, respectively.
On the moduli $\boldsymbol{\rho}$ and $\tilde{\boldsymbol{\rho}}$, they correspond to the shifts $\tau\to \tau+\sigma$ and $\tilde{\tau}\to\tilde{\tau}+\tilde{\sigma}$, respectively.
From \eqref{eq:p-moduli} and \eqref{eq:p-moduli-tilde}, it follows that the combined action leaves $\hat{\boldsymbol{\rho}}$ invariant, and the claim follows.

Let us end with the special case $f=\mathcal{O}$, corresponding to the Hopf surface $\mathcal{M}_{(0,-1)}(\hat{\boldsymbol{\rho}})$ with topology $S^2\times T^2$.
Consider the first relation in \eqref{eq:combined-geom-action} for $\tilde{h}=\mathcal{O}h\mathcal{O}$:
\begin{equation}
    \mathcal{O}\to h\mathcal{O} \tilde{h}^{-1}=\mathcal{O}\,.
\end{equation}
This action leads to the statement that:
\begin{equation}
    M_{\mathcal{O}}(h\boldsymbol{\rho},\mathcal{O}h\boldsymbol{\rho})=M_{\mathcal{O}}(\boldsymbol{\rho},\mathcal{O}\boldsymbol{\rho}) \,.
\end{equation}
Given the fact that in this case $\boldsymbol{\rho}=\hat{\boldsymbol{\rho}}$, we see that this specialization of the combined action reduces to the statement that $\mathcal{H}$ is the group of large diffeomorphisms and gauge transformations of $\mathcal{M}_{(0,-1)}(\hat{\boldsymbol{\rho}})$, which is indeed correct.

\section{Modular factorization of lens indices}\label{sec:mod-fac-lens-indices}

In this section, we consider supersymmetric partition functions on secondary Hopf surfaces, also known as lens indices.
We give a comprehensive review of the holomorphic block factorization of lens indices, which is the statement that lens indices respect the Heegaard splitting of a Hopf surface.
We then argue that (a subset of) the ambiguities discussed in Section \ref{ssec:ambig-heegaard} lead to a modular family of holomorphic blocks into which a given lens index can be factorized.
After providing a geometric interpretation of the modular subset, we proceed to prove the conjecture in the context of concrete $\mathcal{N}=1$ gauge theories.
The proof relies in particular on modular properties of the elliptic $\Gamma$ function derived in Appendix \ref{app:mod-props-Gamma}.

\subsection{Towards a conjecture}\label{ssec:towards-conjecture}

It was shown in \cite{Closset:2013vra} (based on \cite{Festuccia:2011ws,Dumitrescu:2012ha}) that $\mathcal{N}=1$ theories with a $U(1)_R$ symmetry can be formulated on the Hopf surfaces $\mathcal{M}_{(p,q)}(\hat{\boldsymbol{\rho}})$ for general values of the complex structure moduli while preserving two real supercharges.\footnote{The same result holds for the $D_2\times T^2(\boldsymbol{\rho})$ geometry \cite{Longhi:2019hdh}.}
An additional requirement is that the R-charges of the fields are quantized as integers for the case $p=0$ and $q=-1$, i.e., $S^2\times T^2$, and as integer multiples of $\frac{2}{q-1}$ for $q>1$.

The supersymmetric background allows a computation of the supersymmetric partition function on $\mathcal{M}_{(p,q)}(\hat{\boldsymbol{\rho}})$, which equals the associated index, i.e., a weighted trace over the Hilbert space of the theory on $L(p,q)$  \cite{Closset:2013vra,Closset:2014uda,Assel:2014paa}.\footnote{\label{fn:susy-cas-en}We define these indices for general $\mathcal{N}=1$ gauge theories in Appendix \ref{app:lens-indices}. The equality between the localized partition function and the index holds up to a proportionality factor associated with the supersymmetric Casimir energy \cite{Assel:2014paa,Assel:2015nca,Bobev:2015kza}.}
The fugacities appearing in the index can be mapped to both the complex structure moduli of the Hopf surface and the holonomies for background gauge fields associated to global symmetries.

It was argued in \cite{Closset:2013vra,Closset:2014uda} that such partition functions depend holomorphically on the complex structure moduli and holonomies, while being insensitive to other details of the background.
This was confirmed in \cite{Closset:2013sxa,Assel:2014paa,Nishioka:2014zpa} through explicit localization computations.
In the following, we will always consider the formulation of the partition functions as indices, ignoring the prefactor associated with the supersymmetric Casimir energy.
We thus collectively refer to the partition functions on secondary Hopf surfaces as \emph{lens indices} and denote them by:
\begin{equation}\label{eq:notation-lens-index}
    \mathcal{I}_{(p,q)}(\hat{\boldsymbol{\rho}})\,,
\end{equation}
where the notation reflects the holomorphic dependence on the complexified moduli $\hat{\boldsymbol{\rho}}$.
Since we are interested in the full collection of lens indices, we will assume that the R-charges are quantized as even integers.
As explained in \cite{Gadde:2020bov}, this can be achieved through an appropriate shift of the R-symmetry by a flavor symmetry.
The resulting R-symmetry typically does not coincide with the superconformal R-symmetry of the IR $\mathcal{N}=1$ SCFT, and therefore the indices are parametrized in a non-standard way.\footnote{The Bethe Ansatz computation of the superconformal index also employs a shifted R-symmetry such that the R-charges are quantized as integers \cite{Benini:2018mlo,Benini:2018ywd}.}

Let us now introduce an alternative notation for the lens index that reflects the Heegaard splitting $M_{f}(\boldsymbol{\rho},\tilde{\boldsymbol{\rho}})$, as defined in \eqref{eq:notation-Mg-split}, of the underlying Hopf surface:\footnote{\label{fn:FI-rho}In general, we should also include an Fayet--Iliopoulos parameter $\xi$ to $\boldsymbol{\rho}$. We suppress this parameter in most of the discussion and comment on its role and transformation under a large diffeomorphism in the example of SQED.}
\begin{equation}\label{eq:defn-Zf-notation}
    \mathcal{Z}_f(\boldsymbol{\rho})\equiv \mathcal{I}_{(p,q)}(\hat{\boldsymbol{\rho}})\,,
\end{equation}
where $f=g_{(p,q)}\,\mathcal{O}$ with $g_{(p,q)}\in SL(2,\mathbb{Z})_{23}$ as in \eqref{eq:gSL2-from-gSL3}, and $\boldsymbol{\rho}$ is related to $\hat{\boldsymbol{\rho}}$ through \eqref{eq:p-moduli}.
Because of the gluing condition, we write $\mathcal{Z}_f$ as a function of $\boldsymbol{\rho}$ only.
As examples of the notation, $\mathcal{Z}_{S_{23}\mathcal{O}}(\boldsymbol{\rho})$ corresponds to the superconformal index, $\mathcal{Z}_{\mathcal{O}}(\boldsymbol{\rho})$ corresponds to the index on $S^2\times T^2$, and $\mathcal{Z}_{g_{(p,1)}\,\mathcal{O}}(\boldsymbol{\rho})$ corresponds to the index on $L(p,1)\times S^1$.
These indices are defined in Appendix \ref{app:lens-indices} in terms of $\hat{\boldsymbol{\rho}}$.

Consider now two distinct Heegaard splittings $M_f(\boldsymbol{\rho},\tilde{\boldsymbol{\rho}})$ and $M_{f'}(\boldsymbol{\rho}',\tilde{\boldsymbol{\rho}}')$ of a Hopf surface, where $f'$ is related to $f$ via the orientation preserving transformations in \eqref{eq:combined-geom-action}:
\begin{equation}\label{eq:f-rho-orient-pres-transf}
    f'=hf\tilde{h}^{-1}\,,\quad \boldsymbol{\rho}'=h\boldsymbol{\rho}\qquad \text{or}\qquad  f'=f^{-1}\,,\quad \boldsymbol{\rho}'=\tilde{\boldsymbol{\rho}}=f^{-1}\boldsymbol{\rho}\,.
\end{equation}
Note that the first transformation implies $\tilde{\boldsymbol{\rho}}'\equiv(f')^{-1}\boldsymbol{\rho}'=\tilde{h}\tilde{\boldsymbol{\rho}}$ as recorded in the first line of \eqref{eq:combined-geom-action}, while the second transformation similarly implies $\tilde{\boldsymbol{\rho}}'\equiv (f')^{-1}\boldsymbol{\rho}'=\boldsymbol{\rho}$ as recorded in the second line of \eqref{eq:combined-geom-action}.
By definition of $\mathcal{Z}_{f}(\boldsymbol{\rho})$, we have for either transformation:\footnote{The behaviour under the orientation reversing transformation in \eqref{eq:combined-geom-action} is more complicated, and we return to it in Section \ref{ssec:1-cocycle-lens-indices}.}
\begin{equation}\label{eq:lens-index-equivalence}
    \mathcal{Z}_{f'}(\boldsymbol{\rho}')=\mathcal{Z}_{f}(\boldsymbol{\rho})\,
\end{equation}
since the Hopf surface itself has not changed.
As already mentioned in Section \ref{sec:intro}, lens indices of certain $\mathcal{N}=1$ gauge theories have a factorization property that reflects the Heegaard splitting of the Hopf surface.
Postponing a detailed review to Section \ref{ssec:hol-blocks}, we here note that while the second transformation in \eqref{eq:f-rho-orient-pres-transf} is trivially satisfied in factorized form, the first transformation predicts:
\begin{equation}\label{eq:schem-Zg-Bh-fact}
    \mathcal{Z}_{hf\tilde{h}^{-1}}(h\boldsymbol{\rho})=e^{i\pi \mathcal{P}'}\sum \mathcal{B}_{h}(\boldsymbol{\rho})\mathcal{B}_{\tilde{h}}(f^{-1}\boldsymbol{\rho})= e^{i\pi \mathcal{P}} \sum \mathcal{B}(\boldsymbol{\rho})\mathcal{B}(f^{-1}\boldsymbol{\rho})=\mathcal{Z}_{f}(\boldsymbol{\rho})\,,
\end{equation}
where we plugged in \eqref{eq:lens-index-equivalence} with the (schematic) factorized form of the indices alluded to in Section \ref{sec:intro}.
The functions $\mathcal{B}(\boldsymbol{\rho})$ and $\mathcal{B}_{h}(\boldsymbol{\rho})$ represent the (supersymmetric) partition functions on $D_2\times T^2$ with moduli $\boldsymbol{\rho}$ and $h\boldsymbol{\rho}$ respectively (see Section \ref{ssec:hopf-surfaces}), and are thus related:
\begin{equation}\label{eq:defn-Bh}
    \mathcal{B}_{h}(\boldsymbol{\rho})\equiv \mathcal{B}(h\boldsymbol{\rho})\,.
\end{equation}
We also included the phases $\mathcal{P}$ and $\mathcal{P}'$ in the factorization, which in general depend non-trivially on $\boldsymbol{\rho}$ and will be derived explicitly later on.
Given the fact that the action of $h,\tilde{h}\in \mathcal{H}$ on $\boldsymbol{\rho},\tilde{\boldsymbol{\rho}}$ is modular, the prediction \eqref{eq:schem-Zg-Bh-fact} \emph{in factorized form} has the flavor of a modular covariance. 
However, let us stress some important distinctions.
First of all, the functions $\mathcal{B}$ by themselves are not (weight $0$) automorphic forms under $\mathcal{H}$: $\mathcal{B}(h\boldsymbol{\rho})\ncong \mathcal{B}(\boldsymbol{\rho})$ for general $h\in \mathcal{H}$, even though $\mathcal{H}$ represents the group of large diffeomorphisms and gauge transformations of $D_2\times T^2$.\footnote{Modular properties of holomorphic blocks in three dimensions are also more subtle than those of ordinary modular forms \cite{Cheng:2018vpl}.}
It follows that the covariance relies on the product.
But even the product cannot be identified with an ordinary modular object, since the covariance is with respect to a \emph{combined action}, acting on both $\boldsymbol{\rho}$ and $f$ (cf. \eqref{eq:f-rho-orient-pres-transf}).
In other words, our proposed covariance does not reflect large diffeomorphisms of an underlying manifold, as familiar from the $T^2$ partition function of CFT$_2$'s.\footnote{Indeed, the large diffeomorphisms of a general Hopf surface certainly do not include $\mathcal{H}$.} 
Instead, it reflects ambiguities in the Heegaard spitting of a Hopf surface and is non-trivial only at the level of the factorized expressions.

In the following sections, we subject this proposal to physical consistency conditions.
This will result in a refined proposal, discussed in Section \ref{ssec:mod-fact-conjecture}, for which we will provide evidence in Section \ref{ssec:evidence}.
In particular, we will see that \eqref{eq:schem-Zg-Bh-fact} does not hold for general pairs $h,\tilde{h}\in\mathcal{H}$.
The subset of allowed pairs $(h,\tilde{h})$ still forms an interesting modular subset of $\mathcal{H}\times \mathcal{H}$, whose geometric interpretation we discuss in Section \ref{ssec:geom-int-univ-blocks}.
The fact that only a subset of $\mathcal{H}\times \mathcal{H}$ is compatible with the factorization of a given index is closely related to the conjecture of \cite{Gadde:2020bov}, which proposes that the normalized parts of lens indices form a non-trivial cohomology class in $H^1(\mathcal{G},N/M)$, as we will show in detail in Section \ref{sec:gen-modularity}.

\subsection{Review of holomorphic block factorization}\label{ssec:hol-blocks}

In this section we review in detail the factorization of lens indices into holomorphic blocks, as anticipated in \eqref{eq:nieri-pasq-result} and \eqref{eq:schem-Zg-Bh-fact}, based on \cite{Nieri:2015yia,Longhi:2019hdh}.
We first focus on structural properties of the formulae for generic theories, and illustrate the claims in Section \ref{sssec:example-free-chiral} and Section \ref{sssec:sqed} with the example of a free chiral multiplet and SQED, respectively.

For three-dimensional gauge theories, factorization properties of supersymmetric partition functions on $S^2\times S^1$, $S^3$ and $L(p,1)$ were first established and studied in \cite{Pasquetti:2011fj,Beem:2012mb,Hwang:2012jh,Imamura:2013qxa}.
The result was obtained through manipulation of a Coulomb branch formula for the partition functions, which takes the form of an integral over gauge holonomies associated to the Cartan torus of the gauge group.
In these works, it was also argued that the factors, dubbed \emph{holomorphic blocks} in \cite{Beem:2012mb}, could be understood in terms of supersymmetric partition functions on the solid tori associated to the Heegaard splitting of the three-manifold.

A more direct, path integral derivation of the factorization properties was given later in \cite{Benini:2012ui,Doroud:2012xw,Fujitsuka:2013fga,Benini:2013yva}.
These works employ a different localization scheme giving rise to a so-called Higgs branch formula for the partition function.
In this scheme, the gauge symmetry is completely broken and the path integral only receives contributions from field configurations localized at the centers of the two solid tori.
The resulting formula takes on the form of a finite sum over Higgs branch vacua of (a mass deformation of) the theory, and the summand is naturally factorized.
The equality between the Higgs and Coulomb branch formulae can be established by performing the contour integrals defining the latter through residues \cite{Fujitsuka:2013fga,Benini:2013yva}.

We focus here on a completely analogous story for the four-dimensional lens indices \cite{Yoshida:2014qwa,Peelaers:2014ima,Nieri:2015yia}.
In particular, either through evaluation of the contour integrals of the Coulomb branch formula \cite{Yoshida:2014qwa,Nieri:2015yia}, or through Higgs branch localization \cite{Peelaers:2014ima}, the lens index of a general $\mathcal{N}=1$ gauge theory can be shown to take the following form:\footnote{\label{fn:higgs-branch}This expression assumes the existence of a $\mathcal{Q}$-exact deformation of the theory such that the theory localizes onto a finite number of Higgs branch vacua. A sufficient condition for the existence of the Higgs branch expression in terms of the contour integral is the vanishing of the residue at the origin \cite{Peelaers:2014ima,Chen:2014rca}. 
For theories with a $U(1)$ factor
in the gauge group, one can ensure this by adding a non-zero Fayet--Iliopoulos term \cite{Benini:2012ui,Doroud:2012xw,Benini:2013yva,Peelaers:2014ima}.}
\begin{equation}\label{eq:lens-higgs-form}
    \mathcal{I}_{(p,q)}(\hat{\boldsymbol{\rho}})=\mathcal{Z}_f(\boldsymbol{\rho})=\sum_\alpha \mathcal{Z}^{\alpha}_{f}(\boldsymbol{\rho})=\sum_\alpha \mathcal{Z}^{\alpha}_{f,\text{cl}}(\boldsymbol{\rho})\mathcal{Z}^{\alpha}_{f,\text{1-loop}}(\boldsymbol{\rho})\mathcal{Z}^{\alpha}_{\text{v}}(\boldsymbol{\rho})\mathcal{Z}^{\alpha}_{\text{v}}(f^{-1} \boldsymbol{\rho})\,,
\end{equation}
where we have used the notation in \eqref{eq:defn-Zf-notation}.
In the language of the localization computation \cite{Benini:2013yva,Peelaers:2014ima}, the summation runs over a finite number of Higgs branch vacua of the (mass deformed) theory, and $\mathcal{Z}^{\alpha}_{f}(\boldsymbol{\rho})$ represents the contribution to the lens index from a given vacuum $\alpha$.
We stress that the summation domain is independent of $f$ \cite{Nieri:2015yia}.
The last equality shows how the contributions $\mathcal{Z}^{\alpha}_{f}(\boldsymbol{\rho})$ split into a classical $\mathcal{Z}^{\alpha}_{f,\text{cl}}(\boldsymbol{\rho})$, perturbative $\mathcal{Z}^{\alpha}_{f,\text{1-loop}}(\boldsymbol{\rho})$ and non-perturbative vortex $\mathcal{Z}^{\alpha}_{\text{v}}(\boldsymbol{\rho})$ and anti-vortex $\mathcal{Z}^{\alpha}_{\text{v}}(f^{-1}\boldsymbol{\rho})$ contributions.
The latter contributions capture codimension-$2$ multi-(anti-)vortex configurations which in terms of the Heegaard splitting $M_f(\boldsymbol{\rho},\tilde{\boldsymbol{\rho}})$ of the Hopf surface wrap the $T^2$ and are localized at the centers of the disks \cite{Peelaers:2014ima}.
``Anti'' refers to the vortices on the orientation reversed $D_2\times T^2$.
Thus, the vortex partition functions are naturally functions of $\boldsymbol{\rho}$ and $\tilde{\boldsymbol{\rho}}=f^{-1}\boldsymbol{\rho}$.
The perturbative contribution combines the one-loop fluctuations around the vortex configurations, which are similarly localized at the centers of the disks.
Finally, $\mathcal{Z}^{\alpha}_{f,\text{cl}}(\boldsymbol{\rho})$ captures the overall classical action, which consists solely of an Fayet--Iliopoulos (FI) term and therefore is only present when the gauge group contains a $U(1)$ factor.

Based on \cite{Spiridonov:2012ww}, it was shown in \cite{Nieri:2015yia} that the classical and perturbative contribution can also be written in a manifestly factorized form:
\begin{equation}
    \mathcal{Z}^{\alpha}_{f,\text{cl}}(\boldsymbol{\rho})\mathcal{Z}^{\alpha}_{f,\text{1-loop}}(\boldsymbol{\rho})=e^{-i\pi \mathcal{P}_f(\boldsymbol{\rho})}b_S^{\alpha}(\boldsymbol{\rho})b_S^{\alpha}(f^{-1} \boldsymbol{\rho})\,,
\end{equation}
where $b_S^{\alpha}(\boldsymbol{\rho})$ ($b_S^{\alpha}(f^{-1}\boldsymbol{\rho})$) captures the overall classical and one-loop fluctuations around the (anti-)vortex configurations.
The subscript $S$ refers to the element $S_{13}\in SL(2,\mathbb{Z})_{13}$, as we will return to momentarily.
Finally, the phase $\mathcal{P}_f(\boldsymbol{\rho})$ is given by a cubic polynomial in $\vec{z}$ and encodes the 't Hooft anomalies of the theory.
It does not depend on $\alpha$, which reflects the fact that the anomalies do not depend on the vacuum in which the theory resides \cite{Nieri:2015yia,Gadde:2020bov}.\footnote{\label{fn:gauge-anom-canc}At a more technical level, this can be derived from the contour integral expression of the index by making use of the gauge anomaly cancellation \cite{Nieri:2015yia}.}
Combining all of the above, one is led to the holomorphic block factorization of lens indices:
\begin{equation}\label{eq:lens-hol-blocks}
    \mathcal{I}_{(p,q)}(\hat{\boldsymbol{\rho}})=e^{-i\pi \mathcal{P}_f(\boldsymbol{\rho})}\sum_\alpha \norm{\mathcal{B}_S^{\alpha}(\boldsymbol{\rho})}^2_f\,,\quad \norm{\mathcal{B}_S^{\alpha}(\boldsymbol{\rho})}^2_f\equiv  \mathcal{B}_S^{\alpha}(\boldsymbol{\rho}) \mathcal{B}_S^{\alpha}(f^{-1}\boldsymbol{\rho})\,,
\end{equation}
where again $f=g_{(p,q)}\,\mathcal{O}$ with $g_{(p,q)}\in SL(2,\mathbb{Z})_{23}$ as in \eqref{eq:gSL2-from-gSL3}.
Note that the presence of the phase prevents full factorization of the summand. 
The holomorphic blocks $\mathcal{B}_S^{\alpha}(\boldsymbol{\rho})$ combine a factor of the perturbative part with a vortex partition function so that:\footnote{Here we have used invariance $\mathcal{Z}^{\alpha}_{\text{v}}(\boldsymbol{\rho})$ under $S_{13}$, a point we will discuss in more detail in Sections \ref{sssec:sqed} and \ref{sssec:gen-gauge-th}.}
\begin{equation}
    \mathcal{B}_S^\alpha(\boldsymbol{\rho})=b_S^{\alpha}(\boldsymbol{\rho})\,\mathcal{Z}^{\alpha}_{\text{v}}(S_{13}\boldsymbol{\rho})\,.
\end{equation}
If we write $\mathcal{B}^{\alpha}(\boldsymbol{\rho})$ for the supersymmetric partition function on a $D_2\times T^2$ geometry with moduli $\boldsymbol{\rho}=(\vec{z};\tau,\sigma)$ \cite{Longhi:2019hdh}, the holomorphic block $\mathcal{B}^{\alpha}_S$ used in \cite{Nieri:2015yia} is given by:
\begin{equation}\label{eq:defn-Bs}
    \mathcal{B}^{\alpha}_S(\boldsymbol{\rho})=\mathcal{B}^{\alpha}(S_{13}\boldsymbol{\rho})=\mathcal{B}^{\alpha}\left(\tfrac{\vec{z}}{\sigma};\tfrac{\tau}{\sigma},-\tfrac{1}{\sigma}\right)\,,\qquad S_{13}\in SL(2,\mathbb{Z})_{13}\subset \mathcal{H} \,,
\end{equation}
as we will see explicitly in the examples below.
It follows that the factorization expressed in \eqref{eq:lens-hol-blocks} reflects a particular Heegaard splitting $M_{f'}(\boldsymbol{\rho}',\tilde{\boldsymbol{\rho}}')$ of the Hopf surface $\mathcal{M}_{(p,q)}(\hat{\boldsymbol{\rho}})$ with $f'$ and $\boldsymbol{\rho}'$ given by:
\begin{equation}\label{eq:lens-index-equiv-S}
    f'=S_{13}f S_{13}^{-1}\,,\qquad\boldsymbol{\rho}'=S_{13}\boldsymbol{\rho}\,,\quad \tilde{\boldsymbol{\rho}}'=(f')^{-1}\boldsymbol{\rho}'=S_{13}f^{-1}\boldsymbol{\rho}\,.
\end{equation}
From our geometric discussion in Section \ref{ssec:ambig-heegaard}, it is natural to wonder why this specific Heegaard splitting is singled out.
This question motivated in part the present work, and we will indeed show that there exist more factorized expressions for the index where $S_{13}$ is replaced by general elements $h\in \mathcal{H}$.

\paragraph{Boundary conditions:}

It was suggested in \cite{Nieri:2015yia}, and later confirmed in \cite{Longhi:2019hdh}, that apart from a choice of Higgs vacuum $\alpha$ for the entire theory, there exist (at least) two $\frac{1}{2}$-BPS boundary conditions on $D_2\times T^2$ for a given $\mathcal{N}=1$ multiplet of the theory.
In particular, a chiral multiplet can either have Dirichlet (D) or Robin-like (R) boundary conditions, while a vector multiplet admits Neumann (N) or Dirichlet boundary conditions.

In the following, we will assume Neumann boundary conditions for the vector multiplet.
Let us assume that all chiral multiplets in the gauge theory obey the same boundary condition, Dirichlet or Robin-like.
For gauge anomaly cancellation on the boundary,\footnote{As observed in \cite[Section 6]{Longhi:2019hdh}, the relevant supersymmetric $T^3$ partition function of the 3d boundary theory is formally equivalent to the superconformal index of a 2d $(0,2)$ theory.
The anomaly cancellation, then, refers to the cancellation of the gauge anomaly in the relevant 2d theory.} the anti-chiral multiplets have to satisfy the opposite boundary condition, i.e., Robin-like or Dirichlet, respectively \cite{Nieri:2015yia,Longhi:2019hdh}.
We denote the holomorphic block of the full theory, including the vector multiplets, by $\mathcal{B}^{\alpha}(\boldsymbol{\rho})$ and $\mathcal{C}^{\alpha}(\boldsymbol{\rho})$, respectively.
One naturally expects that the compact space partition function $\mathcal{Z}_f(\boldsymbol{\rho})$ is independent of the boundary condition.
This independence was indeed observed in \cite{Nieri:2015yia}, where it was shown that the respective products of holomorphic blocks are equal, up to a phase:
\begin{equation}\label{eq:equiv-hol-block-facts}
    \mathcal{I}_{(p,q)}(\hat{\boldsymbol{\rho}})=e^{-i\pi \mathcal{P}_f(\boldsymbol{\rho})}\sum_\alpha \norm{\mathcal{B}_S^{\alpha}(\boldsymbol{\rho})}^2_f=e^{-i\pi( \mathcal{P}_f(\boldsymbol{\rho})+\mathcal{P}_f^{3d}(\boldsymbol{\rho}))}\sum_\alpha \norm{\mathcal{C}_S^{\alpha}(\boldsymbol{\rho})}^2_f\,,
\end{equation}
where $\mathcal{C}_S^{\alpha}(\boldsymbol{\rho})$ is defined in terms of $\mathcal{C}^\alpha(\boldsymbol{\rho})$ as in \eqref{eq:defn-Bs}.

We now summarize the physical arguments that lead to the equality, following \cite{Longhi:2019hdh}, which will be confirmed mathematically in the examples below.
First of all, the boundary conditions on the (anti-)chiral multiplets can be changed from D to R (or vice versa) through a coupling of theory to degrees of freedom living on the boundary $T^3=\partial D_{2}\times T^2$.
At the level of the partition functions, it can be shown that:
\begin{equation}\label{eq:dir-block-rob-block-reln}
    \mathcal{B}^{\alpha}(\boldsymbol{\rho})=Z^{\alpha}_\partial(\vec{z};\tau) \mathcal{C}^{\alpha}(\boldsymbol{\rho})\,,
\end{equation}
where we recall that $\boldsymbol{\rho}=(\vec{z};\tau,\sigma)$ and $Z^{\alpha}_\partial(\vec{z};\tau)$ captures the contribution of the boundary degrees of freedom.
Note that it does not depend on $\sigma$, the modulus of the non-contractible $T^2$.
In addition, it turns out that $Z^{\alpha}_\partial(\vec{z};\tau)$ is invariant, up to a phase polynomial quadratic in $\vec{z}$, under the usual action of $SL(2,\mathbb{Z})\ltimes \mathbb{Z}^{2r}$:
\begin{equation}
    (z_a;\tau)\to \left(\tfrac{z_a}{m\tau+n};\tfrac{k\tau+l}{m\tau+n}\right)\,,\quad  (z_a;\tau)\to \left(z_a+i\tau+j;\tau\right)\,.
\end{equation}
This action is generated by $\lbrace S_{12},T_{21}, t^{(a)}_{1},t^{(a)}_2\rbrace$ when viewed as a subgroup of $\mathcal{G}$.
As mentioned in the footnote above, this shows that the contribution of the boundary degrees of freedom behaves essentially as the elliptic genus of a 2d $(0,2)$ theory.
We will assume that this is a general feature of the relation between the blocks $\mathcal{B}^{\alpha}$ and $\mathcal{C}^{\alpha}$, and leave a more detailed investigation to future work.

Given the equation \eqref{eq:dir-block-rob-block-reln} and the assumed modular property of $Z^{\alpha}_\partial(\vec{z};\tau)$, we can now derive \eqref{eq:equiv-hol-block-facts}.
First, we write the product of the blocks $\mathcal{B}^{\alpha}_S(\boldsymbol{\rho})$ as:
\begin{equation}
   \norm{\mathcal{B}_S^{\alpha}(\boldsymbol{\rho})}^2_f=Z^{\alpha}_\partial(\vec{z}\,';\tau')Z^{\alpha}_\partial\left(\mathcal{O}_2g_{2}^{-1}(\vec{z}\,';\tau')\right) \norm{\mathcal{C}_S^{\alpha}(\boldsymbol{\rho})}^2_f\,.
\end{equation}
Here, we have defined $(\vec{z}\,';\tau')$ through (cf.\ \eqref{eq:defn-Bs}):
\begin{equation}
    \boldsymbol{\rho}'\equiv (\vec{z}\,';\tau',\sigma')=S_{13}\boldsymbol{\rho}\,.
\end{equation}
Furthermore, we used that $f=g_{(p,q)}\,\mathcal{O}$ with $g_{(p,q)}$ given as in \eqref{eq:gSL2-from-gSL3}:
\begin{equation}\label{eq:g-sl2-in-sl3}
    g_{(p,q)}=\begin{pmatrix}
   1&0&0\\
   0& -s & -r \\
    0&    -p & -q
    \end{pmatrix}\,.
\end{equation}
Since $SL(2,\mathbb{Z})_{12}=S_{13} \,SL(2,\mathbb{Z})_{23} \,S_{13}^{-1}$, the action of $f$ on $(\vec{z}\,';\tau')$ can be written in terms of the standard $g_2\in SL(2,\mathbb{Z})$ action.
We thus arrive at the definition of $g_2$:
\begin{equation}
    g_{2}=\begin{pmatrix}
    -q & -p \\
    -r & -s
    \end{pmatrix}\,,\qquad \mathcal{O}_{2}=\begin{pmatrix}
    1 & 0 \\
    0 & -1
    \end{pmatrix}\,,
\end{equation}
where we have included $\mathcal{O}_2$ as the restriction of $\mathcal{O}$ to $(\vec{z}\,';\tau')$ as well.
Finally, we require one additional property of $Z^{\alpha}_{\partial}$:\footnote{\label{footnote:Zpartial}In examples, we will see that $Z^{\alpha}_{\partial}(z;\tau)$ consists of a product of $q$-$\theta$ functions. This property then follows from the extension and elliptic properties of the $q$-$\theta$ function (see \eqref{eq:ext-theta} and \eqref{eq:elliptic-theta}).}
\begin{equation}
    Z^{\alpha}_\partial\left(\mathcal{O}_2g_{2}^{-1}(\vec{z}\,';\tau')\right)\cong \frac{1}{Z^{\alpha}_\partial\left(g_{2}^{-1}(\vec{z}\,';\tau')\right)}\,,
\end{equation}
where the equality holds up to a phase.
The modular properties of $Z^{\alpha}_\partial(\vec{z};\tau)$ mentioned above now imply:
\begin{equation}\label{eq:consistency-cond}
    \norm{\mathcal{B}_S^{\alpha}(\boldsymbol{\rho})}^2_f\cong  \norm{\mathcal{C}_S^{\alpha}(\boldsymbol{\rho})}^2_f\,,
\end{equation}
where the equality again holds up to a phase.
We arrive at the final claim \eqref{eq:equiv-hol-block-facts} as long as the relative phase, $\mathcal{P}^{3d}(\boldsymbol{\rho})$, is independent of $\alpha$.
This phase can be interpreted as the anomaly polynomial of the effectively 2d $(0,2)$ boundary theory and is independent of $\alpha$ for the same reasons as $\mathcal{P}_f(\boldsymbol{\rho})$ \cite{Longhi:2019hdh}.

To summarize, the holomorphic blocks $\mathcal{B}^{\alpha}_S(\boldsymbol{\rho})$ or $\mathcal{C}^{\alpha}_S(\boldsymbol{\rho})$ of \cite{Nieri:2015yia} correspond to a non-standard Heegaard splitting of the (secondary) Hopf surface, as expressed in \eqref{eq:lens-index-equiv-S}.
Up to a phase, both holomorphic blocks produce upon gluing the same compact space partition function.

\subsubsection{Example: free chiral multiplet}\label{sssec:example-free-chiral}
	
In this section, we illustrate the above using the free chiral multiplet.
The extension to SQED and general gauge theories is discussed in Sections \ref{sssec:sqed} and \ref{sssec:gen-gauge-th}, respectively.
	
\paragraph{Lens indices:}
	
Let us collect from Appendix \ref{app:lens-indices} the free chiral multiplet indices on $S^2\times T^2$, $S^3\times S^1$, and $L(p,1)\times S^1$.
Since the chiral multiplet has a global $U(1)$ flavor symmetry, we may define the indices with respect to an R-symmetry such that fields have R-charges quantized as even integers.
In addition, for the $S^2\times T^2$ index, we allow $\mathbf{n}\in \mathbb{Z}$ units of flavor symmetry flux through the $S^2$.
Finally, for simplicity of notation we do not include holonomies for the flavor symmetry along the non-contractible cycle in $L(p,1)$.\footnote{The more general expressions are recorded in Appendix \ref{app:lens-indices}.}
We then have:
\begin{align}\label{eq:lens-indices-chiral}
	\begin{split}
		I^{R}_{(0,-1),g}(\hat{\boldsymbol{\rho}})&=\begin{cases}
			\hat{p}^{\frac{\mathbf{R}}{12}}\hat{x}^{-\frac{\mathbf{R}}{2}}\prod^{\frac{|\mathbf{R}|-1}{2}}_{m=-\frac{|\mathbf{R}|-1}{2}}\theta(\hat{z}+m\hat{\tau};\hat{\sigma})^{\text{sgn}(\mathbf{R})}\,,\quad &\text{for}\quad \mathbf{R}\neq 0\,,\\
			1\,,\quad &\text{for}\quad \mathbf{R}=0\,,
		\end{cases}\\
		I^{R}_{(1,0)}(\hat{\boldsymbol{\rho}})&=\Gamma(\hat{z}+\tfrac{R}{2}(\hat{\tau}+\hat{\sigma});\hat{\tau},\hat{\sigma})\,,\\
		I^{R}_{(p,1)}(\hat{\boldsymbol{\rho}})&=\Gamma\left(\hat{z}+\tfrac{R}{2}(\hat{\tau}+\hat{\sigma})+p\hat{\sigma};\hat{\tau}+\hat{\sigma},p\hat{\sigma}\right)\Gamma\left(\hat{z}+\tfrac{R}{2}(\hat{\tau}+\hat{\sigma});\hat{\tau}+\hat{\sigma},p\hat{\tau}\right) \,,
	\end{split}
\end{align}
where $\mathbf{R}\equiv R+\mathbf{n}-1$ and $R\in 2\mathbb{Z}$.
The functions $\theta(z;\sigma)$ and $\Gamma(z;\tau,\sigma)$ represent the $q$-$\theta$ and elliptic $\Gamma$ function, respectively, defined in Appendix \ref{app:defs}.
Using properties of the functions recorded there, one easily verifies that for both $R$ and $\mathbf{R}$ even integers these functions are invariant (up to a phase) under symmetries of the respective Hopf surfaces recorded in Section \ref{ssec:hopf-surfaces}.
	
Recall that the moduli of the Hopf surface are related to the $D_2\times T^2$ moduli as:
\begin{equation}\label{eq:p-moduli-2}
		(z;\tau,\sigma)=\begin{cases}
		(\hat{z}+\tfrac{\mathbf{n}}{2}\hat{\tau};\hat{\tau},\hat{\sigma})\,, \; &\text{for}\;p=r= 0\,,\,q=s= -1\,,\\
		(\hat{z};\hat{\tau}+s\hat{\sigma},p\hat{\sigma})\,, \; &\text{for}\;p\neq 0                    \,,    
	\end{cases}
\end{equation}
where for later convenience we have included a shift of $\hat{z}$ in the first line to reflect a non-trivial flavor flux through the $S^2$.
We then use the notation \eqref{eq:defn-Zf-notation} to write:
\begin{equation}
	Z_{t_2^{\mathbf{n}}\mathcal{O}}(z;\tau,\sigma)=I^{R}_{(0,-1),\mathbf{n}}(z+\tfrac{\mathbf{n}}{2}\tau;\tau,\sigma)\,,\qquad Z_{S_{23}\mathcal{O}}(z;\tau,\sigma)=I^{R}_{(1,0)}(z;\tau,\sigma)\,,
\end{equation}
where we suppress the label $R$ and use that the gluing transformation $t_2^{\mathbf{n}}\mathcal{O}$, which acts by large gauge transformations on $z$ (see \eqref{eq:calG-action}), produces an $S^2\times T^2$ geometry with $\mathbf{n}$ units of flavor symmetry flux through the $S^2$ \cite{Gadde:2020bov}.
Similarly, for the index on $L(p,1)\times S^1$ we write:
\begin{equation}\label{eq:Zchiralmultiplet}
	Z_{g_{(p,1)}\mathcal{O}}(z;\tau,\sigma)=I^{R}_{(p,1)}(z;\tau-\tfrac{1}{p}\sigma,\tfrac{1}{p}\sigma)\,.
\end{equation}
To see how these indices factorize into holomorphic blocks, let us also record the partition function on $D_2\times T^2$ with moduli $\boldsymbol{\rho}=(z;\tau,\sigma)$.
This was computed in~\cite{Longhi:2019hdh} through localization for two types of boundary conditions: Dirichlet and Robin-like boundary conditions.
The associated partition functions are given by:\footnote{\label{fn:sim-hol-block-index-chiral}Note the similarity between the $D_2\times T^2$ partition functions and the index on  $S^3\times S^1$ in \eqref{eq:lens-indices-chiral}.}%
\begin{align}\label{eq:hol-blocks-bc-free-chiral}
	\begin{split}
		B(z;\tau,\sigma)&\equiv  \Gamma(z+\tfrac{R}{2}\tau+\sigma;\tau,\sigma)\,,\\
		C(z;\tau,\sigma)&\equiv\Gamma(z+\tfrac{R}{2}\tau;\tau,\sigma)\,.
	\end{split}
\end{align}
Note that we do not write the label $\alpha$, since for the free chiral multiplet it assumes only a single value.
Furthermore, we define the blocks without a phase prefactor, as opposed to \cite{Longhi:2019hdh}.
This is justified because holomorphic block factorization of the lens indices uniquely fixes the overall phase due to properties of the elliptic $\Gamma$ function.
We also note that:
\begin{equation}\label{eq:BC-reln-free-chiral}
	B(\boldsymbol{\rho})=\theta(z+\tfrac{R}{2}\tau;\tau)C(\boldsymbol{\rho})\,,
\end{equation}
where we have used the shift property \eqref{eq:basic-shift-gamma-app} of the elliptic $\Gamma$ function.
As anticipated in general above, we now see explicitly that $B(\boldsymbol{\rho})$ and $C(\boldsymbol{\rho})$ are related through multiplication by a function that is invariant, up to a phase, under $SL(2,\mathbb{Z})_{12}\ltimes \mathbb{Z}^{2+2}\subset \mathcal{G}$.
Finally, unlike the indices on closed manifolds, the functions $B(\boldsymbol{\rho})$ and $C(\boldsymbol{\rho})$ do not have automorphic properties under the full group of large diffeomorphisms and gauge transformations $\mathcal{H}$ of $D_2\times T^2$.
However, we note that they are periodic under $z\to z+1$, $\tau\to\tau+1$ and $\sigma\to \sigma+1$, which again relies on the fact that the R-charges are quantized as even integers.

\paragraph{Holomorphic block factorization:}

We are now ready to describe the explicit factorization of the indices \eqref{eq:lens-indices-chiral} into holomorphic blocks, following \cite{Nieri:2015yia}.
We start with the superconformal index $Z_{S_{23}\mathcal{O}}(\boldsymbol{\rho})$.
The crucial property of the elliptic $\Gamma$ function underlying the factorization of this index is \cite{Felder_2000} (see also Appendix \ref{app:defs}):
\begin{align}\label{eq:gamma-ids-for-hol-blocks}
    \begin{split}
    \Gamma(z;\tau,\sigma)&=e^{-i\pi Q(z;\tau,\sigma)}\Gamma\left(\tfrac{z}{\sigma};\tfrac{\tau}{\sigma},-\tfrac{1}{\sigma}\right)\Gamma\left(\tfrac{z}{\tau};\tfrac{\sigma}{\tau},-\tfrac{1}{\tau}\right)\,,
    \end{split}
\end{align}
where $Q(z;\tau,\sigma)$ is a cubic polynomial in $z$.
Using this property, we may write:
\begin{align}
\begin{split}
    \Gamma(z+\tfrac{R}{2}(\tau+\sigma);\tau,\sigma)&=e^{-i\pi Q(z+\frac{R}{2}(\tau+\sigma)-1;\tau,\sigma)}\\
    & \times\Gamma\left(\tfrac{z+\frac{R}{2}\tau-1}{\sigma};\tfrac{\tau}{\sigma},-\tfrac{1}{\sigma}\right)\Gamma\left(\tfrac{z+\frac{R}{2}\sigma-1}{\tau};\tfrac{\sigma}{\tau},-\tfrac{1}{\tau}\right)\,,
    \end{split}
\end{align}
where we have made use of the periodicity of the elliptic $\Gamma$ function under $z\to z+1$ and the fact that $R$ is quantized as an even integer.
One now easily checks that the index can be written as anticipated in \eqref{eq:lens-hol-blocks}:
\begin{equation}\label{eq:hol-block-S-fact-sci-chiral}
    Z_{S_{23}\mathcal{O}}(\boldsymbol{\rho})=e^{-i \pi P_{S_{23}}(\boldsymbol{\rho};R)}\norm{B_S(\boldsymbol{\rho})}^2_{S_{23}\mathcal{O}}  \,,
\end{equation}
where the holomorphic block $B_S(\boldsymbol{\rho})$ is given in terms of $B(\boldsymbol{\rho})$ in \eqref{eq:hol-blocks-bc-free-chiral} as follows:
\begin{equation}
    B_S(\boldsymbol{\rho})=B(S_{13}\boldsymbol{\rho})=\Gamma\left(\tfrac{z+\frac{R}{2}\tau-1}{\sigma};\tfrac{\tau}{\sigma},-\tfrac{1}{\sigma}\right)\,,
\end{equation}
and we define the corresponding phase by:
\begin{equation}\label{eq:PS23-chiral}
    P_{S_{23}}(\boldsymbol{\rho};R)=Q(z+\tfrac{R}{2}(\tau+\sigma)-1;\tau,\sigma)\,.
\end{equation}

To factorize $Z_{g_{(p,1)}\mathcal{O}}(\boldsymbol{\rho})$, we use a closely related property for the first $\Gamma$ function in the expression \eqref{eq:lens-indices-chiral}:
\begin{align}\label{eq:gamma-ids-for-hol-blocks-2}
    \begin{split}
    \Gamma(z+\sigma;\tau,\sigma)&=e^{-i\pi Q(z+\sigma;\tau,\sigma)}\frac{\Gamma\left(\frac{z}{\sigma};\frac{\tau}{\sigma},-\frac{1}{\sigma}\right)}{\Gamma\left(\frac{z}{\tau};-\frac{\sigma}{\tau},-\frac{1}{\tau}\right)}\,,
    \end{split}
\end{align}
while we use \eqref{eq:gamma-ids-for-hol-blocks} on the second $\Gamma$ function.
Two of the four resulting $\Gamma$ functions cancel, and one easily verifies that in this case again:
\begin{equation}\label{eq:hol-block-gp-fact-sci-chiral}
    Z_{g_{(p,1)}\mathcal{O}}(\boldsymbol{\rho})=e^{-i \pi P_{g_{(p,1)}}(\boldsymbol{\rho};R)}\norm{B_S(\boldsymbol{\rho})}^2_{g_{(p,1)}\mathcal{O}}  \,,
\end{equation}
where now the phase is given by:
\begin{equation}\label{eq:Pgp-chiral}
    P_{g_{(p,1)}}(\boldsymbol{\rho};R)=Q(z+\tfrac{R}{2}\tau+\sigma-1;\tau,\sigma)+Q(z+\tfrac{R}{2}\tau-1;p\tau-\sigma,\tau)\,.
\end{equation}

Finally, for factorization of the $S^2\times T^2$ index, we first use the shift property of the elliptic $\Gamma$ function \eqref{eq:basic-shift-gamma-app} to rewrite the index as follows:
\begin{align}
    \begin{split}
        Z_{t_2^{\mathbf{n}}\mathcal{O}}(\boldsymbol{\rho})&=p^{\frac{\mathbf{R}}{12}}q^{\frac{\mathbf{n}\mathbf{R}}{2}}x^{-\frac{\mathbf{R}}{2}}\frac{\Gamma(z+\frac{R}{2}\tau;\tau,\sigma)}{\Gamma(z-(\frac{R}{2}+\mathbf{n}-1)\tau;\tau,\sigma)}\\
        &=p^{\frac{\mathbf{R}}{12}}q^{\frac{\mathbf{n}\mathbf{R}}{2}}x^{-\frac{\mathbf{R}}{2}}\Gamma(z+\tfrac{R}{2}\tau;\tau,\sigma)\Gamma(z-(\tfrac{R}{2}+\mathbf{n})\tau;-\tau,\sigma)\,,
    \end{split}
\end{align}
where in the second line we have made use of the extension property \eqref{eq:extend} of the elliptic $\Gamma$ function.
The latter equality already takes the form of a factorization:
\begin{equation}
    Z_{t_2^{\mathbf{n}}\mathcal{O}}(\boldsymbol{\rho})=p^{\frac{\mathbf{R}}{12}}q^{\frac{\mathbf{n}\mathbf{R}}{2}}x^{-\frac{\mathbf{R}}{2}}\norm{C(\boldsymbol{\rho})}^2_{t_2^{\mathbf{n}}\mathcal{O}}\,.
\end{equation}
We may also factorize $Z_{t_2^{\mathbf{n}}\mathcal{O}}(\boldsymbol{\rho})$ in terms of $C_S(\boldsymbol{\rho})\equiv C(S_{13}\boldsymbol{\rho})$ by making use of:
\begin{align}\label{eq:gamma-ids-for-hol-blocks-3}
    \begin{split}
    \theta\left(\tfrac{z}{\sigma};-\tfrac{1}{\sigma} \right) &= e^{i\pi B_2(z,\sigma)} \theta(z;\sigma)\,,
    \end{split}
\end{align}
where $B_2(z;\sigma)$ is a quadratic polynomial in $z$ (see Appendix \ref{app:defs}).
Using this transformation, we can write:
\begin{equation}
    Z_{t_2^{\mathbf{n}}\mathcal{O}}(\boldsymbol{\rho})=p^{\frac{\mathbf{R}}{12}}q^{\frac{\mathbf{n}\mathbf{R}}{2}}x^{-\frac{\mathbf{R}}{2}}e^{i\pi \tilde{P}_{t_2^{\mathbf{n}}}(\boldsymbol{\rho};R)}\norm{C_S(\boldsymbol{\rho})}^2_{t_2^{\mathbf{n}}\mathcal{O}}\,,
\end{equation}
where we indicate with the tilde that the phase is associated to $C_S(\boldsymbol{\rho})$, as opposed to $B_S(\boldsymbol{\rho})$, and is given by:
\begin{equation}\label{eq:P1-chiral}
    \tilde{P}_{t_2^{\mathbf{n}}}(\boldsymbol{\rho};R)=
        \text{sgn}(\mathbf{R})\sum^{\frac{|\mathbf{R}|-1}{2}}_{m=-\frac{|\mathbf{R}|-1}{2}}B_2(z+(m-\tfrac{\mathbf{n}}{2})\tau;\sigma)\,.
\end{equation}
All in all, we have seen that the chiral multiplet indices can be factorized, up to a phase, in terms of the holomorphic blocks $B_S(\boldsymbol{\rho})$ or $C_S(\boldsymbol{\rho})$.
In fact, given the relation between the blocks \eqref{eq:BC-reln-free-chiral} and our general arguments above, it follows that the indices can be factorized in terms of both blocks, as we show explicitly below.
Before getting there, let us first show that the relative phase captures the 't Hooft anomalies of the theory \cite{Spiridonov:2012ww,Nieri:2015yia,Gadde:2020bov}.

\paragraph{Anomaly polynomials:}

A convenient parametrization of the 't Hooft anomalies of a general gauge theory is as follows \cite{Gadde:2020bov}:
\begin{align}\label{eq:anomaly-pol-gen-th}
    \begin{split}
        \mathcal{P}(\vec{Z};x_i)\equiv& \frac{1}{3x_1x_2x_3}\left(k_{abc}Z_aZ_cZ_c+3k_{abR}Z_aZ_bX+3k_{aRR}Z_aX^2-k_aZ_a\tilde{X}\right.\\
        &\left.+k_{RRR}X^3-k_RX\tilde{X}\right)\,,
    \end{split}
\end{align}
where $\vec{Z}$ and $x_i$ represent the homogeneous moduli defined in \eqref{eq:rho-homog}, and
\begin{equation}\label{eq:X-and-Xt}
X\equiv \frac{1}{2}\sum^3_{i=1}x_i \,,\qquad \tilde{X}\equiv \frac{1}{4}\sum^3_{i=1}x^2_i \,.
\end{equation}
The coefficients encode anomalies.
For example, $k_{abc}=\mathrm{Tr}\, F_aF_bF_c$ represents a cubic anomaly for the flavor symmetry generators $F_a$, which would be diagonal for fundamental representations.
The label $R$ refers to the R-symmetry generator instead, and $k_a=\mathrm{Tr}\, F_a$ and $k_R=\mathrm{Tr}\, R$ capture the mixed-gravitational anomalies.
In our conventions, the relation between the anomaly polynomial as parametrized by \eqref{eq:anomaly-pol-gen-th} and the phase polynomial $P_{S_{23}}(\boldsymbol{\rho};R)$, is:
\begin{equation}\label{eq:phase-anomaly-pol-reln-chiral}
    P_{S_{23}}(\tfrac{Z+\frac{R}{2}x_1}{x_1},\tfrac{x_2}{x_1},\tfrac{x_3}{x_1};R)=\mathcal{P}_{\chi_R}(\vec{Z};x_i)\,,
\end{equation}
where $\mathcal{P}_{\chi_R}(\vec{Z};x_i)$ captures the anomalies of a chiral multiplet with R-charge $R$.
Expanding $P^R_{S_{23}}(\frac{Z+\frac{R}{2}x_1}{x_1},\frac{x_2}{x_1},\frac{x_3}{x_1};R)$, one easily reads off the anomalies:
\begin{equation}\label{eq:anomaly-coefficient-R}
    k_{F^3}=1\,,\quad k_{F^2R}=R-1\,,\quad k_{FR^2}=(R-1)^2\,,\quad k_F=1\,,\quad k_{R^3}= (R-1)^3\,,\quad k_R=R-1\,,
\end{equation}
as indeed appropriate for a free chiral multiplet.

To understand how the phases on the other backgrounds connect to the anomaly polynomial, we note that $P_{g_{(p,1)}}(\boldsymbol{\rho};R)$ can be written as follows in terms of $\hat{\boldsymbol{\rho}}$:
\begin{equation}\label{eq:Lenspolynomial-Q+dQ}
   P_{g_{(p,1)}}(\boldsymbol{\rho};R)= \frac{1}{p} Q\left(\hat{z} + \frac{R}{2}(\hat{\tau}+\hat{\sigma})-1, \hat{\tau},\hat{\sigma} \right)  +\frac{p^2-1}{12p} (2\hat{z} + (R-1)(\hat{\tau}+\hat{\sigma})-1)\,,
\end{equation}
where we have made use of \eqref{eq:p-moduli-2}.
This phase relates to a more general parametrization of the anomalies.
In particular, let us write an analogue of \eqref{eq:anomaly-pol-gen-th} with the same anomaly coefficients \eqref{eq:anomaly-coefficient-R} as follows:
\begin{eqnarray}\nonumber
        \mathcal{P}^{(p)}(\vec{Z};\hat{x}_i)&\equiv \frac{1}{3p\hat{x}_1\hat{x}_2\hat{x}_3}\left(k_{abc}Z_aZ_bZ_c+3k_{abR}Z_aZ_bX +3k_{aRR}Z_a X^2-k_aZ_a\tilde{X}^{(p,1)}\right.\\ \label{eq:anomaly-pol-gen-th-lens}
        &\left.+k_{RRR} X^3-k_R X\tilde{X}^{(p,1)}\right)\,,
\end{eqnarray}
where $X$ is as before in terms of $\hat{x}_i$ and we have defined $\tilde{X}^{(p,1)}$ as follows:
\begin{equation}
    \tilde{X}^{(p,1)}=\frac{1}{4}\left(\hat{x}_1^2+\hat{x}_2^2+\hat{x}_3^2-2(p^2-1)\hat{x}_2\hat{x}_3\right)\,.
\end{equation}
In this parametrization, $P_{g_{(p,1)}}(\boldsymbol{\rho};R)$ is related to the anomaly polynomial as before:
\begin{equation}
    P_{g_{(p,1)}}(\tfrac{Z+\frac{R}{2}x_1}{x_1},\tfrac{x_2}{x_1},\tfrac{x_3}{x_1};R)=\mathcal{P}^{(p)}_{\chi_R}(\vec{Z};\hat{x}_i)\,,
\end{equation}
where we understand the $x_i$ on the left hand side as functions of $\hat{x}_i$ through \eqref{eq:p-moduli-2}.
Note that for $p=1$, this correctly reduces to the $S^3\times S^1$ case \eqref{eq:phase-anomaly-pol-reln-chiral} as a function of $\hat{x}_i$.
We will derive the phase polynomial for general $g_{(p,q)}$ in Section \ref{sssec:mod-fact-lens}.

Finally, let us expand $\tilde{P}_{t_2^{\mathbf{n}}}(\boldsymbol{\rho})$ in terms of $Z=\hat{Z}+\frac{\mathbf{n}}{2}x_2$ as follows:
\begin{eqnarray}\nonumber
    \tilde{P}_{t_2^{\mathbf{n}}}(\tfrac{Z-x_1+\frac{R}{2}(x_1+x_3)}{x_1},\tfrac{x_2}{x_1},\tfrac{x_3}{x_1};R)&=&\frac{\mathbf{R}}{x_1x_3}\Big(\hat{Z}^2+(R-1)(x_1+x_3)\hat{Z}+\tfrac{(R-1)^2}{4}(x_1+x_3)^2\\ \label{eq:Pt2}
    && -\tfrac{1}{12}(x_1^2+x_3^2-(\mathbf{R}^2-1)x^2_2)\Big)\,.
\end{eqnarray}
Note that there are no cubic terms in this case.
This is consistent with the fact that the twisted theory on $S^2\times T^2$ behaves effectively as a two-dimensional $(0,2)$ theory on $T^2$.
In particular, it consists of $\mathbf{R}$ Fermi multiplets for $\mathbf{R}>0$ and $|\mathbf{R}|$ chiral multiplets for $\mathbf{R}<0$ \cite{Closset:2013sxa}.
The phase polynomial is therefore again consistent with an interpretation in terms of the anomaly polynomial.

\paragraph{Independence on boundary condition:}

Due to the relation between $B(\boldsymbol{\rho})$ and $C(\boldsymbol{\rho})$ in \eqref{eq:BC-reln-free-chiral}, it follows from our general arguments that  up to a phase both holomorphic blocks lead to the same compact space partition function upon gluing.
Explicitly, one may verify that:
\begin{equation}\label{eq:hol-block-fact-gen-index-2}
    Z_{f}(\boldsymbol{\rho})=e^{-i \pi P_g(\boldsymbol{\rho})}\norm{B_S(\boldsymbol{\rho})}^2_f=e^{-i \pi \tilde{P}_g(\boldsymbol{\rho})}\norm{C_S(\boldsymbol{\rho})}^2_f \,,
\end{equation}
where the phase polynomial $\tilde{P}_g$ is related to $P_g$ through:
\begin{align}
    \begin{split}
        \tilde{P}_g(z;\tau,\sigma)&=P_{g}(z+1;\tau,\sigma)\,.
    \end{split}
\end{align}
The difference between these phases takes on a similar form as the phase polynomial associated to the $(0,2)$ multiplets in \eqref{eq:Pt2}.
For example, for $g=g_{(p,1)}$, we have:
\begin{align}
\begin{split}
\tilde{P}_g(z;\tau,\sigma)-P_{g}(z;\tau,\sigma)
= \frac{1}{p} B_2\left(\frac{\hat{z}}{\hat{\tau}} + \frac{R(\hat{\sigma}+\hat{\tau})}{2\hat{\tau}}-1, \frac{\hat{\sigma}}{\hat{\tau}} \right) +\frac{p^2-1}{6p} \,,
\end{split}
\end{align}
where $B_2(z;\tau)$ is a quadratic polynomial in $\hat{z}$ and is defined in Appendix \ref{app:defs}.
As mentioned in the general discussion, we may interpret this phase as capturing the anomalies of the effectively two-dimensional boundary theory, due to a coupling to the bulk, which changes the boundary condition on the bulk fields from Dirichlet to Robin-like \cite{Longhi:2019hdh}.

\subsubsection{Example: SQED}\label{sssec:sqed}

In this section, we consider holomorphic block factorization for SQED with $N_f$ flavors.
For clarity of exposition, we focus on the superconformal index and refer to \cite{Nieri:2015yia} for both the $L(p,1)\times S^1$ and $S^2\times T^2$ indices.

The contour integral expression for the SQED index can be obtained from the general gauge theory index collected in Appendix \ref{app:lens-indices}:
\begin{align}\label{eq:explicit-index-sqed}
    \begin{split}
        I^{\mathrm{SQED}}_{(1,0)}(\hat{\boldsymbol{\rho}})=(\hat{p};\hat{p})_\infty(\hat{q};\hat{q})_{\infty}\oint \frac{dv}{2\pi i v} v^{\xi}\prod^{N_f}_{\beta=1}&\Gamma(-u+z_\beta+\tfrac{R}{2}(\hat{\tau}+\hat{\sigma});\hat{\tau},\hat{\sigma})\\
        &\times\Gamma(u+z_\beta+\tfrac{R}{2}(\hat{\tau}+\hat{\sigma});\hat{\tau},\hat{\sigma}) \,,
    \end{split}
\end{align}
where we added a non-zero FI parameter $\xi$ to ensure that the residue at $v=0$ vanishes.
This is a necessary condition for a Higgs branch expression to exist (see footnote \ref{fn:higgs-branch}).
We note that $\xi$ is integer quantized, as argued in \cite{Imamura:2011uw}, and therefore the index can depend on it.
We also introduced the chemical potentials $z_\alpha$ for the diagonal subgroup of the $SU(N_f)\times SU(N_f)$ flavor symmetry, satisfying $\sum_\alpha z_\alpha=0$.
Furthermore, vanishing of the mixed-gauge anomalies requires $R=1$.

Performing the contour integral, one obtains an expression for the index of the form \eqref{eq:lens-higgs-form} \cite{Yoshida:2014qwa,Peelaers:2014ima}:
\begin{equation}\label{eq:lens-higgs-form-sqed}
    I^{\mathrm{SQED}}_{(1,0)}(\hat{\boldsymbol{\rho}})=Z^{\mathrm{SQED}}_{S_{23}\mathcal{O}}(\boldsymbol{\rho})=\sum^{N_f}_{\alpha=1} Z^{\alpha}_{S_{23}\mathcal{O},\text{cl}}(\boldsymbol{\rho})Z^{\alpha}_{S_{23}\mathcal{O},\text{1-loop}}(\boldsymbol{\rho})Z^{\alpha}_{\text{v}}(\boldsymbol{\rho})Z^{\alpha}_{\text{v}}(\mathcal{O}S_{23}^{-1} \boldsymbol{\rho})\,,
\end{equation}
where we employ the notation \eqref{eq:defn-Zf-notation} and recall that in this case $\hat{\boldsymbol{\rho}}=\boldsymbol{\rho}$.
Furthermore, the factors in the summand can be written as:\footnote{Note that had we not included the FI parameter, the vortex contribution $Z_{\mathrm{v}}^{\alpha}(\boldsymbol{\rho})$ would have been divergent.}
\begin{align}\label{eq:S3S1sqed}
    \begin{split}
        Z^{\alpha}_{S_{23}\mathcal{O},\text{cl}}(\boldsymbol{\rho})&= x^{\xi}_{\alpha}(pq)^{\xi/2}\,, \\
        Z^{\alpha}_{S_{23}\mathcal{O},\text{1-loop}}(\boldsymbol{\rho})&= \frac{1}{\Gamma(0;\tau,\sigma)} \prod_{\beta=1}^{N_f} \frac{\Gamma(z_\beta -z_\alpha;\tau,\sigma)}{\Gamma(-z_\beta-z_\alpha;\tau,\sigma)}\,, \\
        Z_{\mathrm{v}}^{\alpha}(\boldsymbol{\rho}) &= \sum_{\kappa=0}^\infty q^{\kappa\xi}\prod_{\beta=1}^{N_f} \prod^\kappa_{j=1}\frac{\theta(z_\alpha+z_\beta+j\tau;\sigma) }{\theta(z_\alpha-z_\beta+j\tau;\sigma)}\,,
    \end{split}
\end{align}
where we have included the formal prefactor in $Z^{\alpha}_{S_{23}\mathcal{O},\text{1-loop}}(\boldsymbol{\rho})$ to cancel the $\beta=\alpha$ factor in the numerator of the product.
We also used $\Gamma(z;\tau,\sigma)=\frac{1}{\Gamma(\tau+\sigma-z;\tau,\sigma)}$ and define the $\kappa=0$ term in $Z_{\mathrm{v}}^{\alpha}(\boldsymbol{\rho})$ as $1$.
The summation index $\kappa$ can be thought of as the vortex charge.
Pulling out the factor $(pq)^{\xi/2}$, we will now show how the remaining expression factorizes into holomorphic blocks, following \cite{Nieri:2015yia}.

First of all, one can use properties of the $q$-$\theta$ function to show that each coefficient in the sum of $Z_{\mathrm{v}}^{\alpha}(\boldsymbol{\rho})$ is invariant under the action of $\mathcal{H}=SL(2,\mathbb{Z})_{13}\ltimes \mathbb{Z}^{2+2N_f}$ on $\boldsymbol{\rho}$.
For example, invariance under $S_{13}$ follows from \eqref{eq:gamma-ids-for-hol-blocks-3}:
\begin{equation}\label{eq:invarianceof-Vortexpartition-function}
    \prod_{\beta=1}^{N_f} \prod^\kappa_{j=1}\frac{\theta(\frac{z_\alpha+z_\beta+j\tau}{\sigma};-\tfrac{1}{\sigma})}{\theta(\frac{z_\alpha-z_\beta+j\tau}{\sigma};-\tfrac{1}{\sigma})}=\prod_{\beta=1}^{N_f}\prod^\kappa_{j=1} \frac{\theta(z_\alpha+z_\beta+j\tau;\sigma) }{\theta(z_\alpha-z_\beta+j\tau;\sigma)}\,,
\end{equation}
where we have used the $SU(N)$ condition $\sum_\alpha z_\alpha=0$ to show that:
\begin{equation}
    \prod_{\beta=1}^{N_f} \prod^\kappa_{j=1}e^{i\pi( B_{2}(z_\alpha+z_\beta+j\tau;\sigma)-B_{2}(z_\alpha-z_\beta+j\tau;\sigma))}=1\,.
\end{equation}
Invariance under the other generators of $\mathcal{H}$ follows from periodicity of $\theta(z;\sigma)$ under $z\to z+1$ and $\sigma\to \sigma+1$ in combination with \eqref{eq:invarianceof-Vortexpartition-function}.

Furthermore, up to a prefactor independent of $z_{\alpha}$, the overall classical contribution can be factorized as:
\begin{equation}\label{eq:FI-S-fact}
    x_{\alpha}^{\xi}=e^{\frac{\pi i}{6}\left(\frac{\tau}{\sigma}+\frac{\sigma}{\tau}+3\right)}\frac{\theta\left(\frac{z_\alpha}{\sigma};\frac{\tau}{\sigma}\right)\theta\left(\tau\xi;\frac{\tau}{\sigma}\right)}{\theta\left(\frac{z_\alpha}{\sigma}+\tau\xi;\frac{\tau}{\sigma}\right)}\times\frac{\theta\left(\frac{z_\alpha}{\tau};\frac{\sigma}{\tau}\right)\theta\left(\sigma\xi;\frac{\sigma}{\tau}\right)}{\theta\left(\frac{z_\alpha}{\tau}+\sigma\xi;\frac{\sigma}{\tau}\right)}\,,
\end{equation}
One may easily verify the identity making use of the modular and extension property of the $q$-$\theta$ function, as recorded in Appendix \ref{app:defs}.
We also note that expansion parameters in the vortex partition functions for fixed vortex charges $(\kappa_1,\kappa_2)$, $q^{\kappa_1\xi}$ and $p^{\kappa_2\xi}$ respectively, are simply obtained by shifting $z_{\alpha}\to z_{\alpha}+\kappa_1\tau+\kappa_2\sigma$.
This is trivial from the point of view of the left hand side, but follows on the right hand side from the shift property \eqref{eq:elliptic-theta} of the $q$-$\theta$ function.

Finally, since the perturbative part of the index simply consists of a product of elliptic $\Gamma$ functions, i.e., (anti-)chiral multiplet indices, we can now use the results from Section \ref{sssec:example-free-chiral}, specifically \eqref{eq:gamma-ids-for-hol-blocks}, to factorize the entire SQED index.
In particular, we formally define the holomorphic block as:\footnote{Since the FI parameter is integer quantized, this expression vanishes as it stands. However, for the non-trivial large diffeomorphisms $S_{13}$, we will see below that $\mathcal{B}^{\alpha}_S(\boldsymbol{\rho})$ does not vanish. This will be true for more general large diffeomorphisms too, as we will see in Section \ref{sssec:proof-sqed}.}
\begin{align}\label{eq:defn-hol-block-sqed}
\begin{split}
    \mathcal{B}^{\alpha}(\boldsymbol{\rho})&=\frac{\theta\left(z_\alpha;\tau\right)\theta\left(\tau\xi;\tau\right)}{\theta\left(z_\alpha+\tau\xi;\tau\right)}\frac{1}{\Gamma(0;\tau,\sigma)} \prod_{\beta=1}^{N_f} \frac{\Gamma(z_\beta -z_\alpha;\tau,\sigma)}{\Gamma(-z_\beta-z_\alpha;\tau,\sigma)}\\
    \times &\sum_{\kappa=0}^{\infty}q^{\kappa \xi}\prod_{\beta=1}^{N_f} \prod^{\kappa}_{j=1}\frac{\theta(z_\alpha+z_\beta+j\tau;\sigma) }{\theta(z_\alpha-z_\beta+j\tau;\sigma)}\,,
\end{split}
\end{align}
where in this case we let 
\begin{equation}
\boldsymbol{\rho}\equiv (\vec{z};\tau,\sigma;\tau\xi)
\end{equation} 
to emphasize the dependence on the FI parameter $\xi$ as well (see footnote \ref{fn:FI-rho}).
Furthermore, the asymmetric appearance between chiral and anti-chiral multiplet contributions reflects the D and R boundary conditions required for gauge anomaly cancellation (see above \eqref{eq:equiv-hol-block-facts}).
Up to phase prefactors, the index can now be written in the factorized form
\begin{equation}
    I^{\mathrm{SQED}}_{(1,0)}(\hat{\boldsymbol{\rho}})=e^{\frac{\pi i}{6}\left(\frac{\tau}{\sigma}+\frac{\sigma}{\tau}+3\right)}(pq)^{\xi/2}e^{-i\pi P^{\mathrm{SQED}}_{S_{23}}(\boldsymbol{\rho})} \sum_{\alpha=1}^{N_f} \norm{\mathcal{B}_{S}^\alpha (\boldsymbol{\rho})}^2_{S_{23} \mathcal{O}}\,,\qquad \hat{\boldsymbol{\rho}}=\boldsymbol{\rho}\,,
\end{equation}
where we recall the definition of the norm from \eqref{eq:lens-hol-blocks} and extend the action of $S_{13}$ as before to include its action on $ \tau\xi$ as well
\begin{equation}
    \mathcal{B}^{\alpha}_S(\boldsymbol{\rho})\equiv \mathcal{B}^{\alpha}(S_{13}\boldsymbol{\rho})=\mathcal{B}^{\alpha}\left(\tfrac{\vec{z}}{\sigma};\tfrac{\tau}{\sigma},-\tfrac{1}{\sigma};\tau\xi\right)\,.
\end{equation}
That is, for the factorization to hold we need to keep $\tau\xi$ fixed under the action of $S_{13}$, as implicitly observed in \cite{Nieri:2015yia} as well.
This implies that $\xi$ by itself should transform as
\begin{equation}
    \xi\to \sigma\xi \,.
\end{equation}
This is indeed a natural transformation, since the quantization of $\xi$ follows from invariance under large gauge transformations along the temporal circle \cite{Imamura:2011uw} and $S_{13}$ precisely implements the exchange of temporal circle $1\leftrightarrow \sigma$.

In addition, the phase polynomial is given by:
\begin{equation}
    P^{\mathrm{SQED}}_{S_{23}}(\boldsymbol{\rho})=-Q(0;\tau,\sigma)+\sum_{\beta=1}^{N_f}\Big[Q(z_{\beta}-z_{\alpha};\tau,\sigma)-Q(-z_{\beta}-z_{\alpha};\tau,\sigma)\Big] \,.
\end{equation}
Similar to the case of the chiral multiplet, we note that $P^{\mathrm{SQED}}_{S_{23}}(\vec{z};\tau,\sigma)$ parametrizes the anomalies of the theory as in \eqref{eq:anomaly-pol-gen-th}.
In particular, because of the $SU(N)$ condition $\sum_\beta z_\beta=0$, the anomaly polynomial is independent of $\alpha$, as it should be.
Another way to understand this independence is from the integrand in \eqref{eq:explicit-index-sqed}.
If we had reversed the order of operations, namely first factorizing the $\Gamma$ functions in the integrand of \eqref{eq:explicit-index-sqed}, the total phase polynomial would have been manifestly $u$-independent because of gauge anomaly cancellation \cite{Nieri:2015yia}.
This concludes our review of holomorphic block factorization for SQED.

\subsubsection{General gauge theories}\label{sssec:gen-gauge-th}

We briefly point out the extension to general $\mathcal{N}=1$ gauge theories.
As we have learned from the SQED example, there are five ingredients that go into holomorphic block factorization as we have presented it:
\begin{itemize}
    \item The existence of a Higgs branch expression for the relevant index. A sufficient condition is the vanishing of the residue at $v=0$ (see footnote \ref{fn:higgs-branch}) \cite{Peelaers:2014ima,Chen:2014rca}.\footnote{We note that there are concrete expressions only for $L(p,q)\times S^1$ indices with $q=0,1$. We propose a general formula for the index with $q>1$ in Section \ref{ssec:gen-formula}.}
    \item Invariance of the vortex partition function for fixed vortex charge, i.e. the coefficients in the ``grand canonical'' vortex partition function $Z^{\alpha}_{\text{v}}(\boldsymbol{\rho})$, to be invariant under $S_{13}\in SL(2,\mathbb{Z})_{13}\subset \mathcal{H}$.
    \item Factorization of the FI term if present.
    \item Factorization of the free chiral multiplet index.
    \item Independence of the phase polynomial on the Higgs branch vacuum $\alpha$.
\end{itemize}
The first point follows from the contour integral expression of the gauge theory lens indices and the pole structure of their integrands \cite{Nieri:2015yia}.
The second point holds for SQED (Section \ref{sssec:sqed}) and also SQCD with $SU(N)$ gauge group and an arbitrary number of flavors \cite{Nieri:2015yia,Longhi:2019hdh}.
A general argument for this should follow from gauge anomaly cancellation in the four-dimensional theory, which would prohibit 't Hooft anomalies in the vortex worldsheet theory where the 4d gauge symmetries would appear as global symmetries.
The third and fourth point follow from the SQED and chiral multiplet examples respectively.
The final point follows from gauge anomaly cancellation, as we argued for SQED in the final paragraph of the previous section.

\subsection{Consistency condition and its solution}\label{ssec:consistency-cond}

In this section, we impose the consistency condition \eqref{eq:consistency-cond}, namely that the blocks $\mathcal{B}^{\alpha}_S(\boldsymbol{\rho})$ and $\mathcal{C}^{\alpha}_S(\boldsymbol{\rho})$ lead to the same compact space partition function, on the more general proposal in Section \ref{ssec:towards-conjecture}.
That is, we require:
\begin{equation}\label{eq:consistency-cond-h}
	\mathcal{B}_h^{\alpha}(\boldsymbol{\rho})\mathcal{B}_{\tilde{h}}^{\alpha}(f^{-1}\boldsymbol{\rho})\cong  \mathcal{C}_h^{\alpha}(\boldsymbol{\rho})\mathcal{C}_{\tilde{h}}^{\alpha}(f^{-1}\boldsymbol{\rho})\,,
\end{equation}
where equality may hold up to multiplication by a phase (independent of $\alpha$).
This constrains $(h,\tilde{h})$ in a way that depends on $f$, and we will denote the set of pairs that solve the constraints by $S_f$.
This will lead to our conjecture for the modular factorization of lens indices in Section \ref{ssec:mod-fact-conjecture}.

As we have seen in Section \ref{ssec:hol-blocks}, the holomorphic blocks $\mathcal{B}^\alpha(\boldsymbol{\rho})$ and $\mathcal{C}^\alpha(\boldsymbol{\rho})$ are related through multiplication by a function $Z^{\alpha}_\partial(\vec{z};\tau)$ that is invariant under $SL(2,\mathbb{Z})\ltimes \mathbb{Z}^{2r}$ up to a phase.
Plugging in this relation for $\mathcal{B}^\alpha_{h}(\boldsymbol{\rho})$ and $\mathcal{B}^\alpha_{\tilde{h}}(\tilde{\boldsymbol{\rho}})$, we obtain:
\begin{equation}
    \mathcal{B}^\alpha_{h}(\boldsymbol{\rho})=Z^\alpha_\partial (z'_a;\tau')\mathcal{C}^\alpha_{h}(\boldsymbol{\rho})\,,\qquad \mathcal{B}^\alpha_{\tilde{h}}(f^{-1}\boldsymbol{\rho})=Z^\alpha_\partial (\tilde{z}'_a;\tilde{\tau}')\mathcal{C}^\alpha_{\tilde{h}}(f^{-1}\boldsymbol{\rho}) \,,
\end{equation}
where now $(z'_a;\tau')$ and $(\tilde{z}'_a;\tilde{\tau}')$ are defined by:
\begin{equation}\label{eq:rhop-and-tilderhop}
    \boldsymbol{\rho}'\equiv (z'_a;\tau',\sigma')=h\boldsymbol{\rho}\,,\qquad \tilde{\boldsymbol{\rho}}'\equiv (\tilde{z}'_a;\tilde{\tau}',\tilde{\sigma}')=\tilde{h}f^{-1}\boldsymbol{\rho}\,.
\end{equation}
The condition \eqref{eq:consistency-cond-h} is satisfied as long as:
\begin{equation}\label{eq:ind-index-bc-gen}
    Z^\alpha_\partial (z'_a;\tau') \tilde{Z}^\alpha_\partial (\tilde{z}'_a;\tilde{\tau}')\cong 1\,,
\end{equation}
where the equality may hold up to a phase.
Using the fact that $Z^{\alpha}_{\partial}(z;\tau)$ consists of a product of $q$-$\theta$ functions --- see footnote \ref{footnote:Zpartial} --- this equation is satisfied if:
\begin{equation}\label{eq:constraint-h1,2}
    \tilde{z}'_a+\tilde{\mu}_a\tilde{\tau}'+\tilde{\nu}_a= \frac{z'_a+\mu_a\tau'+\nu_a}{\gamma\tau'+\delta}\,,\qquad \tilde{\tau}'=-\frac{\alpha\tau'+\beta}{\gamma\tau'+\delta}\,,\qquad \alpha\delta-\beta\gamma=1\,,
\end{equation}
where $\mu_a,\nu_a\in \mathbb{Z}$ and $\tilde{\mu}_a,\tilde{\nu}_a\in \mathbb{Z}$ can be arbitrary due to the ellipticity of $Z^{\alpha}_\partial(z_a;\tau)$.
Furthermore, the ``$-$" sign reflects the fact that $\tau'$ and $\tilde{\tau}'$ are related through an orientation reversal transformation.
These equations should be read as constraints on $h,\tilde{h}\in \mathcal{H}$ for an appropriate choice of $(\alpha,\beta,\gamma,\delta)$ and $(\mu_a,\nu_a;\tilde{\mu}_a,\tilde{\nu}_a)$. 
For simplicity, we will assume that $h,\tilde{h}$ do not contain large gauge transformations, in which case we only have to solve the second constraint.
We also assume that $f$ does not contain factors of $t_{2}^{(a)}$, but comment at the end of this section on the more general case.

To proceed, we take a generic ansatz for $h,\tilde{h}\in H$ and solve \eqref{eq:constraint-h1,2}.
Explicitly, let $h$ and $\tilde{h}$ be given by:
\begin{equation}\label{eq:h1,2-ansatz}
    h=\begin{pmatrix}
    n & 0 & m\\
    b & 1 & a \\
    l & 0 & k
    \end{pmatrix}\,,\qquad \tilde{h}=\begin{pmatrix}
    \tilde{n} & 0 & \tilde{m}\\
    \tilde{b} & 1 & \tilde{a} \\
    \tilde{l} & 0 & \tilde{k}
    \end{pmatrix}\,,
\end{equation}
with $k n-l m=1$ and $\tilde{k}\tilde{n}-\tilde{l}\tilde{m}=1$.
We will assume periodicity of the holomorphic blocks $\mathcal{B}_h(\boldsymbol{\rho})$ in its arguments, as encountered in Section \ref{sssec:example-free-chiral}.
This implies that the second and third rows of $h$ and $\tilde{h}$ are defined up to integer multiples of the first row:
\begin{equation}\label{eq:periodicity-h}
    h\sim T_{21} h\sim T_{31}h\,,
\end{equation}
and similarly for $\tilde{h}$.
In particular, this fixes $(k,l)$ and $(\tilde{k},\tilde{l})$ for a given $(m,n)$ and $(\tilde{m},\tilde{n})$, respectively.
Furthermore, it implies that we may consider $b$ $\mod n$ and $\tilde{b}$ $\mod \tilde{n}$ and view $(a,\tilde{a})$ as free integers.
Plugging in the constraint \eqref{eq:constraint-h1,2} with \eqref{eq:rhop-and-tilderhop}, it follows that $\alpha\delta-\beta\gamma=1$ requires:
\begin{equation}
    \tilde{m}=m\,.
\end{equation}
The remaining constraints are solved if:
\begin{equation}\label{eq:sl2-constraints}
  \begin{alignedat}{2}
     \alpha&=-q-p\tilde{a} \,, \qquad  & \delta &=-s-pa \,,\\
     m\beta &=r+qa+s\tilde{a}+pa\tilde{a} \,, \qquad  & \gamma&=pm\,,
  \end{alignedat}
\end{equation}
and:
\begin{equation}\label{eq:b-and-n-constraints}
  \begin{alignedat}{2}
     \tilde{b}&=-\alpha b-\beta n \,, \qquad  & \tilde{n}&=\delta n+\gamma b \,,\\
     b &=-\delta \tilde{b}-\beta \tilde{n} \,, \qquad  & n&=\alpha \tilde{n}+\gamma \tilde{b}\,.
  \end{alignedat}
\end{equation}
Let us make some comments.
First of all, using the fact that $qs-pr=1$, it follows immediately that $\alpha\delta-\beta\gamma=1$.
However, we need to ensure $\beta\in \mathbb{Z}$, which imposes a constraint on $(m;a,\tilde{a})$.
Secondly, note that the equations in the first line of \eqref{eq:b-and-n-constraints} are equivalent to those in the second line.
Their solution is immediate either in terms of $(b,n)$ or $(\tilde{b},\tilde{n})$.
Thirdly, our ansatz requires the following coprime conditions on $m$:
\begin{equation}\label{eq:coprime-conds}
    \gcd(m,n)=\gcd(m,\tilde{n})=1\,.
\end{equation}
Finally, note that the (redundant) set of constraints is invariant under the inversion symmetry in \eqref{eq:main-geom-equiv}, which effectively exchanges the untilded and tilded variables and $q\leftrightarrow s$.
We will see this symmetry, and also the other symmetries of the Hopf surface, reflected in the set of solutions.

Let us first analyze the coprime conditions.
If we choose to solve the equations in terms of $(b,n)$, we find that $\gcd(m,\tilde{n})=1$ can be written as:
\begin{equation}\label{eq:coprime-cond-2}
    \gcd(m,(s+pa)n)=1\,.
\end{equation}
Since $\gcd(p,s)=1$, this constraint automatically implies the other coprime condition $\gcd(m,n)=1$, apart from the special cases $p=\pm a=\mp s=1$ and $a=s=0$.\footnote{These cases will be treated separately in the examples.}
Solving in terms of $(\tilde{b},\tilde{n})$ instead, we similarly find that \eqref{eq:coprime-conds} can be captured in a single condition:
\begin{equation}\label{eq:coprime-cond-3}
    \gcd(m,(q+p\tilde{a})\tilde{n})=1\,.
\end{equation}
We continue to solve for $\beta\in\mathbb{Z}$.
Combining \eqref{eq:sl2-constraints} and \eqref{eq:b-and-n-constraints}, one finds that there exists an integral solution for $\beta$ as long as:
\begin{equation}\label{eq:beta-int-cond}
    m\tilde{b}-\tilde{n}\tilde{a}=q(mb-na)-rn\,,
\end{equation}
where we have used $\gcd(m,n)=1$.
Alternatively, using the inversion symmetry we can also find an integral solution for $\beta$ as long as:
\begin{equation}\label{eq:beta-int-cond-2}
    mb-na=s(m\tilde{b}-\tilde{n}\tilde{a})-r\tilde{n}\,,
\end{equation}
where now we used $\gcd(m,\tilde{n})=1$.
Since we may take $0\leq b<n$ and $0\leq \tilde{b}<\tilde{n}$, it follows that the pairs $(a,b)$ and $(\tilde{a},\tilde{b})$ uniquely parametrize the single integers $c$ and $\tilde{c}$, respectively, through:
\begin{equation}\label{eq:ab-solns}
  \begin{alignedat}{2}
     c&=bm-an\,, \qquad & \tilde{c}&=\tilde{b}m-\tilde{a}\tilde{n}\,,\\
     \Longleftrightarrow \quad (a,b)&=(-ck+\kappa m,-cl+\kappa n)\,, \qquad & (\tilde{a},\tilde{b})&=(-\tilde{c}\tilde{k}+\tilde{\kappa}m,-\tilde{c}\tilde{l}+\tilde{\kappa}\tilde{n})\,,
  \end{alignedat}
\end{equation}
where we inverted the relation in the second line, and $\kappa,\tilde{\kappa}\in \mathbb{Z}$ are fixed by the domain of $b$ and $\tilde{b}$.
The conditions \eqref{eq:beta-int-cond} and \eqref{eq:beta-int-cond-2} can now be written, respectively, as:
\begin{equation}\label{eq:c-soln}
    \tilde{c}=qc-rn\,,\qquad c=s\tilde{c}-r\tilde{n}  \,.
\end{equation}
These equations are consistent due to $qs-pr=1$ and the relation between $n$ and $\tilde{n}$.
In conclusion, we find that $\beta\in\mathbb{Z}$ if $\tilde{c}$ is solved in terms of $c$ as in \eqref{eq:c-soln} (or vice versa).

Let $S_f\subset H\times H$ denote the set of pairs $(h,\tilde{h})$ that solve $\tilde{h}$ in terms of $h$.\footnote{\label{fn:subgroup}We caution that in general $S_f$ can not be thought of as a subgroup of $H\times H$. 
This follows from the fact that the coprime conditions \eqref{eq:coprime-cond-2} and \eqref{eq:coprime-cond-3} in general do not respect the semi-direct product structure of $H$.}
Taking into account periodicity \eqref{eq:periodicity-h}, the solution set is a right coset parametrized by three integers $(m,n,c)$:
\begin{equation}\label{eq:Sf-mod}
    \Gamma'_{\infty}\times \Gamma'_{\infty}\backslash S_f\,,\qquad 
    \Gamma'_{\infty}=\langle T_{21},T_{31}\rangle \subset H\,.
\end{equation}
Explicitly, an element in this set can be written as:
\begin{equation}\label{eq:h-ht-Lens-gen}
    h=\begin{pmatrix}
    n & 0 & m\\
    -cl & 1 & -ck\\
    l & 0 & k
    \end{pmatrix}\,,\quad \tilde{h}=\begin{pmatrix}
    -sn+pc & 0 & m\\
    (qc-rn)\tilde{l} & 1 & (qc-rn)\tilde{k}\\
    \tilde{l} & 0 & \tilde{k}
    \end{pmatrix}\,,
\end{equation}
where we have chosen $\kappa=\tilde{\kappa}=0$ as a representative, and recall that $(k,l)$ and $(\tilde{k},\tilde{l})$ are also fixed by periodicity.
To describe \eqref{eq:Sf-mod} more concisely, let us take take $n,c\in\mathbb{Z}$ general and choose $m$ such that it obeys the coprime condition \eqref{eq:coprime-cond-2}.
For fixed $c$ (i.e., for fixed $(a,b)$ in the coset), this set of integers $(m,n)$ has an elegant description in terms of the quotient $\Gamma_\infty \backslash \Gamma_0(s+pa)$, where $\Gamma_0(n)\subset SL(2,\mathbb{Z})$ is the Hecke congruence subgroup:
\begin{equation}
    \Gamma_0(n)= \left\lbrace \begin{pmatrix}
        \textrm{a} & \textrm{b}\\
        \textrm{c} & \textrm{d}
    \end{pmatrix}\quad \left|\; \textrm{a}\textrm{d}-\textrm{b}\textrm{c}=1\,,\; \textrm{c}=0\mod n \right.\right\rbrace\,,
\end{equation}
and $\Gamma_{\infty}\subset SL(2,\mathbb{Z})$ is generated by the (upper triangular) $T$ matrix.
It follows that, as a set, \eqref{eq:Sf-mod} can be described as:
\begin{equation}\label{eq:Sf-gen}
     \Gamma'_{\infty}\times \Gamma'_{\infty}\backslash S_f \cong \bigcup_{\substack{a\in \mathbb{Z}}}\Gamma_\infty\backslash  \Gamma_0(s+pa)\,.
\end{equation}
We thus find the existence of an interesting modular set of holomorphic blocks $\mathcal{B}^{\alpha}_h(\boldsymbol{\rho})$ and $\mathcal{B}^{\alpha}_{\tilde{h}}(f^{-1}\boldsymbol{\rho})$ consistent with the factorization of a general lens index $\mathcal{I}_{(p,q)}(\hat{\boldsymbol{\rho}})$.
We will study some examples below to make this more concrete, and turn to a geometric interpretation in Section \ref{ssec:geom-int-univ-blocks}.
For now, let us similarly describe the solution set $\widetilde{S}_f$ where $h$ is solved in terms of the $\tilde{h}$ parameters:
\begin{equation}\label{eq:tildeSf-gen}
     \Gamma'_{\infty}\times \Gamma'_{\infty}\backslash\widetilde{S}_f\cong \bigcup_{\tilde{a}\in \mathbb{Z}} \Gamma_\infty\backslash \Gamma_0(q+p\tilde{a})\,,
\end{equation}
where we have made use of \eqref{eq:coprime-cond-3}.
As explained above, we see that: $S_f= \widetilde{S}_{f^{-1}}$.
Notice also that both sets are separately invariant under the other symmetries of the Hopf surface, including $s\to s+p$ and $q\to q+p$.

Finally, it will be useful to have a description of the solution set $S_{f'}$ for $f'=g'\,\mathcal{O}$ with general $g'\in SL(3,\mathbb{Z})$.
This is easily obtained from $S_f$ by recalling that there always exist $h,\tilde{h}\in H$ such that $f'=hf\tilde{h}^{-1}$ with $f=g_{(p,q)}\,\mathcal{O}$ and $g_{(p,q)}\in SL(2,\mathbb{Z})_{23}$ as in \eqref{eq:gSL2-from-gSL3}.
This leads us to the following description of $S_{f'}$:
\begin{equation}\label{eq:Sf'-gen}
    S_{f'}=\left\lbrace (h_fh^{-1},\tilde{h}_f\tilde{h}^{-1})\quad  |\quad (h_f,\tilde{h}_f)\in S_f \right\rbrace\,, 
\end{equation}
where $S_f$ is as described above.

\subsubsection*{Examples of \texorpdfstring{$\bm{S_f}$}{Sf}}\label{sssec:examples-Sf}

Here, we describe the solution set $S_f$ for some simple gluing transformations $f$.
We will denote by $S^{(a,b)}_{f}$ the subset of $S_f$ for fixed $(a,b)$.
We also point out in which cases $S_f$ or $S_f^{(a,b)}$ can be understood as a \emph{subgroup} of $H\times H$.

\paragraph{$\bm{S^2\times T^2}$:}

The simplest example is obtained by taking $f=\mathcal{O}$, i.e., $p=r=0$ and $q=s=-1$.
The resulting manifold has topology $S^2\times T^2$.
In this case, the constraints simplify significantly.
In particular, the $a$ parameter decouples from the modular part of the solution set.
We thus obtain the direct product:
\begin{equation}
     \Gamma'_{\infty}\times \Gamma'_{\infty}\backslash S_{\mathcal{O}}\cong \Gamma_\infty \backslash SL(2,\mathbb{Z})\times \mathbb{Z}\cong  \Gamma'_{\infty}\times \Gamma'_{\infty}\backslash\widetilde{S}_{\mathcal{O}}\,.
\end{equation}
More explicitly, $S_\mathcal{O}$ is parametrized by the matrices:
\begin{equation}\label{eq:h-ht-S2xT2}
    h=\begin{pmatrix}
    n & 0 & m\\
    b & 1 & a \\
    l & 0 & k
    \end{pmatrix}\,,\quad \tilde{h}=\begin{pmatrix}
    n & 0 & m\\
    -b & 1 & -a \\
    l & 0 & k
    \end{pmatrix}\,.
\end{equation}
It follows that $S_\mathcal{O}$ embeds (almost) diagonally into $H\times H$:
\begin{equation}\label{eq:SO}
     S_{\mathcal{O}}=\left\lbrace (h,\mathcal{O}h\mathcal{O})\;|\; h\in H \right\rbrace \subset H\times H\,,
\end{equation}
and it clearly forms a subgroup.
The closely related case $f=t_2^g\,\mathcal{O}$ can be checked to satisfy the consistency condition for the above pair $(h,\tilde{h})$ as long as $a=b=0$.
This follows from the elliptic properties of $Z^{\alpha}_{\partial}(z;\tau)$ discussed around \eqref{eq:constraint-h1,2}.
Concluding, there exists a consistent family of holomorphic blocks for the $S^2\times T^2$ index parametrized by $\Gamma'_{\infty}\backslash H$.
This is not surprising given that this index has ordinary modular properties under $H$, as discussed in Section \ref{ssec:hol-blocks}.

\paragraph{$\bm{S^3\times S^1}$:}

Another basic example corresponds to $f=S_{23}\,\mathcal{O}$, i.e., $p=-r=1$ and $q=s=0$, which has topology $S^3\times S^1$.
The constraints again simplify significantly.
In particular, one finds:
\begin{equation}\label{eq:c-soln-s3xs1}
    \tilde{n}=c\,, \qquad   \tilde{c}=n\,.
\end{equation}
The condition on the integers $(m,n,c)$ is now $\gcd(m,n)=\gcd(m,c)=1$, which cannot be reduced to a single coprime condition.
The solution set is written as:
\begin{equation}\label{eq:Sf-S3xS1}
     \Gamma'_{\infty}\times \Gamma'_{\infty}\backslash S_{S_{23}\mathcal{O}}\cong \bigcup_{a\in \mathbb{Z}}  \Gamma_\infty\backslash\Gamma_0(a)\cong  \Gamma'_{\infty}\times \Gamma'_{\infty}\backslash\widetilde{S}_{S_{23}\mathcal{O}}\,.
\end{equation}
The explicit matrices $(h,\tilde{h})$ are given by:
\begin{equation}\label{eq:h-ht-S3xS1}
    h=\begin{pmatrix}
    n & 0 & m\\
    -\tilde{n}l+\kappa n & 1 & -\tilde{n}k+\kappa m \\
    l & 0 & k
    \end{pmatrix}\,,\quad \tilde{h}=\begin{pmatrix}
    \tilde{n} & 0 & m\\
    -n\tilde{l}+\tilde{\kappa}\tilde{n} & 1 & -n\tilde{k}+\tilde{\kappa}m \\
    \tilde{l} & 0 & \tilde{k}
    \end{pmatrix}\,,
\end{equation}
where we have written $\tilde{n}$ as opposed to $c$ and $\kappa,\tilde{\kappa}\in \mathbb{Z}$ are free integers in $S_{S_{23}\mathcal{O}}$, but can be set to zero in the coset.
The meaning of the factor $\Gamma_{\infty}\backslash \Gamma_0(0)$ can be understood from the explicit matrices $(h,\tilde{h})$ and corresponds to a single element $(S_{13},S_{13})$.

In the case $\tilde{n}=n$, $\tilde{k}=k$ and $\tilde{l}=l=\tilde{\kappa}=\kappa$, the solution set turns out to have a particularly nice interpretation.
First, note that the $(h,\tilde{h})$ can now be written as:
\begin{equation}
    h=\tilde{h}=\begin{pmatrix}
    n & 0 & m\\
    0 & 1 & -1 \\
    l & 0 & k
    \end{pmatrix}\,.
\end{equation}
That is, $S^{(-1,0)}_{S_{23}\mathcal{O}}$ almost embeds as a diagonal $SL(2,\mathbb{Z})$ subgroup of $H\times H$.
In fact, we can modify $f$ slightly to make this embedding exact.
For this, we make use of the relation \eqref{eq:Sf'-gen} between two solution sets $S_{f'}$ and $S_f$.
It follows from the expression of $S_{ij}$ in terms of $T_{ij}$ in \eqref{eq:defn-SL2ij} that:
\begin{equation}\label{eq:T32-as-S23}
    T_{32}^{-1}\mathcal{O}=T_{23}^{-1}\,S_{23}\,\mathcal{O}\,T_{23}\,.
\end{equation}
Since $T_{23}\in H$, we can apply \eqref{eq:Sf'-gen} and find that $S^{(0,0)}_{T_{32}^{-1}\mathcal{O}}$ does embed as a diagonal $SL(2,\mathbb{Z})$ subgroup of $H\times H$:
\begin{equation}
     S^{(0,0)}_{T_{32}^{-1}\mathcal{O}}=\left\lbrace (h,h)\;|\; h\in SL(2,\mathbb{Z})_{13} \right\rbrace \subset H\times H\,.
\end{equation}
This example reappears below as a special case of $L(p,-1)\times S^1$ for $p=1$.
The connection with $S^3\times S^1$ follows from the symmetry of Hopf surfaces under $\boldsymbol{\rho}\to T_{23}^{-1}\boldsymbol{\rho}$ and $f\to T_{23}^{-1} fT_{23}$, as mentioned in Section \ref{ssec:ambig-heegaard}.

Concluding, the superconformal index admits a consistent factorization in terms of a family of holomorphic blocks containing a diagonal $SL(2,\mathbb{Z})_{13}\subset H\times H$.
We turn to a geometric interpretation of this potentially surprising fact in Section \ref{ssec:geom-int-univ-blocks}.

\paragraph{$\bm{L(p,\pm 1)\times S^1}$:}

Finally, we turn to $f_{(p,\pm 1)}=S_{23}T_{23}^{-p}S_{23}^{\pm 1}\mathcal{O}$.
The associated Hopf surface has topology $L(p,\pm 1)\times S^1$, and the constraints are solved by:
\begin{equation}
    \tilde{n}=\mp n+pc\,,\qquad \tilde{c}=\pm c\,.
\end{equation}
The solution set can be described as:
\begin{equation}
    \Gamma'_{\infty}\times \Gamma'_{\infty}\backslash S_{f_{(p,\pm 1)}} \cong\bigcup_{a\in \mathbb{Z}} \Gamma_\infty\backslash\Gamma_0(\pm 1+pa)\cong \Gamma'_{\infty}\times \Gamma'_{\infty}\backslash \widetilde{S}_{f_{(p,\pm 1)}}\,.
\end{equation}
In terms of the matrices $(h,\tilde{h})$ we have:
\begin{equation}\label{eq:h-ht-Lens}
    h=\begin{pmatrix}
    n & 0 & m\\
    -cl+\kappa n & 1 & -ck+\kappa m\\
    l & 0 & k
    \end{pmatrix}\,,\quad \tilde{h}=\begin{pmatrix}
    \mp n+pc & 0 & m\\
    \mp c \tilde{l}+\tilde{\kappa}(\mp n+pc) & 1 & \mp c \tilde{k}+\tilde{\kappa}m \\
    \tilde{l} & 0 & \tilde{k}
    \end{pmatrix}\,.
\end{equation}
A simple and interesting example is the case when $c=\kappa=\tilde{\kappa}=0$.
In this case, the set $S_{f_{(p,- 1)}}^{(0,0)}$ embeds as a diagonal subgroup for any $p\in \mathbb{Z}$:
\begin{align}
    \begin{split}
        S_{f_{(p,-1)}}^{(0,0)}&= \left\lbrace (h,h)\;|\; h\in SL(2,\mathbb{Z})_{13} \right\rbrace \subset H\times H\,.
    \end{split}
\end{align}
This reproduces our previous example for $p=1$ since $f_{(1,-1)}=T_{32}^{-1}\,\mathcal{O}$.
On the other hand, $S_{f_{(p, 1)}}^{(0,0)}$ does not immediately define a subgroup.
This follows from the fact that $\tilde{h}=S_{23}^2\,h\,S_{12}^2$, which is not a conjugation of $h$.
There is again a simple fix.
We instead look at the solution set for:
\begin{equation}
    f'_{(p,1)}=S_{13}\, f_{(p,1)} \, S_{13}^{-1}\,.
\end{equation}
In this case, we find that $S^{(0,0)}_{f'_{(p,1)}}$ embeds as follows:
\begin{equation}
    S^{(0,0)}_{f'_{(p,1)}}=\left\lbrace (h,S_{23}^2\,h\,S_{23}^2)\;|\; h\in SL(2,\mathbb{Z})_{13} \right\rbrace\subset H\times H\,.
\end{equation}
This defines a subgroup since $S_{23}^4=\mathbbm{1}$.
Concluding, we see that also the indices on $L(p,\pm 1)\times S^1$, for any $p$, admit a consistent factorization in terms of a family of holomorphic blocks that contains an (almost) diagonal $SL(2,\mathbb{Z})_{13}\subset H\times H$.

\subsection{Modular factorization conjecture}\label{ssec:mod-fact-conjecture}

We can now state our \emph{modular factorization conjecture}.
Provided that a Higgs branch expression for the index exists (see Section \ref{ssec:hol-blocks}), we claim that a given lens index can be factorized in a variety of ways parametrized by $S_f$
\begin{equation}\label{eq:mod-fact-conj}
    \mathcal{I}_{(p,q)}(\hat{\boldsymbol{\rho}})=e^{-i\pi \mathcal{P}^{\mathbf{m}}_f(\boldsymbol{\rho})} \sum_{\alpha} \mathcal{B}^\alpha_{h}(\boldsymbol{\rho})\mathcal{B}^\alpha_{\tilde{h}}(f^{-1}\boldsymbol{\rho})\,,\qquad  (h,\tilde{h})\in S_f\subset H\times H\,.
\end{equation}
Without loss of generality, we take $f=g_{(p,q)}\,\mathcal{O}$ with $g_{(p,q)}\in SL(2,\mathbb{Z})_{23}$ as in \eqref{eq:gSL2-from-gSL3}.\footnote{The extension to more general gluing transformations follows from our observation in \eqref{eq:gSL2-from-gSL3} and the relation between $S_{f'}$ and $S_f$ in \eqref{eq:Sf'-gen}.}
Furthermore, we claim that the phase polynomial $\mathcal{P}^{\mathbf{m}}_f(\boldsymbol{\rho})$ includes the 't Hooft anomalies of the theory, which will be seen to depend on both $f$ and $(h,\tilde{h})$, the latter dependence indicated through $\mathbf{m}\equiv (m;n,c)$.
The constraint $(h,\tilde{h})\in S_f$ ensures that the index is independent of the boundary conditions imposed on the blocks.
That is, the factorization also holds with respect to $\mathcal{C}^{\alpha}_{h}(\boldsymbol{\rho})$ and $\mathcal{C}^{\alpha}_{\tilde{h}}(f^{-1}\boldsymbol{\rho})$.
The use of the word ``modular'' is motivated by the fact that $S_f$ contains modular (congruence sub-)groups.

Before turning to a geometric interpretation in Section \ref{ssec:geom-int-univ-blocks} and evidence in Section \ref{ssec:evidence}, let us examine the conjecture in more detail for the $L(p,-1)\times S^1$ index.
This includes the $S^2\times T^2$ index for $p=0$ and the $S^3\times S^1$ index for $p=1$, up to a change of parameters: 
\begin{equation}\label{eq:f(-1,1)-S23reln}
    \mathcal{Z}_{f_{(1,-1)}}(T_{23}^{-1}\boldsymbol{\rho})=\mathcal{Z}_{S_{23}\mathcal{O}}(\boldsymbol{\rho})\equiv \mathcal{I}_{(1,0)}(\hat{\boldsymbol{\rho}})\,,\qquad \boldsymbol{\rho}=\hat{\boldsymbol{\rho}}\,.
\end{equation}
As mentioned at the end of Section \ref{ssec:consistency-cond}, the set $S_{f_{(p,- 1)}}$ contains a diagonal $SL(2,\mathbb{Z})_{13}$ subgroup of $H\times H$.
Therefore, the modular factorization conjecture implies:
\begin{equation}
    e^{-i\pi\mathcal{P}^{\mathbf{m}}_{f}(\boldsymbol{\rho})}\sum_{\alpha}\mathcal{B}^\alpha(h\boldsymbol{\rho})\mathcal{B}^\alpha(hf^{-1}\boldsymbol{\rho})=e^{-i\pi\mathcal{P}^{\mathbf{1}}_{f}(\boldsymbol{\rho})} \sum_{\alpha}\mathcal{B}^\alpha(\boldsymbol{\rho})\mathcal{B}^\alpha(f^{-1}\boldsymbol{\rho})\,,\quad  h\in SL(2,\mathbb{Z})_{13}\,,
\end{equation}
since both sides represent the same index.
In fact, the equality holds at the level of the summands, as will become clear in Section \ref{ssec:evidence}.
This equation appears similar to an ordinary modular covariance, but, as also stressed in Section \ref{ssec:towards-conjecture}, it is a covariance with respect to a combined action:
\begin{equation}\label{eq:geom-action-conj}
    \boldsymbol{\rho}\to h\boldsymbol{\rho}\,,\qquad f\to hfh^{-1}\,.
\end{equation}
As a result, unlike an ordinary modular covariance, it does not behave nicely under group multiplication:
\begin{equation}\label{eq:h1h2-property}
    \mathcal{B}^\alpha(h_1h_2\boldsymbol{\rho})\mathcal{B}^\alpha(h_1h_2f^{-1}\boldsymbol{\rho})\ncong \mathcal{B}^\alpha(h_2\boldsymbol{\rho})\mathcal{B}^\alpha(h_2f^{-1}\boldsymbol{\rho})\,,
\end{equation}
unless $h_2$ commutes with $f^{-1}$.
The commutant of $f^{-1}$ inside $SL(2,\mathbb{Z})_{13}\subset SL(3,\mathbb{Z})$ is trivial unless $p=0$.
In the latter case, $f=\mathcal{O}$ and the commutant equals $H$.
This is indeed expected since the $S^2\times T^2$ index has modular properties under $H$.
Since a general lens index should not have such modular properties, \eqref{eq:h1h2-property} should be viewed as a feature rather than a bug.
A general lens index only transforms under $H$ in the generalized sense \eqref{eq:geom-action-conj}.
In Section \ref{sec:gen-modularity}, we will nonetheless see that a natural modular object can be constructed from the lens indices.
This object respects the multiplication of \emph{gluing elements} $g\in\mathcal{G}$ in an interesting way, as opposed to the multiplication of large diffeomorphisms of the $D_{2}\times T^2$ geometries.

\subsection[Geometric interpretation of \texorpdfstring{$S_f$}{Sf} and universal blocks]{Geometric interpretation of \texorpdfstring{$\bm{S_f}$}{Sf} and universal blocks}\label{ssec:geom-int-univ-blocks}

Modular factorization asserts that a given lens index can be factorized in terms of a (modular) family $S_f\subset H\times H$ of holomorphic blocks.
The subset $S_f$ arises from the physical constraint that the compact space partition function should not depend on the boundary conditions imposed on the holomorphic blocks.
In this section, we will provide a geometric interpretation of $S_f$ in terms of ``compatible'' Heegaard splittings of a Hopf surface.

We start with an observation about the set $S_f$ with $f=g\,\mathcal{O}$ and $g\in SL(3,\mathbb{Z})$.
First, we define the subgroup $F\subset SL(3,\mathbb{Z})$ as follows:
\begin{equation}\label{eq:def-F1}
\begin{aligned}
F\equiv SL(2,\mathbb{Z})_{12}\ltimes \mathbb{Z}^{2} \,, \quad \textrm{with} \quad \mathbb{Z}^{2} &= \langle T_{31}\,,T_{32}\rangle\,,
\end{aligned}
\end{equation}
where $SL(2,\mathbb{Z})_{ij}$ and $T_{ij}$ were defined in Section \ref{ssec:top-aspects}.
Note that this subgroup takes on the same form as $H$, but corresponds to a different embedding in $SL(3,\mathbb{Z})$.
It turns out the following two statements are equivalent:
\begin{equation}\label{eq:Sf-consequence-f'}
    (h,\tilde{h})\in S_{g\,\mathcal{O}}\quad \text{if and only if}\quad g'=h\,g\,\mathcal{O}\tilde{h}^{-1}\mathcal{O}\in F\,.
\end{equation}
The right implication is easily verified for $g=g_{(p,q)}$, as in \eqref{eq:gSL2-from-gSL3}, by plugging in the associated solutions \eqref{eq:h-ht-Lens-gen} for $(h,\tilde{h})$.\footnote{For general $g\in SL(3,\mathbb{Z})$, the claim follows from the relation \eqref{eq:Sf'-gen} and the observation in \eqref{eq:gSL2-from-gSL3}.}
One finds that $g'$ is of the form: 
\begin{equation}\label{eq:g-in-F1}
    g'=\begin{pmatrix}
        * & * & 0\\
        * & * & 0\\
        * & * & 1
    \end{pmatrix}\,,
\end{equation}
which indeed corresponds to an element in $F\subset SL(3,\mathbb{Z})$.
The converse can be proved by solving the condition $h\,g_{(p,q)}\,\mathcal{O}\tilde{h}^{-1}\mathcal{O}\in F$ in terms of $(h,\tilde{h})$ and noting that the resulting constraints are identical to those specifying $S_f$, as derived in Section \ref{ssec:consistency-cond}.\footnote{In particular, one can check that the inverse of the $SL(2,\mathbb{Z})$ matrix given by $\alpha$, $\beta$, $\gamma$ and $\delta$ in \eqref{eq:constraint-h1,2} is equal to the upper left $2\times 2$ block of $g'$, for $g= g_{(p,q)}$. We will see this explicitly later in \eqref{eq:h1,2-sl3-constr} with \eqref{eq:eq:h1,2-sl3-constr-hprime}.}

Consider now a Hopf surface with a Heegaard splitting in terms of some gluing transformation $f=g\,\mathcal{O}$ and $g\in SL(3,\mathbb{Z})$.
Recall from Section \ref{ssec:top-aspects} the notation:
\begin{equation}\label{eq:Mg-defn-lambda-mu-2}
    \begin{pmatrix}\lambda'&\; \mu &\; \lambda\end{pmatrix}=\begin{pmatrix}\tilde{\lambda}'&\; \tilde{\mu} &\; \tilde{\lambda}\end{pmatrix}\,f^{-1}\,,
\end{equation}
which summarizes how the cycles of the two $D_2\times T^2$ geometries are identified.
More general Heegaard splittings of the same Hopf surface are labeled by $(h,\tilde{h})\in H\times H$, associated to the gluing transformation $f'=hf\tilde{h}^{-1}$, which we similarly write as:
\begin{equation}\label{eq:Mg-defn-lambda-mu-3}
    \begin{pmatrix}\lambda'_h&\; \mu_h &\; \lambda_h\end{pmatrix}=\begin{pmatrix}\tilde{\lambda}'_{\tilde{h}}&\; \tilde{\mu}_{\tilde{h}} &\; \tilde{\lambda}_{\tilde{h}}\end{pmatrix}\,(f')^{-1}\,,
\end{equation}
where we defined:
\begin{equation}
    \begin{pmatrix}\lambda'_h&\; \mu_h &\; \lambda_h\end{pmatrix}\equiv \begin{pmatrix}\lambda'&\; \mu &\; \lambda\end{pmatrix} h^{-1}\,,\qquad \begin{pmatrix}\tilde{\lambda}'_{\tilde{h}}&\; \tilde{\mu}_{\tilde{h}} &\; \tilde{\lambda}_{\tilde{h}}\end{pmatrix}\equiv \begin{pmatrix}\tilde{\lambda}'&\; \tilde{\mu} &\; \tilde{\lambda}\end{pmatrix}\,\tilde{h}^{-1}\,.
\end{equation}
We define the compatible Heegaard splittings of this Hopf surface to be labeled by the subset $(h,\tilde{h})\in S_f\subset H\times H$.
This subset has a nice geometric characterization. 
Indeed, it follows from the form of $g'$ that the set $S_f$ parametrizes all Heegaard splittings of the associated Hopf surface such that the gluing transformation $g'$ fixes the cycle $\tilde{\lambda}_{\tilde{h}}=\lambda_h$.
This cycle is to be identified with the $S^1$ inside the Hopf surface $\mathcal{M}_{(p,q)}(\hat{\boldsymbol{\rho}})$. 
The remaining $L(p,q)\subset \mathcal{M}_{(p,q)}(\hat{\boldsymbol{\rho}})$ is then glued by the appropriate $SL(2,\mathbb{Z})$ block in $g'$ acting on $(\lambda'_h,\,\mu_h)$.

We can thus think of the large diffeomorphisms $(h,\tilde{h})\in S_f$ as the embeddings of the ``time circle'' into the $D_2\times T^2$ geometries such that the gluing transformation of the associated Heegaard splitting leaves it fixed.
This should be viewed as the four-dimensional analogue of how one can choose any embedding of the time circle in a $T^2$ for a CFT$_2$; for the latter, this choice is a consequence of $SL(2,\mathbb{Z})$ invariance of the CFT on the torus.
It provides a more transparent interpretation of $S_f$.
Namely, modular factorization becomes the statement that a given index can be factorized in terms of a pair of holomorphic blocks only if the blocks are defined with respect to a common time circle.

So far, we have seen that for a given Hopf surface the associated set $S_f$ parametrizes all \emph{compatible} ways in which the time circle can be embedded in the Heegaard splitting.
Instead, we could also fix the embedding of the time circle, i.e. fix a general pair $(h,\tilde{h})\in H\times H$, and consider all gluing transformations compatible with this embedding.
Clearly, this set of gluing transformations is in bijection with $F$, since any $f=g\,\mathcal{O}$ with:
\begin{equation}
    g\in F_{h,\mathcal{O}\tilde{h}\mathcal{O}}\,,\quad F_{h,\mathcal{O}\tilde{h}\mathcal{O}}\equiv h^{-1}\,F\,\mathcal{O}\tilde{h}\mathcal{O}\,,
\end{equation}
solves (the right hand side of) \eqref{eq:Sf-consequence-f'} for a general pair $(h,\tilde{h})$.
Note that only $F_h\equiv F_{h,h}$ corresponds to a subgroup of $SL(3,\mathbb{Z})$, while for general $\tilde{h}$ the gluing transformations form a subset.
We conclude that the Heegaard splittings associated to gluing transformations $F_{h,\mathcal{O}\tilde{h}\mathcal{O}}$ are compatible with the $(h,\tilde{h})$ embedding of the time circle into the $D_{2}\times T^2$ geometries, and that this is the maximal compatible subset in $SL(3,\mathbb{Z})$.

To illustrate the above, let us look at some specializations of $(h,\tilde{h})$, starting with $h=\tilde{h}=\mathbbm{1}$.
The associated blocks are the partition functions $D_2\times T^2(\boldsymbol{\rho})$, which we denoted in Section \ref{ssec:towards-conjecture} by $\mathcal{B}^{\alpha}(\boldsymbol{\rho})$.
In this case, it is clear that any $f=g\,\mathcal{O}$ with $g\in F$ solves the constraint in \eqref{eq:Sf-consequence-f'}.
It follows that the indices associated to any $g\in F$ can be consistently factorized in terms of $\mathcal{B}^{\alpha}(\boldsymbol{\rho})$ and that $F$ is the maximal subgroup of $SL(3,\mathbb{Z})$ with this property:
\begin{equation}
    \mathcal{Z}_{f}(\boldsymbol{\rho})\cong \sum_{\alpha}\norm{\mathcal{B}^{\alpha}(\boldsymbol{\rho})}^2_{f}\,,\qquad g\in F\,,
\end{equation}
where we omit the relative phase.
Note that $F$ can be used to construct arbitrary topologies $L(p,q)\times S^1$ since $SL(2,\mathbb{Z})_{12}\subset F$ (see Section \ref{ssec:top-aspects}).
In this sense, $\mathcal{B}^{\alpha}(\boldsymbol{\rho})$ is the unique universal block for indices associated the subgroup $F$.

However, we could have made a different choice of embedding of the time circle.
For example, let us choose $h=\tilde{h}=S_{13}$ so that the relevant geometry is $D_2\times T^2(S_{13}\boldsymbol{\rho})$.
As in Section \ref{ssec:hol-blocks}, we denote the associated partition function by $\mathcal{B}^{\alpha}_S(\boldsymbol{\rho})=\mathcal{B}^{\alpha}(S_{13}\boldsymbol{\rho})$ and, as explained there, this corresponds to the block used in the original work \cite{Nieri:2015yia}.
The set of gluing transformations which solve \eqref{eq:Sf-consequence-f'} is now given by:
\begin{equation}\label{eq:def-FS}
\begin{aligned}
F_S\equiv S_{13}^{-1}\,F\,S_{13}= SL(2,\mathbb{Z})_{23}\ltimes \mathbb{Z}^{2}\,,\quad \textrm{with} \quad \mathbb{Z}^{2} &= \langle T_{12}\,,T_{13}\rangle\,.
\end{aligned}
\end{equation}
Similarly to before, the index associated to any element in $F_S$ can be consistently factorized in terms of $\mathcal{B}_S^{\alpha}(\boldsymbol{\rho})$, and $F_S$ is the maximal subgroup of $SL(3,\mathbb{Z})$ with this property:\footnote{This was also observed in \cite{Gadde:2020bov}.}
\begin{equation}
    \mathcal{Z}_{f}(\boldsymbol{\rho})\cong \sum_{\alpha}\norm{\mathcal{B}_S^{\alpha}(\boldsymbol{\rho})}^2_{f}\,,\qquad g\in F_S\,.
\end{equation}
Since the conventional definition of the Hopf surfaces is with respect to $SL(2,\mathbb{Z})_{23}\subset F_S$, as discussed in Section \ref{ssec:hopf-surfaces}, this explains why the lens indices considered in \cite{Nieri:2015yia} were all factorized in terms of $\mathcal{B}^{\alpha}_S(\boldsymbol{\rho})$.
More generally, the holomorphic block $\mathcal{B}^{\alpha}_h(\boldsymbol{\rho})$ for any $h\in H$ is the unique universal holomorphic block for indices associated to $f=g\,\mathcal{O}$ and $g\in h^{-1}Fh$:
\begin{equation}\label{eq:Zf-Bh}
    \mathcal{Z}_{f}(\boldsymbol{\rho})\cong \sum_{\alpha}\norm{\mathcal{B}_h^{\alpha}(\boldsymbol{\rho})}^2_{f}\,,\qquad g\in F_h\equiv h^{-1}\, F \,h\,.
\end{equation}
Finally, the factorization of indices in terms of the most general pair of holomorphic blocks reads:
\begin{equation}\label{eq:Zf-Bh-Bht}
    \mathcal{Z}_{f}(\boldsymbol{\rho})\cong \sum_{\alpha}\mathcal{B}_h^{\alpha}(\boldsymbol{\rho})\mathcal{B}_{\tilde{h}}^{\alpha}(f^{-1}\boldsymbol{\rho})\,,\qquad g\in F_{h,\mathcal{O}\tilde{h}\mathcal{O}}\,.
\end{equation}
In the remainder of this paper, we will only make use of holomorphic block factorizations corresponding to the cases $F_h$, i.e. when the set of gluing transformations forms a subgroup of $SL(3,\mathbb{Z})$.

Let us now look at a restricted set of topologies.
As in Section \ref{ssec:consistency-cond}, we consider the example of $f_{(p,-1)}=g_{(p,-1)}\,\mathcal{O}$ with $p$ arbitrary, associated to the topologies $L(p,-1)\times S^1$.
Explicitly, $g_{(p,-1)}$ is given by:
\begin{equation}
    g_{(p,-1)}=\begin{pmatrix}
        1 & 0 & 0\\
        0 & 1 & 0\\
        0 & -p & 1
    \end{pmatrix}\,.
\end{equation}
We immediately see that it is an element in $F$ for any $p$, which means that the associated indices can all be factorized in terms of $\mathcal{B}^{\alpha}(\boldsymbol{\rho})$.
However, for this subset of gluing transformations, the condition \eqref{eq:Sf-consequence-f'} is preserved for arbitrary $h=\tilde{h}\in SL(2,\mathbb{Z})_{13}$.\footnote{Note that $\mathcal{O}$ commutes with general $\tilde{h}\in SL(2,\mathbb{Z})_{13}$.}
Therefore, there is an $SL(2,\mathbb{Z})$ family of holomorphic blocks $\mathcal{B}_h^{\alpha}(\boldsymbol{\rho})$ with $h\in SL(2,\mathbb{Z})_{13}$ compatible with the factorization of indices associated to $g_{(p,-1)}$.
Geometrically, this follows from the fact that $g_{(p,-1)}$ fixes both $\tilde{\lambda}=\lambda$ and $\tilde{\lambda}'=\lambda'$, and therefore any combination of cycles $m\lambda+n\lambda'$ with $\gcd(m,n)=1$ could serve as a time circle.

Finally, we note that there does not exist a holomorphic block for which indices associated to the full subgroup $H\subset SL(3,\mathbb{Z})$ can be consistently factorized.
For example, one may check that for $f=T_{13}\,\mathcal{O}$ and $f=T_{31}\,\mathcal{O}$, there exist no $h,\tilde{h}\in H$ such that \eqref{eq:Sf-consequence-f'} holds for both transformations.
This is consistent with the geometric interpretation: $H$ does not fix any combination of the non-contractible cycles $\lambda$ and $\lambda'$ and therefore there is no invariant embedding of the time circle.

\subsection{Evidence for conjecture}\label{ssec:evidence}

In this section, we prove the modular factorization conjecture for the free chiral multiplet and SQED, and indicate how the proof extends to more general $\mathcal{N}=1$ gauge theories.

\subsubsection{Free chiral multiplet}\label{sssec:proof-chiral}

For convenience, let us recall the relevant indices from Section \ref{sssec:example-free-chiral}:
\begin{align}\label{eq:free-chiral-ind-proof}
    \begin{split}
        Z_{\mathcal{O}}(\boldsymbol{\rho})&=\frac{1}{\theta(z;\sigma)}\,,\qquad Z_{S_{23}\mathcal{O}}(\boldsymbol{\rho})=\Gamma\left(z;\tau,\sigma\right)\\
        Z_{g_{(p,1)}\mathcal{O}}(\boldsymbol{\rho})&=\Gamma\left(z+\sigma;\tau,\sigma\right)\Gamma\left(z;p\tau-\sigma,\tau\right)\,,
    \end{split}
\end{align}
where we have taken the R-charge to be vanishing for notational convenience.\footnote{We will indicate below how the formulae generalize to $R\in 2\mathbb{Z}$.}
Furthermore, the $D_2\times T^2$ partition functions associated with Dirichlet and Robin-like boundary conditions are given by:
\begin{align}\label{eq:hol-blocks-bc-free-chiral-2}
    \begin{split}
        B(\boldsymbol{\rho})&= \Gamma(z+\sigma;\tau,\sigma) \qquad \textrm{and}\qquad         C(\boldsymbol{\rho})= \Gamma(z;\tau,\sigma)\,,%
    \end{split}
\end{align}
which satisfy:
\begin{equation}
          B(\boldsymbol{\rho})=\theta(z;\tau)\, C(\boldsymbol{\rho})\,.  
\end{equation}

Let us start with the superconformal index $Z_{S_{23}\mathcal{O}}(\boldsymbol{\rho})$.
The most general modular property involving three elliptic $\Gamma$ functions is derived in Appendix \ref{sapp:3-Gamma}.
It can be written as:
\begin{align}\label{eq:gen-mod-prop-3Gamma}
    \begin{split}
        \Gamma(z;\tau,\sigma)&=e^{-i\pi Q_{\mathbf{m}}(z;\tau,\sigma)}\Gamma\left(\tfrac{z}{m\sigma+n};\tfrac{\tau-\tilde{n}(k\sigma+l)}{m\sigma+n},\tfrac{k\sigma+l}{m\sigma+n}\right)\Gamma\left(\tfrac{z}{m\tau+\tilde{n}};\tfrac{\sigma-n(\tilde{k}\tau+\tilde{l})}{m\tau+\tilde{n}},\tfrac{\tilde{k}\tau+\tilde{l}}{m\tau+\tilde{n}}\right)\,,
    \end{split}
\end{align}
where $\mathbf{m}=(m,n,\tilde{n})$, $kn-lm=1$ and $\tilde{k}\tilde{n}-\tilde{l}m=1$, and the phase prefactor can be written as:
\begin{equation}\label{eq:phase-gen-S23}
	Q_{\mathbf{m}}(z;\tau,\sigma)=\tfrac{1}{m}Q(mz;m\tau+\tilde{n},m\sigma+n)+f_{\mathbf{m}}\,,
\end{equation}
with $f_{\mathbf{m}}$ a constant independent of $\boldsymbol{\rho}$:
\begin{equation}
    f_{\mathbf{m}} \equiv 2 \sigma_1(n,\tilde{n},1;m)\,,
\end{equation}
where $\sigma_1(n,\tilde{n},1;m)$ denotes a Fourier--Dedekind sum defined in \eqref{eq:definition-sigmat}, and can be written in terms of an ordinary Dedekind sum $s(n,m)$ for $\tilde{n}=n$:\footnote{For details of this derivation, we refer again to Appendix \ref{sapp:3-Gamma}.}
\begin{equation}\label{eq:fm-dedekind}
    f_{\mathbf{m}} = 2s(n,m) - \frac{(m-1)(m-5)}{12m}\,.
\end{equation}
This Dedekind sum was also observed to play a role in the subleading order of generalized Cardy-like limits of the index \cite{ArabiArdehali:2021nsx}.
We now note that the transformed variables on the right hand side precisely take the form $h\boldsymbol{\rho}$ and $\tilde{h}\tilde{\boldsymbol{\rho}}$ for the solutions $(h,\tilde{h})\in S_{S_{23}\mathcal{O}}$ given in \eqref{eq:h-ht-S3xS1}, and with $\tilde{\boldsymbol{\rho}}=\mathcal{O}S_{23}^{-1}\boldsymbol{\rho}$.
It follows that the modular property matches with the prediction of modular factorization:
\begin{align}
    \begin{split}
        Z_{S_{23}\mathcal{O}}(\boldsymbol{\rho})&=e^{-i\pi P^{\textbf{m}}_{S_{23}}\left(\boldsymbol{\rho}\right)}B_{h}(\boldsymbol{\rho})B_{\tilde{h}}(\mathcal{O}S_{23}^{-1}\boldsymbol{\rho})\\
        &=e^{-i\pi \tilde{P}^{\textbf{m}}_{S_{23}}\left(\boldsymbol{\rho}\right)}C_{h}(\boldsymbol{\rho})C_{\tilde{h}}(\mathcal{O}S_{23}^{-1}\boldsymbol{\rho})\,,
    \end{split}
\end{align}
where we recall that $B_h(\boldsymbol{\rho})\equiv B(h\boldsymbol{\rho})$ and similarly for $C_h(\boldsymbol{\rho})$, and define the phases in terms of $P_{S_{23}}(\boldsymbol{\rho};R)$ (see \eqref{eq:PS23-chiral}), which is:
\begin{equation}
    P_{S_{23}}(z;\tau,\sigma;R)=Q(z+\tfrac{R}{2}(\tau+\sigma)-1;\tau,\sigma)\,,
\end{equation}
as follows:
\begin{align}\label{eq:PS23-chiral-2}
    \begin{split}
P^{\textbf{m}}_{S_{23}}\left(z;\tau,\sigma\right)&=\tfrac{1}{m}P_{S_{23}}\left(mz;m\tau+\tilde{n},m\sigma+n;0 \right)+f_\mathbf{m}'\,,\\
       \tilde{P}^{\textbf{m}}_{S_{23}}\left(z;\tau,\sigma\right)&=\tfrac{1}{m}P_{S_{23}}\left(mz+1;m\tau+\tilde{n},m\sigma+n;0 \right)+f_{\mathbf{m}}\,,
    \end{split}
\end{align}
where $f_\mathbf{m}'$ is a constant involving Gauss floor function, which might be determined similarly to $f_{\mathbf{m}}$, as discussed in Appendix \ref{sapp:3-Gamma}.
However, in Appendix \ref{sapp:3-Gamma} we only provide the analytic expression for $f_{\mathbf{m}}$ while the analytic form of $f_\mathbf{m}'$  is still less clear. 
The equality between the two factorizations, up to a change in phase, follows from the relation between $B(\boldsymbol{\rho})$ and $C(\boldsymbol{\rho})$ and the fact that the pair $(h,\tilde{h})\in S_{S_{23}\mathcal{O}}$ solves the consistency condition \eqref{eq:consistency-cond-h}.
We also note that the phase prefactor still clearly encodes the 't Hooft anomalies, since it is related to $P_{S_{23}}(z;\tau,\sigma;R)$ up to a change of variables.
We do not have an interpretation of the additional constants, but note that the Dedekind sum for the case $\tilde{n}=n$ also appeared in the subleading parts of the Cardy-like limits studied in \cite{ArabiArdehali:2021nsx}.

Let us briefly comment on the generalization to $R\in 2\mathbb{Z}$.
Using the formulae for the index in \eqref{eq:lens-indices-chiral} and the $D_2\times T^2$ partition functions in \eqref{eq:hol-blocks-bc-free-chiral}, the factorization would follow from a property of the form:
\begin{align}\label{eq:ZS23-hol-blocks-gen}
    \begin{split}
        &\Gamma(z+\tfrac{R}{2}(\tau+\sigma);\tau,\sigma)=e^{-i\pi Q_{\mathbf{m}}(z;\tau,\sigma;R)}\\
        &\times\Gamma\left(\tfrac{z+\frac{R}{2}(\tau-\tilde{n}(k\sigma+l))}{m\sigma+n};\tfrac{\tau-\tilde{n}(k\sigma+l)}{m\sigma+n},\tfrac{k\sigma+l}{m\sigma+n}\right)\Gamma\left(\tfrac{z+\frac{R}{2}(\sigma-n(\tilde{k}\tau+\tilde{l}))}{m\tau+\tilde{n}};\tfrac{\sigma-n(\tilde{k}\tau+\tilde{l})}{m\tau+\tilde{n}},\tfrac{\tilde{k}\tau+\tilde{l}}{m\tau+\tilde{n}}\right)\,,
    \end{split}
\end{align}
for some $Q_{\mathbf{m}}(z;\tau,\sigma;R)$.
This formula reduces to \eqref{eq:gen-mod-prop-3Gamma} through repeated use of the shift property \eqref{eq:basic-shift-gamma-app} of elliptic $\Gamma$ functions and the elliptic and modular property in \eqref{eq:elliptic-theta} and \eqref{eq:result} of the $q$-$\theta$ function.
In the process, one picks up phases, which define $Q_{\mathbf{m}}(z;\tau,\sigma;R)$ in terms of $Q_{\mathbf{m}}(z;\tau,\sigma)$.
Note that $R\in 2\mathbb{Z}$ is crucial for this to work.

We now turn to the lens index $Z_{g_{(p,1)}\mathcal{O}}(\boldsymbol{\rho})$.
In this case, we need a modular property that involves four elliptic $\Gamma$ functions, as derived in Appendix \ref{sapp:4-Gamma}:
\begin{align}\label{eq:holomorphic-block-lensLp1}
    \begin{split}
        \Gamma\left(z+\sigma;\tau,\sigma\right)&\Gamma\left(z;p\tau-\sigma,\tau\right)=e^{-i\pi Q_{\mathbf{m}_p}(z;\tau,\sigma)}\Gamma\left(\tfrac{z}{m\sigma+n_1};\tfrac{\tau-c(k_1\sigma+l_1)}{m\sigma+n_1},\tfrac{k_1\sigma+l_{1}}{m\sigma+n_1}\right)\\
        &\times \Gamma\left(\tfrac{z}{m(p\tau-\sigma)+\tilde{n}_2};\tfrac{\tau-c(\tilde{k}_2(p\tau-\sigma)+\tilde{l}_2)}{m(p\tau-\sigma)+\tilde{n}_2},\tfrac{\tilde{k}_2(p\tau-\sigma)+\tilde{l}_2}{m(p\tau-\sigma)+\tilde{n}_2}\right)\,,
    \end{split}
\end{align}
where $k_1n_1-l_1m=1$, $\tilde{k}_2\tilde{n}_2-\tilde{l}_2m=1$, $\tilde{n}_2=-n_1+pc$, $c\in \mathbb{Z}$ is a free integer and $\mathbf{m}_p=(m,n_1,c;p)$.
Furthermore, the phase is given by: 
\begin{align}
    \begin{split}
        Q_{\mathbf{m}_p}(z;\tau,\sigma)
        &= \tfrac{1}{m p} Q\left(m z, \tfrac{m(p\tau-\sigma)+\tilde{n}_2}{p}, \tfrac{m\sigma+n_1}{p} \right) + \tfrac{p^2-1}{12p}(2z-\tau) + f_{\mathbf{m}_{p}}
    \end{split}
\end{align}
with $f_{\mathbf{m}_p}$ a constant.
Setting $(n_1,k_1,l_1)\equiv (n,k,l)$ and $(\tilde{n}_2,\tilde{k}_2,\tilde{l}_2)\equiv (\tilde{n},\tilde{k},\tilde{l})$, we compare the transformed variables on the right hand side to $h\boldsymbol{\rho}$ and $\tilde{h}\tilde{\boldsymbol{\rho}}$ with $(h,\tilde{h})\in S_{g_{(p,1)}\mathcal{O}}$ given in \eqref{eq:h-ht-Lens} and $\tilde{\boldsymbol{\rho}}=\mathcal{O}g_{(p,1)}^{-1}\boldsymbol{\rho}$.
Sure enough, this property again precisely matches the factorization conjecture for the lens index of the free chiral multiplet:
\begin{align}
    \begin{split}
        Z_{g_{(p,1)}\mathcal{O}}(\boldsymbol{\rho})&=e^{-i\pi P^{\mathbf{m}_p}_{g_{(p,1)}}\left(\boldsymbol{\rho}\right)}B_{h}(\boldsymbol{\rho})B_{\tilde{h}}(\mathcal{O}g_{(p,1)}^{-1}\boldsymbol{\rho})\\
        &=e^{-i\pi \tilde{P}^{\mathbf{m}_p}_{g_{(p,1)}}\left(\boldsymbol{\rho}\right)}C_{h}(\boldsymbol{\rho})C_{\tilde{h}}(\mathcal{O}g_{(p,1)}^{-1}\boldsymbol{\rho})\,,
    \end{split}
\end{align}
where the phases can now be defined in terms of $P_{g_{(p,1)}}(\boldsymbol{\rho};R)$:
\begin{equation}
    P_{g_{(p,1)}}(\boldsymbol{\rho};R)= \tfrac{1}{p} Q\left(z+\tfrac{R}{2}\tau-1,\tfrac{p\tau-\sigma}{p}, \tfrac{\sigma}{p}\right) + \tfrac{p^2-1}{12p}(2z-1+(R-1)\tau)  \,,
\end{equation}
which we have rewritten as compared to its first appearance in \eqref{eq:Pgp-chiral}.
In terms of this function, the phases are defined as follows: 
\begin{align}
    \begin{split}
        P^{\mathbf{m}_p}_{g_{(p,1)}}\left(z;\tau,\sigma\right)&=\tfrac{1}{m}P_{g_{(p,1)}}\left(mz;m\tau+c,m\sigma+n_1;0\right)+f_{\mathbf{m}_p}\,,\\
        \tilde{P}^{\mathbf{m}_p}_{g_{(p,1)}}\left(z;\tau,\sigma\right)&=\tfrac{1}{m}P_{g_{(p,1)}}\left(mz+1;m\tau+c,m\sigma+n_1;0\right)+f_{\mathbf{m}_p}'\,.
    \end{split}
\end{align}
The constants $f_{\mathbf{m}_p}$ and $f_{\mathbf{m}_p}'$ are again distinct.
It follows that the overall phase captures again the anomalies, since it is related to $P_{g_{(p,1)}}(\boldsymbol{\rho};R)$ through a change of variables.
The extension to $R\in 2\mathbb{Z}$ follows along similar lines as in the case of the $S^3\times S^1$ index.

Finally, we verify our factorization conjecture for the $S^2\times T^2$ index of the free chiral multiplet.
As in Section \ref{sssec:example-free-chiral}, we first note that:
\begin{equation}\label{eq:ZO-van-R}
    Z_{\mathcal{O}}(\boldsymbol{\rho})=\frac{1}{\theta(z;\sigma)}=\Gamma(z;\tau,\sigma)\Gamma(z;-\tau,\sigma)\,,
\end{equation}
in the case of vanishing R-charge.
We now make use of the general modular property of the $q$-$\theta$ function:
\begin{align}
    \begin{split}
    \theta\left(\tfrac{z}{ m\sigma+n};\tfrac{k\sigma+l}{m\sigma+n} \right)  &= e^{i\pi B^{\textbf{m}}_2(z;\sigma)}\theta(z;\sigma)  \,.
    \end{split}
\end{align}
Using this transformation, we find:
\begin{equation}\label{eq:ZO-gen-fact}
    \frac{1}{\theta(z;\sigma)}=e^{-i\pi B^{\textbf{m}}_2(z;\sigma)}\Gamma\left(\tfrac{z}{m\sigma+n};\tfrac{\tau+a\sigma+b}{m\sigma+n},\tfrac{k\sigma+l}{m\sigma+n}\right)\Gamma\left(\tfrac{z}{m\sigma+n};-\tfrac{\tau+a\sigma+b}{m\sigma+n},\tfrac{k\sigma+l}{m\sigma+n}\right)\,,
\end{equation}
We note that the left hand side of \eqref{eq:ZO-van-R} is independent of $\tau$ because of the vanishing R-charge.
In general, however, it will depend on $\tau$ (cf.\  \eqref{eq:lens-indices-chiral}).
This constrains the form of the second variable of the $\Gamma$ functions on the right hand side of \eqref{eq:ZO-gen-fact} as indicated.
Comparing with the factorization conjecture for $Z_\mathcal{O}(\boldsymbol{\rho})$, with $(h,\tilde{h})\in S_{\mathcal{O}}$ as in \eqref{eq:h-ht-S2xT2}, we see again that the property above exactly matches with the conjecture:
\begin{align}
    \begin{split}
        Z_{\mathcal{O}}(\boldsymbol{\rho})&=e^{-i\pi P^{\textbf{m}}_{1}(\boldsymbol{\rho})}B_h(\boldsymbol{\rho})B_{\tilde{h}}(\mathcal{O}\boldsymbol{\rho})\\
        &=e^{-i\pi \tilde{P}^{\textbf{m}}_{1}(\boldsymbol{\rho})}C_h(\boldsymbol{\rho})C_{\tilde{h}}(\mathcal{O}\boldsymbol{\rho})
    \end{split}
\end{align}
where the phases are given by:
\begin{align}
    \begin{split}
        P^{\textbf{m}}_{1}(\boldsymbol{\rho})&=B^{\textbf{m}}_2(z-\tfrac{1}{m};\sigma) +\tfrac{1-m}{m} \,,\\
        \tilde{P}^{\textbf{m}}_{1}(\boldsymbol{\rho})&=B^{\textbf{m}}_2(z;\sigma)\,.
    \end{split}
\end{align}
These phases can again be understood as changes of variable of the basic case $ P_{t_2^{\mathbf{n}}}(\boldsymbol{\rho};R)$, discussed in \eqref{eq:P1-chiral}, for $R=\mathbf{n}=0$, which is the anomaly polynomial for the $(0,2)$ theory associated with the twisted reduction of the chiral multiplet on $S^2$.
The extension to general R-charge (and flavor symmetry fluxes) is straightforward and left to the reader.

We thus see that the modular properties of the elliptic $\Gamma$ functions precisely match with our modular factorization conjecture in the context of the free chiral multiplet.
We find this agreement between general physical arguments and mathematically rigorous properties of the elliptic $\Gamma$ function remarkable.

\subsubsection{SQED}\label{sssec:proof-sqed}

In this section, we show how modular factorization can also be proved for a non-trivial gauge theory.
As in Section \ref{sssec:sqed}, we focus on the superconformal index of SQED for simplicity.
To this end, we recall the Higgs branch expression \eqref{eq:lens-higgs-form-sqed} for SQED index:
\begin{equation}
    I^{\mathrm{SQED}}_{(1,0)}(\hat{\boldsymbol{\rho}})=Z^{\mathrm{SQED}}_{S_{23}\mathcal{O}}(\boldsymbol{\rho})=\sum^{N_f}_{\alpha=1} Z^{\alpha}_{S_{23}\mathcal{O},\text{cl}}(\boldsymbol{\rho})Z^{\alpha}_{S_{23}\mathcal{O},\text{1-loop}}(\boldsymbol{\rho})Z^{\alpha}_{\text{v}}(\boldsymbol{\rho})Z^{\alpha}_{\text{v}}(\mathcal{O}S_{23}^{-1} \boldsymbol{\rho})\,,
\end{equation}
where the factors in the summand can be written as:
\begin{align}
    \begin{split}
        Z^{\alpha}_{S_{23}\mathcal{O},\text{cl}}(\boldsymbol{\rho})&= x^{\xi}_{\alpha}(pq)^{\xi/2}\,, \\
        Z^{\alpha}_{S_{23}\mathcal{O},\text{1-loop}}(\boldsymbol{\rho})&= \frac{1}{\Gamma(0;\tau,\sigma)} \prod_{\beta=1}^{N_f} \frac{\Gamma(z_\beta -z_\alpha;\tau,\sigma)}{\Gamma(-z_\beta-z_\alpha;\tau,\sigma)}\,, \\
        Z_{\mathrm{v}}^{\alpha}(\boldsymbol{\rho}) &= \sum_{\kappa=0}^\infty q^{\kappa\xi}\prod_{\beta=1}^{N_f} \prod^{\kappa}_{j=1}\frac{\theta(z_\alpha+z_\beta+j\tau;\sigma) }{\theta(z_\alpha-z_\beta+j\tau;\sigma)}\,,
    \end{split}
\end{align}
Modular factorization can be arrived at as follows.

First of all, we use the invariance of each coefficient in the sum of $Z_{\mathrm{v}}^{\alpha}(\boldsymbol{\rho})$ under $\mathcal{H}=SL(2,\mathbb{Z})_{13}\ltimes \mathbb{Z}^{2+2N_f}$ to write:
\begin{equation}
    \prod_{\beta=1}^{N_f}\prod^\kappa_{j=1} \frac{\theta(z_\alpha+z_\beta+j\tau;\sigma) }{\theta(z_\alpha-z_\beta+j\tau;\sigma)}=\prod_{\beta=1}^{N_f} \prod^\kappa_{j=1}\frac{\theta\left(\frac{z_\alpha+z_\beta+j(\tau-\tilde{n}(k\sigma+l))}{m\sigma+n};\tfrac{k\sigma+l}{m\sigma+n}\right)}{\theta\left(\frac{z_\alpha-z_\beta+j(\tau-\tilde{n}(k\sigma+l))}{m\sigma+n};\tfrac{k\sigma+l}{m\sigma+n}\right)}\,,
\end{equation}
where $kn-lm=1$ as in Section \ref{ssec:consistency-cond} and in this equation $\tilde{n}$ is an integer coprime with $m$ (although this is irrelevant for this specific identity).

Furthermore, we find that for $m\neq 0$ the overall classical contribution can be factorized, up to a prefactor, as (cf. \eqref{eq:FI-S-fact}):
\begin{align}\label{eq:FI-fact-m}
    \begin{split}
         x_{\alpha}^{\xi}=&\,e^{\frac{\pi i}{6m}\left(\frac{m\tau+\tilde{n}}{m\sigma+n}+\frac{m\sigma+n}{m\tau+\tilde{n}}+3-2\sigma_1(k\tilde{n},1;m)\right)}\frac{\theta\left(\frac{z_\alpha}{m\sigma+n};\frac{\tau-\tilde{n}(k\sigma+l)}{m\sigma+n}\right)\theta\left(\frac{(m\tau+\tilde{n})\xi}{m};\frac{\tau-\tilde{n}(k\sigma+l)}{m\sigma+n}\right)}{\theta\left(\frac{z_\alpha}{m\sigma+n}+\frac{(m\tau+\tilde{n})\xi}{m};\frac{\tau-\tilde{n}(k\sigma+l)}{m\sigma+n}\right)}\\
         \times &\, \frac{\theta\left(\frac{z_\alpha}{m\tau+\tilde{n}};\frac{\sigma-n(\tilde{k}\tau+\tilde{l})}{m\tau+\tilde{n}}\right)\theta\left(\frac{(m\sigma+n)\xi}{m};\frac{\sigma-n(\tilde{k}\tau+\tilde{l})}{m\tau+\tilde{n}}\right)}{\theta\left(\frac{z_\alpha}{m\tau+\tilde{n}}+\frac{(m\sigma+n)\xi}{m};\frac{\sigma-n(\tilde{k}\tau+\tilde{l})}{m\tau+\tilde{n}}\right)}\,,
    \end{split}
\end{align}
where now also $\tilde{k}\tilde{n}-\tilde{l}m=1$ and we made use of
\begin{equation}
    \theta\left(\tfrac{z}{m\sigma+n};\tfrac{\tau-\tilde{n}(k\sigma+l)}{m\sigma+n}\right)\theta\left(\tfrac{z}{m\tau+\tilde{n}};\tfrac{\sigma-n(\tilde{k}\tau+\tilde{l})}{m\tau+\tilde{n}}\right)=e^{-\pi i \left(\frac{1}{m}B\left(\frac{mz}{m\sigma+n}-1;\frac{m\tau+\tilde{n}}{m\sigma+n}\right)+2\sigma_1(k\tilde{n},1;m)\right)}\,.
\end{equation}
where the constant $\sigma_1$ is defined in \eqref{eq:def-sigma1}.
This property can be derived from the extension and general modular property of the $q$-$\theta$ function in Appendix \ref{app:defs} and the fact that
\begin{equation}
	\frac{\sigma-n(\tilde{k}\tau+\tilde{l})}{m\tau+\tilde{n}}=\begin{pmatrix}
		-\tilde{k}n & -l-\tilde{l}-l\tilde{l}m \\
		 m& k\tilde{n} 
	\end{pmatrix}\frac{\tau-\tilde{n}(k\sigma+l)}{m\sigma+n}\,,
\end{equation}
where the action of the $SL(2,\mathbb{Z})$ matrix is defined in the usual projective manner.
In particular, note that the transformation exchanges $m\sigma+n\leftrightarrow m\tau+\tilde{n}$.
For completeness, we separately define the $m=0$ and $n=\tilde{n}=1$ case of \eqref{eq:FI-fact-m}
\begin{align}\label{eq:FI-m=0}
    \begin{split}
         x_{\alpha}^{\xi}&=\frac{\theta\left(z_\alpha;\tau-\sigma\right)\theta\left(\tau\xi;\tau-\sigma\right)}{\theta\left(z_\alpha+\tau\xi;\tau-\sigma\right)}\times \frac{\theta\left(z_\alpha;\sigma-\tau\right)\theta\left(\sigma\xi;\sigma-\tau\right)}{\theta\left(z_\alpha+\sigma\xi;\sigma-\tau\right)}\,,
    \end{split}
\end{align}
where we made use of the fact that $\xi$ is quantized as an integer and the shift property \eqref{eq:elliptic-theta} of the $q$-$\theta$ function.

Finally, for the perturbative part we can use the results from Section \ref{sssec:proof-chiral}, specifically \eqref{eq:gen-mod-prop-3Gamma}, to achieve factorization.
We  thus find that the index can be expressed for $m\neq 0$\footnote{Note that only the factorization of the FI term needs to be defined separately when $m=0$, as we have done in \eqref{eq:FI-m=0}.} in the following form
\begin{align}
    \begin{split}
         I^{\mathrm{SQED}}_{(1,0)}(\hat{\boldsymbol{\rho}})&=e^{\frac{\pi i}{6m}\left(\frac{m\tau+\tilde{n}}{m\sigma+n}+\frac{m\sigma+n}{m\tau+\tilde{n}}+3-2\sigma_1(k\tilde{n},1;m)\right)}(pq)^{\xi/2}e^{-i\pi P^{\mathrm{\textbf{m},SQED}}_{S_{23}}(\boldsymbol{\rho})} \\
         &\times\sum_{\alpha=1}^{N_f} \mathcal{B}_{h}^\alpha (\boldsymbol{\rho})\mathcal{B}_{\tilde{h}}^\alpha (\mathcal{O}S_{23}^{-1}\boldsymbol{\rho})\,,\qquad \hat{\boldsymbol{\rho}}=\boldsymbol{\rho}\,.
    \end{split}
\end{align}
In this expression, we used the formal definition of $\mathcal{B}^{\alpha}(\boldsymbol{\rho})$ in \eqref{eq:defn-hol-block-sqed}, but now with a more general action of $(h,\tilde{h})\in S_{S_{23}\mathcal{O}}$ as in \eqref{eq:h-ht-S3xS1}:
\begin{align}
    \begin{split}
        \mathcal{B}^{\alpha}_h(\boldsymbol{\rho})\equiv \mathcal{B}^{\alpha}(h\boldsymbol{\rho})=\mathcal{B}^{\alpha}\left(\tfrac{\vec{z}}{m\sigma+n};\tfrac{\tau-\tilde{n}(k\sigma+l)}{m\sigma+n},\tfrac{k\sigma+l}{m\sigma+n};\tfrac{(m\tau+\tilde{n})\xi}{m}\right)\,,
    \end{split}
\end{align}
where we recall from Section \ref{sssec:sqed} that for SQED we extended $\boldsymbol{\rho}$ to include the FI parameter as $\boldsymbol{\rho}=(\vec{z};\tau,\sigma;\tau\xi)$.
Similarly, since the action of $\mathcal{O}S_{23}^{-1}$ on $\boldsymbol{\rho}$ exchanges $\tau\leftrightarrow\sigma$ we have:
\begin{equation}
    \mathcal{B}_{\tilde{h}}^\alpha (\mathcal{O}S_{23}^{-1}\boldsymbol{\rho})=\mathcal{B}^{\alpha}\left(\tfrac{\vec{z}}{m\tau+\tilde{n}};\tfrac{\sigma-n(\tilde{k}\tau+\tilde{l})}{m\tau+\tilde{n}},\tfrac{\tilde{k}\tau+\tilde{l}}{m\tau+\tilde{n}};\tfrac{(m\sigma+n)\xi}{m}\right)
\end{equation}
It may seem as if $\tau\xi$ transforms non-trivially under $h$, in which case modular factorization actually does not hold.
However, this is only apparent since
\begin{align}
    \begin{split}
        \tau \xi &\xrightarrow{h}\frac{(m\tau+\tilde{n})\xi}{m}=\tau \xi\mod 1\,,
    \end{split}
\end{align}
as long as we assume that $\xi$ is quantized as a multiple of $m$.
This should be closely related to the fact that for $L(p,1)\times S^1$ indices $\xi$ is quantized as a multiple of $p$ \cite{Nieri:2015yia} and the relation of the $m\neq 1$ modular properties to lens-like quotients \cite{Ardehali:2021irq,Jejjala:2021hlt}.\footnote{This connection will be substantiated in upcoming work \cite{Fujiwara:2023}.} 
Furthermore, we note that since $(1,1)\notin S_{S_{23}\mathcal{O}}$, we never run into the issue that the formal expression vanishes \eqref{eq:defn-hol-block-sqed} given the invariance of $\tau\xi$.

Finally, the phase polynomial is given by:
\begin{equation}
    P^{\mathrm{SQED}}_{S_{23}}(\boldsymbol{\rho})=-\tilde{P}^{\textbf{m}}_{S_{23}}\left(0;\tau,\sigma\right)+\sum_{\beta=1}^{N_f}\Big[\tilde{P}^{\textbf{m}}_{S_{23}}(z_{\beta}-z_{\alpha};\tau,\sigma)-\tilde{P}^{\textbf{m}}_{S_{23}}(-z_{\beta}-z_{\alpha};\tau,\sigma)\Big] \,,
\end{equation}
with $\tilde{P}^{\textbf{m}}_{S_{23}}(z;\tau,\sigma)$ defined in \eqref{eq:PS23-chiral-2}.
It follows from the discussions in Section \ref{sssec:sqed} and Section \ref{sssec:proof-chiral} that this polynomial captures the 't Hooft anomalies of SQED.

In conclusion, we see that, up to phase prefactors, the SQED index can be factorized in a variety of ways precisely matching with our modular factorization conjecture.

\subsubsection{General gauge theories}\label{sssec:proof-gen-gauge-th}

Let us briefly indicate how our proof can be extended to more general $\mathcal{N}=1$ gauge theories.
As observed in Section \ref{sssec:gen-gauge-th} in the context of the factorization in terms of $\mathcal{B}^{\alpha}_{S}(\boldsymbol{\rho})$, there were five main ingredients that go into the factorization of a lens index.
The last four points require an update in the context of the more general factorization properties:
\begin{itemize}
    \item Invariance of the vortex partition function at fixed vortex charge under the full group $H=SL(2,\mathbb{Z})_{13}\ltimes \mathbb{Z}^2$.
    \item Modular factorization of the FI term, if present.
    \item Modular factorization of the free chiral multiplet index.
    \item Independence of the phase polynomial of the Higgs branch vacuum $\alpha$.
\end{itemize}
The first point requires full modular invariance and ellipticity of the vortex partition function.
This should follow from the same arguments mentioned in \ref{sssec:gen-gauge-th}, namely that 't Hooft anomalies for global symmetries on the vortex worldsheet should vanish. 
The second and third point we have demonstrated in Sections \ref{sssec:proof-chiral} and \ref{sssec:proof-sqed}.
Finally, the fourth point also follows from the arguments made in \ref{sssec:gen-gauge-th}, since the phase polynomials for the general factorization properties are simply changes of variables of the phases there.
In particular, independence of $\alpha$ follows from gauge anomaly cancellation.
All in all, we believe that this provides strong evidence of modular factorization for more general $\mathcal{N}=1$ gauge theories, as long as a Higgs branch expression for the index exists.

\section{\texorpdfstring{$\bm{SL(3,\mathbb{Z})}$}{SL(3,Z)} one-cocycle condition and lens indices}\label{sec:gen-modularity}

In this section, we begin with a gentle introduction to the mathematical framework of group cohomology, geared towards a generalization of the notion of a modular form.
The insight of Gadde \cite{Gadde:2020bov} relates lens indices to a non-trivial cohomology class in $H^1(\mathcal{G},N/M)$ with $\mathcal{G}=SL(3,\mathbb{Z})\ltimes \mathbb{Z}^{3r}$.
We establish this connection systematically and rigorously, assuming the conjectured properties of lens indices in Section \ref{sec:mod-fac-lens-indices}.
Our approach supplies a physical interpretation of the fact that lens indices are related to a non-trivial cohomology class.
Finally, we briefly comment on a cohomological perspective on the modular factorization of lens indices.

\subsection{Modular group cohomology}\label{ssec:mod-group-cohomology}

For concreteness, let us consider the Jacobi group $\mathcal{J}=SL(2,\mathbb{Z})\ltimes \mathbb{Z}^2$.
This group acts by large diffeomorphisms on the complex structure $\tau$ of a torus and by large gauge transformations on a line bundle modulus $z$:
\begin{equation}\label{eq:action-J-chem-pots}
    (z;\tau)\to \left(\frac{z}{m\tau+n};\frac{k\tau+l}{m\tau+n}\right)\,,\quad  (z;\tau)\to \left(z+a\tau+b;\tau\right)\,.
\end{equation}
A (weight $0$) automorphic form with respect to $\mathcal{J}$ transforms as follows\footnote{The elliptic genera of two-dimensional SCFTs are examples of such forms (see, e.g., \cite{Kawai:1993jk}).}:
\begin{align}\label{eq:defn-autom-form}
	\chi\left(z;\tau\right)&=\phi_g(z;\tau)\chi(g^{-1}(z;\tau))\,,
\end{align}
where $\phi_g(z;\tau)$ is a pure phase and known as the automorphic factor. 
It satisfies:
\begin{equation}\label{eq:group-homom}
    \phi_{g_1g_2}(z;\tau)=\phi_{g_1}(z;\tau)\phi_{g_2}(g_1^{-1}\cdot(z;\tau))\,.
\end{equation}
In general, $\chi(z;\tau)$ is a meromorphic function of $(z;\tau)$.
Following~\cite{Felder_2000} (see also~\cite{Gadde:2020bov}),  we will now set up a general framework that allows us to identify such automorphic forms as elements of the zeroth group cohomology $H^0(\mathcal{J},N/M)$ of the Jacobi group $\mathcal{J}$, valued in $N/M$ to be defined momentarily.

To this end, let us look at the group of $k$-cochains $C^{k}(G,A)$, where $G$ will be a modular group such as $\mathcal{J}$ and with $A$ a multiplicative abelian group of functions.
The group $C^{k}(G,A)$ consists of $k$-cocycles $\alpha:G^k\to A$ such that $\alpha_{g_1,\ldots,g_k}=1$ if $g_j=1$ for some $j$.
Furthermore, one defines $C^{0}(G,A)=A$.
Let us denote by $Y$ a complex manifold endowed with an action of $G$. 
We parametrize the manifold in terms of a set of complex variables collectively denoted by $\boldsymbol{\rho}$.
For example, when $G=\mathcal{J}$ we take $\boldsymbol{\rho}=(z;\tau)$ with $z\in \mathbb{C}$ and $\tau\in \mathbb{H}$ and the action is as in \eqref{eq:action-J-chem-pots}.
The other relevant example will be $G=\mathcal{G}$, discussed in Section \ref{ssec:hopf-surfaces}, in which case we have $\boldsymbol{\rho}=(z;\tau,\sigma)$ with $z\in \mathbb{C}$ and $\tau,\sigma\in \mathbb{H}$, and the action is as in \eqref{eq:calG-action}.
In addition, $A$ will take three concrete forms: the group of meromorphic functions $N$ on $Y$, the group of holomorphic, nowhere vanishing functions $M$ on $Y$, or the quotient $N/M$. 
Note that $M$ is nothing but the set of (holomorphic) phases.
These three groups fit into a short exact sequence:
\begin{equation}\label{eq:short-exact-seq}
    1\to M\to N\to N/M \to 1\,.
\end{equation}
The $G$-action on $Y$ induces an action on $\alpha_{g_1,\ldots,g_k}(\boldsymbol{\rho})\in A$ as follows:
\begin{equation}
    g\cdot \alpha_{g_1,\ldots,g_k}(\boldsymbol{\rho})\equiv \alpha_{g_1,\ldots,g_k}(g^{-1}\boldsymbol{\rho})\,.
\end{equation}

\begin{figure}[t]
     \centering 
     \begin{subfigure}[t]{0.49\textwidth}
         \centering
         \begin{tikzpicture}[scale=0.7]
            \begin{scope}[decoration={                        markings,mark=at position 0.57 with {\arrow{Latex[scale=1.15]}}}] 
            \draw[{Circle[]}-{Circle[]},postaction={decorate}] (0,0) -- node[below=1.5mm] {$g$} (2,0);
            \end{scope}
        \end{tikzpicture}
        \caption{$(\delta \alpha)_g(\boldsymbol{\rho})=\frac{\alpha(\boldsymbol{\rho})}{\alpha(g^{-1}\boldsymbol{\rho})}$}
         \label{fig:alpha0}
     \end{subfigure}
     \hfill
     \begin{subfigure}[t]{0.49\textwidth}
         \centering
         \begin{tikzpicture}[scale=0.7]
            \begin{scope}[decoration={                        markings,mark=at position 0.57 with {\arrow{Latex[scale=1.15]}}}]
            \draw [{Circle[]}-{Circle[]},postaction={decorate}] (0,0) -- node[below=1.5mm] {$g_1$} (2,0) ;
            \draw [{}-{Circle[]},postaction={decorate}] (2,0) -- node[right=1.5mm] {$g_2$} (1,{sqrt(3)});
            \draw [postaction={decorate}] (0,0) -- node[left=1.5mm] {$g_1g_2$} (1,{sqrt(3)})  ;
            \end{scope}
        \end{tikzpicture}
        \caption{$(\delta\alpha)_{g_1,g_2}(\boldsymbol{\rho})=\frac{\alpha_{g_1}(\boldsymbol{\rho})\alpha_{g_2}(g_{1}^{-1}\boldsymbol{\rho})}{\alpha_{g_1g_2}(\boldsymbol{\rho})}$}
         \label{fig:alpha1}
     \end{subfigure}
     \par
     \begin{subfigure}[t]{1\textwidth}
         \centering
         \begin{tikzpicture}[scale=0.9]
            \begin{scope}[decoration={                        markings,mark=at position 0.5 with {\arrow{Latex[scale=1.15]}}}]
            \draw [{Circle[]}-{Circle[]},postaction={decorate}] (0,0) -- node[below=1.5mm] {$g_1$} (4,0) ;
            \draw [{}-{Circle[]},postaction={decorate}] (4,0) -- node[right=1.5mm] {$g_2$} (2,{sqrt(12)});
            \draw [postaction={decorate}] (0,0) -- node[left=1.5mm] {$g_1g_2$}  (2,{sqrt(12)});
            \draw[dashed,{}-{Circle[]},postaction={decorate}] (2,{sqrt(12)}) -- node[left=0.5mm] {$g_3$} (2,{0.3*sqrt(12)}) ;
            \draw[dashed,postaction={decorate}] (0,0) -- node[above=1mm] {$g_1g_2g_3$} (2,{0.3*sqrt(12)}) ;
            \draw[dashed,postaction={decorate}] (4,0) -- node[above=1mm] {$g_2g_3$} (2,{0.3*sqrt(12)}) ;
            \end{scope}
        \end{tikzpicture}
        \caption{$(\delta\alpha)_{g_1,g_2,g_3}(\boldsymbol{\rho})=\frac{\alpha_{g_1,g_2}(\boldsymbol{\rho})\alpha_{g_1g_2,g_3}(\boldsymbol{\rho})}{\alpha_{g_1,g_2g_3}(\boldsymbol{\rho})\alpha_{g_2,g_3}(g_1^{-1}\boldsymbol{\rho})}$}
         \label{fig:alpha2}
     \end{subfigure}
        \caption{In the graphical representation, the right hand side of the equation reflects the boundary components of the left hand side. The arrows indicate the orientation of the boundary components and determine whether $\alpha_{g_1,\ldots,g_k}(\boldsymbol{\rho})$ ends up in the numerator or denominator.
        In the above, the arrows are directed such that there is a single point/edge/face not anchored to the same point as all other points/edges/faces. This is reflected by the action of $g_{1}$ on the relevant $\alpha_{g_1,\ldots,g_k}(\boldsymbol{\rho})$.
        An arrow can be flipped at the expense of inverting the group element, its orientation and anchoring point: $\alpha_{g_1,\ldots,g_k}(\boldsymbol{\rho})=1/\alpha_{g_1^{-1},g_1g_2,\ldots,g_1g_k}(g_1^{-1}\boldsymbol{\rho})$.}
        \label{fig:group-coh}
\end{figure}

\noindent To construct the cohomology groups, one defines a coboundary operator $\delta\equiv\delta_k$ from $C^{k}(G,A)\to C^{k+1}(G,A)$ via:
\begin{align}
    \begin{split}
        &(\delta\alpha)_{g_1,\ldots,g_{k+1}}(\boldsymbol{\rho})=\alpha_{g_1,\ldots,g_k}(\boldsymbol{\rho})\\
    &\left(\alpha_{g_2,\ldots,g_{k+1}}(g_{1}^{-1}\boldsymbol{\rho})\prod_{j=1}^{k}\alpha_{g_1,\ldots,g_jg_{j+1},\ldots,g_{k+1}}(\boldsymbol{\rho})^{(-1)^{j}}\right)^{(-1)^{k+1}}\,,
    \end{split}
\end{align}
where we recall that $A$ is defined multiplicatively.
Furthermore, for $k=0$:
\begin{equation}
    (\delta \alpha)_{g}(\boldsymbol{\rho})=\frac{\alpha(\boldsymbol{\rho})}{\alpha(g^{-1}\boldsymbol{\rho})}\,.
\end{equation}
We have illustrated this equation in Figure \ref{fig:group-coh} for $k=0,1,2$.
One may verify that $\delta^2=1$, as appropriate for a coboundary operator.
We can thus define cohomology groups as:
\begin{equation}
    H^{k}(G,A)=\frac{\ker \delta_{k}}{\text{im}\; \delta_{k-1}}\, ,\, k\geq 1\,,\quad H^{0}(G,A)=\ker \delta_{0}\,.
\end{equation}
Having set up the basic language, we can now describe the property \eqref{eq:defn-autom-form}.
Consider an element $\chi\in C^{0}(\mathcal{J},N)=N$.
This function is a weight $0$ automorphic form if:
\begin{equation}\label{eq:autom-form-defn}
    (\delta \chi)_g(\boldsymbol{\rho})=\frac{\chi(\boldsymbol{\rho})}{\chi(g^{-1}\boldsymbol{\rho})}=\xi_g(\boldsymbol{\rho})\,,\qquad \xi_g(\boldsymbol{\rho})\in C^1(G,M)\,,
\end{equation}
with $\xi_g$ the factor of automorphy.
Note that the cohomological structure automatically ensures the condition \eqref{eq:group-homom}:
\begin{equation}
    (\delta \xi)_{g_1,g_2} =1\quad \Rightarrow \quad \frac{\xi_{g_1} (\boldsymbol{\rho})\xi_{g_2} (g^{-1}_1\boldsymbol{\rho})}{\xi_{g_1g_2} (\boldsymbol{\rho})}=1\,.
\end{equation}
It follows that such a $\chi$ can be thought of as an element in $H^0(\mathcal{J},N/M)$, since it is annihilated by $\delta$ modulo $M$.
This brings us to the claim at the beginning of this section.

The equivalence class $[\chi]$ of an automorphic form $\chi$ modulo $M$ corresponds to a cohomology class in $H^1(\mathcal{J},M)$.
To see this, consider the product of the automorphic form $\chi$ with a phase $\phi\in M$.
Let $\xi_g=(\delta\chi)_g$ and $\psi_g=(\delta(\chi \phi))_g$.
Since $\psi_g=\xi_g (\delta \phi)_g$, it follows that $\xi_g$ and $\psi_g$ sit in the same cohomology class in $H^1(\mathcal{J},M)$.
As in \cite{Felder_2000}, we call this map from $H^0(\mathcal{J},N/M)$ to $H^1(\mathcal{J},M)$: $\delta_*[\chi]\equiv[\delta\chi]$.

Note that the definition of $\delta_*$ generalizes to higher degree.
Together with the short exact sequence~\eqref{eq:short-exact-seq}, it induces a long exact sequence:
\begin{equation}
   \cdots  H^{k-1}(G,N/M)  \xrightarrow{\delta_*} H^k(G,M) \xrightarrow{i_*} H^k(G,N)\xrightarrow{p_*} H^k(G,N/M) \xrightarrow{\delta_*} H^{k+1}(G,M) \cdots \,,
\end{equation}
where $i_*$ and $p_*$ lift the inclusion and projection of the short exact sequence to the cohomology groups.
For exactness at the node $H^k(G,M)$, we first note that the trivial class in $H^k(G,N)$ is of the form $\delta C^{k-1}(G,N)$.
Therefore, $\ker \, i_*$ contains all cocycles $[\xi_{g_1,\ldots,g_{k}}]\in H^k(G,M)$ that can be written as: 
\begin{equation}
    [\xi_{g_1,\ldots,g_{k}}]=[\delta(\chi)_{g_1,\ldots,g_{k}}]=\delta_* ([\chi])_{g_1,\ldots,g_{k}}\,,\quad \chi_{g_1,\ldots,g_{k-1} }\in  C^{k-1}(G,N)\,,
\end{equation}
where we have used the definition of $\delta_*$.
We thus see that $\text{im}\; \delta_*=\ker i_*$.
This implies, in particular, that $\ker i_*\subset H^1(G,M)$ classifies automorphic forms modulo $M$.
Exactness at the other nodes follows similarly.

Let us now focus on $H^1(\mathcal{G},N/M)$.
The elliptic $\Gamma$ function can be understood as part of a class $H^1(\mathcal{G},N/M)$, as first discussed in \cite{Felder_2000} (see also \cite{Jejjala:2021hlt} for a recent review).
From the above, a class $[X_g]\in H^1(\mathcal{G},N/M)$ satisfies a $1$-cocycle condition:
\begin{equation}\label{eq:basic-mod-prop-H1}
    \delta(X)_{g_1,g_2}(\boldsymbol{\rho})=\frac{X_{g_1}(\boldsymbol{\rho})X_{g_2}(g_1^{-1}\boldsymbol{\rho})}{X_{g_1g_2}(\boldsymbol{\rho})}=\xi_{g_1,g_2}(\boldsymbol{\rho})\,,\quad \xi_{g_1,g_2}(\boldsymbol{\rho})\in C^2(\mathcal{G},M)\,.
\end{equation}
We view this property as the degree $1$ analogue of the automorphic property~\eqref{eq:autom-form-defn}.
As stressed in~\cite{Jejjala:2021hlt}, the properties of $X_g$ are associated to relations in the relevant modular group.
This should be contrasted with the degree $0$ case, where modular properties are labeled by elements of the modular group.

Note that the case $g_2=g_1^{-1}$ in particular implies:
\begin{equation}\label{eq:inverse-cocycle}
    X_{g}(\boldsymbol{\rho})=\frac{1}{X_{g^{-1}}(g^{-1}\boldsymbol{\rho})}\,,
\end{equation}
where we use the choice $\xi_{g,g^{-1}}=1$ \cite{Felder_2000}.
Moreover, note that $\xi_{g_1,g_2}$ satisfies:
\begin{equation}\label{eq:2-cocycle-cond}
    (\delta \xi)_{g_1,g_2,g_3}=1 \quad \Rightarrow \quad \frac{\xi_{g_1,g_2} (\boldsymbol{\rho})\xi_{g_1g_2,g_3} (\boldsymbol{\rho})}{\xi_{g_1,g_2g_3} (\boldsymbol{\rho})\xi_{g_2,g_3} (g^{-1}_1\boldsymbol{\rho})}=1\,.
\end{equation}
This is a $2$-cocycle condition and the analogue of \eqref{eq:group-homom}.

It will be important in the following that at degree $1$ there is a notion of exact or trivializable elements.
Indeed, such classes can be written as:
\begin{equation}\label{eq:triv-Xg}
    [X_{g}]=[ (\delta B)_{g}]=\left[\frac{B(\boldsymbol{\rho})}{B(g^{-1}\boldsymbol{\rho})}\right]\,,
\end{equation}
with $B\in C^{0}(\mathcal{G},N)=N$.
Any $[X_{g}]$ of this form trivially satisfies \eqref{eq:basic-mod-prop-H1}, as can be easily verified.
Similar to ordinary differential forms, it will always be possible to find a locally exact expression for $[X_g]\in H^1(\mathcal{G},N/M)$, i.e., for some $g\in \mathcal{G}$.
But if $[X_g]\in H^1(\mathcal{G},N/M)$ corresponds to a non-trivial class, there exists no function $B$ such that~\eqref{eq:triv-Xg} holds for all $g$.

The $1$-cocycle condition implies that a class $[X_g]\in H^{1}(\mathcal{G}, N/M)$ is defined by its values on the generators of $\mathcal{G}$.
Consider now the free group formed by the generators $T_{ij}$ and $t_i$ of $\mathcal{G}$.\footnote{See Section \ref{sec:heegaard-splitting} for definitions. We omit the label $(a)$ on $t_i$ for notational convenience.} 
Any set of functions in $N/M$ associated to the generators provides a $1$-cocycle $X_g$ for the free group.
This set of functions descends to a $1$-cocycle for $\mathcal{G}$ if and only if the relations in $\mathcal{G}$ are sent to one \cite{Felder_2000}.
The $SL(3,\mathbb{Z})$ relations~\eqref{eq:sl3-relns} require:
\begin{align}\label{eq:Felder-relations}
    \begin{split}
        &X_{T_{ij}}\left(\boldsymbol{\rho}\right)X_{T_{kl}}\left(T_{ij}^{-1}\boldsymbol{\rho}\right)\cong X_{T_{kl}}\left(\boldsymbol{\rho}\right)X_{T_{ij}}\left(T_{kl}^{-1}\boldsymbol{\rho}\right)\,, \qquad i\neq l \,,\quad j\neq k \,,  \\
		&X_{T_{ij}}\left(\boldsymbol{\rho}\right)X_{T_{jk}}\left(T_{ij}^{-1}\boldsymbol{\rho}\right)\cong X_{T_{ik}}\left(\boldsymbol{\rho}\right)X_{T_{jk}}\left(T_{ik}^{-1}\boldsymbol{\rho}\right)	 X_{T_{ij}}\left(T_{jk}^{-1}T_{ik}^{-1}\boldsymbol{\rho}\right) \,, \\
		&X_{S_{ij}}\left(\boldsymbol{\rho}\right)X_{S_{ij}}\left(S_{ij}^{-1}\boldsymbol{\rho}\right)X_{S_{ij}}\left(S_{ij}^{-2}\boldsymbol{\rho}\right)X_{S_{ij}}\left(S_{ij}^{-3}\boldsymbol{\rho}\right)\cong 1\,,
    \end{split}
\end{align}
where the $\cong$ sign indicates that these equations should hold up to multiplication by a phase, i.e., an element in $C^2(\mathcal{G},M)$.
In addition, $X_g$ must satisfy the mixed relations between the $T_{ij}$ and $t_i$ given in \eqref{eq:Z3-relns}:
\begin{align}\label{eq:Felder-relations-2}
    \begin{split}
        &X_{T_{ij}}\left(\boldsymbol{\rho}\right)X_{t_k}\left(T_{ij}^{-1}\boldsymbol{\rho}\right)\cong X_{t_k}\left(\boldsymbol{\rho}\right)X_{T_{ij}}\left(t_k^{-1}\boldsymbol{\rho}\right)\,, \qquad i\neq k \,,  \\
		&X_{T_{ij}}\left(\boldsymbol{\rho}\right)X_{t_i}\left(T_{ij}^{-1}\boldsymbol{\rho}\right)\cong X_{t_i}\left(\boldsymbol{\rho}\right)X_{t_j^{-1}}\left(t_i^{-1}\boldsymbol{\rho}\right)	 X_{T_{ij}}\left(t_jt_i^{-1}\boldsymbol{\rho}\right)\,,\\
		&X_{t_i}\left(\boldsymbol{\rho}\right)X_{t_j}\left(t_i^{-1}\boldsymbol{\rho}\right)\cong X_{t_j}\left(\boldsymbol{\rho}\right)X_{t_i}\left(t_j^{-1}\boldsymbol{\rho}\right)\,.
    \end{split}
\end{align}
We now turn to the physical relevance of this construction.

\subsection{The 1-cocycle condition and lens indices}\label{ssec:1-cocycle-lens-indices}

In this section, we systematically prove that the candidate $1$-cocycle of \cite{Gadde:2020bov}, constructed from the collection of lens indices, realizes a non-trivial class in $H^1(\mathcal{G},N/M)$ for general $\mathcal{N}=1$ gauge theories, as long as their index admits a Higgs branch expression (see Section \ref{ssec:hol-blocks} and they have a global symmetry.\footnote{This requirement follows from the requirement of (even) integral R-charges. See Section \ref{ssec:towards-conjecture} for more details. In addition, due to the non-renormalization of indices the proof also applies to IR SCFTs that can be reached through supersymmetric flows from the $\mathcal{N}=1$ gauge theories.}
Our proof is based on three main results from Section \ref{sec:mod-fac-lens-indices}: 
\begin{enumerate}
    \item Ambiguities in the Heegaard splitting lead to the same lens index:
    \begin{equation}\label{eq:ambig-heegaard-index}
        \mathcal{I}_{(p,q)}(\hat{\boldsymbol{\rho}})=\mathcal{Z}_{f}(\boldsymbol{\rho})=\mathcal{Z}_{hf\tilde{h}^{-1}}(h\boldsymbol{\rho})\,.
    \end{equation}
    \item Invariance of the lens indices under the action of their respective groups of large diffeomorphisms and gauge transformations, up to a phase. 
    More specifically, we use periodicity of the superconformal index $\mathcal{Z}_{S_{23}\mathcal{O}}(\boldsymbol{\rho})$ under $z_a\to z_a+1$, $\tau\to\tau+1$ and $\sigma\to \sigma +1$ and the covariance of $\mathcal{Z}_{t_2^{n}\mathcal{O}}$ under an entire copy of $\mathcal{H}$.
    \item Modular factorization of lens indices:
    \begin{equation}\label{eq:mod-fact-conj-proof}
        \mathcal{I}_{(p,q)}(\hat{\boldsymbol{\rho}})=e^{-i\pi \mathcal{P}^{\mathbf{m}}_f(\boldsymbol{\rho})} \sum_{\alpha} \mathcal{B}^\alpha_{h}(\boldsymbol{\rho})\mathcal{B}^\alpha_{\tilde{h}}(f^{-1}\boldsymbol{\rho})\,,\qquad  (h,\tilde{h})\in S_f\subset H\times H\,.
    \end{equation}
\end{enumerate}
The first two points were discussed in Section \ref{ssec:towards-conjecture} and Section \ref{ssec:hol-blocks}, respectively, and the last point in Section \ref{ssec:mod-fact-conjecture}.
Let us summarize the strategy of our proof here.
\begin{enumerate}
    \item After constructing the candidate $1$-cocycle for $\mathcal{G}$ we evaluate it on the generators, which using \eqref{eq:ambig-heegaard-index} can be expressed in terms of $\mathcal{Z}_{S_{23}\mathcal{O}}$ and $\mathcal{Z}_{\mathcal{O}}$.
    \item Using these expressions, we show that it satisfies all basic relations in $\mathcal{G}$, i.e., we verify \eqref{eq:Felder-relations}, \eqref{eq:Felder-relations-2} and also the inverse relation \eqref{eq:inverse-cocycle}.
    The relations separate into three classes that require distinct properties of the lens indices to prove:
    \begin{enumerate}
        \item Relations that only involve generators in $\mathcal{H}\subset \mathcal{G}$.
        These relations can be proven making use of the first result above.
        \item Relations that involve strictly one element in $\lbrace T_{12},T_{32},t_{2}\rbrace$.
        These relations can be proven making use of both the first and second result.
        \item Relations that involve more than one element in $\lbrace T_{12},T_{32},t_{2}\rbrace$.
        These relations can be proven making use of the third result.
        \item The relation between an element and its inverse is also proven using the third result.
    \end{enumerate}
    \item Finally, we show that the $1$-cocycle defines a non-trivial class in $H^1(\mathcal{G},N/M)$ and provide a physical interpretation in terms of the results of Section \ref{ssec:geom-int-univ-blocks}.
\end{enumerate}
Our proof can be viewed as a generalization to arbitrary gauge theories of the proof in \cite{Felder_2000} that the elliptic $\Gamma$ function is part of a $1$-cocycle for $\mathcal{G}$, which in the physical context corresponds to the example of the free chiral multiplet \cite{Gadde:2020bov,Jejjala:2021hlt}.

\paragraph{A candidate $\mathbf{1}$-cocycle:}

The first result above implies that there always exists a lens index $\mathcal{I}_{(p,q)}(\hat{\boldsymbol{\rho}})$ that can be associated to $\mathcal{Z}_f(\boldsymbol{\rho})$ with $f=g\,\mathcal{O}$ and $g\in \mathcal{G}$ arbitrary.
This follows from the standard Heegaard splitting of a Hopf surface \eqref{eq:notation-Mg-split} in terms $f=g_{(p,q)}\,\mathcal{O}$ and our observation in \eqref{eq:gSL2-from-gSL3}.
In addition, recall that there is a natural $\mathcal{G}$ action \eqref{eq:calG-action} on $\boldsymbol{\rho}$.
These are two key features shared with a $1$-cocycle $X_g(\boldsymbol{\rho})$ of $\mathcal{G}$.
However, a $1$-cocycle has two other basic properties not shared by the indices: first of all, it obeys $X_1(\boldsymbol{\rho})= 1$, and secondly it is defined multiplicatively.
In contrast, the index $\mathcal{Z}_\mathcal{O}(\boldsymbol{\rho})$ is not trivial, and moreover the Higgs branch expression for a lens index mixes sums and products.

However, the factorization of a lens index is uniform over the Higgs branch vacua $\alpha$, since the relative phase (the anomaly polynomial) does not depend on $\alpha$. Taken together, this suggests a natural candidate $1$-cocycle \cite{Gadde:2020bov}:\footnote{We label $ \hat{\mathcal{Z}}^{\alpha}_g(\boldsymbol{\rho})$ by $g$ because it will turn out that the dependence on the orientation reversed moduli drops out.}
\begin{equation}\label{eq:defn-hatZ}
    \hat{\mathcal{Z}}^{\alpha}_g(\boldsymbol{\rho})\equiv \frac{\mathcal{Z}^{\alpha}_f(\boldsymbol{\rho})}{\mathcal{Z}^{\alpha}_\mathcal{O}(g^{-1}\boldsymbol{\rho})}\,,\qquad f=g\,\mathcal{O}\,,
\end{equation}
where $\mathcal{Z}^{\alpha}_f(\boldsymbol{\rho})$ is the summand in the expression \eqref{eq:lens-higgs-form} for the index $\mathcal{Z}_f(\boldsymbol{\rho})$.
This ratio clearly satisfies the first property, and we will see that it also fits into a multiplicative structure.
As such, this object defines a $1$-cocycle for the free group of generators of $\mathcal{G}$.
Before turning to the $\mathcal{G}$ relations, we first show that the $1$-cocycle evaluated on the generators can be expressed in terms of $\mathcal{Z}_{S_{23}\mathcal{O}}$ and $\mathcal{Z}_{\mathcal{O}}$.%

Using the first result \eqref{eq:ambig-heegaard-index} for $f=\mathcal{O}$, $h\in \mathcal{H}$ general and $\tilde{h}=\mathbbm{1}$, we have:
\begin{equation}
    \mathcal{Z}_{h\mathcal{O}}^{\alpha}(h\boldsymbol{\rho})= \mathcal{Z}_{\mathcal{O}}^{\alpha}(\boldsymbol{\rho})\quad \Rightarrow\quad \hat{\mathcal{Z}}_{h}^{\alpha}(\boldsymbol{\rho})=1\,.
\end{equation}
We thus find for all generators in $\mathcal{H}$:
\begin{align}\label{eq:hatZ-Tij-jneq2}
    \begin{split}
        \hat{\mathcal{Z}}_{T_{ij}}^{\alpha}(\boldsymbol{\rho})= 1\,,\quad j\neq 2\,,\qquad \hat{\mathcal{Z}}_{t_i}^{\alpha}(\boldsymbol{\rho})= 1\,,\quad i\neq 2\,,
    \end{split}
\end{align}
and similarly for their inverses.
This establishes a similar claim in \cite{Gadde:2020bov} rigorously.
It follows that \eqref{eq:inverse-cocycle} is trivially satisfied for elements in $\mathcal{H}$.
We now compute $\hat{\mathcal{Z}}^{\alpha}_{(\cdot)}$ for the remaining generators $T_{i2}$ and $t_2$, again making use of \eqref{eq:ambig-heegaard-index}.
For example, since $T_{32}^{-1}\mathcal{O}=T_{23}^{-1}S_{23}\mathcal{O}\,T_{23}$ and $T_{23}\in \mathcal{H}$ (see \eqref{eq:T32-as-S23}), we find:
\begin{equation}\label{eq:hatZ-T32}
    \hat{\mathcal{Z}}^{\alpha}_{T_{32}^{-1}}(\boldsymbol{\rho})\cong  \hat{\mathcal{Z}}^{\alpha}_{S_{23}}(T_{23}\boldsymbol{\rho})\,,
\end{equation}
where we have made use of the second result listed above for the $S^2\times T^2$ index:
\begin{equation}\label{eq:S2xT2-mod-cov}
    \mathcal{Z}_{\mathcal{O}}^{\alpha}(h\boldsymbol{\rho})\cong \mathcal{Z}_{\mathcal{O}}^{\alpha}(\boldsymbol{\rho})\,,\qquad h\in \mathcal{H}\,.
\end{equation}
Similarly, since $T_{12}^{-1}\mathcal{O}=S_{13}\,T_{32}^{-1}\mathcal{O}\,S_{13}^{-1}$ we find:
\begin{equation}\label{eq:hatZ-T12}
    \hat{\mathcal{Z}}^{\alpha}_{T_{12}^{-1}}(\boldsymbol{\rho})\cong  \hat{\mathcal{Z}}^{\alpha}_{S_{23}}(T_{23}S_{13}^{-1}\boldsymbol{\rho})\,,
\end{equation}
where we have again made use of \eqref{eq:S2xT2-mod-cov}.
So far, we have evaluated $\hat{\mathcal{Z}}^{\alpha}_{(\cdot)}$ on $T_{12}^{-1}$ and $T_{32}^{-1}$.
For now, we only consider relations involving these inverses, and later show that $\hat{\mathcal{Z}}^{\alpha}_{(\cdot)}$ on $T_{12}$ and $T_{32}$ correctly reflects \eqref{eq:inverse-cocycle}.
Finally, the index associated to $t_2$ is an $S^2\times T^2$ index with an additional unit of magnetic flux for the flavor symmetry associated to $t_2$.\footnote{We suppress the superscript $t_i^{(a)}$ with $a=1,\ldots,r$ and $r$ the rank of the flavor symmetry for notational convenience.}
We simply write out the definition in this case:
\begin{equation}\label{eq:hatZ-t2}
    \hat{\mathcal{Z}}_{t_2}^{\alpha}(\boldsymbol{\rho})=\frac{\mathcal{Z}^{\alpha}_{t_2\mathcal{O}}(\boldsymbol{\rho})}{\mathcal{Z}^{\alpha}_\mathcal{O}(\boldsymbol{\rho})}\,,
\end{equation}
where we recall that the second result above holds for both numerator and denominator.
Having expressed the candidate $1$-cocycle on all generators of $\mathcal{G}$, we now turn to check that it satisfies the basic relations of $\mathcal{G}$.

\paragraph{Basic relations (a):}

Due to \eqref{eq:hatZ-Tij-jneq2}, all relations that only involve $T_{ij},t_i\in\mathcal{H}$ are trivially satisfied.
In particular, the relation associated to $S_{13}^4=\mathbbm{1}$ is satisfied.
The relations $S_{12}^4=S_{23}^4=\mathbbm{1}$ follow from the former as long as the first and second line of \eqref{eq:Felder-relations} are satisfied as well.
Therefore, the non-trivial relations to be checked are the first two relations in \eqref{eq:Felder-relations} and the relations in \eqref{eq:Felder-relations-2} that involve at least one element in $\lbrace T_{12}^{-1},T_{32}^{-1},t_2 \rbrace$.

\paragraph{Basic relations (b):}

We now turn to the relations that involve one element in $\lbrace T_{12}^{-1},T_{32}^{-1},t_{2}\rbrace$.
For the $SL(3,\mathbb{Z})$ part of these relations, we have to check:
\begin{equation}\label{eq:sl3-reln-1-T32}
  \begin{alignedat}{2}
     T_{32}^{-1}T_{31}&=T_{31}T_{32}^{-1}\,,\qquad & T_{12}^{-1}T_{13}&=T_{13}T_{12}^{-1}\,,\\
        T_{32}^{-1}T_{21}&=T_{31}^{-1}T_{21}T_{32}^{-1}\,,\qquad & T_{12}^{-1}T_{23}&=T_{13}^{-1}T_{23}T_{12}^{-1}\,.
  \end{alignedat}
\end{equation}
Plugging in the expressions \eqref{eq:hatZ-Tij-jneq2}, \eqref{eq:hatZ-T32} and \eqref{eq:hatZ-T12}, it follows that $\hat{\mathcal{Z}}_{(\cdot)}$ satisfies these relations as long as:
\begin{align}\label{eq:periodicity-constr-hatZ-S23}
    \begin{split}
        \hat{\mathcal{Z}}^{\alpha}_{S_{23}}(T_{21}\boldsymbol{\rho})\cong \hat{\mathcal{Z}}^{\alpha}_{S_{23}}(\boldsymbol{\rho})\,,\\
        \hat{\mathcal{Z}}^{\alpha}_{S_{23}}(T_{31}\boldsymbol{\rho})\cong \hat{\mathcal{Z}}^{\alpha}_{S_{23}}(\boldsymbol{\rho})\,.
    \end{split}
\end{align}
Since $\tilde{\boldsymbol{\rho}}=\mathcal{O}S_{23}^{-1}T_{21}\boldsymbol{\rho} =T_{31}\mathcal{O}S_{23}^{-1}\boldsymbol{\rho}$ on the left hand side of the first line, and similarly on the second line but with $T_{21}\leftrightarrow T_{31}$, we recognize this action as a large diffeomorphism on the Hopf surface $\mathcal{M}_{(1,0)}(\hat{\boldsymbol{\rho}})$ under which $\hat{\tau}\to \hat{\tau}+1$ and $\hat{\sigma}\to \hat{\sigma} +1$, respectively, described in \eqref{eq:largediffeo-matrixrelation}.
By the second result listed above, it follows that the relations \eqref{eq:sl3-reln-1-T32} are indeed satisfied by $\hat{\mathcal{Z}}^{\alpha}_{(\cdot)}$.

The relations that only involve $t_2$ and elements in $\mathcal{H}$ have the structure $t_2h=h't_2$ for $h,h'\in\mathcal{H}$.
These relations are also satisfied by the second result listed above, which implies that $\hat{\mathcal{Z}}^{\alpha}_{t_2}(\boldsymbol{\rho})$ is invariant under the action of $\mathcal{H}$ up to a phase.
Furthermore, the relations that involve $T_{i2}^{-1}$ and $t_{1,3}$, but not $t_2$, are yet again satisfied due the second result.
In this case, it follows from periodicity of $\mathcal{Z}^{\alpha}_{S_{23}\mathcal{O}}(\boldsymbol{\rho})$ under $z\to z+1$.

\paragraph{Basic relations (c):}

We continue with the relations involving more than one element in $\lbrace T_{12}^{-1},T_{32}^{-1},t_{2}\rbrace$.
For the $SL(3,\mathbb{Z})$ part of the relations, we need to check:
\begin{align}\label{eq:relns-incl-both-T12T32}
    \begin{split}
         T_{13}T_{32}^{-1}&=T_{12}^{-1}T_{32}^{-1}T_{13}\,,\quad T_{31}T_{12}^{-1}=T_{32}^{-1}T_{12}^{-1}T_{31}\,,\quad 
       T_{32}^{-1}T_{12}^{-1}=T_{12}^{-1}T_{32}^{-1}\,.
    \end{split}
\end{align}
Let us first note that, since $\hat{\mathcal{Z}}^{\alpha}_{T_{13}}(\boldsymbol{\rho})=\hat{\mathcal{Z}}^{\alpha}_{T_{31}}(\boldsymbol{\rho})=1$, the third relation follows from the first two as long as $\hat{\mathcal{Z}}^{\alpha}_{(\cdot)}$ satisfies in addition:
\begin{equation}
    T_{13}T_{32}^{-1}=T_{31}T_{12}^{-1}\,.
\end{equation}
One easily checks that this relation is implied by \eqref{eq:periodicity-constr-hatZ-S23}.
Therefore, we can focus on the first two relations in \eqref{eq:relns-incl-both-T12T32}.
The first relation requires us to show that:
\begin{equation}\label{eq:T13T32-reln}
    \hat{\mathcal{Z}}^{\alpha}_{T_{13}}(\boldsymbol{\rho})\hat{\mathcal{Z}}^{\alpha}_{T_{32}^{-1}}(T_{13}^{-1}\boldsymbol{\rho})\cong \hat{\mathcal{Z}}^{\alpha}_{T_{12}^{-1}}(\boldsymbol{\rho})\hat{\mathcal{Z}}^{\alpha}_{T_{32}^{-1}}(T_{12}\boldsymbol{\rho})\hat{\mathcal{Z}}^{\alpha}_{T_{13}}(T_{32}T_{12}\boldsymbol{\rho})\,.
\end{equation}
Note that in this case we have not plugged in \eqref{eq:hatZ-Tij-jneq2}, \eqref{eq:hatZ-T32}, and \eqref{eq:hatZ-T12}, which turns out to be convenient.
In order to prove this relation, we need the third result listed above.
More specifically, consider the modular factorization of both $\mathcal{Z}^{\alpha}_{T_{ij}\mathcal{O}}(\boldsymbol{\rho})$ and $\mathcal{Z}^{\alpha}_{\mathcal{O}}(\boldsymbol{\rho})$ as parametrized by $S_{T_{ij}\mathcal{O}}$ and $S_{\mathcal{O}}$ respectively.
Let $(h,\tilde{h})\in S_{T_{ij}\mathcal{O}}\cap S_{\mathcal{O}}$, which is non-empty for all $T_{ij}$.\footnote{This follows for example from the fact that the generators of $F$ and $F_S$, defined in Section \ref{ssec:geom-int-univ-blocks}, together comprise all the $T_{ij}$.}
For such a pair $(h,\tilde{h})$, it follows that:
\begin{equation}
    \hat{\mathcal{Z}}^{\alpha}_{T_{ij}}(\boldsymbol{\rho})=\frac{\mathcal{Z}^{\alpha}_{T_{ij}\mathcal{O}}(\boldsymbol{\rho})}{\mathcal{Z}^{\alpha}_{\mathcal{O}}(T_{ij}^{-1}\boldsymbol{\rho})}\cong \frac{\mathcal{B}^{\alpha}_h(\boldsymbol{\rho})}{\mathcal{B}^{\alpha}_{h}(T_{ij}^{-1}\boldsymbol{\rho})}\,.
\end{equation}
We thus see that the factorization of $\mathcal{Z}^{\alpha}_{T_{ij}\mathcal{O}}(\boldsymbol{\rho})$ and $\mathcal{Z}^{\alpha}_{\mathcal{O}}(\boldsymbol{\rho})$ into a common set of holomorphic blocks corresponds to a trivialization of $\hat{\mathcal{Z}}^{\alpha}_{T_{ij}}(\boldsymbol{\rho})$ (cf.\ \eqref{eq:triv-Xg}), as first suggested in \cite{Gadde:2020bov}.
This observation allows us to prove that $\hat{\mathcal{Z}}^{\alpha}_{(\cdot)}$ satisfies any relation that involves elements $g$ whose solution sets $S_f$ have a non-empty intersection.

For the relation \eqref{eq:T13T32-reln}, one may observe that the relevant $T_{ij}$ are all elements of $F_S= SL(2,\mathbb{Z})_{23}\ltimes \mathbb{Z}^2\subset SL(3,\mathbb{Z}) $ defined in Section \ref{ssec:geom-int-univ-blocks}.
It follows from the discussion there that the relevant $\mathcal{Z}^{\alpha}_{T_{ij}\mathcal{O}}(\boldsymbol{\rho})$ and $\mathcal{Z}^{\alpha}_{\mathcal{O}}(\boldsymbol{\rho})$ can be factorized in terms of the holomorphic block $\mathcal{B}^{\alpha}_S(\boldsymbol{\rho})$.
By the above, we find that $\hat{\mathcal{Z}}^{\alpha}_{g}(\boldsymbol{\rho})$ for $g\in F_S$ can be written as:
\begin{equation}\label{eq:triv-FS}
    \hat{\mathcal{Z}}^{\alpha}_{g}(\boldsymbol{\rho})\cong \frac{\mathcal{B}^{\alpha}_S(\boldsymbol{\rho})}{\mathcal{B}^{\alpha}_S(g^{-1}\boldsymbol{\rho})}\,.
\end{equation}
Therefore, the relation \eqref{eq:T13T32-reln} is trivially satisfied, as one easily verifies by plugging in \eqref{eq:triv-FS}.
Notice in particular that:
\begin{equation}
    \hat{\mathcal{Z}}^{\alpha}_{T_{13}}(\boldsymbol{\rho})\cong \frac{\mathcal{B}^{\alpha}_S(\boldsymbol{\rho})}{\mathcal{B}^{\alpha}_S(T_{13}^{-1}\boldsymbol{\rho})}=\frac{\mathcal{B}^{\alpha}(S_{13}\boldsymbol{\rho})}{\mathcal{B}^{\alpha}(T_{31}S_{13}\boldsymbol{\rho})}=1\,,
\end{equation}
where in the last equation we have made use of periodicity of the holomorphic blocks (see, e.g., Section \ref{sssec:example-free-chiral}).
This is indeed consistent with \eqref{eq:hatZ-Tij-jneq2}.

Let us apply the same strategy to the second relation in \eqref{eq:relns-incl-both-T12T32}.
The relation we have to prove reads:
\begin{equation}\label{eq:T31T12-reln}
    \hat{\mathcal{Z}}^{\alpha}_{T_{31}}(\boldsymbol{\rho})\hat{\mathcal{Z}}^{\alpha}_{T_{12}^{-1}}(T_{31}^{-1}\boldsymbol{\rho})\cong \hat{\mathcal{Z}}^{\alpha}_{T_{32}^{-1}}(\boldsymbol{\rho})\hat{\mathcal{Z}}^{\alpha}_{T_{12}^{-1}}(T_{32}\boldsymbol{\rho})\hat{\mathcal{Z}}^{\alpha}_{T_{31}}(T_{12}T_{32}\boldsymbol{\rho})\,.
\end{equation}
In this case, all elements involved belong to $F=SL(2,\mathbb{Z})_{12}\ltimes \mathbb{Z}^2$.
The indices associated to this subgroup can all be factorized in terms of $\mathcal{B}^{\alpha}(\boldsymbol{\rho})$, as again discussed in Section \ref{ssec:geom-int-univ-blocks}.
Thus, for any $g\in F$ we have:
\begin{equation}\label{eq:triv-F1}
    \hat{\mathcal{Z}}^{\alpha}_{g}(\boldsymbol{\rho})\cong \frac{\mathcal{B}^{\alpha}(\boldsymbol{\rho})}{\mathcal{B}^{\alpha}(g^{-1}\boldsymbol{\rho})}\,.
\end{equation}
For the same reason as above, it follows that also \eqref{eq:T31T12-reln} is satisfied.

The remaining relations are given by:
\begin{equation}\label{eq:t2-T12-T32-relns}
  \begin{alignedat}{2}
     T_{12}^{-1}t_2&=t_2T_{12}^{-1}\,,\qquad &  T_{32}^{-1}t_2&=t_2T_{32}^{-1}\,,\\
       T_{12}^{-1}t_1&=t_1t_2^{-1}T_{12}^{-1}\,,\qquad & T_{32}^{-1}t_3&=t_3t_2^{-1}T_{32}^{-1}\,.
  \end{alignedat}
\end{equation}
Recall that $\mathcal{Z}_{t_2\mathcal{O}}(\boldsymbol{\rho})$ can be factorized in terms of pairs $(h,h) \in SL(2,\mathbb{Z})_{13}\subset S_{\mathcal{O}}$ (see the comment below \eqref{eq:SO}).
On the other hand, one can make use of the relation between $S_{f'}$ and $S_f$ \eqref{eq:Sf'-gen} for $f'=hf\tilde{h}^{-1}$ to determine the sets $S_{t_1\mathcal{O}}$ and $S_{t_3\mathcal{O}}$ in terms of $S_{\mathcal{O}}$.
One finds that the relevant intersections are non-empty:
\begin{align}
    \begin{split}
         (S_{13},S_{13})&\in S_{T_{32}\mathcal{O}}\cap S_{t_2\mathcal{O}}\cap S_{t_3\mathcal{O}}\cap S_{\mathcal{O}}\,,\\
        (\mathbbm{1},\mathbbm{1})&\in S_{T_{12}\mathcal{O}}\cap S_{t_2\mathcal{O}}\cap S_{t_1\mathcal{O}}\cap S_{\mathcal{O}} \,.
    \end{split}
\end{align}
This shows that there exist a trivializations for each of the relations in \eqref{eq:t2-T12-T32-relns}, and consequently they are satisfied by $\hat{\mathcal{Z}}^{\alpha}_{(\cdot)}$.

Let us briefly pause to note that if there were a global trivialization, i.e., a function $\mathcal{B}^{\alpha}_h(\boldsymbol{\rho})$ such that all generators of $\mathcal{G}$ are trivialized, we could have proven all relations in one go.
However, as we have explained in Section \ref{ssec:geom-int-univ-blocks}, the maximal subset of $SL(3,\mathbb{Z})$ that can be factorized into a common set of holomorphic blocks is isomorphic to $SL(2,\mathbb{Z})\ltimes \mathbb{Z}^2$.
Two indices that can never be simultaneously trivialized, for example, are $\hat{\mathcal{Z}}^{\alpha}_{T_{13}}(\boldsymbol{\rho})$ and $\hat{\mathcal{Z}}^{\alpha}_{T_{31}}(\boldsymbol{\rho})$.
A relation that involves both elements, such as $S_{13}^4=\mathbbm{1}$, has to be proven with an alternative method, as we have used in point \textbf{(a)}.
We will comment in more detail about the absence of a global trivialization below.

\paragraph{Basic relation (d):}

To conclude that $\hat{\mathcal{Z}}^{\alpha}_{(\cdot)}$ descends to a $1$-cocycle for $\mathcal{G}$, we still need to show that $\hat{\mathcal{Z}}^{\alpha}_{(\cdot)}$ for $T_{12}^{-1}$ and $T_{12}$ correctly reflects \eqref{eq:inverse-cocycle}, and similarly for $T_{32}^{-1}$ and $t_2$.
To start, let us note that the orientation reversing symmetry of the Hopf surface, as described in \eqref{eq:combined-geom-action}, naively seems to imply:
\begin{equation}\label{eq:inv-reln-suggestion}
    \mathcal{Z}_{g\mathcal{O}}(\boldsymbol{\rho})\stackrel{?}{=}\mathcal{Z}_{g^{-1}\mathcal{O}}(g^{-1}\boldsymbol{\rho})\,.
\end{equation}
However, an equation of this sort would not lead to the desired relation \eqref{eq:inverse-cocycle}.
To see what goes wrong, we first note that the solution sets for $T_{i2}^{-1}$ and $T_{i2}$ with $i=1,3$ are identical:
\begin{equation}
    S_{T_{i2}^{-1}\mathcal{O}}= S_{T_{i2}\mathcal{O}}\,.
\end{equation}
It follows that any function $\mathcal{B}^{\alpha}_h(\boldsymbol{\rho})$ that trivializes $\hat{\mathcal{Z}}^{\alpha}_{T_{i2}^{-1}}(\boldsymbol{\rho})$ can also be used to trivialize $\hat{\mathcal{Z}}^{\alpha}_{T_{i2}}(\boldsymbol{\rho})$.
For example, we can use $\mathcal{B}^{\alpha}_S(\boldsymbol{\rho})$ to trivialize $\hat{\mathcal{Z}}^{\alpha}_{T_{i2}}$.
One immediately observes:
\begin{align}
    \begin{split}
        \hat{\mathcal{Z}}^{\alpha}_{T_{i2}}(\boldsymbol{\rho})&\cong \frac{\mathcal{B}^{\alpha}_S(\boldsymbol{\rho})}{\mathcal{B}^{\alpha}_S(T_{i2}^{-1}\boldsymbol{\rho})} \cong \frac{1}{\hat{\mathcal{Z}}^{\alpha}_{T_{i2}^{-1}}(T_{i2}^{-1}\boldsymbol{\rho})} \,.
    \end{split}
\end{align}
Comparing with \eqref{eq:inverse-cocycle}, this is indeed the right behaviour for inverses.
This argument also applies to the relation between $\hat{\mathcal{Z}}^{\alpha}_{t_2}$ and $\hat{\mathcal{Z}}^{\alpha}_{t^{-1}_2}$.
As a consequence, there is no simple relation such as the one suggested in \eqref{eq:inv-reln-suggestion}.
We expect this to be related to the fact that orientation reversal changes the preserved supersymmetry algebra.

\paragraph{Conclusion:}

This concludes our proof of the statement that $\hat{\mathcal{Z}}^{\alpha}_{(\cdot)}$ is a $1$-cocycle for $\mathcal{G}$.
In particular, it follows that it satisfies the defining $1$-cocycle condition:
\begin{equation}\label{eq:hatZ-mod-prop}
    \hat{\mathcal{Z}}^{\alpha}_{g_1g_2}(\boldsymbol{\rho})\cong \hat{\mathcal{Z}}^{\alpha}_{g_1}(\boldsymbol{\rho})\hat{\mathcal{Z}}^{\alpha}_{g_2}(g_{1}^{-1}\boldsymbol{\rho})\,,\qquad g_{1,2}\in\mathcal{G}\,,
\end{equation}
where the equality holds up to a phase in $C^2(\mathcal{G},M)$.
An explicit description of this phase is in terms of the anomaly polynomial of the theory (see, e.g., Section \ref{ssec:evidence}).
For consistency of \eqref{eq:hatZ-mod-prop}, this phase should really by thought of as a class in $H^2(\mathcal{G},M)$, and in particular should satisfy the $2$-cocycle condition \eqref{eq:2-cocycle-cond}.
For the free chiral multiplet, or rather a single elliptic $\Gamma$ function, this was shown in \cite{Felder_2000}.
A quick way to argue that it holds more generally for $\mathcal{N}=1$ gauge theories is as follows.
First, note that the vortex contributions to a lens index are automatically factorized (cf.\ \eqref{eq:lens-higgs-form}).
This implies that its contribution to $\hat{\mathcal{Z}}^{\alpha}_{g}(\boldsymbol{\rho})$ is cohomologically trivial \cite{Gadde:2020bov}:
\begin{equation}\label{eq:hatZ-explicit}
    \hat{\mathcal{Z}}^{\alpha}_{g}(\boldsymbol{\rho})=\frac{\mathcal{Z}^{\alpha}_{g\mathcal{O}}(\boldsymbol{\rho})}{\mathcal{Z}^{\alpha}_{\mathcal{O}}(g^{-1}\boldsymbol{\rho})}=\frac{\mathcal{Z}^{\alpha}_{g\mathcal{O},\text{1-loop}}(\boldsymbol{\rho})}{\mathcal{Z}^{\alpha}_{\mathcal{O},\text{1-loop}}(g^{-1}\boldsymbol{\rho})}\frac{\mathcal{Z}^{\alpha}_{\text{v}}(\boldsymbol{\rho})}{\mathcal{Z}^{\alpha}_{\text{v}}(g^{-1}\boldsymbol{\rho})}\,,
\end{equation}
where we have plugged in \eqref{eq:lens-higgs-form}.
As such, this contribution drops out of the $1$-cocycle condition, and in particular does not contribute to the phase.
Since $\mathcal{Z}^{\alpha}_{\text{1-loop},\,g\mathcal{O}}(\boldsymbol{\rho})$ consists of a product of elliptic $\Gamma$ functions, and the cohomology groups are defined multiplicatively, it follows that also the phase for a general gauge theory satisfies the $2$-cocycle condition.
It would be interesting to verify this more explicitly.

\paragraph{Non-triviality of the class:}

We have seen that not all elements in $\mathcal{G}$ are simultaneously trivializable.
This implies, by definition, that $\hat{\mathcal{Z}}^{\alpha}_{(\cdot)}$ defines a non-trivial class in $H^1(\mathcal{G},N/M)$.
The underlying physical reason follows from the connection between trivialization and holomorphic block factorization.

To see this, recall from Section \ref{ssec:geom-int-univ-blocks} that only indices associated to a maximal subset of $SL(3,\mathbb{Z})$, isomorphic to $SL(2,\mathbb{Z})\ltimes \mathbb{Z}^2$, admit a factorization in terms of a given pair of holomorphic blocks.
Physically, this follows from the condition that the factorization of a given index is only compatible with a pair of holomorphic blocks, labeled by $(h,\tilde{h})$, when the embedding of the time circle into the Heegaard splitting is fixed by the gluing transformation $f'$.
Furthermore, recall that $\hat{\mathcal{Z}}^{\alpha}_{g}(\boldsymbol{\rho})$ can only be trivialized in terms of a function $\mathcal{B}^{\alpha}_h(\boldsymbol{\rho})$ if both $\mathcal{Z}_{f}(\boldsymbol{\rho})$ and $\mathcal{Z}_{\mathcal{O}}(\boldsymbol{\rho})$ admit a factorization in terms of a common pair of holomorphic blocks.
This constrains $\tilde{h}=\mathcal{O}h\mathcal{O}$, since only the subgroups $F_h\subset SL(3,\mathbb{Z})$, defined in \eqref{eq:Zf-Bh}, contain the identity element.
Therefore, $F_h$ for any $h\in H$ is a maximal subgroup of $SL(3,\mathbb{Z})$ on which $\hat{\mathcal{Z}}^{\alpha}_{g}(\boldsymbol{\rho})$ can be trivialized, and the trivializing function is $\mathcal{B}^{\alpha}_h(\boldsymbol{\rho})$.
This implies, by definition, that $\hat{\mathcal{Z}}^{\alpha}_{g}(\boldsymbol{\rho})$ is non-trivial in cohomology.

We can extend this statement to maximal subgroups of $\mathcal{G}$.
Since we have seen that $\mathcal{Z}_{t_{2,3}\mathcal{O}}(\boldsymbol{\rho})$ also admit a factorization in terms of $\mathcal{B}^{\alpha}_S(\boldsymbol{\rho})$, it follows that:
\begin{equation}\label{eq:def-calFS}
\begin{aligned}
\mathcal{F}_S\equiv SL(2,\mathbb{Z})_{23}\ltimes \mathbb{Z}^{2+2r} \,, \quad \textrm{with} \quad \mathbb{Z}^{2+2r} &= \langle T_{12}\,,T_{13}\,, t^{(a)}_2\,,t^{(a)}_3\rangle\,,
\end{aligned}
\end{equation}
is a maximal subgroup of $\mathcal{G}$ which can be trivialized in terms of $\mathcal{B}^{\alpha}_S(\boldsymbol{\rho})$.
Similarly, $\hat{\mathcal{Z}}^{\alpha}_{g}(\boldsymbol{\rho})$ can be trivialized in terms of $\mathcal{B}^{\alpha}(\boldsymbol{\rho})$ for $g\in \mathcal{F}$ with:
\begin{equation}\label{eq:def-calF1}
\begin{aligned}
\mathcal{F}\equiv SL(2,\mathbb{Z})_{12}\ltimes \mathbb{Z}^{2+2r} \,, \quad \textrm{with} \quad \mathbb{Z}^{2+2r} &= \langle T_{31}\,,T_{32}\,, t^{(a)}_1\,,t^{(a)}_2\rangle\,.
\end{aligned}
\end{equation}
In general, a trivialization of $\hat{\mathcal{Z}}^{\alpha}_{g}(\boldsymbol{\rho})$ for $g\in \mathcal{F}_h\equiv h^{-1}\mathcal{F}h$ is in terms of the function $\mathcal{B}^{\alpha}_{h}(\boldsymbol{\rho})$.
Finally, indices associated to $\mathcal{H}\subset \mathcal{G}$ can never be simultaneously trivialized.

\subsection{Cohomological perspective on modular factorization}\label{ssec:coh-perspective}

In this section, we provide a cohomological perspective on modular factorization.

We recall from Section \ref{ssec:geom-int-univ-blocks} that $hf\tilde{h}^{-1}\mathcal{O}\in F$ for any gluing transformation $f$ and $(h,\tilde{h})\in S_f$.
Let us write out this fact more explicitly for $f=g_{(p,q)}\,\mathcal{O}$ as in \eqref{eq:gSL2-from-gSL3}.
Using the explicit pair $(h,\tilde{h})\in S_f$ in \eqref{eq:h-ht-Lens-gen}, we have:\footnote{This relation generalizes the relations studied in \cite{Gadde:2020bov,Jejjala:2021hlt}, associated to order $3$ elements in $SL(3,\mathbb{Z})$ for $g_{(1,0)}=S_{23}$.}
\begin{equation}\label{eq:h1,2-sl3-constr}
    h\,g_{(p,q)}\,\tilde{h}^{-1}_{\mathcal{O}}=S_{23}\, h'\, S_{23}^{-1}\,,\qquad \tilde{h}_{\mathcal{O}}\equiv\mathcal{O}\,\tilde{h}\,\mathcal{O}\,,\qquad (h,\tilde{h})\in S_f\,,
\end{equation}
with $h'\in H$ given by:
\begin{equation}\label{eq:eq:h1,2-sl3-constr-hprime}
    h'=\begin{pmatrix}
        \alpha & 0 &-\gamma\\
        k\tilde{l}\alpha-\tilde{k}(l-p\tilde{b}) & 1 & kp \\
        -\beta&0 & \delta
    \end{pmatrix}\,.
\end{equation}
Here, $\alpha, \beta, \gamma$, and $\delta$ refer to the combinations of parameters defined in \eqref{eq:sl2-constraints} and satisfy $\alpha\delta-\beta\gamma=1$.
One easily checks that the right hand side of \eqref{eq:h1,2-sl3-constr} is indeed an element of $F$.

Using the $1$-cocycle condition \eqref{eq:hatZ-mod-prop}, we can evaluate $\hat{\mathcal{Z}}^{\alpha}_{(\cdot)}$ on both sides of \eqref{eq:h1,2-sl3-constr} to find:
\begin{equation}\label{eq:hatZS23-triv}
    \hat{\mathcal{Z}}^{\alpha}_{hg\tilde{h}^{-1}_{\mathcal{O}}}(\boldsymbol{\rho})\cong \frac{\hat{\mathcal{Z}}^{\alpha}_{S_{23}}(\boldsymbol{\rho})}{\hat{\mathcal{Z}}^{\alpha}_{S_{23}}(\tilde{h}_{\mathcal{O}}\,g^{-1}\,h^{-1}\boldsymbol{\rho})}\,,\qquad (h,\tilde{h})\in S_f\,,
\end{equation}
where we have made use of the fact that $\hat{\mathcal{Z}}^{\alpha}_{S_{23}^{-1}}(\boldsymbol{\rho})=1/\hat{\mathcal{Z}}^{\alpha}_{S_{23}}(S_{23}\boldsymbol{\rho})$ and $\hat{\mathcal{Z}}^{\alpha}_{h}(\boldsymbol{\rho})=1$ for $h\in\mathcal{H}$.
This is an interesting expression for two main reasons.
First of all, the equation takes on the form of a trivialization.
Note that the trivialization is now in terms of the function $\hat{\mathcal{Z}}^{\alpha}_{S_{23}}(\boldsymbol{\rho})$.
Since the group element $hg\tilde{h}^{-1}_{\mathcal{O}}\in F$, it could also be trivialized in terms of $\mathcal{B}^{\alpha}(\boldsymbol{\rho})$ (see the comment above \eqref{eq:def-calF1}).
This is not too surprising.
Indeed, let us plug in both numerator and denominator of \eqref{eq:hatZS23-triv} with \eqref{eq:hatZ-explicit}:
\begin{equation}
    \hat{\mathcal{Z}}^{\alpha}_{hg\tilde{h}^{-1}_{\mathcal{O}}}(\boldsymbol{\rho})\cong \frac{\mathcal{Z}^{\alpha}_{S_{23}\mathcal{O},\text{1-loop}}(\boldsymbol{\rho})}{\mathcal{Z}^{\alpha}_{S_{23}\mathcal{O},\text{1-loop}}(\tilde{h}_{\mathcal{O}}\,g^{-1}\,h^{-1}\boldsymbol{\rho})}\frac{\mathcal{Z}^{\alpha}_{\text{v}}(\boldsymbol{\rho})}{\mathcal{Z}^{\alpha}_{\text{v}}(\tilde{h}_{\mathcal{O}}\,g^{-1}\,h^{-1}\boldsymbol{\rho})}\,,
\end{equation}
where we have made use of \eqref{eq:h1,2-sl3-constr} and the fact that both $\mathcal{Z}^{\alpha}_{\mathcal{O},\text{1-loop}}(\boldsymbol{\rho})$ and $\mathcal{Z}^{\alpha}_{\text{v}}(\boldsymbol{\rho})$ are invariant, up to a phase, under the action $\boldsymbol{\rho}\to h\boldsymbol{\rho}$ for $h\in H$.
The right hand side now reflects essentially the trivialization in terms of $\mathcal{B}^{\alpha}(\boldsymbol{\rho})$ (see, e.g., Section \ref{sssec:sqed}).\footnote{The word essentially refers to the additional shift in the $z$ argument in the expression for $\mathcal{B}^{\alpha}(\boldsymbol{\rho})$. This shift produces relative $\theta$ functions in both numerator and denominator, which can be checked to cancel (up to a phase).}

A second observation is that we can further rewrite the equation as:
\begin{equation}
    \hat{\mathcal{Z}}^{\alpha}_{g}(\boldsymbol{\rho})\cong \frac{\hat{\mathcal{Z}}^{\alpha}_{S_{23}}(h\boldsymbol{\rho})}{\hat{\mathcal{Z}}^{\alpha}_{S_{23}}(\tilde{h}_{\mathcal{O}}g^{-1}\boldsymbol{\rho})}\,,\qquad (h,\tilde{h})\in S_f\,,
\end{equation}
where we have used \eqref{eq:hatZ-mod-prop} and the fact that $\hat{\mathcal{Z}}^{\alpha}_{h}= 1$ for $h\in\mathcal{H}$.
This equation can be viewed as the analogue of modular factorization for $\hat{\mathcal{Z}}^{\alpha}_{(\cdot)}$.
We conclude that the modular factorization of lens indices follows, in the cohomological language, from the $SL(3,\mathbb{Z})$ relation \eqref{eq:h1,2-sl3-constr}.
This generalizes the $Y^3=1$ relation of \cite{Gadde:2020bov} and its relation to the original holomorphic block factorization of \cite{Nieri:2015yia}, as mentioned in Section \ref{sec:intro}.

\section{Application: general lens space index}\label{sec:gen-lens-index}

In this section, we show how the $1$-cocycle condition \eqref{eq:hatZ-mod-prop} leads to an expression for the general lens index $\mathcal{I}_{(p,q)}(\hat{\boldsymbol{\rho}})$ in terms of the $S^3\times S^1$ and $S^2\times T^2$ indices.
We then evaluate the formula for the free chiral multiplet, and perform two consistency checks.

\subsection{A general formula}\label{ssec:gen-formula}

In Section \ref{ssec:top-aspects}, we discussed the Heegaard splitting of a general lens space $L(p,q)$.
In order to compute the index using the $1$-cocyle condition, we first decompose the associated gluing element $g_{(p,q)}\in SL(2,\mathbb{Z})_{23}$ into the generators $S_{23}$ and $T_{23}$.
This was called a continued fraction expansion in \cite{jeffrey1992chern}, and is given by:
\begin{equation}\label{eq:def-gpqei0k}
\Delta_t\equiv g_{(p,q)}= S_{23}\prod_{i=1}^{t}\left( T_{23}^{-e_{i}}S_{23} \right)\,,\qquad e_i\geq 2 \,.
\end{equation}
Let us explain this decomposition in some detail.
First, define a truncated product $\Delta_i$:
\begin{eqnarray} \label{eq:convergentmatrices}
	\Delta_{i}=  S_{23} \prod_{j=1}^{i}\left( T_{23}^{-e_{j}}S_{23} \right)=\left(
	\begin{array}{ccc}
		1	& 0 & 0 \\
		0 	& -s_{i}  &  -r_{i} \\ 
		0	& -p_{i} & -q_{i}
	\end{array}	
	\right) \,.
\end{eqnarray}
where the matrix entries of $\Delta_i$ are defined in terms of $e_j$.
We also define $\Delta_0\equiv S_{23}$ for later convenience.
The recurrence relation $\Delta_i = \Delta_{i-1} T_{23}^{-e_i} S_{23}$ can be written in terms of the matrix entries as: 
\begin{equation}\label{eq:recursive-pq}
	\begin{aligned}
    &	p_{i} = e_{i} p_{i-1}- p_{i-2}\,, \qquad   \, q_i=p_{i-1}\,, \\
    &s_{i} = e_{i} s_{i-1}- s_{i-2} \,,\qquad   r_i=s_{i-1}\,,\\
    & p_0=1\,,\quad p_1= e_1 \,, \quad	s_0=0\,, \quad s_1 =1 \,.
	\end{aligned}
\end{equation}
Let us also define negatively indexed parameters consistent with the initial conditions:
\begin{equation}\label{eq:minusone-pquv}
	p_{-1}=0, \qquad s_{-1}= -1\,.
\end{equation}
The first line in \eqref{eq:recursive-pq} implies that the solutions obey:
\begin{align}\label{eq:poveruandpoverq}
	\begin{split}
\frac{p_i}{q_i}&=	 [e_i,e_{i-1},\ldots,e_{1}]^-\equiv e_i-\cfrac{1}{e_{i-1}-\frac{1}{\cdots-\frac{1}{e_1}}}\,,
\end{split} 
\end{align}
which is known as the Hirzebruch--Jung continued fraction expansion. 
This expansion is unique for $e_j\ge 2$ \cite{jeffrey1992chern}. 
Similarly:
\begin{align}\label{eq:poveruandsoverp}
	\begin{split}
    -\frac{s_i}{p_i} &= [0,e_1,\ldots, e_i]^- \,.
    \end{split} 
\end{align}
Note that the continued fraction expansions imply that $1 \le s_i,q_i < p_i$.
For any coprime pair $(p,q)$ defining a lens space $L(p,q)$ with $1 \le q < p$, there exists a $t$ such that \cite{jeffrey1992chern}:
\begin{equation}\label{eq:definitionptqt=pq}
    g_{(p,q)}=\Delta_t\,,\quad \text{with}\quad p_t = p\,, \quad q_t = q\,,\quad s_t =s\,, \quad r_t = r\,.
\end{equation}
This establishes the claim in \eqref{eq:def-gpqei0k}.
Geometrically, the $e_i$ parametrize a $-p/q$ surgery on the unknot in $S^3$, which provides an alternative construction of $L(p,q)$ \cite{jeffrey1992chern,tange2010complete,bleiler1989lens,2005math6432P}. 

We can now write a formula for the $L(p,q)\times S^1$ index of a general gauge theory:
\begin{align}\label{eq:lenspartitionfunction}
    \begin{split}
\mathcal{I}_{(p,q)} (\hat{\boldsymbol{\rho}}) \equiv \mathcal{Z}_{g_{(p,q)}\mathcal{O}}(\boldsymbol{\rho})&=\sum_\alpha \mathcal{Z}_{\mathcal{O}}^\alpha(g_{(p,q)}^{-1}\,\boldsymbol{\rho}) \hat{\mathcal{Z}}^{\alpha}_{g_{(p,q)}}(\boldsymbol{\rho})\\
&=\sum_\alpha  \mathcal{Z}_{\mathcal{O}}^\alpha (g_{(p,q)}^{-1} \boldsymbol{\rho}) \prod_{i=0}^t \hat{\mathcal{Z}}_{S_{23}}^\alpha (S_{23}\Delta_i^{-1} \boldsymbol{\rho})\,,
    \end{split}
\end{align}
where we assume a Higgs branch expression for the index and have used the definition \eqref{eq:defn-hatZ} of $\hat{\mathcal{Z}}^{\alpha}_{g}(\boldsymbol{\rho})$ to rewrite the summand.
In the second line, we have used the $1$-cocycle condition \eqref{eq:hatZ-mod-prop} and the fact that $\hat{\mathcal{Z}}^{\alpha}_{T_{23}}(\boldsymbol{\rho})=1$.
Furthermore, the moduli $\hat{\boldsymbol{\rho}}$ are related to $\boldsymbol{\rho}$ through the usual relation \eqref{eq:p-moduli}.
We thus see that the $1$-cocycle condition leads to a concrete formula for the lens index in terms of the superconformal and $S^2\times T^2$ indices, apparently avoiding difficulties with a direct definition of the lens index for $q>1$.\footnote{See a Appendix \ref{sapp:lens-index} for a direct definition when $q=1$ and also \cite{Benini:2011nc,Razamat:2013jxa}.}
We also note that the formula is structurally similar to a proposed formula for the $L(p,q)$ partition function of three-dimensional $\mathcal{N}=2$ theories \cite{Alday:2017yxk}.

\subsection{Consistency checks for the chiral multiplet}\label{ssec:cons-checks-lens}

We now evaluate \eqref{eq:lenspartitionfunction} explicitly for the free chiral multiplet and perform a number of consistency checks on the result.
For simplicity of notation, we focus on vanishing R-charge.
Using the indices collected in Appendix \ref{app:lens-indices} we find:
\begin{align}\label{eq:physical-lens-4d}
    \begin{split}
        Z_{g_{(p,q)} \mathcal{O}} (\boldsymbol{\rho})  =& \prod_{i=0}^{t-1}\Gamma (z+ p_{i-1}\tau-s_{i-1}\sigma;  p_i \tau- s_i \sigma\,, p_{i-1}\tau-s_{i-1} \sigma) \\
        &\times \Gamma (z;  p_t \tau- s_t \sigma\,, p_{t-1}\tau-s_{t-1} \sigma)\\
       =&\Gamma(z;\tau,\sigma)\prod_{i=1}^{t}\Gamma (z+ p_{i}\tau-s_{i}\sigma;  p_i \tau- s_i \sigma\,, p_{i-1}\tau-s_{i-1} \sigma) \,,
    \end{split}
\end{align}
where we have made use of the shift property of the elliptic $\Gamma$ function to absorb $Z_{\mathcal{O}}(\boldsymbol{\rho})$ either in the last or the first $\hat{Z}_{S_{23}}(\boldsymbol{\rho})$ in the product.
Also note that $p_{-1}$ and $s_{-1}$ were defined in \eqref{eq:minusone-pquv}. 
We will perform two types of consistency checks on this formula. 
First, we check that it is invariant under the symmetries of the Hopf surface, as described in Section \ref{ssec:hopf-surfaces}.
Secondly, we will show that it can be factorized into holomorphic blocks consistent with modular factorization.

\subsubsection{Invariance under symmetries Hopf surface}\label{sssec:inv-syms-lens}

We implement the symmetries of the Hopf surface through an action on its Heegaard splitting, as discussed at the end of Section \ref{ssec:ambig-heegaard}.
First of all, the index is obviously invariant under $\tau\to \tau+1$, $\sigma\to \sigma +1$ and $z\to z+1$ due to the periodicity of the elliptic $\Gamma$ function.
This implies that it is invariant under all the symmetries in \eqref{eq:largediffeo-matrixrelation}.
In addition, it should be invariant under:
\begin{align}\label{eq:shift-stos+p}
    \begin{split}
        \tau &\to \tau + \sigma\,, \quad s \to s+p\,,\quad r\to r+q\,,\\
        \tilde{\tau} &\to \tilde{\tau} + \tilde{\sigma}\,, \quad q \to q+p\,,\quad r\to r+s\,,
    \end{split}
\end{align}
where we recall $\tilde{\boldsymbol{\rho}}=\mathcal{O}g_{(p,q)}^{-1}\boldsymbol{\rho}$.
We only have to check the first line, since invariance under the second line is automatic.
Note that $s_i+p_i$ satisfies the same recurrence relation as $s_i$ and leads to $s_t'=s_t+p_t$ and $r_t'=r_t+q_t$.
The transformation in \eqref{eq:shift-stos+p} thus shifts all $s_i$ and $r_i$ by $s_i \to s_i+ p_i$ and $r_i\to r_i+q_i$.  
Combined with $\tau\to\tau+\sigma$, we see that the combinations $p_i \tau -s_i\sigma$ are invariant for all $i$, and therefore the index is invariant too.
Finally, the most non-trivial transformation to check is:
\begin{equation}
    \boldsymbol{\rho}\leftrightarrow \tilde{\boldsymbol{\rho}}\,,\quad q\leftrightarrow s\,.
\end{equation}
Recall that the transformation on the moduli is implemented on $\hat{\boldsymbol{\rho}}$ through the exchange $\hat{\tau}\leftrightarrow\hat{\sigma}$.
Let us denote the transformed lens data as:
\begin{align}\label{eq:transformation-qstausigma}
\begin{split}
  &  s_t' = q_t, \qquad q_t' =s_t, \qquad p_t' =p_t\,.
  \end{split}
\end{align}
It follows that the continued fraction expansions \eqref{eq:poveruandpoverq} for the primed lens data are given in terms of the $e_i$ for $i=1, \cdots , t$ by:
\begin{align}\label{eq:transformation-eidata}
	\begin{split}
     e_{i}' = e_{t-i+1}\quad \Rightarrow\quad \frac{p_i'}{q_i'} = [ e_{t-i+1},\ldots , e_t]^-\,, \qquad-\frac{s_i'}{p_i'} &= [0,e_t,\ldots, e_{t-i+1}]^- \,.
    \end{split} 
\end{align}
To prove invariance, we need to show that:
\begin{equation}\label{eq:qs-inv}
    Z_{g_{(p,q)} \mathcal{O}} (\boldsymbol{\rho})=Z_{g_{(p,s)} \mathcal{O}} (\tilde{\boldsymbol{\rho}})\,.
\end{equation}
Recall from Section \ref{ssec:ambig-heegaard} that:
\begin{equation}\label{eq:p-moduli-3}
    (\tau,\sigma)=(\hat{\tau}-s\hat{\sigma},p\hat{\sigma})\,,\qquad (\tilde{\tau},\tilde{\sigma})=(\hat{\sigma}-q\hat{\tau},p\hat{\tau})\,.
\end{equation}
Plugging in the left and right hand side of \eqref{eq:qs-inv} with the explicit expressions \eqref{eq:physical-lens-4d} in the first and second line, respectively, and writing $(\boldsymbol{\rho},\tilde{\boldsymbol{\rho}})$ in terms of $\hat{\boldsymbol{\rho}}$ as in \eqref{eq:p-moduli-3}, one finds the invariance if:
\begin{equation}\label{eq:transformed-convergent-relation}
    p_{i-1}' = s_t\,p_{t-i} -p_t\,s_{t-i} , \qquad q_t\,p_{i-1}' -p_t\,s_{i-1}'= p_{t-i}\,.
\end{equation}
We can prove these equations as follows.
First, note that both expressions on the left hand side satisfy the recurrence relations \eqref{eq:recursive-pq} with respect to $e_{i}'$, whereas the right hand sides satisfy them with respect to $e_{t-i+1}$.
Since $e_{i}'=e_{t-i+1}$, this is consistent.
Furthermore, the initial conditions in \eqref{eq:recursive-pq} and \eqref{eq:minusone-pquv} on $p_{-1,0}$ and $s_{-1,0}$ are correctly reproduced by the left hand sides, as one may verify by evaluating both equations for $i=t$ and $i=t+1$.
This proves \eqref{eq:transformed-convergent-relation}, and therefore the invariance \eqref{eq:qs-inv}.

\subsubsection{Modular factorization of general lens index}
\label{sssec:mod-fact-lens}

In Section \ref{ssec:evidence}, we have shown that the modular properties of the elliptic $\Gamma$ function lead to the modular factorization of the $S^3\times S^1$, $L(p,1)\times S^1$ and $S^2\times T^2$ index.
In this section, we generalize this result to the $L(p,q)\times S^1$ index in the context of the free chiral multiplet.\footnote{Given this result, the extension to general $\mathcal{N}=1$ gauge theories follows along the same lines as discussed in Section \ref{sssec:proof-gen-gauge-th}.}

Let us first collect the modular property involving $t+3$ elliptic $\Gamma$ functions from Appendix \ref{sapp:t+3-Gamma}, where $t$ is the length of the continued fraction expansion of $p/q$ (cf.\ \eqref{eq:poveruandpoverq}).
This generalizes the $t=0$ ($q=0$) and $t=1$ ($q=1$) modular properties used in Section \ref{ssec:evidence}.
The relevant formula is given by:
\begin{eqnarray} \nonumber
       & &\left( \prod_{i=0}^{t-1}\Gamma (z+ p_{i-1}\tau-s_{i-1}\sigma;  p_i \tau- s_i \sigma\,, p_{i-1}\tau-s_{i-1} \sigma)\right) \Gamma (z;  p_t \tau- s_t \sigma\,, p_{t-1}\tau-s_{t-1} \sigma)\\ \label{eq:mod-prop-Gamma-t+3}
       &=&e^{-i\pi \tilde{P}_{g_{(p,q)}}^{\mathbf{m}}(z,\tau,\sigma)}\Gamma\left(\tfrac{z}{m\sigma+n_1}; \tfrac{\tau-c(k_1\sigma+l_1)}{m\sigma+n_1}, \tfrac{k_1\sigma +l_1}{m\sigma+n_1} \right) \\
       \nonumber
        &&\qquad \qquad \qquad \times  \Gamma\left(\tfrac{z}{m(p\tau
        -s \sigma) +{\tilde{n}}_{t+1}}; \tfrac{q\tau-r\sigma-n_{t+1}(\tilde{k}_{t+1} (p \tau-s \sigma) +\tilde{l}_{t+1})}{m(p\tau-s \sigma) +{\tilde{n}}_{t+1}} , \tfrac{\tilde{k}_{t+1} (p \tau-s \sigma) +\tilde{l}_{t+1}}{m(p\tau-s\sigma) +{\tilde{n}}_{t+1}}\right)\,,
\end{eqnarray}
where $k_1n_1-l_1m=1$, $
\tilde{k}_{t+1}\tilde{n}_{t+1}-\tilde{l}_{t+1}m=1$ and:
\begin{equation}
    n_{t+1}=qc-rn_1\,,\qquad \tilde{n}_{t+1}=-sn_1+pc\,.
\end{equation}
In addition, the phase polynomial is given by:
\begin{align}\label{eq:physical-lens-phasepolynomial}
\begin{split}
\tilde{P}_{g_{(p,q)}}^{\mathbf{m}}(\boldsymbol{\rho})  &= \frac{1}{m p} Q \left(m z, \frac{m(p \tau - s \sigma) + \tilde{n}_{t+1}}{p} , \frac{m \sigma+n_1}{p} \right)   +\delta \tilde{P}_{g_{(p,q)}}^{\mathbf{m}}(\boldsymbol{\rho}) \\
    \delta \tilde{P}_{g_{(p,q)}}^{\mathbf{m}}(\boldsymbol{\rho})  &=  \frac{(\eta_t+3)p -3}{6p}z  - \frac{(p^2-1)(p \tau-s\sigma+\sigma)}{12  p^2}+f_{\mathbf{m};(p,q)}\,,
    \end{split}
\end{align}
where we let $\mathbf{m}$ denote the various modular parameters and $f_{\mathbf{m},(p,q)}$ is a constant.
The constant $\eta_t$ is the continued fraction representation of the Dedekind sum $s(s,p)$:
\begin{equation}\label{eq:kappa-definition}
	\eta_t  =   \frac{q +s}{p} -3t + \sum_{i=1}^t e_i = 12s(s,p) \,.
\end{equation}
The appearance of the Dedekind sum in the context of $L(p,q)$ is not too surprising.
In particular, if two lens spaces $L(p,q)$ and $L(p,q')$ are related to each other by an orientation preserving diffeomorphism, then the Dedekind sums $s(q,p)$ and $s(q',p)$ are equal, namely, $(q-q')(qq'-1) \equiv 0 \,\mod \, p$. 
Note that the converse does not hold \cite{katase1990classifying}.

Setting $(n_1,k_1,l_1)\equiv (n,k,l)$ and $(\tilde{n}_{t+1},\tilde{k}_{t+1},\tilde{l}_{t+1})\equiv (\tilde{n},\tilde{k},\tilde{l})$, one easily checks that the modular property \eqref{eq:mod-prop-Gamma-t+3} can be written as:
\begin{align}
    \begin{split}
        Z_{g_{(p,q)}\mathcal{O}}(\boldsymbol{\rho})&=e^{-i\pi P^{\mathbf{m}}_{g_{(p,q)}}\left(\boldsymbol{\rho}\right)}B_{h}(\boldsymbol{\rho})B_{\tilde{h}}(\mathcal{O}g_{(p,q)}^{-1}\boldsymbol{\rho})\\
        &=e^{-i\pi \tilde{P}^{\mathbf{m}}_{g_{(p,q)}}\left(\boldsymbol{\rho}\right)}C_{h}(\boldsymbol{\rho})C_{\tilde{h}}(\mathcal{O}g_{(p,q)}^{-1}\boldsymbol{\rho})\,,
    \end{split}
\end{align}
where now $(h,\tilde{h})\in S_{g_{(p,q)}\mathcal{O}}$ was given in \eqref{eq:h-ht-Lens-gen}, and the holomorphic blocks $B(\boldsymbol{\rho})$ and $C(\boldsymbol{\rho})$ were given in \eqref{eq:hol-blocks-bc-free-chiral-2}.
The relative phase can again be interpreted in terms of the anomaly polynomial of the theory.
As in Section \ref{sssec:example-free-chiral}, 
we find that the phase can be written in terms of:
\begin{equation}
P_{g_{(p,q)}} \left(z;\tau,\sigma \right)=
\tfrac{1}{p} Q(z,\tfrac{p\tau-s\sigma}{p},\tfrac{\sigma}{p}) + \tfrac{(\eta_t+3)p-3}{12p} (2z+1) -\tfrac{p^2-1}{12p^2}(p\tau-s\sigma+\sigma) \,,
\end{equation}
as follows:
\begin{align}\label{eq:twoP-constant}
\begin{split}
& P_{g_{(p,q)}}^{\mathbf{m}} (z,\tau,\sigma) = \frac{1}{m } P_{g_{(p,q)}} (m z; m \tau+c,m\sigma+n) +\text{const} ~,
\\
&    \tilde{P}_{g_{(p,q)}}^{\mathbf{m}} (z,\tau,\sigma) = \frac{1}{m } P_{g_{(p,q)}} (m z+1; m \tau+c,m\sigma+n) +\text{const} ~.
    \end{split}
\end{align}
We have not found a general formula for the constants in \eqref{eq:twoP-constant}, although for any fixed set of integers $\mathbf{m}$ we can compute it (see Appendix \ref{sapp:t+3-Gamma}). 

Let us introduce the following parametrization of the anomalies:
\begin{eqnarray}\nonumber
        \mathcal{P}^{(p,q)}(\vec{Z};\hat{x}_i)&\equiv& \frac{1}{3p\hat{x}_1\hat{x}_2\hat{x}_3}\left(k_{abc}Z_aZ_bZ_c+3k_{abR}Z_aZ_bX +3k_{aRR}Z_a X^2-k_aZ_a\tilde{X}^{(p,q)}\right.\\
        \label{eq:anomaly-pol-gen-th-lenspq}
        &&\left.+k_{RRR}X^3-k_R X\tilde{X}^{(p,1)}\right)\,,
\end{eqnarray}
where $X$ is given in terms of the $\hat{x}_i$ as in the case of the $L(p,1)\times S^1$ (cf.\ \eqref{eq:X-and-Xt}), and we use a modified definition for $\tilde{X}$:
\begin{equation}
    \tilde{X}^{(p,q)} = \frac{1}{4} (\hat{x}_1^2+\hat{x}_2^2+\hat{x}_3^2-2(p \eta_t+3p -3)\hat{x}_2\hat{x}_3)\,.
\end{equation}
One may verify that again the phase polynomial correctly captures the anomalies for a free chiral with $R=0$ (cf.\ \eqref{eq:anomaly-coefficient-R}) by noting that:
\begin{equation}\label{eq:relation-Lpq}
P_{g_{(p,q)}}(\tfrac{Z+x_1}{x_1},\tfrac{x_2}{x_1},\tfrac{x_3}{x_1};0) = \mathcal{P}^{(p,q)} (\vec{Z},\hat{x}_i)+\text{const.}
\end{equation}
where we view $x_i$ as functions of $\hat{x}_i$ according to \eqref{eq:p-moduli-3}, and have not obtained an analytic formula for the constant.
We stress that the form of $\mathcal{P}^{(p,q)}(\vec{Z};\hat{x}_i)$ is not preserved under the shift $Z_a\to Z_a+X$ for $q>1$.
It is therefore not entirely clear how to reinstate a non-vanishing R-charge consistent with the anomaly polynomial.
We hope to come back to a better understanding of this point in future work. 

In conclusion, we see that a general $L(p,q)\times S^1$ index can be factorized in terms of a family of holomorphic blocks that are consistent with modular factorization.
We view this as an additional consistency check of our proposed formula for the general $L(p,q)\times S^1$ index \eqref{eq:lenspartitionfunction}.

\section{Summary and future directions}\label{sec:sum-future}

In this work, we have argued that a subset $S_f\subset H\times H$ of the ambiguities in the Heegaard splitting of a (secondary) Hopf surface of toplogy $L(p,q)\times S^1$ leads to a modular family of factorization properties for the lens indices of four-dimensional $\mathcal{N}=1$ gauge theories, provided the index of the theory has a Higgs branch expression.
We proved the claim for the free chiral multiplet and SQED with a non-zero FI parameter, and indicated how those proofs can be extended to more general gauge theories.
Furthermore, we have shown that $S_f$ can be geometrically characterized as capturing all embeddings of the time circle into the Heegaard splitting such that the associated gluing transformation fixes this circle.
The embedding is labeled by two large diffeomorphisms $(h,\tilde{h})$, which also label the pair of holomorphic blocks compatible with the factorization:
\begin{equation}
    \mathcal{I}_{(p,q)}(\hat{\boldsymbol{\rho}})=e^{-i\pi \mathcal{P}^{\mathbf{m}}_f(\boldsymbol{\rho})} \sum_{\alpha} \mathcal{B}^\alpha_{h}(\boldsymbol{\rho})\mathcal{B}^\alpha_{\tilde{h}}(f^{-1}\boldsymbol{\rho})\,,\qquad  (h,\tilde{h})\in S_f\subset H\times H\,.
\end{equation}
That there exists a modular family of such embeddings can be understood from the fact that there is a $T^2\subset D_2\times T^2$, and the fact that a general (secondary) Hopf surface admits a Heegaard splitting already in terms of an $SL(2,\mathbb{Z})\subset SL(3,\mathbb{Z})$ subset of gluing transformations.
In this family, $\tilde{h}$ is completely determined by $h$ and $h\in H$ itself cannot be completely arbitrary.
For example, if the topology of the lens space $L(p,q)$ has $|q|>1$ or if $h$ takes non-trivial value in the $\mathbb{Z}^2$ part of $H$, $h$ is constrained to take value in a congruence subgroup $\Gamma_0(q+ap)\subset SL(2,\mathbb{Z})\subset H$ for $a\in \mathbb{Z}$, as discussed in detail in Section \ref{ssec:consistency-cond}.
An interesting subcase is the $SL(2,\mathbb{Z})$ family of holomorphic blocks into which $L(p,\pm 1)\times S^1$ indices can be factorized for any $p$, generalizing the factorization of such indices in terms $\mathcal{B}^{\alpha}_{S}(\boldsymbol{\rho})$ originally discovered in \cite{Nieri:2015yia}.

The proof of modular factorization involves modular properties of the elliptic $\Gamma$ function and the $q$-$\theta$ function, which are the building blocks of any $\mathcal{N}=1$ gauge theory index.
Because of the non-renormalization of indices, the statement also applies to the IR SCFTs obtained from (supersymmetric) RG flows.\footnote{It has been suggested by Razamat on various occasions that any $\mathcal{N}=1$ SCFT could be reached by such flows (see, e.g., \cite{Razamat:2022gpm} and references therein).}

These results provide a clear physical basis to systematically prove that the normalized part of the collection of lens indices $\hat{\mathcal{Z}}^{\alpha}_{g}(\boldsymbol{\rho})$ obeys a $1$-cocycle condition associated to the group $\mathcal{G}=SL(3,\mathbb{Z})\ltimes \mathbb{Z}^{3r}$, as first proposed in \cite{Gadde:2020bov}.
In particular, the non-triviality of the cohomology class of $\hat{\mathcal{Z}}^{\alpha}_{g}(\boldsymbol{\rho})$ in $H^1(\mathcal{G},N/M)$ is explained by the fact only a subset of indices, associated to an $SL(2,\mathbb{Z})\ltimes \mathbb{Z}^{2(1+r)}\subset \mathcal{G}$, can be simultaneously factorized in terms of a common holomorphic block.

Finally, as an application of the $1$-cocycle condition, we derived a formula for the $L(p,q)\times S^1$ index, generalizing \cite{Benini:2011nc,Razamat:2013opa}.

\bigskip

There are many interesting directions for future research.
First of all, the analysis in this paper was originally motivated by its applications to the black holes in the gravitational dual.
A number of questions in this direction deserve further study:
\begin{itemize}
    \item As mentioned in the introduction, the modular properties of elliptic $\Gamma$ functions can be used to compute asymptotic, Cardy-like limits of indices.
    In an upcoming work \cite{Jejjala:20222023}, we apply modular factorization in the context of the $\mathcal{N}=4$ theory to study a family of such Cardy-like limits, which is parametrized by the right coset $\Gamma_{\infty}'\times \Gamma_{\infty}'\backslash S_f$.
    The interpretation of this coset on the gravitational side will also be discussed there.
    \item Apart from the superconformal index, we can also study Cardy-like limits of more general lens indices.
    It would be interesting to understand the gravitational interpretation, in particular the existence of supersymmetric black lenses in AdS.\footnote{Such solutions are not available in the literature, in spite of the existence of asymptotically \emph{flat} black lenses in minimal gauged supergravity or in the $U(1)^3$ gauged supergravity \cite{Kunduri:2014kja,Kunduri:2016xbo,Tomizawa:2016kjh,Breunholder:2017ubu,Breunholder:2018roc,Tomizawa:2019yzb}.}
    \item The fact that $S_f$ can be expressed in terms of modular groups makes it tempting to compare the situation to AdS$_3$/CFT$_2$ and the associated Farey tail expansion of the elliptic genus \cite{Dijkgraaf:2000fq,Manschot:2007ha}.
    However, modular factorization of lens indices $\mathcal{Z}_{f}(\boldsymbol{\rho})$ follows from a combined action $\boldsymbol{\rho}\to h\boldsymbol{\rho}$ and $f\to hf\tilde{h}^{-1}$ for $(h,\tilde{h})\in S_f$.
    It is not clear to us whether this more general type of covariance can lead to Farey tail-like formulas for the index. 
    \item The holographic duals of the original holomorphic blocks were dubbed gravitational blocks \cite{Hosseini:2019iad} and have played a role in a number of follow-ups \cite{Hosseini:2020mut,Hosseini:2021mnn,Hosseini:2022vho}.
    Our work predicts a modular family of these gravitational blocks in AdS$_5$, and it will be interesting to see if this sheds light on the previous point.
\end{itemize}
There are also implications of this work within the context of SCFTs:
\begin{itemize}
    \item The Schur limit of the superconformal index of $\mathcal{N}=2$ SCFTs is known to have modular properties \cite{Razamat:2012uv}, which can be explained in terms of the underlying chiral algebras \cite{Beem:2013sza,Beem:2017ooy,Pan:2021mrw,Beem:2021zvt,Hatsuda:2022xdv}.
    We expect that the Schur limit of modular factorization provides a geometric explanation for the modular properties of the Schur index.
    \item The fact that the elliptic genus of a CFT$_2$ is a Jacobi form has a clear interpretation (see, e.g., \cite{Kawai:1993jk}).
    We would like to have a similarly transparent physical argument for the relevance of degree $1$ automorphic forms in the context of lens indices.
    Part of the argument certainly includes modular factorization, but the fact that the normalized part of a lens index, $\hat{\mathcal{Z}}^{\alpha}_{g}(\boldsymbol{\rho})$, plays a crucial role obscures the argument because physically this is a somewhat unnatural object.
    More recent mathematical work \cite{Felder_2008} adopts the language of gerbes and stacks to investigate the elliptic $\Gamma$ function, which could potentially be relevant as well.
    \item Finally, we would like to have a physical interpretation of the constant $f_{(m,n,\tilde{n})}$ appearing in the phase polynomials studied in Section \ref{ssec:evidence}.
    The case $\tilde{n}=n$, where the constant reduces to a Dedekind sum $s(n,m)$, was studied in \cite{ArabiArdehali:2021nsx} from a different perspective.
\end{itemize}
We hope to report on these topics in future work.

\section*{Acknowledgements}
We are grateful to Miranda Cheng, Abhijit Gadde, and Finn Larsen for conversations about this work.
We also thank an anonymous referee for several critical remarks on v2 of this paper.
VJ is supported by the South African Research Chairs Initiative of the Department of Science and Innovation and the National Research Foundation.
YL was supported by the UCAS program of special research associate, the internal funds of the KITS, and the
Chinese Postdoctoral Science Foundation.
YL is also supported by a Project Funded by the Priority Academic Program Development of Jiangsu Higher Education Institutions (PAPD).
WL is supported by NSFC No.\ 11875064, No.\ 11947302, and the Max-Planck Partnergruppen fund; she is also grateful for the hospitality of AEI Potsdam, where part of this work was done.

\appendix

\section{Definitions and properties of special functions}
\label{app:defs}

In this appendix we collect definitions and mathematical properties of the $q$-$\theta$ function and the elliptic $\Gamma$ function.

\subsection[\texorpdfstring{The $q$-$\theta$ function}{The q-theta function}]{The \texorpdfstring{$\bm{q}$-$\bm{\theta}$}{q-theta} function}
\label{append:q-theta}

The $q$-$\theta$ function $\theta(z;\tau)$ is defined as:
\begin{equation}
\theta(z;\tau) = \exp \left(- \sum_{m=1}^\infty \frac{x^m+x^{-m}q^m}{m(1-q^m)} \right) = \prod_{n=0}^{\infty} (1-x q^n) (1-x^{-1}q^{n+1}) \,,
\end{equation}
where $q=e^{2\pi i \tau}$ and $x=e^{2\pi i z}$.
The elliptic and extension properties of the $\theta(z;\tau)$ function are given by:
\begin{eqnarray}\label{eq:elliptic-theta}
&&		\theta(z+m\tau+n; \tau) = (-x)^{-m} q^{ - \frac{m(m-1)}{2}} \theta(z;\tau) \,, \\
&& \label{eq:ext-theta}
		\theta(-z;\tau) = \theta(z+\tau;\tau) \,, \qquad 	\theta(z;-\tau)\theta(z;\tau) = -x \,.
\end{eqnarray}
In addition, the $q$-$\theta$ function satisfies a multiplication formula:
\begin{equation}\label{eq:multiplication-qtheta}
    \theta(z;\tau) = \prod_{j=0}^{m-1} \theta(z+j\tau,m\tau)\,.
\end{equation}
Finally, it also satisfies a modular property under $SL(2,\mathbb{Z})$ transformation:
\begin{align}\label{eq:result}
	\theta\left(\frac{z}{ m\tau+n};\frac{k\tau+l}{m\tau+n} \right)  &=e^{i\pi B_2^\mathbf{m}(z;\tau)} \theta(z;\tau)  \,, \quad \mathbf{m}=(m,n)\,,
\end{align}
where the phase is given by \cite{Felder_2008}:
\begin{align}\label{eq:theB-polynomial}
\begin{split}
	B_2^\mathbf{m}(z;\tau) =& \tfrac{m z^2}{m\tau+n} + z\left(\tfrac{1}{m\tau+n}-1 \right) +	\tfrac{1}{6} \left(\tau+ \tfrac{1}{m(m\tau+n)} \right)+ \tfrac{n}{6m} -\tfrac{1}{2} -2s(n,m)\,,
	 \end{split}
\end{align}
and the Dedekind sum is defined below.
We also define $B_2(z,\tau)\equiv B_2^{(1,0)}(z,\tau)$.
One can check that: 
\begin{equation}\label{eq:Bpolynomial-cB-constrant}
    B_2^{\mathbf{m}}(z,\tau)= \tfrac{1}{m} B_2(mz,m\tau+n) + 2\sigma_1(n,1;m)\,.
\end{equation}
where $\sigma_1(n,1;m)$ is the Fourier--Dedekind sum defined as \cite{beck2002frobenius}:
\begin{equation}\label{eq:def-sigma1}
    \sigma_1(n,1;m) = \frac{1}{m}\sum_{\mu=1}^{m-1} \frac{\xi^\mu}{(\xi^{n\mu}-1)(\xi^\mu-1)}, \qquad \xi= e^{\frac{2\pi i}{m}}
\end{equation}

\subsection{Dedekind sum and its generalizations}\label{appendix:Dedekinsum}
\subsection*{Dedekind sum}
The ordinary Dedekind sum is defined for two coprime integers $(n,m)$ as follows:
	\begin{equation}\label{eq:def-Dedekindsum}
	s(n,m) =
	\frac{1}{4m} \sum_{\mu=1}^{m-1} \cot \frac{\pi \mu}{m} \cot \frac{\pi n \mu}{m}\,=- \sigma_1 (n,1;m) +\frac{1-m}{4m} \,.
\end{equation}
Let the Hirzebruch--Jung continued fraction expansion of $m/n$ be given by $[e_t;...,e_1]^-$, see \eqref{eq:poveruandpoverq}.
The Dedekind sum can alternatively be written as \cite{holzapfel1988chern,girstmair2006continued,Urza2007ArrangementsOC}:
\begin{equation}
    s(n,m) = \frac{n+n'}{12m}-\frac{t}{4} +\frac{1}{12}\sum_{i=1}^t e_i\,,
\end{equation}
where $n' n \equiv 1 \mod m$ and $0<n'<m$.
One of its best known properties is the reciprocal relation:
\begin{equation}\label{eq:Dedekindsum-reciprocal}
	s(n,m)+s(m,n) = -\frac{1}{4} + \frac{1}{12} \left(\frac{m}{n} + \frac{n}{m} + \frac{1}{mn} \right)\,.
\end{equation}
See also the lectures \cite{rademacher1972dedekind}. 

\subsection*{Generalized Dedekind sum}
There are various kinds of generalizations of the Dedekind sum \cite{beck2002frobenius,rademacher1972dedekind,rademacher1954generalization,carlitz1954note}.
One such generalization takes the following form:
\begin{equation}\label{eq:definition-sigmat}
	\sigma_t (n_1,n_2 \cdots,n_r;m) = \frac{1}{m} \sum_{\xi^m=1 \neq \xi} \frac{\xi^t}{(\xi^{n_1}-1) \cdots (\xi^{n_r} -1)}
\end{equation} 
It generalizes the object $\sigma_1(n,1;m)$, which is related to the Dedekind sum as in \eqref{eq:def-Dedekindsum}. 
As we will see in Appendix \ref{app:mod-props-Gamma}, $\sigma_1(n_1,n_2,1;m)$ will play a role in modular properties of the elliptic $\Gamma$ function.

\subsection[\texorpdfstring{The elliptic $\Gamma$ function}{The elliptic Gamma function}]{The elliptic \texorpdfstring{$\bm{\Gamma}$}{Gamma} function}

The elliptic $\Gamma$ function is defined as follows \cite{Felder_2000}:
\begin{align}\label{eq:defn-ell-gamma-app}
	\begin{split}
		\Gamma(z;\tau,\sigma)=\exp\left(\sum^{\infty}_{\ell=1}\frac{x^\ell-(x^{-1}pq)^\ell}{\ell(1-p^\ell)(1-q^\ell)}\right)=\prod^{\infty}_{m,n=0}\frac{1-x^{-1}p^{m+1}q^{n+1}}{1-x p^{m}q^{n}} \,,
	\end{split}
\end{align}
where $q=e^{2\pi i \tau}$, $p=e^{2\pi i \sigma}$, and $x=e^{2\pi i z}$.
Note that the formulae are convergent for $\mathrm{Im}(\tau)>0$, $\mathrm{Im}(\sigma)>0$ and $0<\mathrm{Im}(z)<\mathrm{Im}(\tau)+\mathrm{Im}(\sigma)$.
The following basic properties follow from the definition:
\begin{align}
	\begin{split}
		\Gamma(z;\tau,\sigma)&=\Gamma(z;\sigma,\tau) \,, \\
		\Gamma(z+1;\tau,\sigma)&=\Gamma(z;\tau+1,\sigma)=\Gamma(z;\tau,\sigma+1)=\Gamma(z;\tau,\sigma)\\
    	\Gamma(z;\tau,\sigma)&=\frac{1}{\Gamma(\tau+\sigma-z;\tau,\sigma)}\,.
	\end{split}
\end{align} 
The elliptic $\Gamma$ function also satisfies the shift property:
\begin{align}\label{eq:basic-shift-gamma-app}
	\begin{split}
		\Gamma(z+\tau;\tau,\sigma)&=\theta(z;\sigma)\Gamma(z;\tau,\sigma) \,,\\
		\Gamma(z+\sigma;\tau,\sigma)&=\theta(z;\tau)\Gamma(z;\tau,\sigma) \,,
	\end{split}
\end{align}
where $\theta(z;\tau)$ was defined above.
The $\Gamma$ function satisfies a reflection property: 
\begin{equation}\label{eq:gamma-theta-id-app}
	\Gamma(z;\tau,\sigma)\Gamma(-z;\tau,\sigma)=\frac{1}{\theta(z;\sigma)\theta(-z;\tau)}\,.
\end{equation}
In addition, it can be extended to $\tau,\sigma \in \mathbb{C}\setminus \mathbb{R}$ as follows \cite{Felder_2000}:
\begin{align}\label{eq:extend}
	\begin{split}
		& \Gamma(z;-\tau,\sigma) = \frac{1}{\Gamma(z+\tau;\tau,\sigma)} = \Gamma(\sigma-z;\tau,\sigma) \\
		& \Gamma(z;\tau,-\sigma) = \frac{1}{\Gamma(z+\sigma;\tau,\sigma)} = \Gamma(\tau-z;\tau,\sigma) \,.
	\end{split}
\end{align}
Two interesting relations involving three elliptic $\Gamma$ functions are
\begin{equation}\label{eq:ts3-reln-2}
    \Gamma(z;\tau,\sigma)=\Gamma(z;\tau-\sigma,\sigma)\Gamma(z;\sigma-\tau,\tau)
\end{equation}
and
\begin{equation}\label{eq:Y3-prop-2}
    \Gamma(z;\tau,\sigma)=e^{-i\pi Q(z;\tau,\sigma)}\Gamma\left(\tfrac{z}{\sigma};\tfrac{\tau}{\sigma},-\tfrac{1}{\sigma}\right)\Gamma\left(\tfrac{z}{\tau};\tfrac{\sigma}{\tau},-\tfrac{1}{\tau}\right)\,.
\end{equation}
The function $Q(z;\tau,\sigma)$ is defined as:
\begin{align}
\label{eq:feldermoduality}
	\begin{split}
		Q(z;\tau,\sigma) =& \frac{z^3}{3\tau\sigma} -\frac{\tau+\sigma-1}{2\tau\sigma} z^2 + \frac{\tau^2+\sigma^2 +3\tau \sigma -3\tau-3\sigma+1}{6\tau\sigma} z \\
		&+ \frac{(\tau+\sigma-1)(\tau^{-1}+\sigma^{-1}-1)}{12} \,.
	\end{split}
\end{align}

\section{Hopf surfaces and their Heegaard splitting}\label{app:hopf-surfaces}

In this appendix, we review how the manifolds $D_2\times T^2$, $S^2\times T^2$, $S^3\times S^1$, and finally $L(p,q)\times S^1$ can be endowed with complex structure moduli.
For each of the closed manifolds, we also indicate their Heegaard splitting including a mapping of the complex structure moduli.
We employ the notation introduced in Section \ref{ssec:hopf-surfaces}.

\subsection[\texorpdfstring{${D_2\times T^2}$}{D2xT2}]{\texorpdfstring{$\bm{D_2\times T^2}$}{D2xT2}}\label{sapp:D2T2}

Consider a metric on $D_2\times \mathbb{C}$ in terms of complex coordinates $(z,\bar{z})$ and $(w,\bar{w})$:\footnote{The following discussion is based on \cite{Closset:2013sxa,Longhi:2019hdh}.}
\begin{equation}\label{eq:metric-solid-t3}
    ds^2=\frac{4\,dz\,d\bar{z}}{(1+|z|^2)^2}+dw\,d\bar{w}\,,
\end{equation}
where we take the disc to lie within $|z|\leq 1$.
We can now obtain a complex manifold of topology $D_2\times T^2$ through the following quotient:
\begin{equation}\label{eq:solid-torus-ids}
    (z,w)\sim(e^{2\pi i\alpha}z,w+2\pi)\,,\qquad (z,w)\sim (e^{2\pi i\beta}z,w+2\pi\sigma)\,. 
\end{equation}
Here, $\sigma\in \mathbb{H}$ is the standard complex structure parameter of the $T^2$, and $\alpha$ and $\beta$ represent two real parameters.
Note that the metric \eqref{eq:metric-solid-t3} descends to the quotient.
The resulting manifold can be viewed as a disc fibration over the torus.
It will also be useful to write the metric in terms of real coordinates:
\begin{equation}
  \begin{alignedat}{2}
     z&=\tan\frac{\theta}{2}e^{i(\phi+\alpha x+\beta y)}\,,\qquad  & \bar{z}&=\tan \frac{\theta}{2}e^{-i(\phi+\alpha x+\beta y)} \,,\\
        w&=x+\sigma y\,, &  \bar{w}&=x+\bar{\sigma}y\,.
  \end{alignedat}
\end{equation}
Here, $\theta\in [0,\frac{\pi}{2}]$ and the other coordinates $\phi$, $x$ and $y$ are identified modulo $2\pi$.
In these coordinates, the metric takes the form of a torus fibration over the disc:
\begin{equation}\label{eq:metric-solid-t3-2}
    ds^2=d\theta^2+\sin^2\theta(d\phi+\alpha dx+\beta dy)^2+(dx+\mathrm{Re}(\sigma)dy)^2+\mathrm{Im}(\sigma)^2dy^2\,.
\end{equation}
In Section \ref{ssec:top-aspects}, we described how the group of large diffeomorphisms $H$ of $D_2\times T^2$ acts on the cycles $(\lambda',\mu,\lambda)$.
The same group reemerges through an action on $(\alpha,\beta;\sigma)$ that keeps the identifications \eqref{eq:solid-torus-ids} invariant \cite{Closset:2013sxa}.
Its action is generated by the following transformations:
\begin{equation}
\begin{array}{cclcccl}
        (\alpha,\beta;\sigma)&\to& (\alpha,\beta+\alpha;\sigma+1)\,, &\qquad& (\alpha,\beta;\sigma)&\to&(\beta,-\alpha;-\frac{1}{\sigma})\,, \cr
        (\alpha,\beta;\sigma)&\to& (\alpha+1,\beta;\sigma)\,, &\qquad& (\alpha,\beta;\sigma)&\to& (\alpha,\beta+1;\sigma)\,,
\end{array}
\label{eq:large-diffeo-action}
\end{equation}
where the first line covers the $SL(2,\mathbb{Z})$ part of $H$, and the second line the $\mathbb{Z}^2$ part.

In addition to large diffeomorphisms, there may also be large gauge transformations associated to global symmetries of the theory in question.
Accordingly, we can introduce complex background holonomies $\vec{z}=(z_a)$:
\begin{equation}\label{eq:chem-pots-D2xT2}
    z_a=z^{(x)}_a\hat{\sigma}-z^{(y)}_a\,,
\end{equation}
where $z^{(x)}_a$ and $z^{(y)}_a$ represent the holonomies of the gauge field along the cycles of $T^2$, and $a=1,\ldots,r$ for the Cartan components of a background gauge field of a rank $r$ global symmetry.
Geometrically, the holonomies $\vec{z}$ parametrize a rank $r$ complex vector bundle over $D_2\times T^2$.
This yields a $\mathbb{Z}^{2r}$ group of large gauge transformations:
\begin{equation}\label{eq:large-gauge-H}
    z_a\to z_a+1\,,\quad z_a\to z_a+\sigma\,.
\end{equation}
It will be convenient to also combine the parameters $\alpha$ and $\beta$ into a single complex parameter \cite{Closset:2013vra,Closset:2013sxa,Longhi:2019hdh}:
\begin{equation}
    \tau\equiv\alpha\sigma -\beta\,.
\end{equation}
As introduced in Section \ref{ssec:hopf-surfaces}, we combine the above moduli into:
\begin{equation}\label{eq:rho-homog-app}
   \boldsymbol{\rho}\equiv(\vec{z}; \tau, \sigma)=\left(\tfrac{\vec{Z}}{x_1};\tfrac{x_2}{x_1},\tfrac{x_3}{x_1}\right) \,.
\end{equation}
In terms of $\boldsymbol{\rho}$, one can check that the action \eqref{eq:large-diffeo-action} turns into the action action of $\mathcal{H}\subset \mathcal{G}$ described in \eqref{eq:calG-action}.

\subsection[\texorpdfstring{$S^2\times T^2$}{S2xT2}]{\texorpdfstring{$\bm{S^2\times T^2}$}{S2xT2}}\label{sapp:S2T2}

Complex manifolds of topology $S^2\times T^2$ can also be obtained from the metric \eqref{eq:metric-solid-t3} by simply allowing $z\in \mathbb{C}$.
Indeed, the $(z,\bar{z})$ part of this metric is then nothing but the Fubini--Study metric on $S^2$.
Similar to before, we impose the identifications:
\begin{equation}\label{eq:s2xt2-ids}
    (z,w)\sim(e^{2\pi i\hat{\alpha}}z,w+2\pi)\,,\qquad (z,w)\sim (e^{2\pi i\hat{\beta}}z,w+2\pi\hat{\sigma})\,,
\end{equation}
where we use hats to distinguish the parameters of the closed manifold with those of $D_2\times T^2$.
This leads to a manifold of topology $S^2\times T^2$ parametrized by two complex structure moduli $(\hat{\tau},\hat{\sigma})$ with $\hat{\tau}=\hat{\alpha}\hat{\sigma}-\hat{\beta}$ \cite{Closset:2013vra,Closset:2013sxa}.
Including holonomies for the global symmetries as in \eqref{eq:chem-pots-D2xT2}, we write the combined set of moduli as: $\hat{\boldsymbol{\rho}}=(\hat{z}_a;\hat{\tau},\hat{\sigma})$.
We employ the notation \eqref{eq:notation-hopf-surface} to write this manifold, including the complex vector bundle, as $\mathcal{M}_{(0,-1)}(\hat{\boldsymbol{\rho}})$.
The combined group of large diffeomorphisms and gauge transformations is again a copy of $\mathcal{H}$, and it acts in an identical fashion on $\hat{\boldsymbol{\rho}}$ as in the case of $D_{2}\times T^2$.

The relation between the moduli $\hat{\boldsymbol{\rho}}$ and the $D_2\times T^2$ moduli $\boldsymbol{\rho}$ and $\tilde{\boldsymbol{\rho}}$ associated to the Heegaard splitting is as follows.
By construction, the geometry of $\mathcal{M}_{(0,-1)}(\hat{\boldsymbol{\rho}})$ is equivalent to the $D_2\times T^2$ geometry for $|z|\leq 1$.
Therefore, the complex structure moduli of the first $D_2\times T^2$ in \eqref{eq:notation-Mg-split} map trivially onto those of $\mathcal{M}_{(0,-1)}(\hat{\boldsymbol{\rho}})$:
\begin{equation}
    (z_a;\tau,\sigma)=(\hat{z}_a;\hat{\tau},\hat{\sigma})\,.
\end{equation}
To describe $S^2\times T^2$ around $z=\infty$ instead, we transform coordinates via $z'=1/z$.
It follows that the moduli of the second $D_2\times T^2$ geometry are given by:
\begin{equation}
    (\tilde{z}_a;\tilde{\tau},\tilde{\sigma})=(\hat{z}_a;-\hat{\tau},\hat{\sigma})\,.
\end{equation}
Note that the moduli are correctly related through the gluing condition \eqref{eq:gluing-condition}: $\tilde{\boldsymbol{\rho}}=\mathcal{O}\boldsymbol{\rho}$.
We may thus write the Heegaard splitting of $\mathcal{M}_{(0,-1)}(\hat{\boldsymbol{\rho}})$ as:
\begin{equation}\label{eq:M1-split}
    \mathcal{M}_{(0,-1)}(\hat{\boldsymbol{\rho}})\cong M_{\mathcal{O}}(\boldsymbol{\rho},\mathcal{O}\boldsymbol{\rho})\,,\qquad \hat{\boldsymbol{\rho}}=\boldsymbol{\rho}\,,
\end{equation}
where we use the notation \eqref{eq:notation-Mg-split}.

\subsection[\texorpdfstring{$S^3\times S^1$}{S3xS1}]{\texorpdfstring{$\bm{S^3\times S^1}$}{S3xS1}}\label{sapp:S3S1}

We now consider the manifold $S^3\times S^1$, which can be viewed as the special case of the lens space geometries for $(p,q)=(1,0)$.
The standard way to endow this manifold with complex structure moduli is to view it as the (primary) Hopf surface (see, e.g., \cite{Closset:2013vra} and references therein).
The Hopf surface is defined as a quotient of $\mathbb{C}^2\setminus \lbrace(0,0)\rbrace$ by the $\mathbb{Z}$-action: 
\begin{equation}\label{eq:hopf-surface-ids-app}
    (z_1,z_2)\sim (\hat{p}z_1,\hat{q}z_2)\,,\qquad  0<|\hat{p}|\leq |\hat{q}|<1\,,
\end{equation}
with $\hat{p}=e^{2\pi i\hat{\sigma}}$ and $\hat{q}=e^{2\pi i\hat{\tau}}$.\footnote{The complex parameters $\hat{p}$ and $\hat{q}$ are not to be confused with the integers $p$ and $q$ defining the lens space $L(p,q)$.}
To see that the Hopf surface is diffeomorphic to $S^3\times S^1$, consider the following parametrization of $z_{1,2}$:%
\begin{equation}
    z_1=\hat{p}^x\cos(\frac{\theta}{2})e^{i\phi_1}\,,\qquad z_2=\hat{q}^x\sin(\frac{\theta}{2})e^{i\phi_2}\,.
\end{equation}
Here, $x\sim x+1$ ensures the identification \eqref{eq:hopf-surface-ids-app}, $0\leq \theta\leq \pi$, and $\phi_{1,2}$ are identified modulo $2\pi$.
Parametrized in this way, it is easy to see that:
\begin{equation}
    \left|\frac{z_1}{\hat{p}^x}\right|^2+\left|\frac{z_2}{\hat{q}^x}\right|^2=1\,.
\end{equation}
For fixed $x$, this represents a (squashed) $S^3$.
Given that the left hand side is a monotonic function of $x$ and since $x\sim x+1$, one establishes the diffeomorphism with $S^3\times S^1$.
Note that this parametrization reflects the picture of $S^3$ as a torus fibration over an interval where the $\phi_1$ cycle shrinks at $\theta=\pi$ and the $\phi_2$ cycle shrinks at $\theta=0$.

We also introduce (real) holonomies $\vec{z}$ along $S^1$ associated to global symmetries, parametrizing a real vector bundle over $S^3\times S^1$.
Together, we capture the moduli by $\hat{\boldsymbol{\rho}}=(\hat{z}_a;\hat{\tau},\hat{\sigma})$.
Note the following symmetries of the Hopf surface:
\begin{equation}\label{eq:syms-s3xs1}
    \hat{z}_a\to \hat{z}_a+1\,,\quad \hat{\tau}\to\hat{\tau}+1\,,\quad \hat{\sigma}\to \hat{\sigma}+1\,,\qquad \hat{\tau}\leftrightarrow \hat{\sigma}\,.
\end{equation}
We now turn to the Heegaard splitting of the Hopf surface. 
To this end, let us first write down a Hermitian metric on $\mathbb{C}^2\setminus\lbrace(0,0)\rbrace$ \cite{Closset:2013vra}:
\begin{equation}\label{eq:hopf-metric}
    ds^2=\frac{\mathrm{Im}(\hat{\tau})}{\mathrm{Im}(\hat{\sigma})}\frac{dz_1d\bar{z}_1}{|\hat{p}|^{2x}}+\frac{\mathrm{Im}(\hat{\sigma})}{\mathrm{Im}(\hat{\tau})}\frac{dz_2d\bar{z}_2}{|\hat{q}|^{2x}}\,.
\end{equation}
This metric is constructed such that it is invariant under the identifications \eqref{eq:hopf-surface-ids-app}, and thus reduces to a metric on the Hopf surface.
For comparison with the $D_2\times T^2$ geometries, it will be convenient to change coordinates via \cite{Closset:2013sxa}:
\begin{equation}\label{eq:coord-transf-hopf-to-solid-t3}
   z=\frac{z_2}{z_1^\frac{\mathrm{Im}(\hat{\tau})}{\mathrm{Im}(\hat{\sigma})}}\,,\qquad  w=i\log z_1\,.
\end{equation}
These coordinates cover $z_1\neq 0$, which is a coordinate patch with topology $D_2\times T^2$.
For $\mathrm{Im}(\hat{\tau})=\mathrm{Im}(\hat{\sigma})$, one may verify that the metric \eqref{eq:hopf-metric} becomes the $D_2\times T^2$ metric \eqref{eq:metric-solid-t3}.
We will however keep $\hat{\tau}$ and $\hat{\sigma}$ general.
In particular, we then note that the identification $(z_1,z_2)= (e^{2\pi i}z_1,e^{2\pi i}z_2)$ and the Hopf surface identifications \eqref{eq:hopf-surface-ids-app} in terms of the $(z,w)$ coordinates become:
\begin{equation}
    (z,w)\sim (e^{2\pi i\alpha}z,w+2\pi)\,,\qquad (z,w)\sim (e^{2\pi i\beta}z,w+2\pi\hat{\sigma})\,,
\end{equation}
with $\hat{\tau}=\alpha\hat{\sigma}-\beta$.
In other words, in a neighbourhood around $z_2=0$ the Hopf surface becomes a $D_2\times T^2$ geometry with complex structure moduli $(\tau,\sigma)$ that map trivially onto $(\hat{\tau},\hat{\sigma})$:
\begin{equation}
    (z_a;\tau,\sigma)=(\hat{z}_a;\hat{\tau},\hat{\sigma})\,.
\end{equation}
Since $\hat{z}_a$ parametrizes real holonomies, for consistency one has to set $u_x^{(a)}=0$ for $z_a$ (see \eqref{eq:chem-pots-D2xT2}). 
The description around $z_1=0$ is analogous.
In this case, we change coordinates according to:
\begin{equation}
   z'=\frac{z_1}{z_2^{\frac{\mathrm{Im}(\hat{\sigma})}{\mathrm{Im}(\hat{\tau})}}}\,,\qquad  w'=i\log z_2\,.
\end{equation}
In these coordinates, the identifications become:
\begin{equation}
    (z',w')\sim (e^{2\pi i\alpha'}z',w'+2\pi)\,,\qquad (z',w')\sim (e^{2\pi i\beta'}z',w'+2\pi\hat{\tau})\,,
\end{equation}
with $\hat{\sigma}=\alpha'\hat{\tau}-\beta'$.
Thus, around $z_1=0$ we find that the Hopf surface becomes a $D_2\times T^2$ geometry with moduli:
\begin{equation}
    (\tilde{z}_a;\tilde{\tau},\tilde{\sigma})=(\hat{z}_a;\hat{\sigma},\hat{\tau})\,.
\end{equation}
Note that these moduli are correctly related through the gluing condition $\tilde{\boldsymbol{\rho}}=\mathcal{O}S_{23}^{-1}\boldsymbol{\rho}$, where we recall that $f=S_{23}\,\mathcal{O}$ leads to $S^3\times S^1$ (see Section \ref{ssec:top-aspects}).
We can thus write the Heegaard splitting of the Hopf surface as: 
\begin{equation}
    \mathcal{M}_{(1,0)}(\hat{\boldsymbol{\rho}})\cong M_{S_{23}\mathcal{O}}(\boldsymbol{\rho},\mathcal{O}S_{23}^{-1}\boldsymbol{\rho})\,,\qquad \hat{\boldsymbol{\rho}}=\boldsymbol{\rho}\,.
\end{equation}

\subsection[\texorpdfstring{$L(p,q)\times S^1$}{L(p,q)xS1}]{\texorpdfstring{$\bm{L(p,q)\times S^1}$}{L(p,q)xS1}}\label{sapp:LpqS1}

Finally, let us discuss the general lens space geometries $ L(p,q)\times S^1$.
These geometries are endowed with complex structure moduli by simply performing the lens quotient on the primary Hopf surface \cite{Closset:2013vra}:
\begin{equation}\label{eq:lens-quotient-2}
     (z_1,z_2)\sim (e^{\frac{ 2\pi i q}{p}  }z_1,e^{\frac{-2\pi i}{p}  }z_2)\qquad \Leftrightarrow \qquad (z_1,z_2)\sim(e^{\frac{2\pi i }{p}}z_1,e^{-\frac{2\pi i s}{p}}z_2)\,,
\end{equation}
where both $q$ and $s$ are identified $\mod p$ and $qs=1\mod p$.\footnote{Note that the second description reflects the diffeomorphism $L(p,q)\cong L(p,s)$ described in Section \ref{ssec:top-aspects}.}
This geometry is also known as a secondary Hopf surface.
Similarly to $S^3\times S^1$, it can be parametrized by complex coordinates $z_{1,2}$ as follows:
\begin{equation}
    z_1=\hat{p}^x\cos(\frac{\theta}{2})e^{i\frac{p\phi_1+q\phi_2}{p}}\,\,\qquad z_2=\hat{q}^x\sin(\frac{\theta}{2})e^{-i\frac{\phi_2}{p}}\,,
\end{equation}
with $x\sim x+1$, $0\leq \theta\leq \pi$, $\phi_{1,2}\sim \phi_{1,2}+2\pi$.
Note that these coordinates, for fixed $x$, make manifest the description of $L(p,q)$ as a torus fibration with a $(1,0)$ cycle shrinking at $\theta=0$ and a $(q,p)$ cycle shrinking at $\theta=\pi$.
Equivalently, we can introduce $\phi_1'=p\phi_1+q\phi_2$ and $\phi_2'=s\phi_1+r\phi_2$ with $qs-pr=1$, such that:
\begin{equation}
    z_1=\hat{p}^x\cos(\frac{\theta}{2})e^{i\frac{\phi_1'}{p}}\,\,\qquad z_2=\hat{q}^x\sin(\frac{\theta}{2})e^{i\frac{p\phi_2'-s\phi_1'}{p}}\,.
\end{equation}
The symmetries of a general secondary Hopf surface were captured in Section \ref{ssec:hopf-surfaces} and we will not repeat them here.

The Heegaard splitting can be made manifest similar to the $S^3\times S^1$ case.
In particular, in a neighbourhood around $z_2=0$ we define:
\begin{equation}\label{eq:coord-transf-lens-to-solid-t3}
    z=\frac{z_2}{z_1^{\frac{\mathrm{Im}(\hat{\tau})}{\mathrm{Im}(\hat{\sigma})}}}\,,\qquad w=ip\log z_1 \,.
\end{equation}
This neighbourhood has the topology of $D_{2}\times T^2$.
The new coordinates allow us to read off the complex structure moduli associated to the solid torus from the identifications for the lens space \eqref{eq:lens-quotient-2} and the Hopf surface \eqref{eq:hopf-surface-ids-app}, which in these coordinates read: 
\begin{equation}\label{eq:zw-identification-alphabeta}
    (z,w)\sim (e^{2\pi i\alpha}z,w+2\pi)\,,\qquad (z,w)\sim (e^{2\pi i\beta}z,w+2\pi \sigma)\,,
\end{equation}
where $(\alpha,\beta)$ are defined through $\tau=\alpha \sigma-\beta$ and:
\begin{equation}\label{eq:lens-space-moduli}
    \tau=\hat{\tau}+s\hat{\sigma}\,,\qquad \sigma=p\hat{\sigma}\,.
\end{equation}
The holonomies for the global symmetries map trivially, $z_a=\hat{z}_a$, as long as one puts $u^{(a)}_x=0$ in the former.
Therefore, the $D_2\times T^2$ moduli $\boldsymbol{\rho}$ map to linear combinations of the Hopf surface moduli $\hat{\boldsymbol{\rho}}$.\footnote{
Note that the expressions of $(\tau,\sigma)$ in terms of the Hopf moduli $(\hat{\tau},\hat{\sigma})$ correspond to the invariant combinations under the fractional shifts in \eqref{eq:syms-lens-geom}.}

To describe the neighbourhood around $z_1=0$, we instead change coordinates through:
\begin{equation}
     z'=\frac{z_1}{z_2^{\frac{\mathrm{Im}(\hat{\sigma})}{\mathrm{Im}(\hat{\tau})}}}\,,\qquad w'=ip\log z_2\,,
\end{equation}
in which case the identifications become:
\begin{equation}
    (z,w)\sim (e^{2\pi i\alpha'}z,w+2\pi)\,,\qquad (z,w)\sim (e^{2\pi i\beta'}z,w+2\pi \tilde{\sigma})\,,
\end{equation}
with $(\alpha',\beta')$ defined through $\tilde{\tau}=\alpha' \tilde{\sigma}-\beta'$ and:
\begin{equation}\label{eq:lens-space-moduli-2}
    \tilde{\tau}=\hat{\sigma}+q\hat{\tau}\,,\qquad \tilde{\sigma}=p\hat{\tau}\,.
\end{equation}
The relation between the solid tori moduli $\boldsymbol{\rho}$ and $\tilde{\boldsymbol{\rho}}$ is given by:
\begin{equation}\label{eq:lens-gluing-cond}
    \tilde{\boldsymbol{\rho}}=\mathcal{O}g^{-1}\boldsymbol{\rho}\,,\qquad g=\begin{pmatrix}
    1&0&0\\
    0& -s & -r\\
    0& -p & -q
    \end{pmatrix} \in SL(2,\mathbb{Z})_{23}\,,
\end{equation}
reflecting the gluing condition \eqref{eq:gluing-condition}.
We summarize the Heegaard splitting of the secondary Hopf surface as follows:
\begin{equation}\label{eq:Mg-split}
    \mathcal{M}_{(p,q)}(\hat{\boldsymbol{\rho}})\cong M_f(\boldsymbol{\rho},f^{-1}\boldsymbol{\rho})\,,
\end{equation}
where $\hat{\boldsymbol{\rho}}$ is related to $\boldsymbol{\rho}$ through \eqref{eq:lens-space-moduli}.

\section{Lens indices for general gauge theories}\label{app:lens-indices}

In this appendix, we collect the known contour integral formulae for the superconformal index $\mathcal{I}_{(1,0)}(\hat{\boldsymbol{\rho}})$ \cite{Romelsberger:2005eg,Kinney:2005ej,Dolan:2008qi}, the lens space index $\mathcal{I}_{(p,1)}(\hat{\boldsymbol{\rho}})$ \cite{Benini:2011nc,Razamat:2013jxa,Razamat:2013opa,Closset:2017bse} and the $S^2\times T^2$ index $\mathcal{I}_{(0,-1)}(\hat{\boldsymbol{\rho}})$ \cite{Closset:2013sxa,Nishioka:2014zpa,Benini:2015noa,Honda:2015yha,Gadde:2015wta} of general $\mathcal{N}=1$ gauge theories.

\subsection{Superconformal index}\label{sapp:sci}

In this section, we loosely follow the recent review~\cite{Gadde:2020yah}, and write explicit indices in the conventions of \cite{Dolan:2008qi}.

The $\mathcal{N}=1$ superconformal algebra contains 4 complex super(conformal) charges  $\lbrace Q_{\alpha},\,S^{\alpha},\,\tilde{Q}_{\dot{\alpha}},\,\tilde{S}^{\dot{\alpha}}\rbrace$, where $S^{\alpha}=Q_{\alpha}^{\dagger}$ and $\tilde{S}^{\dot{\alpha}}=(\tilde{Q}_{\dot{\alpha}})^{\dagger}$.
Here, $\alpha,\dot{\alpha}=\pm$ represent the $SU(2)_{1,2}$ rotation symmetry indices, whose Cartan generators we will denote by $j_{1,2}$.
Furthermore, the algebra contains a $U(1)_r$ R-symmetry whose generator we denote by $r$.
We will define the index with respect to the supercharge $\mathcal{Q}\equiv Q_{-}$, which has charges $j_1=-\frac{1}{2}$, $j_2=0$ and $r=-1$.
Its anti-commutator with $\mathcal{Q}^{\dagger}$ gives:
\begin{equation}
    \delta\equiv 2\lbrace \mathcal{Q},\mathcal{Q}^\dagger\rbrace = \Delta-2j_1+\tfrac{3}{2}r\,.
\end{equation}
The superconformal index can be defined with respect to charges in the commutant of $\mathcal{Q}$ in the superconformal algebra:
\begin{equation}\label{eq:trace-defn-index}
\mathcal{I}_{(1,0)}(\hat{\boldsymbol{\rho}})=\mathrm{tr}_{\mathcal{H}}(-1)^F \hat{p}^{j_1+j_2-\frac{r}{2}}\hat{q}^{j_1-j_2-\frac{r}{2}} \hat{x}_a^{q_a}e^{-\beta \delta}\,,
\end{equation}
where $q_a$ are generators of the Cartan subalgebra of the global symmetry and $\hat{\boldsymbol{\rho}}\equiv (\hat{z}_a;\hat{\tau},\hat{\sigma})$ is shorthand for the chemical potentials, which are related to the fugacities through
\begin{equation}
\hat{p}=e^{2\pi i\hat{\sigma}} \,,\quad \hat{q}=e^{2\pi i \hat{\tau}} \,, \quad \hat{x}_a=e^{2\pi i\hat{z}_a}  \,.
\end{equation}
Furthermore, $\mathcal{H}$ is the Hilbert space of the theory quantized on $S^3$.
Due to the insertion of $(-1)^F$, with $F$ the fermion number operator, the index localizes on $\mathcal{H}_{\text{BPS}}$, the quarter BPS Hilbert space corresponding to the vanishing locus of $\delta$.
Therefore, the index is independent of $\beta$, and in fact any continuous deformation of the theory that preserves $\mathcal{Q}$.
This implies that, for example, one may compute the index of a gauge theory at weak Yang--Mills coupling.

The definition of the superconformal index employs the superconformal R-symmetry $U(1)_r$. 
However, if the theory has a global symmetry, we can define the index with respect to a shifted R-symmetry, under which the fields can have arbitrary R-charge.
The basic building blocks of a gauge theory index are the indices of a free chiral and vector multiplet.
In the case of a free chiral multiplet, there is a $U(1)$ flavor symmetry, under which the elementary fields can be taken to have unit charge.
Upon redefining the associated fugacity $\hat{x}$ as $\hat{x}\to (\hat{p}\hat{q})^{\frac{r}{2}-\frac{R}{2}}\hat{x}$, the index of a chiral multiplet with arbitrary charge R-charge $R$ can be written in terms of the elliptic $\Gamma$ function as:
\begin{equation}
    I^R_{(1,0)}(\hat{\boldsymbol{\rho}})=\Gamma\left(\hat{z}+\tfrac{R}{2}(\hat{\tau}+\hat{\sigma});\hat{\tau},\hat{\sigma}\right)\,.
\end{equation}
For a free vector multiplet, one has instead:
\begin{equation}
    I^V_{(1,0)}(\hat{\boldsymbol{\rho}})=\frac{1}{(1-v^{-1})\Gamma(-u;\hat{\tau},\hat{\sigma})}\,.
\end{equation}
Here, $u$ is a chemical potential for the $U(1)$ Cartan component of the gauge symmetry and $v=e^{2\pi i u}$.

For a general gauge theory with gauge group $G$, the index can now be written as:
\begin{equation}\label{eq:explicit-index-gen-gt}
\mathcal{I}_{(1,0)}(\hat{\boldsymbol{\rho}})=\frac{1}{|W|} \oint \prod_{i=1}^r \frac{dv_i}{2\pi i v_i}\,\Delta_G(\vec{u})\,I^{V_G}_{(1,0)}(\vec{u};\hat{\tau},\hat{\sigma})\prod_{i}I^{R_i}_{(1,0)}(\vec{u},\vec{z};\hat{\tau},\hat{\sigma}) \,,
\end{equation}
where the contour is taken as the unit circle in the complex $v_i$-plane, and the integral over the gauge fugacities associated to the Cartan torus of $G$ ensures projection onto gauge invariant states.
The Vandermonde determinant $\Delta(\vec{u})$ and $I^{V_G}_{(1,0)}(\hat{\boldsymbol{\rho}})$ combine together into:
\begin{equation}
    \Delta_G(\vec{u})\,I^{V_G}_{(1,0)}(\vec{u};\hat{\tau},\hat{\sigma})=\kappa^r\prod_{\alpha\neq 0} \Gamma(\alpha(\vec{u})+\hat{\tau}+\hat{\sigma};\hat{\tau},\hat{\sigma})\,,
\end{equation}
where $\kappa\equiv (\hat{p};\hat{p})_\infty(\hat{q};\hat{q})_{\infty}$ represents the contribution of modes associated to the Cartan torus of the gauge group, $r$ is the rank of $G$, the product runs over the (non-zero) roots of $G$ and we used the extension property of the elliptic $\Gamma$ function \eqref{eq:gamma-theta-id-app}.
Furthermore, let $\rho$ and $\rho'$ be the weight vectors of the gauge group and flavor symmetry representation of the chiral multiplet with R-charge $R_i$.
Then $I^{R_i}_{(1,0)}(\vec{u},\vec{z};\hat{\tau},\hat{\sigma})$ is defined as:
\begin{equation}
    I^{R_i}_{(1,0)}(\vec{u},\vec{z};\hat{\tau},\hat{\sigma}) =\prod_{\rho,\rho'}\Gamma(\rho(\vec{u})+\rho'(\vec{z})+\tfrac{R_i}{2}(\hat{\tau}+\hat{\sigma});\hat{\tau},\hat{\sigma})\,.
\end{equation}

\subsection{Lens index}\label{sapp:lens-index}

In this section, we review how the index on a lens space $L(p,1)$ is computed.

As discussed in Section \ref{ssec:top-aspects}, $L(p,1)$ is a quotient of $S^3$ \eqref{eq:lens-quotient}. 
For a free chiral multiplet of $R$-charge $R$, one obtains the lens index from the ordinary index by projecting onto the invariant states.
This projection can be implemented, while preserving supersymmetry, through the inclusion of a fugacity $e^{\frac{2\pi i j_2}{p}}$ into the trace \eqref{eq:trace-defn-index}.\footnote{A quotient where the phases have the same sign would instead by generated by $j_1$. Since $j_1$ does not commute with $\mathcal{Q}$, the associated index cannot be defined with respect to $\mathcal{Q}$. Instead, one would need to define (an equivalent) index with respect to, e.g., $\mathcal{Q}'\equiv Q_{+}$.}
Its effect is to project onto only states with multiples of $p$ derivatives $\partial^{pm,pn}_{+\pm}$ (see \cite{Kinney:2005ej} for our conventions).
The resulting expression for the index is given by:
\begin{align}
    \begin{split}
        I^R_{(p,1)}(\hat{\boldsymbol{\rho}})=\Gamma\left(\hat{z}+\tfrac{R}{2}(\hat{\tau}+\hat{\sigma})+p\hat{\sigma};\hat{\tau}+\hat{\sigma},p\hat{\sigma}\right)\Gamma\left(\hat{z}+\tfrac{R}{2}(\hat{\tau}+\hat{\sigma});\hat{\tau}+\hat{\sigma},p\hat{\tau}\right)\,.
    \end{split}
\end{align}
In addition, we can add a holonomy for the $U(1)$ flavor symmetry along the non-contractible cycle of the lens space.
Such holonomies are labeled by an integer $m=0,\ldots,p-1$.
Inclusion of this holonomy instead projects onto (single particle) states with $j_2\pm m\mod p$, where $\pm$ depends on the charge of the state under the global symmetry.
The expression for the index becomes:
\begin{align}
    \begin{split}
        I^R_{(p,1)}(\hat{\boldsymbol{\rho}};m)=&I^R_{0}(\hat{\boldsymbol{\rho}};m)\Gamma\left(\hat{z}+\tfrac{R}{2}(\hat{\tau}+\hat{\sigma})+(p+m)\hat{\sigma};\hat{\tau}+\hat{\sigma},p\hat{\sigma}\right)\\
        &\times \Gamma\left(\hat{z}+\tfrac{R}{2}(\hat{\tau}+\hat{\sigma})-m\hat{\tau};\hat{\tau}+\hat{\sigma},p\hat{\tau}\right)\,,
    \end{split}
\end{align}
where $I^R_{0}(\hat{\boldsymbol{\rho}};m)$ represents a ``vacuum energy'' contribution and will not play an important role for us.
We refer to \cite[eq.\ (2.11)]{Razamat:2013opa} for the explicit expression.
The lens index for a free vector multiplet can be parametrized by:
\begin{equation}
     I^V_{(p,1)}(\hat{\boldsymbol{\rho}};m)=\frac{(1-v^{-1})^{-\delta_{m,0}}I^V_{0}(\hat{\boldsymbol{\rho}};m)}{\Gamma\left(-u+(p+m)\hat{\tau};\hat{\tau}+\hat{\sigma},p\hat{\tau}\right) \Gamma\left(-u-m\hat{\sigma};\hat{\tau}+\hat{\sigma},p\hat{\sigma}\right)}\,,
\end{equation}
where again $I^V_{0}(\hat{\boldsymbol{\rho}};m)$ represents a vacuum energy contribution, and we similarly allow for a (gauge) holonomy around the non-contractible lens cycle.

For a general gauge theory with gauge group $G$, the expression for the full index includes a sum over all possible gauge holonomies along the non-contractible cycle of the lens space.
This gives rise to the following expression for the index:
\begin{equation}
    \mathcal{I}_{(p,1)}(\hat{\boldsymbol{\rho}})=\sum_{(m_i)}\mathcal{I}_{(p,1)}(\hat{\boldsymbol{\rho}};m_i)\,,
\end{equation}
where $m_i=0,\ldots, p-1$ for $i=1,\ldots, r_G$ labels the homology class of the gauge holonomy for the Cartan torus of the gauge group with rank $r_G$.
The inclusion of such holonomies breaks the gauge symmetry to the commutant of $\lbrace e^{2\pi i m_1/p},\ldots,e^{2\pi i m_r/p}\rbrace$.
Therefore, each index $\mathcal{I}_{(p,1)}(\hat{\boldsymbol{\rho}};m_i)$ for fixed $(m_i)$ along the Hopf fiber takes the following form:
\begin{align}\label{eq:explicit-lens-index-gen-gt}
    \begin{split}
        \mathcal{I}_{(p,1)}(\hat{\boldsymbol{\rho}};m_i)&=\frac{1}{|W_{m_i}|} \oint  \prod_{i=1}^r \frac{dv_i}{2\pi i v_i}\,\Delta(\vec{u};m_i)\,I^V_{(p,1)}(\vec{u};\hat{\tau},\hat{\sigma};m_i) \prod_{i}I^{R_i}_{(p,1)}(\vec{u},\vec{z};\hat{\tau},\hat{\sigma};m_i) \,.
    \end{split}
\end{align}
The functions $\Delta(\vec{u};m_i)$, $I^V_{(p,1)}(\vec{u};\hat{\tau},\hat{\sigma})$ and $I^{R_i}_{(p,1)}(\vec{u},\vec{z};\hat{\tau},\hat{\sigma};m_i)$ are defined similarly as in the case of the superconformal index, but now with respect to the unbroken gauge group and in terms of $I^V_{(p,1)}$ and $I^R_{(p,1)}$.

\subsection[\texorpdfstring{$S^2\times T^2$ index}{S2xT2 index}]{\texorpdfstring{$\bm{S^2\times T^2}$}{S2xT2} index}\label{sapp:s2xt2-index}

In this section, we review the computation of the $S^2\times T^2$ index.

To preserve supersymmetry on the $S^2\times T^2$ background, it is necessary to turn on a single unit of R-symmetry flux through the $S^2$.
This means that the $R$-charges of the fields have to be quantized as integers.
The $S^2\times T^2$ index can then be defined as (cf.\ \eqref{eq:trace-defn-index}):
\begin{equation}\label{eq:trace-defn-s2xs1-index}
\mathcal{I}=\mathrm{tr}_{\mathcal{H},g_a}(-1)^F \hat{p}^{L_0}\hat{q}^{J_3} x_a^{q_a}\,,
\end{equation}
where now $\mathcal{H}$ is the Hilbert space on $S^2\times S^1$, and $g_a$ indicate fluxes for the gauge fields coupled global symmetries labeled by the $q_a$.
The fugacities $\hat{p}=e^{2\pi i\hat{\sigma}}$, $\hat{q}=e^{2\pi i\hat{\tau}}$ and $x_a=e^{2\pi i\hat{z}_a}$, where $\hat{\sigma}$ parametrizes the complex structure of the $T^2$ and $\hat{\tau}=\alpha\hat{\sigma}-\beta$ parametrizes the twists of $S^2$ over the $T^2$ and $z_a=z^{(x)}_a\hat{\sigma}-z^{(y)}_a$ captures holonomies for the global symmetry along the $T^2$ cycles (see Section \ref{ssec:hopf-surfaces}).
Furthermore,  $J_3$ and $L_0$ are the Cartan generators of rotations on $S^2\times S^1$, respectively.

The $S^2\times T^2$ index for a free chiral multiplet depends on the shifted $R$-charge $\mathbf{R}\equiv R+q_0g-1$, where $q_0$ is the charge and $g$ is the flux for the $U(1)$ flavor symmetry.
Note that $\mathbf{R}$ is also quantized as an integer.
The twisted reduction of a four-dimensional chiral multiplet $S^2$ yields $\mathbf{R}$ two-dimensional $(0,2)$ Fermi multiplets for $\mathbf{R}>0$, $|\mathbf{R}|$ two-dimensional $(0,2)$ chiral multiplets for $\mathbf{R}< 0$ and is trivial for $\mathbf{R}=0$.
The $S^2\times T^2$ index reduces then to the computation of the elliptic genus for these two-dimensional theories.
Explicitly, they are given by:
\begin{align}\label{eq:S2T2-chiral-app}
	\begin{split}
		I^{R}_{(0,-1),g}(\hat{\boldsymbol{\rho}})&=\begin{cases}
			\hat{p}^{\frac{\mathbf{R}}{12}}\hat{x}^{-\frac{\mathbf{R}}{2}}\prod^{\frac{|\mathbf{R}|-1}{2}}_{m=-\frac{|\mathbf{R}|-1}{2}}\theta(\hat{z}+m\hat{\tau};\hat{\sigma})^{\text{sgn}(\mathbf{R})}\,,\quad &\text{for}\quad \mathbf{R}\neq 0\\
			1\,,\quad &\text{for}\quad \mathbf{R}=0\,.
		\end{cases}
	\end{split}
\end{align}
Here, $\theta(\hat{z};\hat{\sigma})$ is given by the $q$-theta function defined in Appendix \ref{append:q-theta}.
In particular, the $q$-$\theta$ function obeys modular properties under the group $SL(2,\mathbb{Z})\ltimes \mathbb{Z}^{2}$.

The four-dimensional vector multiplet for gauge group $G$ reduces to a single two-dimensional $(0,2)$ vector multiplet, whose elliptic genus is given by:
\begin{equation}
    I^{V}_{(0,-1)}(\hat{\boldsymbol{\rho}})=(\hat{p}^{\frac{1}{12}}y^{-\frac{1}{2}})^{\dim(G)-r}\eta(\hat{\sigma})^{2r}\prod_{\alpha\neq 0}\theta(\alpha(\vec{u})+|\alpha(\vec{m})|\hat{\tau};\hat{\sigma})\,,
\end{equation}
where $\eta(\hat{\sigma})$ is the Dedekind $\eta$ function and $r$ is the rank of $G$.
Furthermore, $\vec{m}$ comprises the gauge fluxes associated to the Cartan torus of the gauge group.

For a general gauge theory, the index takes a similar form as the lens index discussed above:
\begin{equation}\label{eq:explicit-s2xt2-index-gen-gt}
\mathcal{I}_{(0,-1)}(\hat{\boldsymbol{\rho}})=\sum_{(m_i)}\,\oint_{\mathrm{J.K.}} \prod_{i=1}^r \frac{dv_i}{2\pi i v_i}\,I^{V}_{(0,-1)}(\vec{u};\hat{\tau},\hat{\sigma};m_i)\prod_{i}I^{R}_{(0,-1),m_i}(\vec{u},\vec{z};\hat{\tau},\hat{\sigma};m_i) \,,
\end{equation}
however in this case the contour is given by the Jeffrey--Kirwan prescription (see, e.g., \cite{Gadde:2015wta}, for an explicit description of the contour for a general gauge theory).
Moreover, the sum over $(m_i)$ is over gauge fluxes for the Cartan torus of the gauge group, and the integrand is with respect to the unbroken gauge symmetry, i.e., the commutant of $\lbrace m_i\rbrace$ in $G$.
Note that all the fluxes $g_a$ for the flavor symmetry have been taken to be vanishing.

\section{Modular properties of the elliptic \texorpdfstring{$\bm{\Gamma}$}{Gamma} function}\label{app:mod-props-Gamma}

The purpose of this appendix is to derive properties of the elliptic $\Gamma$ function that we will use in Section \ref{ssec:evidence} to provide evidence for the modular factorization of lens indices. 
Before we start, let us collect some relevant formulae.
We will make use of the multiplication formula \cite{felder2002multiplication}:
\begin{align}\label{eq:mult-form-2}
    \begin{split}
        \Gamma(z;\tau,\sigma)&=\prod^{m-1,\tilde{m}-1}_{i,j=0}\Gamma(z+i\sigma+j\tau;\tilde{m}\tau,m\sigma)\,,
    \end{split}
\end{align}
and the standard modular property of the elliptic $\Gamma$ function \cite{Felder_2000}:
\begin{equation}\label{eq:ZS23andZp-form-for-hol-blocks-1}
    \Gamma(z;\tau,\sigma)=e^{-i\pi Q(z;\tau,\sigma)}\Gamma\left(\tfrac{z}{\sigma};\tfrac{\tau}{\sigma},-\tfrac{1}{\sigma}\right)\Gamma\left(\tfrac{z}{\tau};\tfrac{\sigma}{\tau},-\tfrac{1}{\tau}\right)\,,
\end{equation}
where $Q(z;\tau,\sigma)$ is a cubic polynomial in $z$, given in Appendix \ref{app:defs}. 
This property can also be written as:
\begin{equation}\label{eq:ZS23andZp-form-for-hol-blocks-2}
    \Gamma(z+\sigma;\tau,\sigma)=e^{-i\pi Q(z+\sigma;\tau,\sigma)}\frac{\Gamma\left(\frac{z}{\sigma};\frac{\tau}{\sigma},-\frac{1}{\sigma}\right)}{\Gamma\left(\frac{z}{\tau};-\tfrac{\sigma}{\tau},-\frac{1}{\tau}\right)}\,.
\end{equation}

\subsection[Three elliptic \texorpdfstring{$\Gamma$}{Gamma} functions]{Three elliptic \texorpdfstring{$\bm{\Gamma}$}{Gamma} functions}\label{sapp:3-Gamma}

In this appendix, we derive the most general modular property involving three elliptic $\Gamma$ functions, of which \eqref{eq:ZS23andZp-form-for-hol-blocks-1} is a special case.
This property will be useful when describing the factorization of the superconformal indices $\mathcal{Z}_{S_{23}\mathcal{O}}(\boldsymbol{\rho})$.
The formula we derive has appeared implicitly before in \cite{Felder_2008}.
Our strategy is to first replace the elliptic $\Gamma$ function using the multiplication formula \eqref{eq:mult-form-2}.
Then, on each factor in the resulting product, we apply \eqref{eq:ZS23andZp-form-for-hol-blocks-1}.
Finally, we use the multiplication formula again, but in the opposite direction, to rewrite the products in terms of two $\Gamma$ functions.

The first step yields:
\begin{align}
    \begin{split}
        \Gamma(z;\tau,\sigma)&=\prod^{m-1,\tilde{m}-1}_{i,j=0}\Gamma(z+i\sigma+j\tau+k_{(i,j)};\tilde{m}\tau+\tilde{n},m\sigma+n)\,,
    \end{split}
\end{align}
where the integers $n,\tilde{n}\in \mathbb{Z}$ and $k_{(i,j)}\in \mathbb{Z}$ can be added due to the periodicity of the elliptic $\Gamma$ function in all its arguments.
We now apply \eqref{eq:ZS23andZp-form-for-hol-blocks-1} to find:
\begin{align}\label{eq:ZS23-hol-blocks-step-2}
    \begin{split}
        \Gamma(z;\tau,\sigma)=&e^{-i\pi Q_{\mathbf{m}}(z;\tau,\sigma)}\prod^{m-1,\tilde{m}-1}_{i,j=0}\Gamma\left(\tfrac{z+i\sigma+j\tau+k_{(i,j)}}{m\sigma+n};\tfrac{\tilde{m}\tau+\tilde{n}}{m\sigma+n},-\tfrac{1}{m\sigma+n}\right)\\
        &\times \Gamma\left(\tfrac{z+i\sigma+j\tau+k_{(i,j)}}{\tilde{m}\tau+\tilde{n}};\tfrac{m\sigma+n}{\tilde{m}\tau+\tilde{n}},-\tfrac{1}{\tilde{m}\tau+\tilde{n}}\right)\,,
    \end{split}
\end{align}
where:
\begin{equation}
    Q_{\mathbf{m}}(z;\tau,\sigma)=\sum^{m-1,\tilde{m}-1}_{i,j=0}Q(z+i\sigma+j\tau+k_{(i,j)};\tilde{m}\tau+\tilde{n},m\sigma+n)\,,
\end{equation}
and $\mathbf{m}$ captures the integers $m,\tilde{m}$ and $n,\tilde{n}$.

The final step is the trickiest, which asks us to reduce the products on the right hand side of \eqref{eq:ZS23-hol-blocks-step-2} to two single elliptic $\Gamma$ functions by using the multiplication formula in the opposite direction.
To make progress, let us first disregard the shifts in the $z$ variable, and consider whether the second and third arguments can be written as multiples of $m,\tilde{m}$ respectively.
For the third argument of either elliptic $\Gamma$ function, this can be achieved in general if and only if:
\begin{equation}\label{eq:coprime}
    \gcd(m,n)=\gcd(\tilde{m},\tilde{n})=1\,,
\end{equation} 
in which case we can use periodicity in the third argument to write:
\begin{equation}
    -\frac{1}{m\sigma+n}=m \frac{k\sigma+l}{m\sigma+n}\;\mod 1\,,
\end{equation}
where $kn-lm=1$ and similarly for $-\frac{1}{\tilde{m}\tau+\tilde{n}}$.
However, to extract a factor of $\tilde{m}$ in the second argument of the first elliptic $\Gamma$ functions in \eqref{eq:ZS23-hol-blocks-step-2}, we must demand that $\tilde{m}=m$
because of the coprime condition \eqref{eq:coprime}.
In this case, we have:
\begin{equation}
    \frac{m\tau+\tilde{n}}{m\sigma+n}=m \frac{\tau-\tilde{n}(k\sigma+l)}{m\sigma+n}\;\mod 1\,.
\end{equation}
And similarly for $\frac{m\sigma+n}{m\tau+\tilde{n}}$ in the second argument of the second elliptic $\Gamma$ functions in \eqref{eq:ZS23-hol-blocks-step-2}.

The above implies that we can use the multiplication formula to write:
\begin{align}
    \begin{split}
        &\Gamma\left(\tfrac{z}{m\sigma+n};\tfrac{\tau-\tilde{n}(k\sigma+l)}{m\sigma+n},\tfrac{k\sigma+l}{m\sigma+n}\right)=\\
        &\qquad \qquad \prod^{m-1}_{i,j=0}\Gamma\left(\tfrac{z+i(k\sigma+l)+j(\tau-\tilde{n}(k\sigma+l))+d_{(i,j)}(m\sigma+n)}{m\sigma+n};\tfrac{m\tau+\tilde{n}}{m\sigma+n},-\tfrac{1}{m\sigma+n}\right)\,,
    \end{split}
\end{align}
where $d_{(i,j)}$ can be an arbitrary integer for each pair $(i,j)$.
Similarly, we have:
\begin{align}
    \begin{split}
        &\Gamma\left(\tfrac{z}{m\tau+\tilde{n}};\tfrac{\sigma-n(\tilde{k}\tau+\tilde{l})}{m\tau+\tilde{n}},\tfrac{\tilde{k}\tau+\tilde{l}}{m\tau+\tilde{n}}\right)=\\
        &\qquad \qquad\prod^{m-1}_{i,j=0}\Gamma\left(\tfrac{z+i(\tilde{k}\tau+\tilde{l})+j(\sigma-n(\tilde{k}\tau+\tilde{l}))+e_{(i,j)}(m\tau+\tilde{n})}{m\tau+\tilde{n}};\tfrac{m\sigma+n}{m\tau+\tilde{n}},-\tfrac{1}{m\tau+\tilde{n}}\right)\,,
    \end{split}
\end{align}
where again $e_{(i,j)}$ can be an arbitrary integer for each pair $(i,j)$.

To finish the computation, we need to show that shifts in $z$ in the previous two formulae are equivalent to the shifts of $z$ in \eqref{eq:ZS23-hol-blocks-step-2}.
We are thus led to the definitions: 
\begin{align}
    \begin{split}
        \tilde{\imath}=k(i-\tilde{n}j)+md_{(i,j)}\,,\quad \tilde{\jmath}=j\,,\quad \tilde{k}_{(\tilde{\imath},\tilde{\jmath})}=l(i-\tilde{n}j)+n d_{(i,j)}\,,
    \end{split}
\end{align}
and:
\begin{align}
    \begin{split}
        \hat{\imath}=j\,,\quad \hat{\jmath}=\tilde{k}(i-nj)+m e_{(i,j)}\,,\quad \hat{k}_{(\hat{\imath},\hat{\jmath})}=\tilde{l}(i-nj)+\tilde{n} e_{(i,j)}\,.
    \end{split}
\end{align}
We will first show that $\tilde{\imath}$ and $\hat{\jmath}$ run for fixed $j$ over all values in $\lbrace 0,\ldots,m-1\rbrace$ as $i$ runs over the same set for appropriate choices of $d_{(i,j)}$ and $e_{(i,j)}$, respectively.
This allows us to relabel $(\tilde{\imath},\tilde{\jmath})\to (i,j)$, and similarly  $(\hat{\imath},\hat{\jmath})\to (i,j)$.
We then show that $\tilde{k}_{(i,j)}=\hat{k}_{(i,j)}$.
Choosing $k_{(i,j)}$ in \eqref{eq:ZS23-hol-blocks-step-2} equal to $\tilde{k}_{(i,j)}$ will lead to our desired result.

First, let us ignore the constant $-k\tilde{n}j$ in the expression for $\tilde{\imath}$.
Then it is clear that $d_{(i,j)}$ can be chosen to bring $ki$ into the domain $\lbrace 0,\ldots,m-1\rbrace$ for all $i$. 
Since $k$ is coprime with $m$, it is guaranteed that for each $i$ there is a unique $\tilde{\imath}$.
The ignored term $-k\tilde{n}j$ represents a constant shift and therefore does not alter the conclusion.
The argument is identical for $\hat{\jmath}$.
Thus, the first part of the claim follows.

To show $\tilde{k}_{(i,j)}=\hat{k}_{(i,j)}$ we note that:
\begin{align}
    \begin{split}
        m\tilde{k}_{(\tilde{\imath},\tilde{\jmath})}&=-i+\tilde{\imath}n+\tilde{\jmath}\tilde{n} \,,\\
        m\hat{k}_{(\hat{\imath},\hat{\jmath})}&=-i+\hat{\imath}n+\hat{\jmath}\tilde{n}\,.
    \end{split}
\end{align}
Clearly, upon relabeling it follows that the constants are equal.
The actual value of $k_{(i,j)}$ we have to choose is given by the integral solution to these equations.
Since $0\leq a<m$ the solution is given by:
\begin{equation}
	k_{(i,j)}=\bigg\lfloor\frac{in+j\tilde{n}}{m}\bigg\rfloor\,,
\end{equation}
where the floor function is defined as:
\begin{equation}
	\bigg\lfloor\frac{p}{q}\bigg\rfloor=r \qquad \text{where}\qquad p=qr+s, \quad  \;0\leq s< q\,.
\end{equation}
Collecting all the results above, we find:
\begin{align}\label{eq:ZS23-hol-blocks-final}
    \begin{split}
        \Gamma(z;\tau,\sigma)&=e^{-i\pi Q_{\mathbf{m}}(z;\tau,\sigma)}\Gamma\left(\tfrac{z}{m\sigma+n};\tfrac{\tau-\tilde{n}(k\sigma+l)}{m\sigma+n},\tfrac{k\sigma+l}{m\sigma+n}\right)\Gamma\left(\tfrac{z}{m\tau+\tilde{n}};\tfrac{\sigma-n(\tilde{k}\tau+\tilde{l})}{m\tau+\tilde{n}},\tfrac{\tilde{k}\tau+\tilde{l}}{m\tau+\tilde{n}}\right)\,,
    \end{split}
\end{align}
where:
\begin{equation}\label{eq:phase-FHRZ-sumabpart}
    Q_{\textbf{m}}(z;\tau,\sigma)=\sum^{m-1}_{i,j=0}Q\left(z+i\sigma+j\tau+\bigg\lfloor\frac{in+j\tilde{n}}{m}\bigg\rfloor;m\tau+\tilde{n},m\sigma+n\right)\,.
\end{equation}
Let us end this section by simplifying the expression for the phase $Q_{\textbf{m}}(z;\tau,\sigma)$.
This summation can be simplified to:
\begin{equation}\label{eq:phase-FHRZ-singleQpart}
	Q_{\mathbf{m}}(z;\tau,\sigma)=\tfrac{1}{m}Q(mz;m\tau+\tilde{n},m\sigma+n)+f_{\mathbf{m}}\,,
\end{equation}
where $\mathbf{m}=(m,n,\tilde{n})$ and $f_{\mathbf{m}}$ can be proved to be the generalized Dedekind sum defined in Appendix \ref{appendix:Dedekinsum}: 
\begin{equation}\label{eq:fm-as-generalizedDS}
    f_{\mathbf{m}} = 2\sigma_1(n,\tilde{n},1;m)\,.
\end{equation}
To obtain \eqref{eq:fm-as-generalizedDS}, we first derive the following expression for $f_{\mathbf{m}}$ by subtracting \eqref{eq:phase-FHRZ-singleQpart} from \eqref{eq:phase-FHRZ-sumabpart}:
\begin{equation}\label{eq:fm-def-quotient}
f_{\mathbf{m}} = -\frac{1}{4m}+ \sum_{i,j=0}^{m-1}    \frac{(2i-m)(2j-m)}{4m^3} \left( 
m-2( in  + j\tilde{n} )+2m \bigg\lfloor\frac{in+j\tilde{n}}{m}\bigg\rfloor
\right)\,.
\end{equation}
The summation involving the quotient functions can be simplified by making use of identities of quotient functions \cite{carlitz1960some}:
\begin{equation}\label{eq:identity-Carlitz-integer}
	\sum_{j=0}^{m-1} \bigg\lfloor \frac{jn}{m}+x\bigg\rfloor = \lfloor m x \rfloor + \frac{1}{2}(m-1)(n-1)\,,  \qquad x \in \mathbb{R}\,.
\end{equation}
The identity \eqref{eq:identity-Carlitz-integer} can simplify \eqref{eq:fm-def-quotient}, except for the term:
\begin{equation}
    S(n,\tilde{n};m) = \sum_{i,j=0}^{m-1} i j \bigg\lfloor\frac{in+j\tilde{n}}{m}\bigg\rfloor\,.
\end{equation}
This quantity was introduced in \cite{carlitz1974inversions}, and
it was shown in \cite{hodel1975note} that the $S(n,\tilde{n};m)$ and $\sigma_1(n,\tilde{n},1;m)$ are related by: 
\begin{equation}\label{eq:Stos-generalized}
S(n,\tilde{n};m) = \frac{1}{12}m(2m-1)(m-1)^2(n+\tilde{n}) - \frac{1}{8}m (m-1)^3 -m^2 \sigma_1(n,\tilde{n},1;m)\,,
\end{equation}
where $\sigma_1(n,\tilde{n},1;m)$ is a generalized Dedekind sum defined in \eqref{eq:definition-sigmat}.
Using \eqref{eq:Stos-generalized} to simplify the summation and combining with other terms summed up by the identity \eqref{eq:identity-Carlitz-integer}, we then get \eqref{eq:fm-as-generalizedDS}.

Finally, when $n =\tilde{n}$, the expression for $S(n,\tilde{n};m)$ simplifies:
\begin{equation}
    S(n,n;m) = \sum_{l=0}^{2m-2} \sum_{i=0}^l i(l-i) \bigg \lfloor 
    \frac{l n }{m}
    \bigg \rfloor\,,\qquad l=i+j
\end{equation}
The sum over $i$ is independent of quotient functions so it can be summed as polynomials, while the sum over $l$ is known to be the alternative form of Dedekind sum $s(n,m)$ \cite{rademacher1972dedekind}. Thus this confirms that $\sigma_1(n,n,1;m)$ is of the form of Dedekind sum, up to rational functions of $m$, as written in \eqref{eq:fm-dedekind}.

\subsection[Four elliptic \texorpdfstring{$\Gamma$}{Gamma} functions]{Four elliptic \texorpdfstring{$\bm{\Gamma}$}{Gamma} functions}\label{sapp:4-Gamma}

In this appendix, we derive a general modular property involving four elliptic $\Gamma$ functions.
In particular, two of these $\Gamma$ functions are taken in the following specific form:
\begin{align}\label{eq:lens-indices-chiral-only}
    \begin{split}
       \Gamma\left(z+\sigma;\tau,\sigma\right)\Gamma\left(z;p\tau-\sigma,\tau\right) \,.
    \end{split}
\end{align}
Indeed, this is precisely the expression for the $L(p,1)\times S^1$ index of the free chiral multiplet with vanishing R-charge.
The property to be derived will be useful when describing the factorization of the indices $\mathcal{Z}_{g_{(p,1)}\mathcal{O}}(\boldsymbol{\rho})$.

To derive the modular property, we employ a similar strategy as in Appendix \ref{sapp:3-Gamma}.
In particular, we first replace both elliptic $\Gamma$ functions in \eqref{eq:lens-indices-chiral-only} using the multiplication formula \eqref{eq:mult-form-2}.
We then apply \eqref{eq:ZS23andZp-form-for-hol-blocks-2} on the product associated to the first factor and \eqref{eq:ZS23andZp-form-for-hol-blocks-1} on the second, giving a total of four products of $\Gamma$ functions.
The new step in this case is to show that a cancellation occurs between two products out of the four.
Finally, the remaining two products of $\Gamma$ functions will be simplified using the multiplication formula again, but in the opposite way, to eventually arrive at an expression involving a total of four elliptic $\Gamma$ functions.

Let us start with the first $\Gamma$ function in \eqref{eq:lens-indices-chiral-only}.
Before we plug in the multiplication formula, we note that the first elliptic $\Gamma$ function has a shift in its $z$ argument.
To use the methods of Appendix \ref{sapp:3-Gamma} in this case, it will be convenient to first replace this $\Gamma$ function using:
\begin{equation}
    \Gamma(z+\sigma;\tau,\sigma)=\frac{1}{\Gamma(z;\tau,-\sigma)}\,.
\end{equation}
The multiplication formula then yields:
\begin{align}\label{eq:Zgp-der-hol-blocks-step1-1}
    \begin{split}
        \Gamma\left(z+\sigma;\tau,\sigma\right)&=\prod^{m-1}_{i,j=0}\frac{1}{\Gamma(z-i\sigma+j\tau+k^{(1)}_{(i,j)};m\tau+n_1,-m\sigma-\tilde{n}_1)}\,,
    \end{split}
\end{align}
where similarly to Appendix \ref{sapp:3-Gamma} we allow for general integers $n_1,\tilde{n}_1$ and $k^{(1)}_{(i,j)}$.
In particular, we stress that at this point there are no coprime constraints on $m$ and $n_1,\tilde{n}_1$.
Note that we have already anticipated $\tilde{m}=m$ in the multiplication formula, which is required for the same reasons as in Appendix \ref{sapp:3-Gamma}.
For the second $\Gamma$ function, we similarly have:
\begin{align}\label{eq:Zgp-der-hol-blocks-step1-2}
    \begin{split}
      &  \Gamma\left(z;p\tau-\sigma,\tau\right)\\
      &=\prod^{m-1}_{i,j=0}\Gamma(z+i\tau+j(p\tau-\sigma)+k^{(2)}_{(i,j)};m(p\tau-\sigma)+\tilde{n}_2,m\tau+n_2)\,.
    \end{split}
\end{align}
Again, we allow for general integers $n_2,\tilde{n}_2$ and $k^{(2)}_{(i,j)}$.
Also for this $\Gamma$ function, we anticipate $\tilde{m}=m$.
In addition, we anticipate that the $m$ parameter here is the same as in \eqref{eq:Zgp-der-hol-blocks-step1-1}.
Indeed, for our purposes this is a necessary requirement.
To avoid unnecessary clutter, we put them equal already at this point.

We now apply \eqref{eq:ZS23andZp-form-for-hol-blocks-2} to  all the factors on the right hand side of \eqref{eq:Zgp-der-hol-blocks-step1-1}.
We then find:
\begin{align}\label{eq:Zgp-hol-blocks-step-2-1}
    \begin{split}
        \Gamma\left(z+\sigma;\tau,\sigma\right)&=e^{i\pi Q^{(1)}_{\mathbf{m}}(z;\tau,\sigma)}\prod^{m-1}_{i,j=0} \frac{\Gamma\left(\frac{z-i\sigma+j\tau+k^{(1)}_{(i,j)}}{m\sigma+n_1};\frac{m\tau+\tilde{n}_1}{m\sigma+n_1},-\frac{1}{m\sigma+n_1}\right)}{\Gamma\left(\frac{z-i\sigma+j\tau+k^{(1)}_{(i,j)}}{m\tau+\tilde{n}_1};-\frac{m\sigma+n_1}{m\tau+\tilde{n}_1},-\frac{1}{m\tau+\tilde{n}_1}\right)} \,,
    \end{split}
\end{align}
where we have used 
$
    \Gamma(-z;-\tau,-\sigma)=1/\Gamma(z;\tau,\sigma)
$.
For the second product of $\Gamma$ functions \eqref{eq:Zgp-der-hol-blocks-step1-2}, we apply \eqref{eq:ZS23andZp-form-for-hol-blocks-1}:
\begin{align}\label{eq:Zgp-hol-blocks-step-2-2}
    \begin{split}
        &\Gamma\left(z;p\tau-\sigma,\tau\right)=e^{-i\pi Q^{(2)}_{\mathbf{m}}(z;\tau,\sigma)} \prod^{m-1}_{i,j=0} \Gamma\left(\tfrac{z+i\tau+j(p\tau-\sigma)+k^{(2)}_{(i,j)}}{m\tau+n_2};\tfrac{m(p\tau-\sigma)+\tilde{n}_2}{m\tau+n_2},-\tfrac{1}{m\tau+n_2}\right)\\
        &\qquad \qquad \times\Gamma\left(\tfrac{z+i\tau+j(p\tau-\sigma)+k^{(2)}_{(i,j)}}{m(p\tau-\sigma)+\tilde{n}_2};\tfrac{m\tau+n_2}{m(p\tau-\sigma)+\tilde{n}_2},-\tfrac{1}{m(p\tau-\sigma)+\tilde{n}_2}\right) \,.
    \end{split}
\end{align}
The associated phases are given respectively by:
\begin{eqnarray}\label{eq:phases-lensp,1}
        Q^{(1)}_{\mathbf{m}}(z;\tau,\sigma)&=&\sum^{m-1}_{i,j=0}Q(z-i\sigma+j\tau+k^{(1)}_{(i,j)};m\tau+\tilde{n}_1,-m\sigma-n_1)\\ \nonumber
        Q^{(2)}_{\mathbf{m}}(z;\tau,\sigma)&=&\sum^{m-1}_{i,j=0}Q(z+i\tau+j(p\tau-\sigma)+k^{(2)}_{(i,j)};m(p\tau-\sigma)+\tilde{n}_2,m\tau+n_2)\,,
\end{eqnarray}
respectively.
As mentioned above, we are looking to cancel two products of elliptic $\Gamma$ functions, such that the remaining two products will eventually lead to the two holomorphic blocks.
Clearly, the candidates for cancellation are the $\Gamma$ functions in the denominator of \eqref{eq:Zgp-hol-blocks-step-2-1} and the $\Gamma$ functions in the first line of \eqref{eq:Zgp-hol-blocks-step-2-2}.
The cancellation occurs when all arguments between these pairs of $\Gamma$ functions agree.
This implies the following constraints:
\begin{equation}
    n_2=\tilde{n}_1\,, \quad \tilde{n}_2=p\tilde{n}_1-n_1\,,
\end{equation}
and in addition: 
\begin{equation}\label{eq:c2ab-constr}
    k^{(1)}_{(i,j)}=k^{(2)}_{(\tilde{\imath},\tilde{\jmath})}+d_{(\tilde{\imath},\tilde{\jmath})}\tilde{n}_1\,,\quad d_{(\tilde{\imath},\tilde{\jmath})}=\frac{j-\tilde{\imath}-\tilde{\jmath}p}{m}\,,
\end{equation}
where $0\leq i,j<m$ and $0\leq \tilde{\imath},\tilde{\jmath}<m$, $\tilde{\jmath}=i$ and $j$ is the (unique) integer such that $d_{(\tilde{\imath},\tilde{\jmath})}$ is an integer.

We now proceed to reduce the two remaining products of $\Gamma$ functions to two single $\Gamma$ functions, following the steps in Appendix \ref{sapp:3-Gamma}.
For the product remaining in \eqref{eq:Zgp-hol-blocks-step-2-1}, we have to enforce the constraint $\gcd(m,n_1)=1$ for the same reasons as in Appendix \ref{sapp:3-Gamma}.
To match with the notation in Section \ref{ssec:consistency-cond}, we define $c\equiv \tilde{n}_1$ and parametrize $c$ as $c=bm-an_{1}$.
We then repeat the analysis of Appendix \ref{sapp:3-Gamma}, keeping in mind a few sign differences.
One finds that for:
\begin{equation}
    k_{(i,j)}^{(1)}=\bigg\lfloor\frac{-in_1+jc}{m}\bigg\rfloor\,,
\end{equation}
the product reduces to a single $\Gamma$ function:
\begin{align}\label{eq:Zgp-hol-blocks-step-2-3}
    \begin{split}
        &\prod^{m-1}_{i,j=0}\Gamma\left(\tfrac{z-i\sigma+j\tau+k^{(1)}_{(i,j)}}{m\sigma+n_1};\tfrac{m\tau+\tilde{n}_1}{m\sigma+n_1},-\tfrac{1}{m\sigma+n_1}\right)=\Gamma\left(\tfrac{z}{m\sigma+n_1};\tfrac{\tau-c(k_1\sigma+l_1)}{m\sigma+n_1},\tfrac{k_1\sigma+l_{1}}{m\sigma+n_1}\right)\,,
    \end{split}
\end{align}
where $k_1n_1-ml_1=1$.

One can similarly work out the reduction of the remaining product in \eqref{eq:Zgp-hol-blocks-step-2-2}.
Now, one has to require $\gcd(m,\tilde{n}_2)=1$.
As in Section \ref{ssec:consistency-cond}, since $\tilde{n}_2=-n_1+pc$, we find that this constraint implies that $m$ has to be both coprime with $n_1$ and $-1+ap$.
In addition, it follows from the analysis in Appendix \ref{sapp:3-Gamma} that we have to take:
\begin{equation}
    k_{(i,j)}^{(2)}=\bigg\lfloor\frac{in_2+j\tilde{n}_2}{m}\bigg\rfloor=\bigg\lfloor\frac{ic+j(-n_1+pc)}{m}\bigg\rfloor\,.
\end{equation}
Note that the explicit expressions for $k_{(i,j)}^{(1)}$ and $k_{(i,j)}^{(2)}$ are consistent with the constraint \eqref{eq:c2ab-constr}. %
We can thus consistently reduce the second product as well to find:
\begin{align}
    \begin{split}
        &\prod^{m-1}_{i,j=0} \Gamma\left(\tfrac{z+i\tau+j(p\tau-\sigma)+k^{(2)}_{(i,j)}}{m(p\tau-\sigma)+\tilde{n}_2};\tfrac{m\tau+n_2}{m(p\tau-\sigma)+\tilde{n}_2},-\tfrac{1}{m(p\tau-\sigma)+\tilde{n}_2}\right)\\
        &\qquad \qquad = \Gamma\left(\tfrac{z}{m(p\tau-\sigma)+\tilde{n}_2};\tfrac{\tau-c(\tilde{k}_2(p\tau-\sigma)+\tilde{l}_2)}{m(p\tau-\sigma)+\tilde{n}_2},\tfrac{\tilde{k}_2(p\tau-\sigma)+\tilde{l}_2}{m(p\tau-\sigma)+\tilde{n}_2}\right)\,,
    \end{split}
\end{align}
where $\tilde{k}_2\tilde{n}_2-m\tilde{l}_2=1$.
Combining the above, we finally have:
\begin{align}
    \begin{split}
        \Gamma\left(z+\sigma;\tau,\sigma\right)&\Gamma\left(z;p\tau-\sigma,\tau\right)=e^{-i\pi Q_{\mathbf{m}_p}(z;\tau,\sigma)}\Gamma\left(\tfrac{z}{m\sigma+n_1};\tfrac{\tau-c(k_1\sigma+l_1)}{m\sigma+n_1},\tfrac{k_1\sigma+l_{1}}{m\sigma+n_1}\right)\\
        &\qquad \qquad \times \Gamma\left(\tfrac{z}{m(p\tau-\sigma)+\tilde{n}_2};\tfrac{\tau-c(\tilde{k}_2(p\tau-\sigma)+\tilde{l}_2)}{m(p\tau-\sigma)+\tilde{n}_2},\tfrac{\tilde{k}_2(p\tau-\sigma)+\tilde{l}_2}{m(p\tau-\sigma)+\tilde{n}_2}\right)\,,
    \end{split}
\end{align}
where $\mathbf{m}_p=(m,n_1,c;p)$ and:
\begin{align}
    \begin{split}
  &      Q_{\mathbf{m}_p}(z;\tau,\sigma)=-\sum^{m-1}_{i,j=0}\bigg[Q\left(z-i\sigma+j\tau+\big\lfloor\tfrac{-in_1+jc}{m}\big\rfloor;m\tau+c,-m\sigma-n_1\right)\\
        &-Q\left(z+i\tau+j(p\tau-\sigma)+\big\lfloor\tfrac{ic+j(-n_1+pc)}{m}\big\rfloor;m(p\tau-\sigma)-n_1+pc,m\tau+c\right)\bigg]\,.
    \end{split}
\end{align}
Similarly to Appendix \ref{sapp:3-Gamma}, this polynomial simplifies significantly.
In particular, we have:
\begin{align}
    \begin{split}
        Q_{\mathbf{m}_p}(z;\tau,\sigma)=&\tfrac{1}{m p} Q\left(m z, \tfrac{m(p\tau-\sigma)+ pc -n_1}{p}, \tfrac{m\sigma+n_1}{p} \right) + \tfrac{p^2-1}{12p}(2z-\tau) + f_{\mathbf{m}_{p}}\,,
    \end{split}
\end{align}
where we have not been able to find an explicit formula for $f_{\mathbf{m}_p}$.

\subsection[\texorpdfstring{$t+3$}{t+3} elliptic \texorpdfstring{$\Gamma$}{Gamma} functions]{\texorpdfstring{$\bm{t+3}$}{t+3} elliptic \texorpdfstring{$\bm{\Gamma}$}{Gamma} functions}\label{sapp:t+3-Gamma}

In this appendix, we derive the most general modular property that involves $t+3$ elliptic $\Gamma$ functions.
The integer $t$ refers to the length of the continued fraction expansion of $p/q$.
On the left hand side of our modular property, we have:
\begin{equation}
    \left( \prod_{i=0}^{t-1}\Gamma (z+ p_{i-1}\tau-s_{i-1}\sigma;  p_i \tau- s_i \sigma\,, p_{i-1}\tau-s_{i-1} \sigma)\right) \Gamma (z;  p_t \tau- s_t \sigma\,, p_{t-1}\tau-s_{t-1} \sigma)\,,
\end{equation}
corresponding to the $L(p,q)\times S^1$ index of the free chiral multiplet in Section \ref{ssec:cons-checks-lens}.

We employ exactly the same strategy as in Appendix \ref{sapp:4-Gamma}.
In particular, we make the replacement \eqref{eq:Zgp-hol-blocks-step-2-1} for $i=0,\cdots,t-1$:
\begin{align}\label{eq:block-3Gamma}
    \begin{split}
    & \Gamma(z+ p_{i-1}\tau-s_{i-1} \sigma\,; p_{i}\tau-s_{i} \sigma \,, p_{i-1}\tau- s_{i-1} \sigma) = e^{i\pi Q^{(i+1)}_m (z,\tau,\sigma)} \times \\
 &\prod_{a,b=0}^{m-1}
\frac{\Gamma\Big(\tfrac{z-a(p_{i-1}\tau-s_{i-1} \sigma)+b(p_{i}\tau-s_{i} \sigma)+c_{(a,b)}^{(i+1)}}{m(p_{i-1}\tau-s_{i-1} \sigma)+n_{i+1}}; \frac{m (p_i\tau-s_i \sigma)+ {\tilde{n}}_{i+1}}{m (p_{i-1}\tau-s_{i-1} \sigma)+ n_{i+1} } , \frac{-1}{m(p_{i-1} \tau- s_{i-1}\sigma) + n_{i+1}} \Big)}{\Gamma\Big(\tfrac{z-a(p_{i-1}\tau-s_{i-1} \sigma)+b(p_{i}\tau-s_{i} \sigma)+c_{(a,b)}^{(i+1)}}{m(p_{i}\tau-s_{i} \sigma)+{\tilde{n}}_{i+1}}; -\frac{m(p_{i-1}\tau-s_{i-1}\sigma) +n_{i+1}}{m(p_{i}\tau-s_{i} \sigma) + {\tilde{n}}_{i+1}} , \frac{-1}{m(p_{i}\tau-s_{i} \sigma) + {\tilde{n}}_{i+1}}\Big)}\,,
    \end{split}
\end{align}
and the phase is defined similarly as $Q^{(1)}_{\mathbf{m}}(z;\tau,\sigma)$ in \eqref{eq:phases-lensp,1}.
For now, the data $n_{i},{\tilde{n}}_{i}$ are unconstrained integers, but we have anticipated that $m$ is the same for each factor. 
Instead, we apply \eqref{eq:Zgp-hol-blocks-step-2-2} for the $t^{\text{th}}$ factor:
\begin{eqnarray}
        &&\Gamma(z; p_{t}\tau-s_{t} \sigma \,, p_{t-1}\tau- s_{t-1} \sigma)=e^{-i\pi Q^{(t+1)}_{\mathbf{m}}(z;\tau,\sigma)}\\ \nonumber
        && \times \prod^{m-1}_{a,b=0} \Gamma\left(\tfrac{z+a(p_{t-1}\tau-s_{t-1} \sigma)+b(p_{t}\tau-s_{t} \sigma)+c_{(a,b)}^{(t+1)}}{m(p_{t-1}\tau-s_{t-1}\sigma)+n_{t+1}};\tfrac{m(p_t\tau-s_t\sigma)+\tilde{n}_{t+1}}{m(p_{t-1}\tau-s_{t-1}\sigma)+n_{t+1}},-\tfrac{1}{m(p_{t-1}\tau-s_{t-1}\sigma)+n_{t+1}}\right)\\ \nonumber
        &&\times\Gamma\left(\tfrac{z+a(p_{t-1}\tau-s_{t-1} \sigma)+b(p_{t}\tau-s_{t} \sigma)+c_{(a,b)}^{(t+1)}}{m(p_t\tau-s_t\sigma)+\tilde{n}_{t+1}};\tfrac{m(p_{t-1}\tau-s_{t-1}\sigma)+n_{t+1}}{m(p_t\tau-s_t\sigma)+\tilde{n}_{t+1}},-\tfrac{1}{m(p_t\tau-s_t\sigma)+\tilde{n}_{t+1}}\right) \,.
\end{eqnarray}
Similar comments for the parameters hold here, except that the phase is defined similarly to $Q^{(2)}_{\mathbf{m}}(z;\tau,\sigma)$ in \eqref{eq:phases-lensp,1}.
Cancellation of the $i^{\text{th}}$ $\Gamma$ function in the denominator with the $(i+1)^{\text{th}}$ in the numerator requires: 
\begin{equation}\label{eq:recursiven1n2i}
    n_{i+1} = {\tilde{n}}_{i}, \qquad {\tilde{n}}_{i+1} = e_i n_{i+1} -n_i \,,
\end{equation}
where we have used the recurrence relations \eqref{eq:recursive-pq}:
\begin{equation}\label{eq:recursive-pq-app}
	\begin{aligned}
    &	p_{i} = e_{i} p_{i-1}- p_{i-2}\,, \quad   s_{i} = e_{i} s_{i-1}- s_{i-2}\,,\quad  q_i=p_{i-1}\,, \quad s_i=r_{i-1}\,.
	\end{aligned}
\end{equation}
Plugging in the first recurrence relation into the second, we can solve for $\tilde{n}_i$ explicitly.
We have:
\begin{equation}\label{eq:tildeni-soln}
 {\tilde{n}}_{i} = -s_{i-1} n_1+p_{i-1}c\,,
\end{equation}
where $n_1$ and $c$ are free integers.
In addition, cancellation requires:
\begin{equation}\label{eq:ciab-constr}
    c^{(i)}_{(a,b)}=c^{(i+1)}_{(\tilde{a},\tilde{b})}+d^{(i)}_{(a,\tilde{b})}\tilde{n}_i\,,\quad d^{(i)}_{(\tilde{a},\tilde{b})}=\frac{b-\tilde{a}s_i-\tilde{b}p_i}{m}\,,
\end{equation}
where $0\leq a,b<m$ and $0\leq \tilde{a},\tilde{b}<m$, $\tilde{b}=a$ and $b$ is the (unique) integer such that $d_{(\tilde{a},\tilde{b})}$ is an integer.
In this case, all cancellations occur and we are left with the final step to reduce the products of the remaining $\Gamma$ functions into a single one, by making use of the multiplication formula in the opposite direction.
This happens when $\gcd(m,n_1)=\gcd(m,\tilde{n}_{t+1})=1$ and in addition:
\begin{equation}
    c^{(i)}_{(a,b)}=\Big\lfloor \tfrac{-a {n}_{i +1} +  b {\tilde{n}}_{i+1}}{m} \Big\rfloor\,,\quad  i=1,\ldots,t\,,\qquad c^{(t+1)}_{(a,b)}=\Big\lfloor \tfrac{a {n}_{t +1} +  b {\tilde{n}}_{t+1}}{m} \Big\rfloor\,,
\end{equation}
which can be checked to satisfy \eqref{eq:ciab-constr}.
All in all, we find the following result:
\begin{eqnarray} \nonumber
       &&\left( \prod_{i=0}^{t-1}\Gamma (z+ p_{i-1}\tau-s_{i-1}\sigma;  p_i \tau- s_i \sigma\,, p_{i-1}\tau-s_{i-1} \sigma)\right) \Gamma (z;  p_t \tau- s_t \sigma\,, p_{t-1}\tau-s_{t-1} \sigma)\\
       &&=e^{-i\pi \tilde{P}_{g_{(p,q)}}^{\mathbf{m}}(z,\tau,\sigma)}\Gamma\left(\tfrac{z}{m\sigma+n_1}; \tfrac{\tau-c(k_1\sigma+l_1)}{m\sigma+n_1}, \tfrac{k_1\sigma +l_1}{m\sigma+n_1} \right) \\  \nonumber
 &&\qquad \qquad \times  \Gamma\left(\tfrac{z}{m(p\tau
-s \sigma) +{\tilde{n}}_{t+1}}; \tfrac{q\tau-r\sigma-n_{t+1}(\tilde{k}_{t+1} (p \tau-s \sigma) +\tilde{l}_{t+1})}{m(p\tau-s \sigma) +{\tilde{n}}_{t+1}} , \tfrac{\tilde{k}_{t+1} (p \tau-s \sigma) +\tilde{l}_{t+1}}{m(p\tau-s\sigma) +{\tilde{n}}_{t+1}}\right)\,,
\end{eqnarray}
where we have used $p\equiv p_t$, $q\equiv q_t=p_{t-1}$, $s\equiv s_t$, and $r\equiv r_t=s_{t-1}$.
Furthermore, $k_1n_1-l_1m=1$, $\tilde{k}_{t+1}\tilde{n}_{t+1}-\tilde{l}_{t+1}m=1$ and finally:
\begin{equation}
    n_{t+1}=qc-rn_1\,,\qquad \tilde{n}_{t+1}=-sn_1+pc\,,
\end{equation}
where we have used \eqref{eq:recursive-pq-app} and \eqref{eq:tildeni-soln}.
Furthermore, the phase polynomial is given by:
\begin{eqnarray}\label{eq:exact-lens-phase-polynomial}
\tilde{P}_{g_{(p,q)}}^{\mathbf{m}}(z,\tau,\sigma)&=& -\left(\sum_{i=0}^{t-1} Q^{(i+1)}_{\mathbf{m}}(z,\tau,\sigma)\right)  + Q^{(t+1)}_{\mathbf{m}} (z,\tau,\sigma) \,, \\ \nonumber
 Q^{(i+1)}_{\mathbf{m}}(z,\tau,\sigma) &=& \sum_{a,b=0}^{m-1} 
Q\Big(z-a(p_{i-1}\tau-s_{i-1}\sigma) + b (p_i\tau-s_i \sigma) +\Big\lfloor \tfrac{-a n_{i+1 } +  b {\tilde{n}}_{i+1}}{m} \Big\rfloor, \\ \nonumber
&& \qquad \qquad \qquad m(p_i\tau-s_i \sigma) +{\tilde{n}}_{i+1}, -m(p_{i-1}\tau-s_{i-1}\sigma) -n_{i+1 }\Big) \,, \\ \nonumber
 Q^{(t+1)}_{\mathbf{m}} (z,\tau,\sigma) &=& \sum_{a,b=0}^{m-1}  Q\Big(z+a(p_{t-1} \tau- s_{t-1}\sigma) + b (p_t\tau-s_t \sigma)+ \Big\lfloor \tfrac{a n_{t+1 } +  b {\tilde{n}}_{t+1}}{m} \Big\rfloor ,\\ \nonumber
&& \qquad \qquad  \qquad  m(p_t\tau-s_t \sigma) +{\tilde{n}}_{t+1} , m(p_{t-1} \tau - s_{t-1} \sigma) +n_{t+1 }\Big) \,.
\end{eqnarray}
The phase polynomial can be further simplified. 
As we have shown in Appendix \ref{sapp:3-Gamma}, the summation over $a,b$ in $Q_{\mathbf{m}}^{(i)}$ results in a single $Q$-polynomial plus a constant (cf.\ \eqref{eq:phase-FHRZ-singleQpart}).
This yields: 
\begin{eqnarray} \nonumber
 \tilde{P}_{g_{(p,q)}}^{\mathbf{m}} (z,\tau,\sigma)&=& \Bigg[ \sum_{i=0}^{t-1} \frac{1}{m} Q\Big(m z+m(p_{i-1} \tau- s_{i-1} \sigma)+n_{i+1}, \\ \label{eq:summation-formulaPt}
&& \qquad \qquad \quad m(p_{i} \tau- s_{i} \sigma)+\tilde{n}_{i+1}\,,m(p_{i-1} \tau- s_{i-1} \sigma)+n_{i+1}\Big)\Bigg] \\ \nonumber
& &+ \frac{1}{m} Q(m z,m(p_{t} \tau- s_{t} \sigma)+\tilde{n}_{t+1},m(p_{t-1} \tau- s_{t-1} \sigma)+n_{t+1})  + \text{const} \,.
\end{eqnarray}
We have not been able to determine a general formula for the constant, but for any specific set of values of the parameters it can be determined by subtracting \eqref{eq:exact-lens-phase-polynomial} from \eqref{eq:summation-formulaPt}.

Finally, we can simplify the summation over $i=0,\dots, t$.
In order to accomplish the summation, we notice the following two facts: 
\footnotesize
\begin{eqnarray} \label{eq:necessary-identities-1}
		&& \tfrac{1}{[m(p_i \tau-s_i \sigma)+ \tilde{n}_{i+1}][m(p_{i-1}\tau-s_{i-1} \sigma) + n_{i+1}]} 
		= \tfrac{1}{m\sigma+n_1} \left(\tfrac{p_i}{m(p_i \tau-s_i \sigma)+ \tilde{n}_{i+1}} - \tfrac{p_{i-1}}{m (p_{i-1} \tau-s_{i-1} \sigma)+n_{i+1}} \right) \\ \nonumber
		&& \tfrac{m(p_i \tau-s_i \sigma)+\tilde{n}_{i+1}}{m (p_{i-1} \tau-s_{i-1} \sigma) + n_{i+1}} +\tfrac{m(p_{i-1} \tau-s_{i-1} \sigma) +n_{i+1}}{m (p_{i} \tau-s_{i} \sigma) + \tilde{n}_{i+1}} 
		= e_i + \left(
		\tfrac{m(p_{i-1} \tau-s_{i-1} \sigma) +n_{i+1}}{m (p_{i} \tau-s_{i} \sigma) + \tilde{n}_{i+1}}  - \tfrac{m(p_{i-2} \tau-s_{i-2} \sigma)+\tilde{n}_{i-1}}{m (p_{i-1} \tau-s_{i-1} \sigma) + n_{i+1}} 
		\right)\,,
\end{eqnarray}
\normalsize
where we have made use of the recurrence relation \eqref{eq:recursive-pq-app} and the fact that $q_is_i-p_ir_i=1$.
These relations make the summation over these terms telescopic and computable due to the recursive relation \eqref{eq:recursiven1n2i}. 
As a result, only the boundary terms ($i=0,t$) in the summation contribute, resulting in the phase polynomial:
\begin{align}\label{eq:final-phase-pq}
\begin{split}
\tilde{P}_{g_{(p,q)}}^{\mathbf{m}}(\boldsymbol{\rho})  &= \frac{1}{m p_t} Q \left(m z, \frac{m(p_t \tau - s_t \sigma) + \tilde{n}_{t+1}}{p_t} , \frac{m \sigma+n_1}{p_t} \right)   +\delta \tilde{P}_{g_{(p,q)}}^{\mathbf{m}}(\boldsymbol{\rho}) \\
    \delta \tilde{P}_{g_{(p,q)}}^{\mathbf{m}}(\boldsymbol{\rho})  &=  \frac{(\eta_t+3)p_t -3}{6p_t}z  - \frac{(p_t^2-1)(p_t \tau-s_t\sigma+\sigma)}{12  p_t^2}+\text{const}\,.
    \end{split}
\end{align}
The constant term in $\delta \tilde{P}_{g_{(p,q)}}^{\mathbf{m}}$ are the sum of a few generalized Dedekind sums of the form $\sigma_1(n_1,n_2,1;m)$.

\bibliographystyle{JHEP}
\bibliography{bib-bh}

\end{document}